\documentclass[12pt]{article}
\usepackage[fontsize=11.7pt]{fontsize}

\usepackage[top=1in,bottom=1in,left=1in,right=1in]{geometry}
\usepackage[doublespacing]{setspace}
\usepackage{amssymb, amsmath, amsthm}
\usepackage{graphicx, hyperref, subcaption, booktabs, enumitem}
\usepackage[sans]{dsfont}
\usepackage[scr=rsfs,cal=boondox]{mathalpha}
\hypersetup{pdfborder = {0 0 0},colorlinks=true,allcolors=blue}
\usepackage[numbers]{natbib}

\newtheorem{assumption}{Assumption}
\newtheorem{thm}{Theorem}

\allowdisplaybreaks[1]

\DeclareMathOperator*{\argmin}{arg\,min}
\DeclareMathOperator{\Tstat}{T}
\DeclareMathOperator{\CI}{I}
\DeclareMathOperator{\setbd}{bd}
\DeclareMathOperator{\setint}{int}

\renewcommand{\P}{\mathbb{P}}
\newcommand{\E}{\mathbb{E}}
\newcommand{\Var}{\mathbb{V}}
\newcommand{\Cov}{\mathbb{C}\mathrm{ov}}
\newcommand{\Indicator}{\mathds{1}}
\newcommand{\En}{\E_n}
\newcommand{\dif}{\mathop{}\!\mathrm{d}}

\newcommand{\bb}{\mathbf{b}}
\newcommand{\be}{\mathbf{e}}
\newcommand{\bk}{\mathbf{k}}
\newcommand{\br}{\mathbf{r}}
\newcommand{\bu}{\mathbf{u}}
\newcommand{\bV}{\mathbf{V}}
\newcommand{\bx}{\mathbf{x}}
\newcommand{\bX}{\mathbf{X}}

\newcommand{\bbeta}{\boldsymbol{\beta}}
\newcommand{\bGamma}{\boldsymbol{\Gamma}}
\newcommand{\bSigma}{\boldsymbol{\Sigma}}
\newcommand{\bUpsilon}{\boldsymbol{\Upsilon}}

\newcommand{\A}{\mathcal{A}}
\newcommand{\B}{\mathcal{B}}
\newcommand{\q}{\mathcal{q}}
\newcommand{\X}{\mathcal{X}}

\begin{document}

\title{\vspace{-.5in}Estimation and Inference in Boundary Discontinuity Designs: Location-Based Methods\thanks{We thank Alberto Abadie, Xiaohong Chen, Boris Hanin, Kosuke Imai, Oliver Linton, Xinwei Ma, Victor Panaretos, J\"org Stoye, and Jeff Wooldridge for comments and discussions. We also thank three anonymous referees for helpful suggestions. Cattaneo and Titiunik gratefully acknowledge financial support from the National Science Foundation (SES-2019432, DMS-2210561, SES-2241575 and SES-2342226). Cattaneo gratefully acknowledges financial support from the National Institute for Food and Agriculture (NIFA) through grant 2024-67023-42704, the Data-Driven Social Science initiative at Princeton University, and a 2026 Guggenheim Fellowship.}\bigskip}

\author{Matias D. Cattaneo\thanks{Department of Operations Research and Financial Engineering, Princeton University.} \and
	    Rocio Titiunik\thanks{Department of Politics, Princeton University.} \and
	    Ruiqi (Rae) Yu\thanks{Department of Operations Research and Financial Engineering, Princeton University.}
	    }
\maketitle

\begin{abstract}
    \begin{onehalfspace}
    Boundary discontinuity designs are used to learn about causal treatment effects along a continuous assignment boundary that splits units into control and treatment groups according to a bivariate location score. We analyze location-based local polynomial treatment effect estimators that directly employ the bivariate score of each unit. We develop pointwise and uniform estimation and inference methods for the \textit{Boundary Average Treatment Effect Curve} (BATEC), as well as for two aggregated causal parameters: the \textit{Weighted Boundary Average Treatment Effect} (WBATE) and the \textit{Largest Boundary Average Treatment Effect} (LBATE). Our results cover both sharp and fuzzy (imperfect compliance) designs. We illustrate the methods with an empirical application, and provide companion general-purpose software. The supplemental appendix includes additional substantive theoretical results, methodological details, and simulation evidence.
    \end{onehalfspace}
\end{abstract}


\thispagestyle{empty}

\clearpage
\tableofcontents
\thispagestyle{empty}

\clearpage
\setcounter{page}{1}

\section{Introduction}\label{sec: Introduction}

We study estimation and inference in boundary discontinuity (BD) designs, where treatment assignment changes discontinuously along a known boundary that separates treatment and control assignment regions according to a bivariate score. This setup is also known as a Multi-Score Regression Discontinuity (RD) design \citep{Papay-Willett-Murnane_2011_JoE,Reardon-Robinson_2012_JREE,Wong-Steiner-Cook_2013_JEBS}, with Geographic RD designs as an important special case \citep{Keele-Titiunik_2015_PA,Keele-Titiunik-Zubizarreta_2015_JRSSA,Keele-Titiunik_2016_PSRM,Keele-etal_2017_AIE,Galiani-McEwan-Quistorff_2017_AIE,Rischard-Branson-Miratrix-Bornn_2021_JASA,Diaz-Zubizarreta_2023_AOAS}. As in the univariate RD design, the discontinuous assignment rule can be used to identify causal effects under continuity restrictions on potential outcomes, even when treatment is targeted to units with the greatest need. This makes BD designs a central tool in non-experimental policy evaluation. See \cite{Cattaneo-Titiunik_2022_ARE} for an overview of the RD literature, \cite[Section 5]{Cattaneo-Idrobo-Titiunik_2024_CUP} for a practical introduction to Multi-Dimensional RD designs, and \cite{Cattaneo-Titiunik-Yu_2026_BookCh} for a review of empirical practice employing BD designs.

\begin{figure}
    \centering
    \begin{subfigure}[b]{0.45\textwidth}
        \centering
        \includegraphics[width=\linewidth]{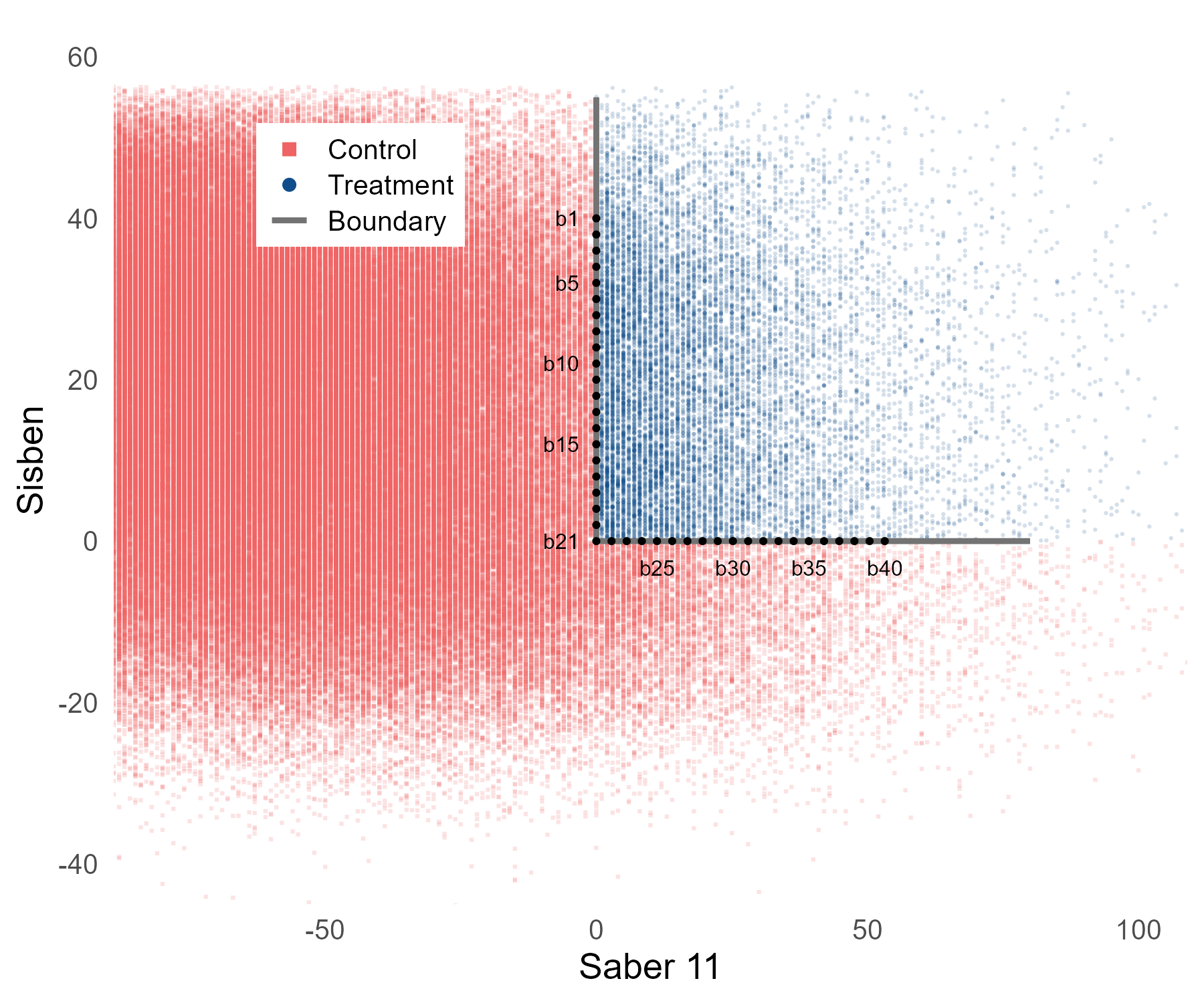}
        \caption{Scatterplot and Boundary.}
        \label{fig:fig1a}
    \end{subfigure}
    \quad
    \begin{subfigure}[b]{0.45\textwidth}
        \centering
        \includegraphics[width=\linewidth]{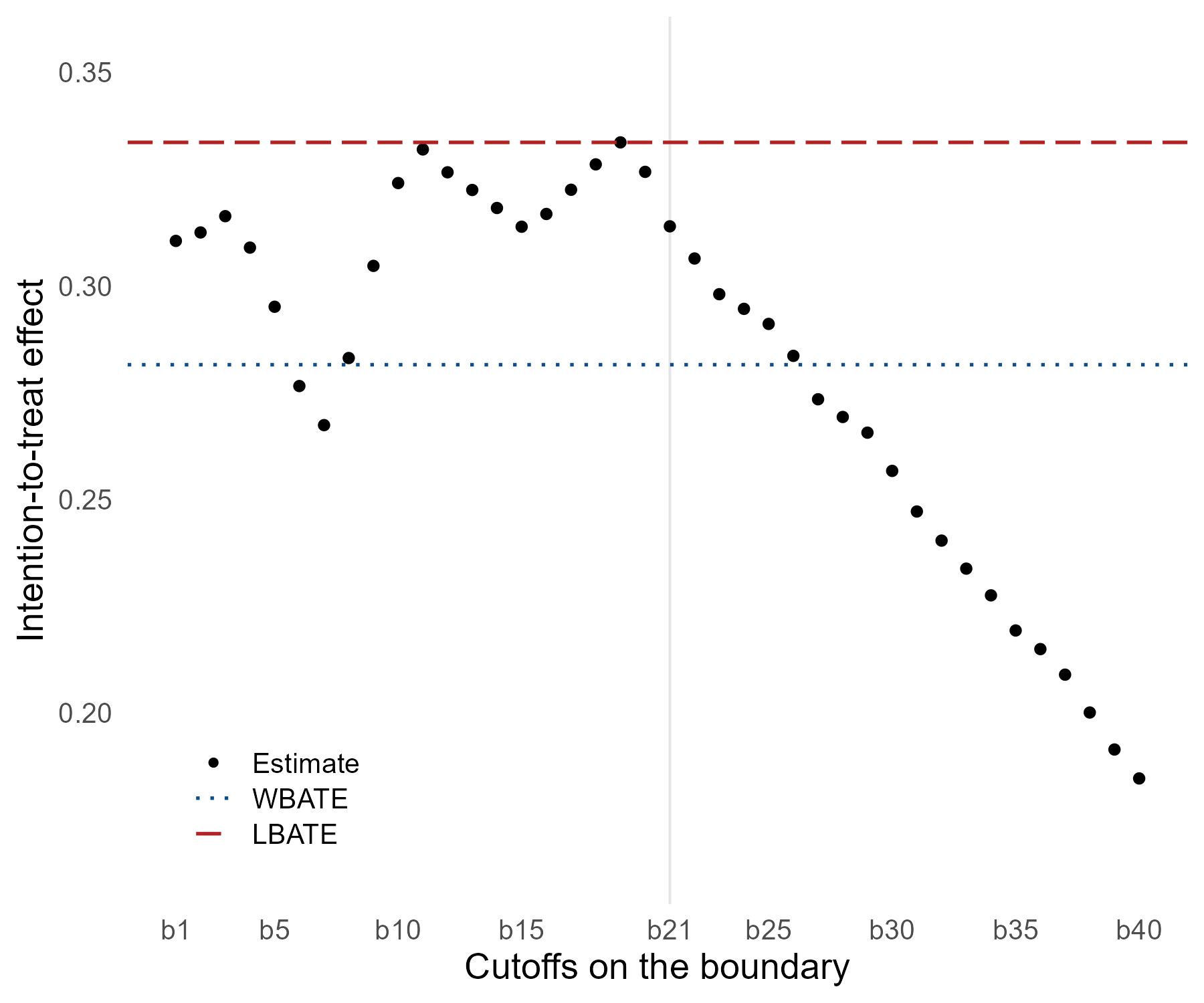}
        \caption{Treatment Effects Along Boundary.}
        \label{fig:fig1b}
    \end{subfigure}

    \caption{Scatterplot, Assignment Boundary, and Treatment Effects Using SPP data.\\ {\footnotesize \textit{Notes}. Panel (a) presents a scatterplot of the bivariate score $\bX_i$ using the SPP data, and also plots the treatment boundary $\B$ with $40$ marked grid points. Panel (b) presents causal treatment effect estimates over the $40$ boundary grid points depicted in Panel (a). Specifically, the black solid dots correspond to $\widehat{\tau}(\bb_j)$ (BATEC), the blue dotted line corresponds to $\widehat{\tau}_{\mathtt{WBATE}}$ (WBATE), and the red dot-dash line corresponds to $\widehat{\tau}_{\mathtt{LBATE}}$ (LBATE).}}
\end{figure}

Our motivating application is the \textit{Ser Pilo Paga} (SPP) program studied by \cite{LondonoVelezRodriguezSanchez_2020_AEJ}. SPP was a Colombian government subsidy that provided tuition support for post-secondary students to attend a government-certified higher education institution. Eligibility was based on both merit and economic need: students had to obtain a high grade in Colombia's national standardized high school exit exam, SABER 11, and they also had to come from economically disadvantaged families, as measured by the survey-based wealth index SISBEN. Thus the assignment rule was deterministic but genuinely bivariate: to qualify for the program, students had to obtain a SABER 11 score in the top $9$ percent of scores and their household's SISBEN index had to be below a region-specific threshold. After recentering both scores at their corresponding thresholds, Figure \ref{fig:fig1a} plots the bivariate score $\bX_i=(\mathtt{SABER11}_i,\mathtt{SISBEN}_i)^\top$ for the $2014$ cohort ($n=363,096$ observations), together with the assignment boundary $\B$ and $40$ evenly-spaced boundary points $\bb_1,\ldots,\bb_{40}$.

The main object of interest is the collection of treatment effects along the assignment boundary. Using standard potential outcomes notation \citep[and references therein]{Hernan-Robins_2020_Book}, we define the \textit{Boundary Average Treatment Effect Curve} (BATEC) as
\begin{align*}
    \tau(\bx) = \E[Y_i(1) - Y_i(0) | \bX_i = \bx],
    \qquad \bx\in\B,
\end{align*}
where $Y_i(0)$ and $Y_i(1)$ are the potential outcomes under control and treatment assignment, $\bX_i=(X_{1i},X_{2i})^\top$ is the bivariate score, and $\B$ is the assignment boundary. In the SPP application, $Y_i$ is an indicator for college enrollment, so $\tau(\bx)$ is the effect of SPP on the probability of college attendance for students at the margin of eligibility at boundary point $\bx\in\B$. For example, in Figure \ref{fig:fig1a}, $\tau(\bb_1)$ corresponds to students with relatively high SISBEN score and low SABER11 score, while $\tau(\bb_{40})$ corresponds to students with relatively low SISBEN score and high SABER11 score. Identification is the natural bivariate analogue of the continuity argument in univariate RD designs \citep{Hahn-Todd-vanderKlaauw_2001_ECMA}: treatment assignment changes abruptly at $\B$, while the potential outcome regression functions are assumed to vary continuously through the boundary. For more discussion, see \cite{Papay-Willett-Murnane_2011_JoE}, \cite{Reardon-Robinson_2012_JREE}, \cite{Wong-Steiner-Cook_2013_JEBS}, \cite{Keele-Titiunik_2015_PA}, \cite{Cattaneo-Keele-Titiunik-VazquezBare_2016_JOP}, and references therein.

We focus on a \textit{location-based} approach to BD designs: local polynomial regressions are fit using the two-dimensional score itself near each boundary point. This distinguishes our approach from analyses that first transform the score into a scalar distance-to-boundary measure, or that use frontier, binned, or pooled one-dimensional RD analyses as approximations \citep[and references therein]{Cattaneo-Titiunik-Yu_2026_BookCh}. Those alternatives can be useful benchmarks, but they do not directly deliver a theory for location-based local polynomial estimation using the full bivariate score. Although pointwise versions of these estimators are natural extensions of familiar univariate local polynomial methods, their statistical properties in BD designs are different from those in univariate RD designs \citep{Cattaneo-Titiunik-Yu_2026_BDD-Pooling,Cattaneo-Titiunik-Yu_2026_JOE}.

The BATEC captures treatment effect heterogeneity along $\B$, but researchers may also want scalar summaries of that heterogeneity. We study two leading examples. The \textit{Weighted Boundary Average Treatment Effect} (WBATE) is
\begin{align*}
    \tau_{\mathtt{WBATE}} = \frac{\int_{\B} \tau(\bx) w(\bx) \dif \bx}
                                 {\int_{\B} w(\bx) \dif \bx},
\end{align*}
where $w(\bx)$ is a weighting function. Motivated by the SPP application, the main paper assumes that $\B$ is a piecewise-linear boundary with finitely many line segments, so the integrals above are line integrals; the supplemental appendix covers more general assignment boundaries, including irregular boundaries sometimes encountered in geographic RD applications. Intuitively, $\tau_{\mathtt{WBATE}}$ averages all boundary treatment effects; see \cite{Reardon-Robinson_2012_JREE} and \cite{Wong-Steiner-Cook_2013_JEBS} for related discussion. The \textit{Largest Boundary Average Treatment Effect} (LBATE) is
\begin{align*}
    \tau_{\mathtt{LBATE}} = \sup_{\bx\in\B} \tau(\bx),
\end{align*}
which captures the largest causal effect along the assignment boundary. See \cite{Andrews-Kitagawa-McCloskey_2024_QJE} for discussion of extreme treatment effects in policy decisions. In the SPP application, WBATE and LBATE summarize, respectively, the average and largest causal effect of receiving the college subsidy across students near the eligibility boundary.

We make four contributions. First, we develop pointwise and uniform estimation and inference methods for the location-based local polynomial estimator of the BATEC. The uniform results are new and allow researchers to form confidence bands for the entire treatment effect curve along $\B$. Second, we provide mean squared error (MSE) expansions, bandwidth selection rules, robust bias-corrected inference, and implementation guidance for the location-based approach. Third, we develop estimation and inference methods for WBATE and LBATE, thereby covering both average and extreme summaries of treatment effect heterogeneity along the boundary. Fourth, we extend the results to BD designs with imperfect compliance, delivering pointwise and uniform inference for fuzzy BATEC, fuzzy WBATE, and fuzzy LBATE parameters. The supplemental appendix gives more general theory for higher-dimensional scores, more general boundaries, derivatives, and coordinate-specific bandwidths, while the companion \texttt{R} package \texttt{rd2d} implements the proposed methods.

Figure \ref{fig:fig1b} previews the empirical value of these methods in the SPP application. It reports estimated treatment effects $\widehat{\tau}(\bb_j)$ at the $40$ boundary points in Figure \ref{fig:fig1a}, together with point estimates $\widehat{\tau}_{\mathtt{WBATE}}$ and $\widehat{\tau}_{\mathtt{LBATE}}$ of WBATE and LBATE, respectively. The pattern suggests meaningful treatment effect heterogeneity: among students near the threshold of academic merit, the program appears to have a roughly constant effect across wealth levels; among students near the threshold of wealth, the effect appears to decline as academic performance improves. This finding suggests that students from underprivileged backgrounds benefit more from SPP than wealthier and high-achieving students. The application also features one-sided imperfect compliance: roughly $41\%$ of eligible students opted not to take up the SPP subsidy. We use the fuzzy methods to account for imperfect compliance in the analysis.

\subsection{Organization and Related Literature}

Section \ref{sec: Setup} introduces the formal causal inference framework and location-based estimator. For a review of the methodological RD literature see \cite{Cattaneo-Titiunik_2022_ARE}, and for practical introductions see \cite{Cattaneo-Idrobo-Titiunik_2020_CUP,Cattaneo-Idrobo-Titiunik_2024_CUP}. For an overview of empirical practice employing BD designs see \cite{Cattaneo-Titiunik-Yu_2026_BookCh}.

Section \ref{sec: BATEC} presents pointwise, integrated mean squared, and uniform estimation and inference methods for the BATEC. The pointwise results generalize well-known results in the nonparametric literature \citep[see][and references therein]{Hardle-etal_2004_Book}, and early BD design results in \cite{Papay-Willett-Murnane_2011_JoE}. The uniform results are new and provide valid inference for the entire treatment effect curve and transformations thereof. We also discuss bandwidth selection based on MSE expansions, robust bias-corrected inference, and implementation details, building on and generalizing state-of-the-art univariate RD methods; see \cite{Calonico-Cattaneo-Farrell_2020_ECTJ} for bandwidth selection, and \cite{Calonico-Cattaneo-Titiunik_2014_ECMA,Calonico-Cattaneo-Farrell_2018_JASA,Calonico-Cattaneo-Farrell_2022_Bernoulli} for robust bias correction.

Sections \ref{Sec: WBATE} and \ref{sec: LBATE} study the aggregated causal parameters $\tau_{\mathtt{WBATE}}$ and $\tau_{\mathtt{LBATE}}$. WBATE is an integral functional over a one-dimensional manifold of a two-dimensional conditional expectation contrast. Related integral functionals on submanifolds have only been studied recently; in concurrent work, \cite{Chen-Gao_2025_arXiv} developed series/sieve methods for such functionals, whereas our results use local polynomial regression as the nonparametric ingredient. See \cite{Reardon-Robinson_2012_JREE} and \cite{Wong-Steiner-Cook_2013_JEBS} for early methodological discussion in the context of BD designs. Our large-sample nonparametric results complement the design-based methods developed in \cite{Keele-Titiunik-Zubizarreta_2015_JRSSA} and \cite{Diaz-Zubizarreta_2023_AOAS}, and the Bayesian methods developed in \cite{Rischard-Branson-Miratrix-Bornn_2021_JASA} for BD designs.

Section \ref{sec: Imperfect Compliance} considers imperfect compliance in BD designs. We present estimation and inference methods for fuzzy BATEC, fuzzy WBATE, and fuzzy LBATE, while noting that the causal interpretation of these parameters is an active research area; see \cite{choi2023complier}, \cite{schwarz2025effect}, \cite{jiang2026extrapolating}, \cite{Cattaneo-Keele-Titiunik-VazquezBare_2016_JOP,Cattaneo-Keele-Titiunik-VazquezBare_2021_JASA} for fuzzy multidimensional RD designs, as well as \cite{Arai-etal_2022_QE} for univariate fuzzy RD designs.

Section \ref{sec: Empirical Application} applies the methods to the SPP data, revising the main results reported in \cite{LondonoVelezRodriguezSanchez_2020_AEJ} and documenting sharp and fuzzy treatment effect heterogeneity along $\B$. Section \ref{sec: Extensions and Future Research} discusses derivative estimation, multidimensional scores, covariate adjustment for efficiency improvements and heterogeneity analysis, bivariate RD Plots, and clustered and spatially correlated data.

The supplemental appendix presents more general theoretical results, reports all proofs, and gives other technical results that may be of independent interest. In particular, it generalizes the main results to (i) $d$-dimensional scores, for $d\geq 2$; (ii) more general assignment boundaries $\B\subseteq\mathbb{R}^{d}$ with effective dimension $d-1$; (iii) derivative estimation and inference; and (iv) different bandwidths for each coordinate of the score. Those results leverage ideas from geometric measure theory \citep{simon1984lectures,federer2014geometric}, thereby accommodating geographic and other BD settings where $\B$ is irregular. From a technical perspective, Section SA-7 leverages ideas in \cite{Cattaneo-Yu_2025_AOS} and presents new strong approximation theorems for empirical processes with polynomially bounded moments. In addition, the supplemental appendix gives practical implementation details in Section SA-9, empirical results in Section SA-10, and simulation evidence in Section SA-11.

Finally, a companion software article \citep{Cattaneo-Titiunik-Yu_2025_rd2d} discusses the general-purpose \texttt{R} software package \texttt{rd2d} (\url{https://rdpackages.github.io/rd2d}) implementing the methods developed in this paper. We also provide complete replication files for our numerical results.

\subsection{Notation}

We employ standard concepts and notations from empirical process theory \citep{van-der-Vaart-Wellner_1996_Book,Gine-Nickl_2016_Book}. For a random variable $Z_i$, we write $\En[g(Z_i)] = n^{-1} \sum_{i = 1}^n g(Z_i)$. For sequences, we write $a_n = o(b_n)$ if $\limsup_{n\to\infty} \frac{|a_n|}{|b_n|} = 0$, and $a_n \lesssim b_n$ if there exist constants $C$ and $N > 0$ such that $n > N$ implies $|a_n| \leq C |b_n|$. For sequences of random variables, we write $a_n = o_{\P}(b_n)$ if $\operatorname{plim}_{n \to \infty}\frac{|a_n|}{|b_n|} = 0$, and $|a_n| \lesssim_\P |b_n|$ if $\limsup_{M \to \infty} \limsup_{n \to \infty} \P[|\frac{a_n}{b_n}| \geq M] = 0$.
For a set $A$, $\setbd(A)$ denotes the boundary of $A$ and $\setint(A)$ denotes the interior of $A$.
Finally, $\Phi(x)$ denotes the standard Gaussian cumulative distribution function.

\section{Setup}\label{sec: Setup}

We begin with sharp BD designs, where all units comply perfectly with their treatment assignment. Let $\A_0$ and $\A_1$ denote the control and treatment regions, respectively, and let $\B=\setbd(\A_0)\cap\setbd(\A_1)$ denote their common assignment boundary. We use the convention that boundary points belong to the treatment region, so $\B\subseteq\A_1$ and $\B\cap\A_0=\emptyset$. Units are assigned to the control or treatment conditions according to the rule $\Indicator(\bX_i \in \A_1)$, and the observed outcome is $Y_i=\Indicator(\bX_i\in\A_0)Y_i(0)+\Indicator(\bX_i\in\A_1)Y_i(1)$. We start by assuming that all units assigned to the treatment condition do receive the treatment, and none of the units assigned to the control condition receive the treatment. Section \ref{sec: Imperfect Compliance} discusses fuzzy BD designs, where treatment status need not coincide with treatment assignment.

We impose the following basic conditions on the underlying data generating process.

\begin{assumption}[Sharp DGP]\label{assump: Sharp DGP}
    Let $t\in\{0,1\}$, $p \geq 1$, and $v \geq 2$.
    \begin{enumerate}[label=\normalfont(\roman*),noitemsep,leftmargin=*]

    \item $(Y_1(t),\bX_1^\top)^\top,\ldots,(Y_n(t),\bX_n^\top)^\top$ are independent and identically distributed random vectors.

    \item The distribution of $\bX_i$ has compact support $\X\subseteq\mathbb{R}^2$ with nonempty interior and a Lebesgue density $f_X(\bx)$ that is continuous and bounded away from zero on $\X$.

    \item $\mu_t(\bx) = \E[Y_i(t)| \bX_i = \bx]$ has a $(p+1)$-times continuously differentiable extension to an open neighborhood of $\X$.

    \item $\sigma^2_t(\bx) = \Var[Y_i(t)|\bX_i = \bx]$ is bounded away from zero and continuous on $\X$.

    \item $\sup_{\bx \in \X}\E[|Y_i(t)|^{2+v}|\bX_i = \bx] < \infty$ for some $v\geq2$.
    \end{enumerate}
\end{assumption}

These conditions are the bivariate analogues of standard continuity-based RD assumptions. Part (i) imposes random sampling; this may be restrictive in some geographic applications with spatial dependence, an issue we discuss in Section \ref{sec: Extensions and Future Research}. Part (ii) imposes compact score support and ensures that local neighborhoods around boundary points contain observations with positive probability. Because Assumption \ref{assump: Boundary and Kernel} places $\B$ in the interior of $\X$, the boundary-local neighborhoods used by the estimators remain away from the edge of the score support for small bandwidths. Parts (iii)--(v) impose standard smoothness, variance, and moment conditions on the potential outcomes.

Under Assumption \ref{assump: Sharp DGP}, the BATEC introduced in Section \ref{sec: Introduction} is identified at every boundary point as $\tau(\bx)=\mu_1(\bx)-\mu_0(\bx)$ for $\bx\in\B$. Because treatment assignment changes discontinuously at $\B$ while $\mu_0(\cdot)$ and $\mu_1(\cdot)$ are continuous, the conditional mean of the observed outcome on side $\A_t$ identifies $\mu_t(\bx)$ as one approaches $\bx\in\B$ from that side. This is the usual RD continuity argument adapted to a bivariate score; see \cite{Papay-Willett-Murnane_2011_JoE}, \cite{Reardon-Robinson_2012_JREE}, \cite{Keele-Titiunik_2015_PA}, \cite{Cattaneo-Keele-Titiunik-VazquezBare_2016_JOP}, and references therein. Identification of $\tau_{\mathtt{WBATE}}$ and $\tau_{\mathtt{LBATE}}$ follows from identification of $\tau(\bx)$ over $\B$. Our focus is on the statistical properties of local polynomial methods for estimating these objects and conducting pointwise, uniform, and aggregate inference.

For each $\bx\in\B$, the location-based estimator fits two local polynomial regressions, one using observations from the treatment side and one using observations from the control side. The resulting BATEC estimator is
\begin{align*}
    \widehat{\tau}(\bx) = \be_{\mathbf{0}}^{\top}\widehat{\bbeta}_1(\bx) - \be_{\mathbf{0}}^{\top}\widehat{\bbeta}_0(\bx),
    \qquad \bx \in \B,
\end{align*}
where, for $t \in \{0,1\}$,
\begin{align}\label{eq: locpoly estimator}
    \widehat{\bbeta}_t(\bx)
    = \argmin_{\bbeta \in \mathbb{R}^{\mathfrak{p}_p}} \En \Big[ \big(Y_i - \br_p(\bX_i - \bx)^{\top}\bbeta \big)^2 K_h(\bX_i - \bx)\Indicator(\bX_i \in \A_t) \Big],
\end{align}
with $\mathfrak{p}_p = (2+p)(1+p)/2$ and $\br_p(\bu)$ collecting all monomials $u_1^{a_1}u_2^{a_2}$ with $a_1+a_2\leq p$, starting with the constant term. Let $\be_{\mathbf{0}}$ denote the unit vector selecting this constant term. The kernel is $K_h(\bu)=K(u_1/h,u_2/h)/h^2$, where $K(\cdot)$ is a bivariate function and $h$ is a bandwidth. We use a common bandwidth for expositional simplicity and because researchers can standardize the two score coordinates before implementation. The supplemental appendix allows for different bandwidths across score coordinates.

We impose the following assumption on the assignment boundary curve and bivariate kernel function.

\begin{assumption}[Boundary and Kernel]\label{assump: Boundary and Kernel}
    \begin{enumerate}[label=\normalfont(\roman*),noitemsep,leftmargin=*]
        \item $\B\subset\setint(\X)$, and $\B$ is piecewise linear with finitely many line segments and positive finite length.
        \item $K: \mathbb{R}^2 \to [0,\infty)$ is compactly supported and Lipschitz continuous, or $K(\bu)=\Indicator(\bu \in [-1,1]^2)$.
        \item $K(\bu)>0$ for all $\bu$ in a neighborhood of the origin.
    \end{enumerate}
\end{assumption}

Assumption \ref{assump: Boundary and Kernel}(i) gives a simple and transparent boundary condition. It ensures that line integrals over $\B$ are well defined, that the boundary can be discretized for uniform inference and aggregate estimands, and that the boundary is away from the edge of the support of the score. Assumption \ref{assump: Boundary and Kernel}(ii) imposes standard regularity conditions on the kernel. Assumption \ref{assump: Boundary and Kernel}(iii) imposes local positivity around the origin; together with the piecewise-linear boundary in Assumption \ref{assump: Boundary and Kernel}(i), this condition implies that, after local rescaling, the kernel assigns nonnegligible mass to each side of the boundary and hence each side contains enough kernel-weighted variation to identify the local polynomial coefficients uniformly along $\B$. Assumption \ref{assump: Boundary and Kernel} may be restrictive in geographic or other applications where the boundary cannot be well approximated by piecewise linear functions. In the supplemental appendix, we give more primitive geometric conditions that allow for irregular boundaries.

Finally, to ensure that $\tau_{\mathtt{WBATE}}$ is well-defined, we impose the following conditions on the weight function.

\begin{assumption}[Weight Function and Boundary]\label{assump: Weight Function and Boundary}
    Let $w:\B \to (0,\infty)$ satisfy $\sup_{\bx\in\B} w(\bx) < \infty$, $\inf_{\bx\in\B} w(\bx) >0$, and $\int_{\B} w(\bx) \dif \bx < \infty$.
\end{assumption}

Weights may be normalized without loss of generality. In particular, $w(\bx)=f_X(\bx)$ satisfies Assumption \ref{assump: Weight Function and Boundary}. In this case, the WBATE reduces to the \textit{boundary average treatment effect} (BATE):
$\tau_{\mathtt{BATE}} =  \int_{\B} \tau(\bx) f(\bx|\bX_i\in\B) \dif \bx$,
with
$f(\bx|\bX_i\in\B) = \frac{f_X(\bx)}{\int_{\B} f_X(\bx) \dif \bx}$.
More formally, the expression $f(\bx|\bX_i\in\B)$ denotes the normalized boundary density with respect to arc length, with Radon--Nikodym derivative proportional to $f_X(\bx)$ on $\B$.
This parameter is the density-weighted average causal effect along the assignment boundary, and is discussed by \cite{Wong-Steiner-Cook_2013_JEBS}. See \cite{Cattaneo-Titiunik-Yu_2026_BDD-Pooling} for an alternative regression-based approach commonly used in practice to estimate the BATE.

\section{Boundary Average Treatment Effect Curve}\label{sec: BATEC}

This section studies pointwise, integrated, and uniform estimation and inference for the BATEC $(\tau(\bx):\bx\in\B)$ in sharp BD designs. We focus on a bivariate score, a scalar bandwidth $h$, and treatment-effect levels; in the supplemental appendix, we generalize this setup to treatment-effect derivatives, diagonal bandwidth matrices, and score of three or more dimensions. Throughout this section, integrals over $\B$ are line integrals with respect to arc length; in the appendix these are written as integrals with respect to a Hausdorff measure \citep{federer2014geometric}. Pointwise results follow from standard local polynomial arguments once the one-sided local design is nonsingular. The integrated MSE expansion and the uniform distribution approximation require additional empirical-process and geometric arguments, all of which are developed in the supplemental appendix.

\subsection{Treatment Effect Estimation}\label{sec: Treatment Effect Estimation}

The first result gives pointwise and uniform convergence rates for the location-based estimator. The stochastic terms are the usual variance component and a polynomial-moment empirical-process remainder, while the smoothing bias is of order $h^{p+1}$.

\begin{thm}[Convergence Rates: Sharp BATEC]\label{thm: Convergence Rates: Sharp BATEC}
    Suppose Assumptions \ref{assump: Sharp DGP} and \ref{assump: Boundary and Kernel} hold. If $n h^2/\log(1/h) \to \infty$ and $h\to0$, then
    \begin{enumerate}[label=\normalfont(\roman*),noitemsep,leftmargin=*]
        \item $|\widehat{\tau}(\bx) - \tau(\bx) | \lesssim_\P \frac{1}{\sqrt{n h^2}} + \frac{1}{n^{\frac{1+v}{2+v}}h^2} + h^{p+1}$ for $\bx \in \B$, and

        \item $\sup_{\bx \in \B} |\widehat{\tau}(\bx) - \tau(\bx)| \lesssim_\P \sqrt{\frac{\log(1/h)}{ n h^2}} + \frac{\log(1/h)}{n^{\frac{1+v}{2+v}}h^2} + h^{p+1}$.
    \end{enumerate}
\end{thm}

Theorem \ref{thm: Convergence Rates: Sharp BATEC} implies consistency of the treatment effect estimator based on the bivariate location score under bandwidth sequences that make the displayed rate vanish; for uniform consistency, it suffices that $n^{(1+v)/(2+v)}h^2/\log(1/h)\to\infty$ and $h\to0$. The bias order is $h^{p+1}$ because smoothing is performed in the two-dimensional score space on each side of the boundary; it does not require reparameterizing the boundary by a scalar running variable. This feature is important when $\B$ has corners or other piecewise-linear changes in direction; see \cite{Cattaneo-Titiunik-Yu_2026_JOE} for more discussion.

Next, we state the conditional MSE expansion. The pointwise expansion has the standard bias--variance form; the integrated expansion averages this local expansion along the one-dimensional assignment boundary. Recall that $t\in\{0,1\}$. The population Gram matrix is
\[
    \bGamma_{t,\bx} =
    \E \Big[\br_p\Big(\frac{\bX_i - \bx}{h}\Big) \br_p \Big(\frac{\bX_i - \bx}{h}\Big)^{\top} K_h(\bX_i - \bx) \Indicator(\bX_i \in \A_t) \Big].
\]
The leading non-random pointwise variance is $V_{\bx} = V_{1,\bx} + V_{0,\bx}$, where $V_{t,\bx} = \be_{\mathbf{0}}^{\top} \bGamma_{t,\bx}^{-1} \bSigma_{t,\bx,\bx} \bGamma_{t,\bx}^{-1}\be_{\mathbf{0}}$ with
\begin{align*}
    \bSigma_{t, \bx,\bx}
    = h^2 \E \bigg[
        \br_p\Big(\frac{\bX_i - \bx}{h}\Big)
        \br_p\Big(\frac{\bX_i - \bx}{h}\Big)^{\top}
        K_h(\bX_i - \bx)^2  \sigma_t^2(\bX_i) \Indicator(\bX_i \in \A_t)
      \bigg].
\end{align*}
Using standard multi-index notation, the leading non-random pointwise bias is $B_{\bx} = B_{1,\bx} - B_{0,\bx}$, where
\begin{align*}
    B_{t,\bx}
    = \be_{\mathbf{0}}^{\top} \bGamma_{t,\bx}^{-1}
      \sum_{|\bk| = p + 1} \frac{\mu_t^{(\bk)}(\bx)}{\bk!}
      \E \bigg[
        \br_p \Big(\frac{\bX_i - \bx}{h}\Big)
        \Big(\frac{\bX_i - \bx}{h}\Big)^{\bk}
        K_h(\bX_i - \bx) \Indicator(\bX_i \in \A_t)
      \bigg].
\end{align*}
The following theorem gives the MSE expansions. Let $\bX = (\bX_1^\top,\ldots,\bX_n^\top)$.

\begin{thm}[MSE Expansions: Sharp BATEC]\label{thm: MSE Expansions: Sharp BATEC}
    Suppose Assumptions \ref{assump: Sharp DGP}, \ref{assump: Boundary and Kernel}, and \ref{assump: Weight Function and Boundary} hold. If $n^{v/(2+v)} h^2/\log(1/h) \to \infty$ and $h\to0$, then
    \begin{enumerate}[label=\normalfont(\roman*),noitemsep,leftmargin=*]
        \item $\E [(\widehat{\tau}(\bx) - \tau(\bx))^2 | \bX]
               = h^{2(p+1)} B_{\bx}^2 + \frac{1}{n h^2} V_{\bx} + o_{\P}(\mathfrak{R})$, and

        \item $\int_\B \E [(\widehat{\tau}(\bx) - \tau(\bx))^2 | \bX] w(\bx) \dif \bx
               = h^{2(p+1)} \int_\B B_{\bx}^2 w(\bx) \dif \bx + \frac{1}{n h^2} \int_\B V_{\bx} w(\bx) \dif \bx + o_{\P}(\mathfrak{R})$,
    \end{enumerate}
    with $\mathfrak{R} = h^{2p+2} + n^{-1} h^{-2}$.
\end{thm}

For a common scalar bandwidth $h$, ignoring higher-order terms gives the approximate MSE-optimal and IMSE-optimal bandwidth choices
\begin{align*}
    h_{\mathtt{MSE},\bx} = \Big(\frac{2 V_{\bx} }{(2p+2) B_{\bx}^2} \frac{1}{n} \Big)^{1/(2p+4)}
    \quad\text{and}\quad
    h_{\mathtt{IMSE}} = \Big(\frac{2 \int_{\B} V_{\bx} w(\bx) \dif \bx}{(2p+2) \int_{\B} B_{\bx}^2 w(\bx) \dif \bx} \frac{1}{n} \Big)^{1/(2p+4)},
\end{align*}
provided that $B_{\bx}\neq0$ and $\int_{\B} B_{\bx}^2 w(\bx) \dif \bx \neq 0$, respectively. These choices are infeasible because a preliminary bandwidth, as well as estimates of the conditional variances and higher-order derivatives of the conditional mean, are needed. We discuss feasible implementation in Section \ref{sec: Implementation}, and in the companion software article \citep{Cattaneo-Titiunik-Yu_2025_rd2d}. The same calculations extend in the appendix to diagonal bandwidth matrices, higher-dimensional scores, derivatives of the BATEC, and more general assignment boundaries.

The pointwise rates and MSE expansion mirror familiar nonparametric local polynomial results \citep[see][for a textbook review]{tsybakov2008introduction}. The integrated MSE expansion in Theorem \ref{thm: MSE Expansions: Sharp BATEC} requires additional care because the criterion integrates pointwise conditional MSE over the one-dimensional manifold $\B$. These point estimation results are also the inputs for the uniform and aggregation results below.

\subsection{Uncertainty Quantification}\label{sec: Uncertainty Quantification}

Given a bandwidth choice, define the feasible $t$-statistic $\widehat{\Tstat}(\bx) = (\widehat{\tau}(\bx) - \tau(\bx))/\widehat{\Omega}^{1/2}_{\bx,\bx}$ where, using standard least squares algebra, the covariance estimator is
\begin{align*}
    \widehat{\Omega}_{\bx_1,\bx_2}
     = \frac{1}{nh^2}\be_{\mathbf{0}}^\top
       \big[ \widehat{\bGamma}^{-1}_{0,\bx_1}\widehat{\bSigma}_{0,\bx_1,\bx_2}\widehat{\bGamma}^{-1}_{0,\bx_2}
           + \widehat{\bGamma}^{-1}_{1,\bx_1}\widehat{\bSigma}_{1,\bx_1,\bx_2}\widehat{\bGamma}^{-1}_{1,\bx_2} \big] \be_{\mathbf{0}}
\end{align*}
with
$\widehat{\bSigma}_{t,\bx_1,\bx_2}
    = h^2 \En \big[\br_p(\tfrac{\bX_i - \bx_1}{h}) \br_p(\tfrac{\bX_i - \bx_2}{h})^{\top} K_h(\bX_i - \bx_1) K_h(\bX_i - \bx_2) \widehat{\varepsilon}_i(\bx_1)\widehat{\varepsilon}_i(\bx_2) \Indicator(\bX_i \in \A_t) \big]
$
and $\widehat{\varepsilon}_i(\bx) = Y_i - \Indicator(\bX_i \in \A_0) \br_p(\bX_i - \bx)^\top\widehat{\bbeta}_0(\bx) - \Indicator(\bX_i \in \A_1) \br_p(\bX_i - \bx)^\top\widehat{\bbeta}_1(\bx)$, for all $\bx_1,\bx_2\in\B$ and $t\in\{0,1\}$.

Wald-type feasible confidence intervals and confidence bands over $\B$ take the form
\begin{align*}
    \widehat{\CI}_\alpha(\bx)
  = \Big[\;\widehat{\tau}(\bx) - \q_{\alpha} \widehat{\Omega}^{1/2}_{\bx,\bx}
         \; , \;
         \widehat{\tau}(\bx) + \q_{\alpha} \widehat{\Omega}^{1/2}_{\bx,\bx}\;\Big],
  \qquad \bx\in\B,
\end{align*}
for any $\alpha \in (0,1)$, where $\q_{\alpha}$ denotes the appropriate quantile for the desired coverage objective. For pointwise inference, the usual normal approximation gives $\sup_{u\in\mathbb{R}} |\P[\widehat{\Tstat}(\bx) \leq u] - \Phi(u)| \to 0$ for each $\bx\in\B$, provided that the ``small bias" condition $n h^{2p+4} \to 0$ holds. This result gives the usual pointwise confidence interval for $\tau(\bx)$.

For uniform inference over $\B$, two challenges arise: (i) the stochastic process $(\widehat{\Tstat}(\bx) : \bx\in\B)$ is not asymptotically tight, and thus it does not converge weakly in the space of uniformly bounded real functions supported on $\B$ and equipped with the uniform norm \citep{van-der-Vaart-Wellner_1996_Book,Gine-Nickl_2016_Book}; and (ii) the geometry of $\B$ can affect the validity of the distributional approximation. We handle both issues by approximating the distribution of the supremum $\sup_{\bx\in\B} |\widehat{\Tstat}(\bx)|$ directly, rather than relying on weak convergence of the full process. This approach enables asymptotically valid confidence bands because
$\P \big[\tau(\bx) \in \widehat{\CI}_\alpha(\bx), \text{ for all } \bx \in \B \big]
 = \P \big[ \sup_{\bx\in\B} | \widehat{\Tstat}(\bx) | \leq  \q_{\alpha} \big]$. The resulting Gaussian approximation for the supremum requires VC-type entropy and moment bounds. The supplemental appendix also gives a stronger process-level coupling under additional technical conditions. These approximation results leverage technical tools from \cite{Cattaneo-Yu_2025_AOS}, \cite{Cattaneo-Chandak-Jansson-Ma_2024_Bernoulli}, \cite{Chernozhukov-Chetverikov-Kato_2014a_AoS,Chernozhukov-Chetverikov-Kato_2014b_AoS}, \cite{Chernozhukov-Chetverikov-Kato-Koike_2022_AoS}, and \cite{dudley2014uniform}. 
Let $\bV = ((Y_1, \bX_1^{\top}), \cdots, (Y_n, \bX_n^{\top}))^\top$.

\begin{thm}[Confidence Intervals and Bands: Sharp BATEC]\label{thm: Confidence Intervals and Bands: Sharp BATEC}
    Suppose Assumptions \ref{assump: Sharp DGP} and \ref{assump: Boundary and Kernel} hold.
    \begin{enumerate}[label=\normalfont(\roman*),noitemsep,leftmargin=*]
        \item If $nh^2 \to \infty$ and $n h^{2p+4} \to 0$, then
              $\P \big[\tau(\bx) \in \widehat{\CI}_\alpha(\bx) \big] \to 1 - \alpha$
              with $\q_{\alpha} = \Phi^{-1}(1-\alpha/2)$.

        \item If $\liminf_{n\to\infty} \log h/\log n > -\infty$, $n^{v/(2+v)} h^2/(\log n)^9 \to \infty$, and $n h^{2p+4}\log n\to0$, then
              $\P \big[\tau(\bx) \in \widehat{\CI}_\alpha(\bx), \text{ for all } \bx \in \B \big] \to 1 - \alpha$,
              with $\q_{\alpha} = \inf \big\{c > 0: \P\big[\sup_{\bx \in \B} |\widehat{Z}(\bx)| \geq c \,\big|\, \bV\big] \leq \alpha\big\}$, where $(\widehat{Z}(\bx): \bx \in \B)$ is a Gaussian process conditional on $\bV$ with $\E[\widehat{Z}(\bx)|\bV]=0$ and
              $\Cov[\widehat{Z}(\bx_1),\widehat{Z}(\bx_2) | \bV]
                = \widehat{\Omega}_{\bx_1,\bx_2}/(\widehat{\Omega}_{\bx_1,\bx_1} \widehat{\Omega}_{\bx_2,\bx_2})^{1/2}$, for all $\bx,\bx_1,\bx_2 \in \B$.
    \end{enumerate}
\end{thm}

This theorem gives asymptotically valid pointwise and uniform uncertainty quantification for $\tau(\bx)$ using the location-based estimator $\widehat{\tau}(\bx)$. The appendix also shows how the result extends to higher-dimensional scores, derivatives, and more general boundaries under the corresponding geometric regularity conditions. As expected in a nonparametric smoothing setting, the pointwise undersmoothing condition $n h^{2p+4} \to 0$ and its $\log(n)$-strengthened uniform analogue rule out the (I)MSE-optimal point estimator of $\tau(\bx)$. Thus, for implementation of both pointwise and uniform inference, we employ robust bias correction \citep{Calonico-Cattaneo-Titiunik_2014_ECMA,Calonico-Cattaneo-Farrell_2018_JASA,Calonico-Cattaneo-Farrell_2022_Bernoulli}; see Section \ref{sec: Implementation} for details.

\subsection{Implementation}\label{sec: Implementation}

It is straightforward to implement local and global bandwidth selectors based on Theorem \ref{thm: MSE Expansions: Sharp BATEC}. In particular, replacing the leading bias and variance quantities, $B_\bx$ and $V_\bx$, with preliminary estimators, we obtain the feasible plug-in bandwidth selectors
\begin{align*}
    \widehat{h}_{\mathtt{MSE},\bx}
    = \Big(\frac{2 \widehat{V}_{\bx}}{(2p+2) \widehat{B}_{\bx}^2} \frac{1}{n}\Big)^{1/(2p+4)}
    \quad\text{and}\quad
    \widehat{h}_{\mathtt{IMSE}}
    = \Big(\frac{2 \int_{\B} \widehat{V}_{\bx} w(\bx) \dif \bx}{(2p+2) \int_{\B} \widehat{B}_{\bx}^2 w(\bx) \dif \bx} \frac{1}{n} \Big)^{1/(2p+4)},
\end{align*}
where, for a preliminary bandwidth choice $a\to0$, $\widehat{B}_{\bx} = \widehat{B}_{1,\bx} - \widehat{B}_{0,\bx}$ is constructed using
$\widehat{B}_{t,\bx}
= \be_{\mathbf{0}}^{\top} \widehat{\bGamma}_{t,\bx}^{-1} \sum_{|\bk| = p + 1} \frac{\widehat{\mu}_t^{(\bk)}(\bx)}{\bk!}
   \En \big[ \br_p (\frac{\bX_i - \bx}{a})(\frac{\bX_i - \bx}{a})^{\bk}K_a(\bX_i - \bx) \Indicator(\bX_i \in \A_t)\big]$,
with $\widehat{\bGamma}_{t,\bx}$ computed using the preliminary bandwidth $a$, and where $\widehat{\mu}_t^{(\bk)}(\bx)$ is a preliminary estimator of $\mu_t^{(\bk)}(\bx)$, and $\widehat{V}_\bx = n a^2 \widehat{\boldsymbol{\Omega}}_{\bx,\bx}$ is constructed using the variance estimator with the preliminary bandwidth $a$. Omitted implementation details are discussed in the companion software article \citep{Cattaneo-Titiunik-Yu_2025_rd2d}. See also \cite{Calonico-Cattaneo-Farrell_2020_ECTJ} for a review on modern bandwidth selection methods in RD designs with univariate score.

The bandwidth choices $\widehat{h}_{\mathtt{MSE},\bx}$ and $\widehat{h}_{\mathtt{IMSE}}$ can be used to implement (I)MSE-optimal $\widehat{\tau}(\bx)$ treatment effect estimators, both pointwise and uniformly over $\B$. Furthermore, leveraging the results in Theorem \ref{thm: Confidence Intervals and Bands: Sharp BATEC}, a simple application of robust bias-corrected inference proceeds by employing the same (I)MSE-optimal bandwidth (for $p$th order point estimation), but then constructing the t-statistic $\widehat{\Tstat}(\bx)$ with a $(p+1)$th polynomial order instead of $p$th polynomial order. The core idea is to simultaneously (i) debias the (I)MSE-optimal point estimator $\widehat{\tau}(\bx)$, and (ii) adjust the variance estimator to incorporate the additional uncertainty introduced by the bias correction. This inference approach has several theoretical advantages \citep{Calonico-Cattaneo-Titiunik_2014_ECMA,Calonico-Cattaneo-Farrell_2018_JASA,Calonico-Cattaneo-Farrell_2022_Bernoulli}, and has been validated empirically \citep{Hyytinen-Tukiainen-etal2018_QE,DeMagalhaes-etal_2025_PA}.

Finally, regarding the computation of the Gaussian process conditional on $\bV$, $(\widehat{Z}(\bx): \bx \in \B)$, there are two methodological issues to consider. First, simulation is implemented over a grid of points forming a discretization of the index set of the continuous stochastic process $\widehat{Z}$; as the number of points in the mesh increases, the approximation becomes more accurate. Second, the estimated (discretized) covariance function may fail to be positive definite in finite samples, but this issue can be easily fixed via regularization; see \cite{Cattaneo-Feng-Underwood_2024_JASA} for a discussion and related technical results.

Section SA-9 in the supplemental appendix provides additional implementation details; see also the companion software package \texttt{rd2d} and article \citep{Cattaneo-Titiunik-Yu_2025_rd2d}.

\section{Weighted Boundary Average Treatment Effect}\label{Sec: WBATE}

Without loss of generality, we normalize the weight function so that $\int_{\B} w(\bx) \dif \bx = 1$, and consider $\tau_{\mathtt{WBATE}} = \int_{\B} \tau(\bx) w(\bx) \dif \bx$, where $w$ satisfies Assumption \ref{assump: Weight Function and Boundary}. The WBATE aggregates the heterogeneous treatment effects $(\tau(\bx):\bx\in\B)$ according to a user-specified weighting scheme. The same construction can be applied to a subset of the boundary, as in \cite{Reardon-Robinson_2012_JREE}, in which case the estimand is the weighted average over that target subset.

The plug-in estimator is $\widehat{\tau}_{\mathtt{WBATE}} = \int_{\B} \widehat{\tau}(\bx) w(\bx) \dif \bx$. When analytic line integration is inconvenient, the integral can be approximated on a boundary grid. For grid points $(\bb_j:j=1,\ldots,J)$ and quadrature weights $(\omega_j:j=1,\ldots,J)$, normalized so that $\sum_{j=1}^J \omega_j w(\bb_j)=1$, $\widehat{\tau}_{\mathtt{WBATE}} \approx \sum_{j=1}^J \omega_j \widehat{\tau}(\bb_j) w(\bb_j)$. In some applications the weights may instead be chosen directly from local empirical information, for example $\omega_j w(\bb_j)=N_j/\sum_{\ell=1}^J N_\ell$ with $N_j = \sum_{i=1}^n \Indicator(\|\bX_i-\bb_j\|\leq h)$. See \cite{Cattaneo-Titiunik-Yu_2025_rd2d} for implementation details.

Our first result establishes a conditional MSE expansion for $\widehat{\tau}_{\mathtt{WBATE}}$. Let $B_{\mathtt{WBATE}} = B_{1,\mathtt{WBATE}} - B_{0,\mathtt{WBATE}}$, with $B_{t,\mathtt{WBATE}} = \int_{\B} B_{t,\bx} w(\bx) \dif \bx$, and let $\Omega_{\mathtt{WBATE}} = \Omega_{1,\mathtt{WBATE}} + \Omega_{0,\mathtt{WBATE}}$, with $\Omega_{t,\mathtt{WBATE}} = \int_{\B} \int_{\B} \Omega_{t,\bx_1,\bx_2} w(\bx_1) w(\bx_2) \dif \bx_1 \dif \bx_2$, for $t\in\{0,1\}$. For the variance-rate and inference results below, we impose the scalar aggregate nondegeneracy condition $\Omega_{\mathtt{WBATE}}\gtrsim(nh)^{-1}$, which rules out cancellation of the local covariance kernel after weighting and integrating over $\B$.

\begin{thm}[MSE Expansion: Sharp WBATE]\label{thm: MSE Expansion: Sharp WBATE}
    Suppose Assumptions \ref{assump: Sharp DGP}, \ref{assump: Boundary and Kernel} and \ref{assump: Weight Function and Boundary} hold, and $\Omega_{\mathtt{WBATE}}\gtrsim(nh)^{-1}$. If $n^{v/(2+v)}h^2/\log(1/h)\to\infty$ and $h\to0$, then
    $\E \big[(\widehat{\tau}_{\mathtt{WBATE}} - \tau_{\mathtt{WBATE}})^2 \big| \bX \big] = \Omega_{\mathtt{WBATE}} + h^{2p+2} B_{\mathtt{WBATE}}^2 + o_{\P}(\mathfrak{R})$, where $(n h)^{-1} \lesssim \Omega_{\mathtt{WBATE}} \lesssim (n h)^{-1}$, and $\mathfrak{R} = (n h)^{-1}+ h^{2p+2}$.
\end{thm}

The theorem implies $\widehat{\tau}_{\mathtt{WBATE}} = \tau_{\mathtt{WBATE}} + o_\P(1)$. It also shows how aggregation changes the effective variance order. The pointwise estimator $\widehat{\tau}(\bx)$ has variance of order $(nh^2)^{-1}$, whereas $\widehat{\tau}_{\mathtt{WBATE}}$ has variance of order $(nh)^{-1}$ because only boundary locations within $O(h)$ of one another have non-negligible covariance. Thus WBATE behaves like a one-dimensional nonparametric estimator.

Writing the leading variance term as $\Omega_{\mathtt{WBATE}}=V_{\mathtt{WBATE}}/(nh)$, the infeasible MSE-optimal scalar bandwidth is $h_{\mathtt{WBATE}} = \big(\frac{V_{\mathtt{WBATE}}}{(2p+2) B_{\mathtt{WBATE}}^2} \frac{1}{n} \big)^{1/(2p+3)}$, provided that $B_{\mathtt{WBATE}}\neq0$. A feasible counterpart can be constructed using plug-in estimators of $B_{\mathtt{WBATE}}$ and $V_{\mathtt{WBATE}}$, as in Section \ref{sec: Implementation}. The exponent differs from the pointwise bandwidth because the WBATE variance is of order $(nh)^{-1}$ rather than $(nh^2)^{-1}$.

For inference, define the feasible $t$-statistic $\widehat{\Tstat}_{\mathtt{WBATE}} = (\widehat{\tau}_{\mathtt{WBATE}} - \tau_{\mathtt{WBATE}})/\widehat{\Omega}^{1/2}_{\mathtt{WBATE}}$, with
$\widehat{\Omega}_{\mathtt{WBATE}} = \widehat{\Omega}_{1,\mathtt{WBATE}} + \widehat{\Omega}_{0,\mathtt{WBATE}}$,
where
$\widehat{\Omega}_{t,\mathtt{WBATE}} = \int_{\B} \int_{\B} \widehat{\Omega}_{t,\bx_1,\bx_2} w(\bx_1) w(\bx_2) \dif \bx_1 \dif \bx_2$,
for $t\in\{0,1\}$. The associated confidence interval estimator is
\begin{align*}
    \widehat{\CI}_{\alpha,\mathtt{WBATE}}
     = \Big[\; \widehat{\tau}_{\mathtt{WBATE}} - \q_{\alpha} \widehat{\Omega}^{1/2}_{\mathtt{WBATE}}
        \; , \;
        \widehat{\tau}_{\mathtt{WBATE}} + \q_{\alpha} \widehat{\Omega}^{1/2}_{\mathtt{WBATE}} \; \Big],
\end{align*}
where $\q_{\alpha} = \Phi^{-1}(1-\alpha/2)$.

\begin{thm}[Confidence Intervals: Sharp WBATE]\label{thm: CLT: Sharp WBATE}
    Suppose Assumptions \ref{assump: Sharp DGP}, \ref{assump: Boundary and Kernel}, and \ref{assump: Weight Function and Boundary} hold, and $\Omega_{\mathtt{WBATE}}\gtrsim(nh)^{-1}$. If $n^{v/(2+v)}h^2/\log(1/h) \to \infty$ and $n h^{2p+3} \to 0$, then
    $\P \big[\tau_{\mathtt{WBATE}} \in \widehat{\CI}_{\alpha,\mathtt{WBATE}} \big] \to 1 - \alpha$.
\end{thm}

For implementation, the MSE-optimal WBATE point estimator can be paired with robust bias correction \citep{Calonico-Cattaneo-Titiunik_2014_ECMA,Calonico-Cattaneo-Farrell_2018_JASA,Calonico-Cattaneo-Farrell_2022_Bernoulli}, following the same logic discussed after Theorem \ref{thm: Confidence Intervals and Bands: Sharp BATEC}. See Section SA-9 in the supplemental appendix, and the companion software package \texttt{rd2d} and article \citep{Cattaneo-Titiunik-Yu_2025_rd2d}, for more details.

\section{Largest Boundary Average Treatment Effect}\label{sec: LBATE}

As an alternative to the WBATE, researchers may be interested in the largest treatment effect along the boundary, $\tau_{\mathtt{LBATE}} = \sup_{\bx\in\B} \tau(\bx)$, with plug-in estimator $\widehat{\tau}_{\mathtt{LBATE}} = \sup_{\bx\in\B} \widehat{\tau}(\bx)$. The smallest boundary treatment effect can be handled analogously by applying the same construction to $-\tau(\bx)$. In practice, the supremum is computed over a sufficiently fine boundary grid, $\widehat{\tau}_{\mathtt{LBATE}} \approx \max_{1\leq j \leq J} \widehat{\tau}(\bb_j)$. If the target set is restricted to a subset of $\B$, then both the estimand and estimator should be interpreted relative to that subset.

Theorem \ref{thm: Convergence Rates: Sharp BATEC} gives $|\widehat{\tau}_{\mathtt{LBATE}} - \tau_{\mathtt{LBATE}}| \leq \sup_{\bx\in\B}|\widehat{\tau}(\bx)-\tau(\bx)|$, and therefore gives the corresponding convergence rate. This rate implies consistency under the same side conditions discussed after Theorem \ref{thm: Convergence Rates: Sharp BATEC}. For uncertainty quantification, we project the uniform confidence band from Theorem \ref{thm: Confidence Intervals and Bands: Sharp BATEC} onto the supremum functional. Define
\begin{align*}
    \widehat{\CI}_{\alpha,\mathtt{LBATE}}
     = \Big[\; \sup_{\bx \in \B} \Big(\widehat{\tau}(\bx) - \q_{\alpha} \widehat{\Omega}^{1/2}_{\bx,\bx}\Big)
            \;,\;
               \sup_{\bx \in \B} \Big(\widehat{\tau}(\bx) + \q_{\alpha} \widehat{\Omega}^{1/2}_{\bx,\bx}\Big) \; \Big]
\end{align*} 
where $\q_{\alpha} = \inf\{c > 0: \P(\sup_{\bx \in \B} |\widehat{Z}(\bx)|\geq c |\bV) \leq \alpha\}$, and $\widehat{Z}$ is the conditionally mean-zero Gaussian process from Theorem \ref{thm: Confidence Intervals and Bands: Sharp BATEC}.

\begin{thm}[Confidence Intervals: Sharp LBATE]\label{thm: Confidence Intervals: Sharp LBATE}
    Suppose the assumptions and conditions in Theorem \ref{thm: Confidence Intervals and Bands: Sharp BATEC} hold. If $\liminf_{n \to \infty} \log h/\log n > - \infty$, $n^{v/(2+v)}h^2/(\log n)^{9} \to \infty$ and $nh^{2p+4}\log n \to 0$, then
    $\P\big[\tau_{\mathtt{LBATE}} \in \widehat{\CI}_{\alpha,\mathtt{LBATE}} \big] \geq 1 - \alpha + o(1)$.
\end{thm}

The resulting interval is generally conservative because it is obtained from simultaneous coverage of the entire BATEC. Implementation follows the same grid and Gaussian-simulation steps outlined for $\widehat{\CI}_{\alpha}(\bx)$ and $\widehat{\CI}_{\alpha,\mathtt{WBATE}}$. In practice, finite-sample regularization may be useful to ensure that $\inf_{\bx \in \B} \widehat{\Omega}_{\bx,\bx}$ is bounded away from zero. See Section SA-9 in the supplemental appendix for more discussion, and the companion software package \texttt{rd2d} and article \citep{Cattaneo-Titiunik-Yu_2025_rd2d} for implementation details.

\section{Imperfect Compliance}\label{sec: Imperfect Compliance}

We extend the results to settings with imperfect treatment compliance, where treatment assignment and treatment status may not be equal for some units; see \cite{Hernan-Robins_2020_Book} for an overview in causal inference. We only give a brief discussion to conserve space, but all omitted details can be found in the supplemental appendix (Section SA-6).

To formalize the fuzzy BD design setup, we need to expand the potential outcomes notation. Let $W_i = \Indicator(\bX_i \in \A_0) \cdot W_i(0) + \Indicator(\bX_i \in \A_1) \cdot W_i(1)$ be the observed treatment status, where $W_i(t)$ denotes the potential treatment status under treatment assignment $t\in\{0,1\}$ for each unit. In addition, the observed outcome is now $Y_i = \Indicator(\bX_i \in \A_0) \cdot Y_i(0,W_i(0)) + \Indicator(\bX_i \in \A_1) \cdot Y_i(1,W_i(1))$, where the potential outcomes are now functions of two arguments: $Y_i(t,w)$ denotes the potential outcome for unit $i$ when this unit is assigned to treatment $t\in\{0,1\}$ and takes treatment status $w\in\{0,1\}$.

To recycle the notation and assumptions from the sharp BD design setup, we redefine $Y_i(t) = Y_i(t,W_i(t))$ for each $t\in\{0,1\}$. In addition, we impose the following assumption on each reduced-form potential outcome/status pair.

\begin{assumption}[Fuzzy DGP]\label{assump: Fuzzy DGP}
    Let $t\in\{0,1\}$.
    \begin{enumerate}[label=\normalfont(\roman*),noitemsep,leftmargin=*]

    \item $(Y_1(t),W_1(t),\bX_1^\top)^\top,\ldots,(Y_n(t),W_n(t),\bX_n^\top)^\top$ are independent and identically distributed random vectors.

    \item $\mu_{W,t}(\bx) = \E[W_i(t)| \bX_i = \bx]$ has a $(p+1)$-times continuously differentiable extension to an open neighborhood of $\X$.

    \item $\sigma^2_{W,t}(\bx) = \Var[W_i(t)|\bX_i = \bx]$ is continuous on $\X$.

    \item $\bUpsilon_{t}(\bx) = \Cov[(Y_i(t),W_i(t))^{\top}|\bX_i = \bx]$ has entries that are continuous on $\X$.

    \end{enumerate}
\end{assumption}

The usual fuzzy estimators and estimand are, respectively,
\begin{align*}
    \widehat{\zeta}(\bx) = \frac{\widehat{\tau}_{Y}(\bx)}{\widehat{\tau}_{W}(\bx)}
    \qquad\text{and}\qquad
    \zeta(\bx) = \frac{\tau_{Y}(\bx)}{\tau_{W}(\bx)},
\end{align*}
for each $\bx \in \B$, where $\widehat{\tau}_{Y}(\bx) = \be_{\mathbf{0}}^{\top}\widehat{\bbeta}_{Y,1}(\bx) - \be_{\mathbf{0}}^{\top}\widehat{\bbeta}_{Y,0}(\bx)$ and $\widehat{\tau}_{W}(\bx) = \be_{\mathbf{0}}^{\top}\widehat{\bbeta}_{W,1}(\bx) - \be_{\mathbf{0}}^{\top}\widehat{\bbeta}_{W,0}(\bx)$, with $\widehat{\bbeta}_{A,t}(\bx)$ denoting the local polynomial fit \eqref{eq: locpoly estimator} using the outcome variable $A \in \{Y,W\}$ and side $t\in\{0,1\}$, and $\tau_{Y}(\bx) = \E[Y_i(1,W_i(1)) - Y_i(0,W_i(0)) | \bX_i = \bx]$ and $\tau_{W}(\bx) = \E[W_i(1) - W_i(0) | \bX_i = \bx]$.

The functional (causal) parameters $\tau_{Y}(\bx)$ and $\tau_{W}(\bx)$ correspond to the intention-to-treat effect and the first-stage effect, at the boundary point $\bx$, respectively. The ratio $\zeta(\bx)$ is the fuzzy BATEC parameter, which can be interpreted as a complier average treatment effect at each boundary point $\bx$ under additional assumptions; see \cite{choi2023complier}, \cite{schwarz2025effect}, \cite{jiang2026extrapolating}, \cite{Cattaneo-Keele-Titiunik-VazquezBare_2016_JOP,Cattaneo-Keele-Titiunik-VazquezBare_2021_JASA} for more discussion of causal identification in fuzzy multidimensional RD designs. See also \cite{Arai-etal_2022_QE} and references therein for an overview of causal identification in univariate fuzzy RD designs.

Fuzzy analogues of WBATE and LBATE can be defined by replacing $\tau(\bx)$ with $\zeta(\bx)$ in the definitions of $\tau_{\mathtt{WBATE}}$ and $\tau_{\mathtt{LBATE}}$, respectively. More precisely, after normalizing the weights so that $\int_{\B} w(\bx)\dif\bx=1$, the fuzzy WBATE is $\zeta_{\mathtt{WBATE}} = \int_{\B} \zeta(\bx) w(\bx) \dif \bx$, and the fuzzy LBATE is $\zeta_{\mathtt{LBATE}} = \sup_{\bx\in\B} \zeta(\bx)$. Plug-in estimators of these parameters are $\widehat{\zeta}_{\mathtt{WBATE}} = \int_{\B} \widehat{\zeta}(\bx) w(\bx) \dif \bx$ and $\widehat{\zeta}_{\mathtt{LBATE}} = \sup_{\bx\in\B} \widehat{\zeta}(\bx)$, respectively.

The properties of the fuzzy estimators can be established using the results for the sharp BD design, along with ``continuous mapping'' and ``probability concentration'' arguments. More precisely, under Assumptions \ref{assump: Sharp DGP}, \ref{assump: Boundary and Kernel} and \ref{assump: Fuzzy DGP}, and bandwidth sequences that make the sharp convergence rates vanish, our results imply that $\widehat{\tau}_{Y}(\bx) = \tau_{Y}(\bx) + o_\P(1)$ and $\widehat{\tau}_{W}(\bx) = \tau_{W}(\bx) + o_\P(1)$, pointwise and uniformly over $\bx\in\B$. For uniform consistency, it suffices that $n^{(1+v)/(2+v)}h^2/\log(1/h)\to\infty$ and $h\to0$. Therefore, under the additional regularity condition $\inf_{\bx\in\B} |\tau_W(\bx)| > 0$, and using the exact second-order ``linearization'' $\widehat{\zeta}(\bx) - \zeta(\bx) = \mathfrak{w}(\bx)^\top \mathfrak{T}(\bx) + \mathfrak{R}_{\mathtt{F}}(\bx)$ with
\begin{align*}
    \mathfrak{w}(\bx)
    = \big[ \frac{1}{\tau_W(\bx)}, -\frac{\tau_Y(\bx)}{\tau_W(\bx)^{2}} \big]^\top
    \quad\text{and}\quad
    \mathfrak{T}(\bx)
    = \big[ \widehat{\tau}_Y(\bx) - \tau_Y(\bx), \widehat{\tau}_W(\bx) - \tau_W(\bx) \big]^\top,
\end{align*}
and $\sup_{\bx\in\B} |\mathfrak{R}_{\mathtt{F}}(\bx)| \lesssim_\P \sup_{\bx\in\B} |\widehat{\tau}_W(\bx) - \tau_W(\bx)|^2 + \sup_{\bx\in\B} |\widehat{\tau}_Y(\bx) - \tau_Y(\bx)| \sup_{\bx\in\B} |\widehat{\tau}_W(\bx) - \tau_W(\bx)|$, which is negligible when the sharp convergence rates vanish. See the supplemental appendix (Section SA-6.1) for details.

Therefore, estimation and inference results for fuzzy BATEC, fuzzy WBATE, and fuzzy LBATE can be established using the linearization argument, and then applying the theoretical results in the supplemental appendix to the specific linear combination $\mathfrak{w}(\bx)^\top \mathfrak{T}(\bx)$. In the remainder of this section we briefly overview the main estimation and inference results for fuzzy BD designs. See the supplemental appendix (Section SA-6) for more results and omitted details.

\subsection{Fuzzy BATEC}\label{sec: Fuzzy BATEC}

Pointwise and uniform convergence rates, AMSE/AIMSE expansions, and bandwidth selectors for the fuzzy BATEC estimator $\widehat{\zeta}(\bx)$ can be established using the linearization argument outlined above, along with the results for the sharp BD design. To state the leading terms, let $B_{A,\bx}=B_{A,1,\bx}-B_{A,0,\bx}$ denote the leading bias constant from Section \ref{sec: Treatment Effect Estimation}, with the outcome variable $A\in\{Y,W\}$ in place of $Y$, and define $B_{\mathtt{F},\bx} = \mathfrak{w}(\bx)^\top (B_{Y,\bx},B_{W,\bx})^\top$.
Similarly, let $V_{A_1,A_2,\bx}$ denote the leading covariance constant for the reduced-form local polynomial estimators using outcomes $A_1,A_2\in\{Y,W\}$, and define
\begin{align*}
    V_{\mathtt{F},\bx}
    = \mathfrak{w}(\bx)^\top
      \begin{pmatrix}
        V_{Y,Y,\bx} & V_{Y,W,\bx}\\
        V_{W,Y,\bx} & V_{W,W,\bx}
      \end{pmatrix}
      \mathfrak{w}(\bx).
\end{align*}
For inference below, we impose $\inf_{\bx\in\B}V_{\mathtt{F},\bx}>0$ directly. This is the scalar nondegeneracy condition for the linearized fuzzy Wald score and allows one-sided compliance because it does not require the treatment-status variances on both sides to be bounded away from zero.

\begin{thm}[Convergence Rates: Fuzzy BATEC]\label{thm: Convergence Rates: Fuzzy BATEC}
    Suppose Assumptions \ref{assump: Sharp DGP}, \ref{assump: Boundary and Kernel}, and \ref{assump: Fuzzy DGP} hold, and $\inf_{\bx\in\B} |\tau_W(\bx)| > 0$. If $n^{(1+v)/(2+v)}h^2/\log(1/h) \to \infty$ and $h\to0$, then
    \begin{enumerate}[label=\normalfont(\roman*),noitemsep,leftmargin=*]
        \item $|\widehat{\zeta}(\bx)-\zeta(\bx)|\lesssim_\P \frac{1}{\sqrt{nh^2}}+\frac{1}{n^{\frac{1+v}{2+v}}h^2}+h^{p+1}$ for $\bx\in\B$, and
        \item $\sup_{\bx\in\B}|\widehat{\zeta}(\bx)-\zeta(\bx)|\lesssim_\P \sqrt{\frac{\log(1/h)}{nh^2}}+\frac{\log(1/h)}{n^{\frac{1+v}{2+v}}h^2}+h^{p+1}$.
    \end{enumerate}
\end{thm}

Using the linearization argument, we focus on the approximate MSE and approximate IMSE, respectively, $\operatorname{AMSE}_{\mathtt{F}}(\bx)=\E[(\mathfrak{w}(\bx)^\top\mathfrak{T}(\bx))^2|\bX]$ and $\operatorname{AIMSE}_{\mathtt{F}}=\int_\B \operatorname{AMSE}_{\mathtt{F}}(\bx)w(\bx)\dif\bx$.

\begin{thm}[Approximate MSE Expansions: Fuzzy BATEC]\label{thm: MSE Expansions: Fuzzy BATEC}
    Suppose Assumptions \ref{assump: Sharp DGP}, \ref{assump: Boundary and Kernel}, \ref{assump: Weight Function and Boundary}, and \ref{assump: Fuzzy DGP} hold, and $\inf_{\bx\in\B} |\tau_W(\bx)| >0$. If $n^{v/(2+v)} h^2/\log(1/h) \to \infty$ and $h\to0$, then
    \begin{enumerate}[label=\normalfont(\roman*),noitemsep,leftmargin=*]
        \item $\operatorname{AMSE}_{\mathtt{F}}(\bx)
               = h^{2p+2} B_{\mathtt{F},\bx}^2+\frac{1}{nh^2}V_{\mathtt{F},\bx}+o_\P(\mathfrak{R})$, and
        \item $\operatorname{AIMSE}_{\mathtt{F}}
               = h^{2p+2}\int_\B B_{\mathtt{F},\bx}^2 w(\bx) \dif \bx+\frac{1}{nh^2}\int_\B V_{\mathtt{F},\bx} w(\bx) \dif \bx+o_\P(\mathfrak{R})$,
    \end{enumerate}
    with $\mathfrak{R}=h^{2p+2}+n^{-1}h^{-2}$.
\end{thm}

Ignoring the higher-order terms, the approximate MSE-optimal and IMSE-optimal bandwidth selectors are obtained from the formulas in Section \ref{sec: Treatment Effect Estimation} by replacing $B_{\bx},V_{\bx}$ with $B_{\mathtt{F},\bx},V_{\mathtt{F},\bx}$. For inference, we consider the usual non-Donsker t-statistic process, leading to the confidence intervals and bands
\begin{align*}
    \widehat{\CI}_{\mathtt{F},\alpha}(\bx)
    = \Big[\;\widehat{\zeta}(\bx) - \q_{\alpha} \widehat{\Omega}^{1/2}_{\mathtt{F},\bx,\bx}
           \; , \;
           \widehat{\zeta}(\bx) + \q_{\alpha} \widehat{\Omega}^{1/2}_{\mathtt{F},\bx,\bx}\;\Big],
\end{align*}
where $\q_{\alpha}$ denotes an appropriate quantile depending on the desired confidence level $\alpha\in(0,1)$ and coverage objective (pointwise for $\bx\in\B$ versus uniform over $\B$), and the variance estimator is derived from the above stochastic linearization using the feasible weights
\begin{align*}
    \widehat{\mathfrak{w}}(\bx)
    = \big[ \frac{1}{\widehat{\tau}_W(\bx)}, -\frac{\widehat{\tau}_Y(\bx)}{\widehat{\tau}_W(\bx)^2} \big]^\top.
\end{align*}
Specifically,
\begin{align*}
    \widehat{\Omega}_{\mathtt{F},\bx_1,\bx_2}
     = \widehat{\mathfrak{w}}(\bx_1)^\top \begin{pmatrix}
        \widehat{\Omega}_{Y,Y,\bx_1, \bx_2} & \widehat{\Omega}_{Y,W,\bx_1, \bx_2} \\
        \widehat{\Omega}_{W,Y,\bx_1, \bx_2} & \widehat{\Omega}_{W,W,\bx_1, \bx_2}
      \end{pmatrix}
    \widehat{\mathfrak{w}}(\bx_2),
    \qquad \bx_1, \bx_2 \in \B,
\end{align*}
where $\widehat{\Omega}_{Y,W,\bx_1,\bx_2} = \frac{1}{n h^2} \be_{\mathbf{0}}^{\top} \big[ \widehat{\bGamma}_{0,\bx_1}^{-1} \widehat{\bSigma}_{Y,W,0,\bx_1,\bx_2} \widehat{\bGamma}_{0,\bx_2}^{-1} + \widehat{\bGamma}_{1,\bx_1}^{-1} \widehat{\bSigma}_{Y,W,1,\bx_1,\bx_2} \widehat{\bGamma}_{1,\bx_2}^{-1} \big] \be_{\mathbf{0}}$ with
$\widehat{\bSigma}_{Y,W,t,\bx_1,\bx_2}
    = h^2 \En \big[\br_p(\tfrac{\bX_i-\bx_1}{h}) \br_p(\tfrac{\bX_i-\bx_2}{h})^{\top} K_h(\bX_i - \bx_1) K_h(\bX_i - \bx_2) \widehat{\varepsilon}_{Y,i}(\bx_1) \widehat{\varepsilon}_{W,i}(\bx_2) \Indicator(\bX_i \in \A_t) \big]
$
and $\widehat{\varepsilon}_{A,i}(\bx) = A_i - \br_p(\bX_i - \bx)^\top[\Indicator(\bX_i \in \A_0) \widehat{\bbeta}_{A,0}(\bx) + \Indicator(\bX_i \in \A_1) \widehat{\bbeta}_{A,1}(\bx)]$, with $A\in\{Y,W\}$. The population covariance $\Omega_{\mathtt{F},\bx_1,\bx_2}$ is defined analogously, replacing feasible weights and covariance components by their population counterparts.

The following result establishes the validity of these confidence intervals and bands. Let $\bV = ((Y_1, W_1, \bX_1^\top)^\top,\ldots, (Y_n, W_n, \bX_n^\top)^\top)$ be the observed data.

\begin{thm}[Confidence Intervals and Bands: Fuzzy BATEC]\label{thm: Confidence Intervals and Bands: Fuzzy BATEC}
    Suppose Assumptions \ref{assump: Sharp DGP}, \ref{assump: Boundary and Kernel} and \ref{assump: Fuzzy DGP} hold, $\inf_{\bx\in\B} |\tau_W(\bx)| > 0$, and $\inf_{\bx\in\B}V_{\mathtt{F},\bx}>0$.
    \begin{enumerate}[label=\normalfont(\roman*),noitemsep,leftmargin=*]
        \item If $n h^2 \to \infty$ and $n h^{2p+4} \to 0$, then
              $\P \big[\zeta(\bx) \in \widehat{\CI}_{\mathtt{F},\alpha}(\bx) \big] \to 1 - \alpha$
              with $\q_{\alpha} = \Phi^{-1}(1-\alpha/2)$.

        \item If $\liminf_{n\to\infty} \log h/\log n > -\infty$, $n^{v/(2+v)} h^2/(\log n)^9 \to \infty$, and $n h^{2p+4}\log n\to0$, then
              $\P \big[\zeta(\bx) \in \widehat{\CI}_{\mathtt{F},\alpha}(\bx), \text{ for all } \bx \in \B \big] \to 1 - \alpha$
              with $\q_{\alpha} = \inf \{c > 0: \P[\sup_{\bx \in \B} |\widehat{Z}_{\mathtt{F}}(\bx)| \geq c | \bV] \leq \alpha\}$, where $(\widehat{Z}_{\mathtt{F}}(\bx): \bx \in \B)$ is a Gaussian process conditional on $\bV$ with $\E[\widehat{Z}_{\mathtt{F}}(\bx_1)|\bV]=0$ and $\Cov[\widehat{Z}_{\mathtt{F}}(\bx_1),\widehat{Z}_{\mathtt{F}}(\bx_2) | \bV] = \widehat{\Omega}_{\mathtt{F},\bx_1,\bx_2}/(\widehat{\Omega}_{\mathtt{F},\bx_1,\bx_1} \widehat{\Omega}_{\mathtt{F},\bx_2,\bx_2})^{1/2}$, for all $\bx_1, \bx_2 \in \B$.
    \end{enumerate}
\end{thm}

See the supplemental appendix (Section SA-6.1) for more results.

\subsection{Fuzzy WBATE}\label{sec: Fuzzy WBATE}

Without loss of generality, we set $\int_{\B} w(\bx) \dif \bx=1$. The fuzzy WBATE and its plug-in estimator are
\begin{align*}
    \zeta_{\mathtt{WBATE}}=\int_{\B} \zeta(\bx) w(\bx) \dif \bx
    \qquad\text{and}\qquad
    \widehat{\zeta}_{\mathtt{WBATE}}=\int_{\B} \widehat{\zeta}(\bx) w(\bx) \dif \bx.
\end{align*}
Let $B_{\mathtt{F,WBATE}}=\int_{\B} B_{\mathtt{F},\bx} w(\bx) \dif \bx$ and
$\Omega_{\mathtt{F,WBATE}} = \int_{\B}\int_{\B}\Omega_{\mathtt{F},\bx_1,\bx_2} w(\bx_1) w(\bx_2) \dif \bx_1 \dif \bx_2$,
and define $\widehat{\Omega}_{\mathtt{F,WBATE}}$ analogously by replacing
$\Omega_{\mathtt{F},\bx_1,\bx_2}$ with $\widehat{\Omega}_{\mathtt{F},\bx_1,\bx_2}$.
For the fuzzy WBATE AMSE and inference results, we impose the aggregate nondegeneracy condition $\Omega_{\mathtt{F,WBATE}}\gtrsim(nh)^{-1}$, which rules out cancellation after applying the fuzzy Wald weights and integrating over $\B$.
The approximate MSE is for the linearized aggregate
\begin{align*}
    \check{\zeta}_{\mathtt{WBATE}}
    = \int_\B\mathfrak{w}(\bx)^\top\mathfrak{T}(\bx)w(\bx)\dif\bx,
    \qquad
    \operatorname{AMSE}_{\mathtt{F,WBATE}}=\E[\check{\zeta}_{\mathtt{WBATE}}^2|\bX].
\end{align*}

\begin{thm}[Approximate MSE Expansion: Fuzzy WBATE]\label{thm: MSE Expansion: Fuzzy WBATE}
    Suppose Assumptions \ref{assump: Sharp DGP}, \ref{assump: Boundary and Kernel}, \ref{assump: Weight Function and Boundary}, and \ref{assump: Fuzzy DGP} hold, $\inf_{\bx\in\B}|\tau_W(\bx)|>0$, $\inf_{\bx\in\B}V_{\mathtt{F},\bx}>0$, and $\Omega_{\mathtt{F,WBATE}}\gtrsim(nh)^{-1}$. If $n^{v/(2+v)}h^2/\log(1/h)\to\infty$ and $h\to0$, then $\operatorname{AMSE}_{\mathtt{F,WBATE}}
        = \Omega_{\mathtt{F,WBATE}}+h^{2p+2}B_{\mathtt{F,WBATE}}^2+o_\P(\mathfrak{R})$,
    where $(nh)^{-1}\lesssim\Omega_{\mathtt{F,WBATE}}\lesssim(nh)^{-1}$ and $\mathfrak{R}=(nh)^{-1}+h^{2p+2}$.
\end{thm}

For inference, define $\widehat{\Tstat}_{\mathtt{F,WBATE}} = (\widehat{\zeta}_{\mathtt{WBATE}}-\zeta_{\mathtt{WBATE}})/\widehat{\Omega}^{1/2}_{\mathtt{F,WBATE}}$ with associated confidence interval
\begin{align*}
    \widehat{\CI}_{\mathtt{F},\alpha,\mathtt{WBATE}}
    =
    \Big[\;\widehat{\zeta}_{\mathtt{WBATE}}-\q_{\alpha} \widehat{\Omega}^{1/2}_{\mathtt{F,WBATE}}
         \; , \;
         \widehat{\zeta}_{\mathtt{WBATE}} + \q_{\alpha} \widehat{\Omega}^{1/2}_{\mathtt{F,WBATE}}\;\Big],
\end{align*}
where $\q_{\alpha} = \Phi^{-1}(1-\alpha/2)$.

\begin{thm}[Confidence Intervals: Fuzzy WBATE]\label{thm: Confidence Intervals: Fuzzy WBATE}
    Suppose Assumptions \ref{assump: Sharp DGP}, \ref{assump: Boundary and Kernel}, \ref{assump: Weight Function and Boundary}, and \ref{assump: Fuzzy DGP} hold, $\inf_{\bx\in\B}|\tau_W(\bx)|>0$, $\inf_{\bx\in\B}V_{\mathtt{F},\bx}>0$, and $\Omega_{\mathtt{F,WBATE}}\gtrsim(nh)^{-1}$. If $n^{v/(2+v)}h^2/\log(1/h)\to\infty$, $n h^{2p+3}\to0$, $\sqrt{n}h^{3/2}/\log(1/h)\to\infty$, and $n^{2v/(2+v)-1/2}h^{7/2}/\log^2(1/h)\to\infty$, then $\P[\zeta_{\mathtt{WBATE}}\in\widehat{\CI}_{\mathtt{F},\alpha,\mathtt{WBATE}}] \to 1-\alpha$.
\end{thm}

See the supplemental appendix (Section SA-6.2) for more results.

\subsection{Fuzzy LBATE}\label{sec: Fuzzy LBATE}

The fuzzy LBATE and its plug-in estimator are
\begin{align*}
    \zeta_{\mathtt{LBATE}}=\sup_{\bx\in\B}\zeta(\bx)
    \qquad\text{and}\qquad
    \widehat{\zeta}_{\mathtt{LBATE}}=\sup_{\bx\in\B}\widehat{\zeta}(\bx).
\end{align*}

Theorem \ref{thm: Convergence Rates: Fuzzy BATEC} gives consistency and convergence rates for $\widehat{\zeta}_{\mathtt{LBATE}}$ by taking suprema over $\B$. For inference, use the Gaussian process from Theorem \ref{thm: Confidence Intervals and Bands: Fuzzy BATEC} and define
\begin{align*}
    \widehat{\CI}_{\mathtt{F},\alpha,\mathtt{LBATE}}
    = \Big[\; \sup_{\bx\in\B}\Big(\widehat{\zeta}(\bx)-\q_{\alpha} \widehat{\Omega}^{1/2}_{\mathtt{F},\bx,\bx}\Big)
           \; , \;
           \sup_{\bx\in\B}\Big(\widehat{\zeta}(\bx)+\q_{\alpha} \widehat{\Omega}^{1/2}_{\mathtt{F},\bx,\bx} \Big)\;\Big],
\end{align*}
where $\q_{\alpha}=\inf\{c>0:\P[\sup_{\bx\in\B}|\widehat{Z}_{\mathtt{F}}(\bx)|\ge c|\bV]\le\alpha\}$.

\begin{thm}[Confidence Intervals: Fuzzy LBATE]\label{thm: Confidence Intervals: Fuzzy LBATE}
    Suppose Assumptions \ref{assump: Sharp DGP}, \ref{assump: Boundary and Kernel} and \ref{assump: Fuzzy DGP} hold, $\inf_{\bx\in\B} |\tau_W(\bx)| > 0$, and $\inf_{\bx\in\B}V_{\mathtt{F},\bx}>0$. If $\liminf_{n\to\infty}\log h/\log n>-\infty$, $n^{v/(2+v)}h^2/(\log n)^9\to\infty$, and $nh^{2p+4}\log n\to0$, then $\P[\zeta_{\mathtt{LBATE}}\in\widehat{\CI}_{\mathtt{F},\alpha,\mathtt{LBATE}}]\ge1-\alpha+o(1)$.
\end{thm}

See the supplemental appendix (Section SA-6.3) for additional results.

\section{The Causal Effects of SPP on College Attendance}\label{sec: Empirical Application}

We illustrate our proposed causal inference methodology for BD designs with the SPP application introduced in Section \ref{sec: Introduction}. Recall that the dataset has $n=363,096$ complete observations for the first cohort of the program ($2014$), where each observation corresponds to one student, and the bivariate score is $\bX_i= (X_{1i},X_{2i})^\top = (\mathtt{SABER11}_i,\mathtt{SISBEN}_i)^\top$, where the first dimension is the student's SABER11 test score (ranging from $-310$ to $172$) and the second dimension is the student's SISBEN wealth index (ranging from $-103.41$ to $127.21$). Without loss of generality, each dimension of the score is recentered at its corresponding cutoff for program eligibility, so that the treatment assignment boundary is as shown in Figure \ref{fig:fig1a}. All the results in this section are implemented using our companion \texttt{R} software package \texttt{rd2d}, and omitted details are given in the supplemental appendix (Sections SA-9 and SA-10), replication files, and companion software article \citep{Cattaneo-Titiunik-Yu_2025_rd2d}.

The outcome variable of interest is \textit{college enrollment}, with $Y_i=1$ if the student enrolled in college and $Y_i=0$ otherwise. In this application, $15,423$ students were eligible for the SPP program, but only $9,165$ took up the benefit. We therefore report both sharp reduced-form effects of eligibility assignment and fuzzy effects that adjust for imperfect compliance. Specifically, we estimate the reduced-form BATEC $\tau_Y(\bx)$, the first-stage BATEC $\tau_W(\bx)$, and the fuzzy BATEC $\zeta(\bx)=\tau_Y(\bx)/\tau_W(\bx)$ at the $40$ evenly spaced grid points $\bb_j\in\B$, $j=1,\ldots,40$, depicted in Figure \ref{fig:fig1a}. We also report the corresponding WBATE and LBATE parameters. The empirical results use the ITT bandwidth selector; in the supplemental appendix, we also report results obtained with fuzzy bandwidths. All local polynomial estimators use data-driven MSE-optimal bandwidths as discussed in Section \ref{sec: Implementation}; the two score coordinates are standardized for bandwidth selection and estimation, while the bandwidths reported in the tables are transformed back to the original SABER11 and SISBEN scales. Uncertainty quantification uses robust bias-corrected inference for pointwise confidence intervals and uniform confidence bands.

\begin{figure}
    \centering
    \begin{subfigure}[b]{0.48\textwidth}
        \centering
        \includegraphics[width=\linewidth]{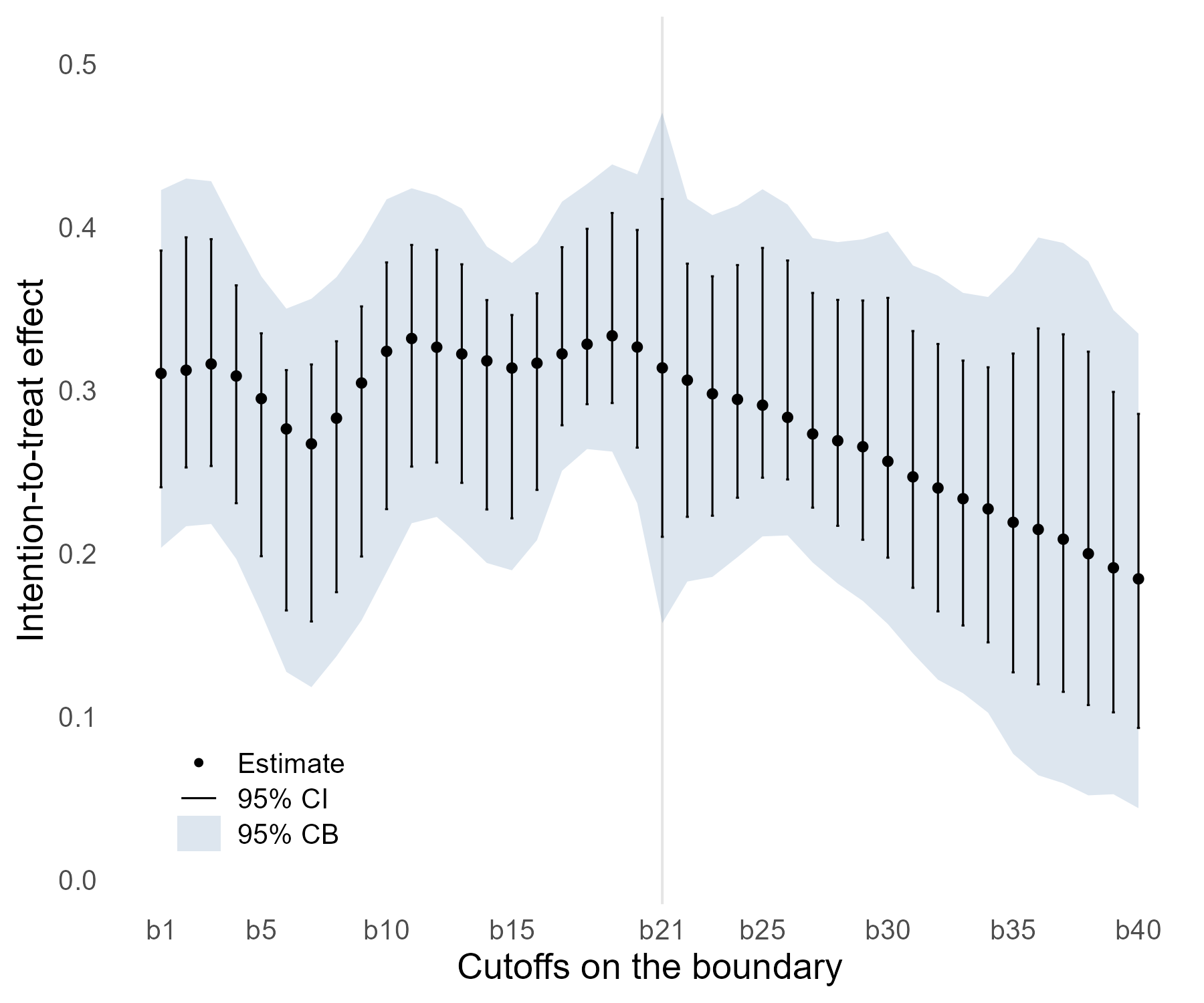}
        \caption{Intention-to-Treat Effects ($\tau_Y(\bx)$).}
        \label{fig:fig2a}
    \end{subfigure}
    \hspace{0.02\linewidth}
    \begin{subfigure}[b]{0.48\textwidth}
        \centering
        \includegraphics[width=\linewidth]{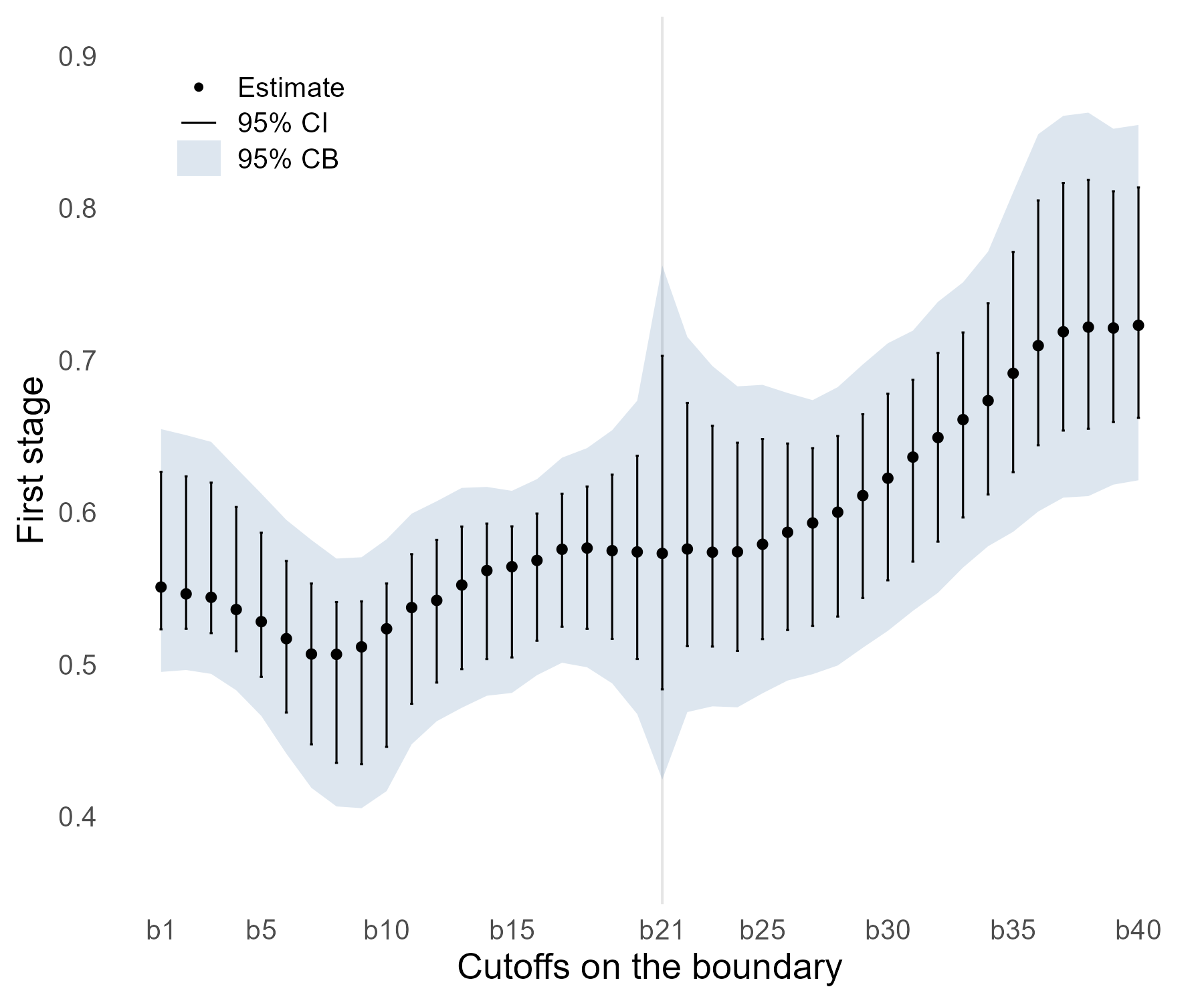}
        \caption{First-Stage Effects ($\tau_W(\bx)$).}
        \label{fig:fig2b}
    \end{subfigure}
    
    \medskip

    \begin{subfigure}[b]{0.48\textwidth}
        \centering
        \includegraphics[width=\linewidth]{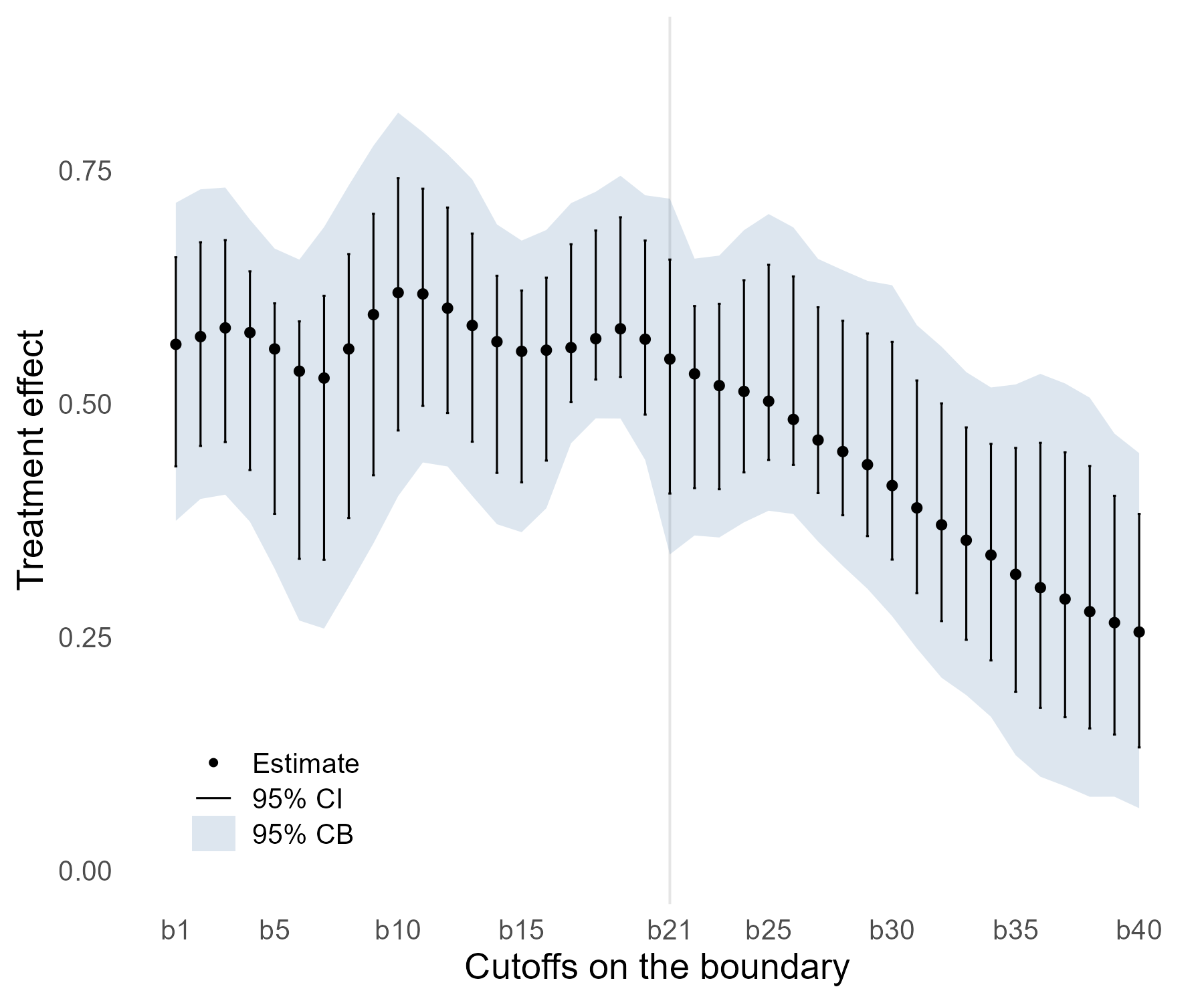}
        \caption{Fuzzy Effects ($\zeta(\bx)$).}
        \label{fig:fig2c}
    \end{subfigure}
    \hspace{0.02\linewidth}
    \begin{subfigure}[b]{0.48\textwidth}
        \centering
        \includegraphics[width=\linewidth]{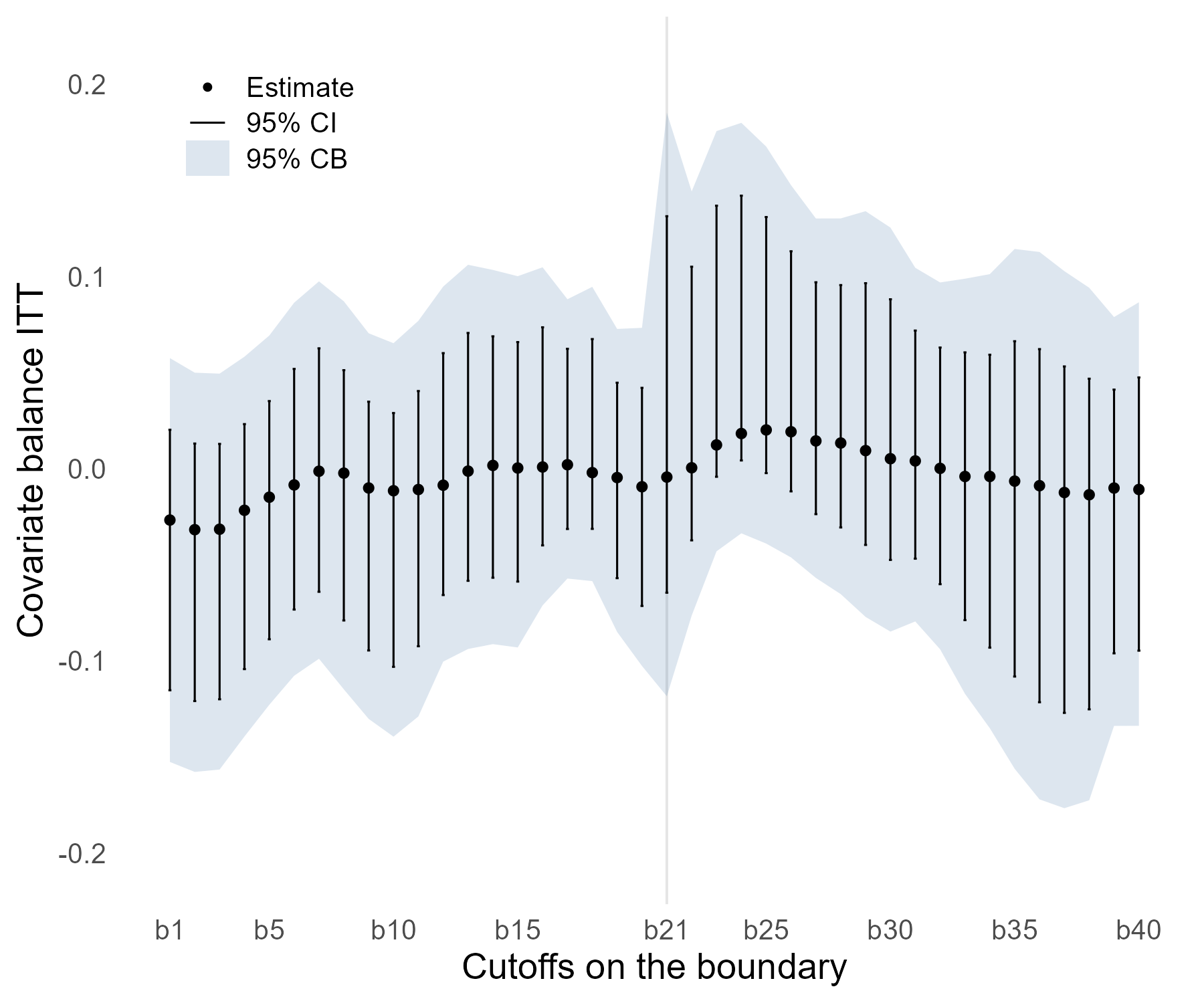}
        \caption{Intention-to-Treat Covariate Balance.}
        \label{fig:fig2d}
    \end{subfigure}

    \medskip

    \caption{Empirical Results for BATEC (SPP Application).\\
            {\footnotesize \textit{Notes}. Panel (a) reports reduced-form effects of eligibility on college enrollment. Panel (b) reports first-stage effects on SPP take-up. Panel (c) reports fuzzy effects on college enrollment. Panel (d) reports a covariate analysis using mother's education as the outcome. Point estimates, robust bias-corrected confidence intervals (CI), and robust bias-corrected confidence bands (CB) using the $40$ boundary grid points shown in Figure \ref{fig:fig1a}. Bandwidths are MSE-optimal, selected separately at each grid point, and based on the ITT bandwidth selector.}}
    \label{fig:fig2}
\end{figure}

Figure \ref{fig:fig2} summarizes the main results for the BATEC graphically. Panel (a) reports reduced-form intention-to-treat effects of eligibility on college enrollment, Panel (b) reports the first-stage effects on SPP take-up, Panel (c) reports the fuzzy effects on college enrollment, and Panel (d) reports covariate balance using mother's education as the outcome. Mother's education is predetermined, so a credible BD design should deliver treatment effects on this covariate that are statistically indistinguishable from zero; see \cite[Section 5]{Cattaneo-Idrobo-Titiunik_2020_CUP} for related discussion of falsification tests in RD designs. The supplemental appendix reports additional heat maps for point estimates and robust bias-corrected $p$-values along the assignment boundary, as well as the complete numerical tables for all $40$ boundary grid points.

The intention-to-treat effects in Panel (a) are statistically significant along the boundary. The pattern indicates roughly homogeneous effects along the portion of the boundary where students are marginal in terms of academic performance but satisfy the minimal poverty requirement ($\bb_1$ through approximately $\bb_{21}$), and more pronounced heterogeneity along the portion where students are marginal in terms of wealth ($\bb_{21}$ through $\bb_{40}$). In particular, the estimated reduced-form effect on college enrollment remains close to $30$ percentage points along the first portion of the boundary and declines for the wealthiest eligible students as their SABER11 scores increase. This pattern is consistent with the program having larger enrollment effects among students for whom the subsidy is more likely to relax a binding financial constraint. The first-stage estimates in Panel (b) are strongly positive along the boundary, confirming substantial but imperfect take-up of the subsidy. Consequently, the fuzzy estimates in Panel (c) are larger than the corresponding intention-to-treat effects. The placebo estimates in Panel (d) are centered near zero and statistically insignificant, which supports the continuity-based identifying assumptions in this application.

The magnitudes are also large relative to the control-side baseline enrollment rates. Section SA-10 of the supplemental appendix reports the estimates of $\mu_0(\bx)$ and the corresponding ratios $\widehat{\tau}_Y(\bx)/\widehat{\mu}_0(\bx)$ and $\widehat{\zeta}(\bx)/\widehat{\mu}_0(\bx)$. The reduced-form WBATE estimate is about $63$ percent of the corresponding control-side baseline, while the fuzzy WBATE estimate is about $109$ percent of that baseline. Across the full boundary grid, the reduced-form effects range from about $28$ to $95$ percent of the local baseline enrollment rate, and the fuzzy effects range from about $38$ to $182$ percent. The results in Table SA-5 show that, in the piece of the boundary where students are near the academic threshold (points $\bb_1$-$\bb_{20}$) the control mean is roughly constant: between 37 and 40 percent of control students enroll in college. In contrast, in the region of the boundary where students are near the wealth threshold (points $\bb_{21}$-$\bb_{40}$), the share of control students who attend college increases steadily as SABER 11 increases; the treatment effect is there reduced in this region of the boundary because more students are already enrolling in college in the absence of the program. Near the marginal level of wealth, students who are highly achieving academically are therefore less constrained by wealth than their less academically achieving peers.  

Table~\ref{tab:empirical-results} presents a summary of the numerical results underlying Figure \ref{fig:fig2}. Due to space limitations, each panel reports selected boundary points, $\bb_j$ with $j=1,5,10,15,20,25,30,35,40$, together with the corresponding WBATE and LBATE estimates. The WBATE estimator uses equal weights over the $40$ boundary grid points. The table also reports MSE-optimal bandwidths in the original score units, point estimates, robust bias-corrected test statistics, $p$-values, and confidence intervals. See Section SA-10 of the supplemental appendix for omitted details.

\begin{table}[p]
    \centering
    \begin{subtable}[t]{0.48\linewidth}
        \centering
        \caption{Intention-to-Treat Effects: $\tau_Y(\bx)$.}
        \label{tab:empirical-results-itt}
        \resizebox{\linewidth}{!}{\begin{tabular}{lccccccc}
  \toprule\toprule
   & $h_1$ & $h_2$ & $N_{\mathrm{Co}}$ & $N_{\mathrm{Tr}}$ & Estimate & $p$-value & 95\% CI \\
  \midrule
   $\bb_{1}$ & 27.6 & 12.2 & 8921 & 3532 & 0.311 & 0.000 & (0.204, 0.423)\\
   $\bb_{5}$ & 23.1 & 10.2 & 9044 & 3997 & 0.295 & 0.000 & (0.164, 0.370)\\
   $\bb_{10}$ & 19.3 & 8.6 & 6925 & 3563 & 0.324 & 0.000 & (0.189, 0.417)\\
   $\bb_{15}$ & 22.2 & 9.8 & 10063 & 4977 & 0.314 & 0.000 & (0.190, 0.378)\\
   $\bb_{20}$ & 19.4 & 8.6 & 8945 & 2680 & 0.327 & 0.000 & (0.231, 0.433)\\
   $\bb_{25}$ & 21.0 & 9.3 & 6188 & 3114 & 0.291 & 0.000 & (0.211, 0.423)\\
   $\bb_{30}$ & 26.3 & 11.7 & 4698 & 4560 & 0.257 & 0.000 & (0.157, 0.398)\\
   $\bb_{35}$ & 28.1 & 12.5 & 2875 & 2950 & 0.219 & 0.000 & (0.077, 0.373)\\
   $\bb_{40}$ & 36.3 & 16.1 & 2916 & 2993 & 0.185 & 0.000 & (0.044, 0.335)\\
  \midrule
   $\mathtt{WBATE}$ &  &  &  &  & 0.282 & 0.000 & (0.248, 0.316)\\
  \midrule
   $\mathtt{LBATE}$ &  &  &  &  & 0.334 &  & (0.263, 0.472)\\
  \bottomrule\bottomrule
\end{tabular}
}
    \end{subtable}
    \hspace{0.02\linewidth}
    \begin{subtable}[t]{0.48\linewidth}
        \centering
        \caption{First-Stage Effects: $\tau_W(\bx)$.}
        \label{tab:empirical-results-first-stage}
        \resizebox{\linewidth}{!}{\begin{tabular}{lccccccc}
  \toprule\toprule
   & $h_1$ & $h_2$ & $N_{\mathrm{Co}}$ & $N_{\mathrm{Tr}}$ & Estimate & $p$-value & 95\% CI \\
  \midrule
   $\bb_{1}$ & 27.6 & 12.2 & 8921 & 3532 & 0.551 & 0.000 & (0.495, 0.655)\\
   $\bb_{5}$ & 23.1 & 10.2 & 9044 & 3997 & 0.528 & 0.000 & (0.466, 0.612)\\
   $\bb_{10}$ & 19.3 & 8.6 & 6925 & 3563 & 0.524 & 0.000 & (0.417, 0.582)\\
   $\bb_{15}$ & 22.2 & 9.8 & 10063 & 4977 & 0.564 & 0.000 & (0.481, 0.614)\\
   $\bb_{20}$ & 19.4 & 8.6 & 8945 & 2680 & 0.574 & 0.000 & (0.467, 0.674)\\
   $\bb_{25}$ & 21.0 & 9.3 & 6188 & 3114 & 0.579 & 0.000 & (0.481, 0.684)\\
   $\bb_{30}$ & 26.3 & 11.7 & 4698 & 4560 & 0.623 & 0.000 & (0.522, 0.711)\\
   $\bb_{35}$ & 28.1 & 12.5 & 2875 & 2950 & 0.691 & 0.000 & (0.587, 0.811)\\
   $\bb_{40}$ & 36.3 & 16.1 & 2916 & 2993 & 0.723 & 0.000 & (0.621, 0.855)\\
  \midrule
   $\mathtt{WBATE}$ &  &  &  &  & 0.592 & 0.000 & (0.567, 0.620)\\
  \midrule
   $\mathtt{LBATE}$ &  &  &  &  & 0.723 &  & (0.624, 0.860)\\
  \bottomrule\bottomrule
\end{tabular}
}
    \end{subtable}

    \medskip
    \begin{subtable}[t]{0.48\linewidth}
        \centering
        \caption{Fuzzy Effects: $\zeta(\bx)$.}
        \label{tab:empirical-results-fuzzy}
        \resizebox{\linewidth}{!}{\begin{tabular}{lccccccc}
  \toprule\toprule
   & $h_1$ & $h_2$ & $N_{\mathrm{Co}}$ & $N_{\mathrm{Tr}}$ & Estimate & $p$-value & 95\% CI \\
  \midrule
   $\bb_{1}$ & 27.6 & 12.2 & 8921 & 3532 & 0.564 & 0.000 & (0.374, 0.715)\\
   $\bb_{5}$ & 23.1 & 10.2 & 9044 & 3997 & 0.559 & 0.000 & (0.323, 0.666)\\
   $\bb_{10}$ & 19.3 & 8.6 & 6925 & 3563 & 0.619 & 0.000 & (0.401, 0.812)\\
   $\bb_{15}$ & 22.2 & 9.8 & 10063 & 4977 & 0.556 & 0.000 & (0.362, 0.675)\\
   $\bb_{20}$ & 19.4 & 8.6 & 8945 & 2680 & 0.569 & 0.000 & (0.440, 0.723)\\
   $\bb_{25}$ & 21.0 & 9.3 & 6188 & 3114 & 0.503 & 0.000 & (0.385, 0.703)\\
   $\bb_{30}$ & 26.3 & 11.7 & 4698 & 4560 & 0.412 & 0.000 & (0.272, 0.627)\\
   $\bb_{35}$ & 28.1 & 12.5 & 2875 & 2950 & 0.317 & 0.000 & (0.123, 0.521)\\
   $\bb_{40}$ & 36.3 & 16.1 & 2916 & 2993 & 0.255 & 0.000 & (0.067, 0.447)\\
  \midrule
   $\mathtt{WBATE}$ &  &  &  &  & 0.487 & 0.000 & (0.435, 0.535)\\
  \midrule
   $\mathtt{LBATE}$ &  &  &  &  & 0.619 &  & (0.487, 0.808)\\
  \bottomrule\bottomrule
\end{tabular}
}
    \end{subtable}
    \hspace{0.02\linewidth}
    \begin{subtable}[t]{0.48\linewidth}
        \centering
        \caption{Intention-to-Treat Covariate Balance.}
        \label{tab:empirical-results-placebo}
        \resizebox{\linewidth}{!}{\begin{tabular}{lccccccc}
  \toprule\toprule
   & $h_1$ & $h_2$ & $N_{\mathrm{Co}}$ & $N_{\mathrm{Tr}}$ & Estimate & $p$-value & 95\% CI \\
  \midrule
   $\bb_{1}$ & 27.4 & 12.2 & 8750 & 3469 & -0.027 & 0.170 & (-0.153, 0.058)\\
   $\bb_{5}$ & 24.2 & 10.7 & 9767 & 4156 & -0.015 & 0.398 & (-0.123, 0.069)\\
   $\bb_{10}$ & 21.0 & 9.3 & 8085 & 3981 & -0.011 & 0.272 & (-0.139, 0.065)\\
   $\bb_{15}$ & 20.3 & 9.0 & 8051 & 4329 & 0.000 & 0.909 & (-0.093, 0.100)\\
   $\bb_{20}$ & 20.2 & 8.9 & 9828 & 2825 & -0.009 & 0.613 & (-0.103, 0.073)\\
   $\bb_{25}$ & 18.7 & 8.3 & 4697 & 2598 & 0.020 & 0.058 & (-0.039, 0.168)\\
   $\bb_{30}$ & 24.5 & 10.9 & 3842 & 3945 & 0.005 & 0.556 & (-0.085, 0.126)\\
   $\bb_{35}$ & 23.1 & 10.3 & 1825 & 1842 & -0.006 & 0.639 & (-0.156, 0.114)\\
   $\bb_{40}$ & 27.8 & 12.3 & 1560 & 1511 & -0.011 & 0.516 & (-0.134, 0.087)\\
  \midrule
   $\mathtt{WBATE}$ &  &  &  &  & -0.004 & 0.938 & (-0.030, 0.028)\\
  \midrule
   $\mathtt{LBATE}$ &  &  &  &  & 0.020 &  & (-0.031, 0.182)\\
  \bottomrule\bottomrule
\end{tabular}
}
    \end{subtable}

    \medskip
    \caption{Selected Empirical Results for BATEC (SPP Application).\\
        {\footnotesize \textit{Notes}. Panel (a) reports reduced-form effects of eligibility on college enrollment. Panel (b) reports take-up. Panel (c) reports fuzzy effects on college enrollment. Panel (d) reports a covariate balance using mother's education as the outcome. Point estimates and robust bias-corrected confidence intervals are computed using the $40$ boundary grid points shown in Figure \ref{fig:fig1a}. MSE-optimal bandwidths for each score coordinate are computed at each grid point using the ITT bandwidth selector. A subset of boundary grid points are reported for exposition. See the supplemental appendix for details and omitted results.}}
    \label{tab:empirical-results}
\end{table}

\section{Extensions and Future Research}\label{sec: Extensions and Future Research}

This section discusses several generalizations of our results that are useful for applied researchers. The theoretical results in the supplemental appendix already accommodate some of these extensions, while in other cases they serve as a foundation for future theoretical and methodological research.

\subsection{Multidimensional Scores and General Assignment Boundaries}

We considered bivariate scores and piecewise linear assignment boundaries with finitely many segments, a setting that covers many empirical BD designs while keeping the assumptions transparent. The supplemental appendix gives pointwise and uniform estimation and inference results when $\bX_i$ is a $d$-dimensional score and $\B$ has effective dimension $d-1$, with $d\geq2$, allowing both diagonal bandwidth matrices and assignment boundaries that are not necessarily piecewise linear. The more general theory requires regularity conditions on the geometry of $\B$, formulated using ideas from geometric measure theory \citep{federer2014geometric}. Practically, these conditions rule out boundary regions that are too irregular or too poorly supported by observations on one side for stable local polynomial estimation. Higher-dimensional scores are possible in principle, but the usual nonparametric curse of dimensionality remains; empirical work with $d>2$ should therefore include careful bandwidth, support, and sensitivity checks.

\subsection{Derivatives and Regression Kink Designs}

The supplemental appendix also studies pointwise and uniform estimation and inference for derivatives of the BATEC, transformations of those derivatives, and their fuzzy counterparts. Derivative estimands are useful when the object of interest is not only the level of the treatment effect at the boundary, but also how that effect changes along $\B$. For example, derivatives can be used to assess monotonicity or other local patterns of treatment effect heterogeneity. The same theory covers regression kink designs \citep{Card-Lee-Pei-Weber_2015_ECMA} in BD settings, where the identifying variation comes from a change in slope rather than a change in level at the assignment boundary.

\subsection{Covariate Adjustment}

In RD designs, pre-treatment covariates can be used as placebo outcomes or balance checks \citep[Section 5]{Cattaneo-Idrobo-Titiunik_2020_CUP}, incorporated in the local polynomial fit to improve precision \citep{Calonico-Cattaneo-Farrell-Titiunik_2019_RESTAT}, or used to study treatment effect heterogeneity \citep{Calonico-Cattaneo-Farrell-Palomba-Titiunik_2026_wp}. These uses carry over naturally to BD designs. The empirical application above reports a placebo analysis using mother's education as the outcome variable, and the supplemental appendix gives a basic formulation for incorporating pre-treatment covariates into location-based estimators. To conserve space, we present the sharp design formulation in the supplemental appendix (Section SA-3.3), while extensions to the fuzzy estimands in Section \ref{sec: Imperfect Compliance} follow the same logic.

\subsection{Multidimensional RD Plots}

Visualization is an important part of univariate RD analysis \citep{Calonico-Cattaneo-Titiunik_2015_JASA,Korting-Lieberman-Matsudaira-Pei-Shen_2023_QJE}, but a single univariate RD plot is generally not enough in BD designs because treatment effects may vary along $\B$. A useful starting point is to display the assignment scores, boundary, and estimation grid, as in Figure \ref{fig:fig1a}, followed by plots of the estimated BATEC and associated uncertainty measures, as in Figure \ref{fig:fig2}. More formal bivariate RD plots can be constructed by extending binscatter methods: univariate RD plots are a particular instance of binscatter methods \citep{Cattaneo-Crump-Farrell-Feng_2024_AER,Cattaneo-Crump-Farrell-Feng_2026_RESTAT}, and analogous BD plots can be based on bins in the assignment space or on binned summaries local to the boundary. Such plots can help assess outcome discontinuities, visualize heterogeneity, and compare the location-based analysis with simpler binned or distance-to-boundary summaries. We plan to develop these visualization tools in future work.

\subsection{Clustered and Spatially Correlated Data}

Many BD applications involve observations that are clustered or spatially correlated, especially when the score contains geographic coordinates. The formal results in this paper are developed under random sampling, and therefore they should be applied directly only when that approximation is credible. When dependence is present and its structure is known or can be consistently approximated, the location-based estimators can in principle be paired with cluster-robust or spatially robust covariance estimators \citep{mackinnon2023cluster}. Pointwise results for nonparametric local-constant regression under spatial dependence have been recently developed \citep{Shimizu_2025_JoE}. Extending those ideas to general local polynomial BD estimators, especially for uniform inference over $\B$, requires additional empirical process and covariance-estimation results. We therefore view dependence-robust pointwise and uniform inference for BD designs as an important direction for future work.

\onehalfspacing
\bibliography{CTY_2026_JASA--bib}

@article{Andrews-Kitagawa-McCloskey_2024_QJE,
  title={Inference on winners},
  author={Andrews, Isaiah and Kitagawa, Toru and McCloskey, Adam},
  journal={Quarterly Journal of Economics},
  volume={139}, number={1}, pages={305--358}, year={2024}
}

@article{Arai-etal_2022_QE,
  title={Testing Identifying Assumptions in Fuzzy Regression Discontinuity Designs},
  author={Arai, Yoichi and Hsu, Yu-Chin and Kitagawa, Toru and Mourifi{\'e}, Ismael and Wan, Yuanyuan},
  journal={Quantitative Economics}, volume={13}, number={1}, pages={1--28}, year={2022}
}

@article{Calonico-Cattaneo-Titiunik_2014_ECMA,
	Author = {Calonico, Sebastian and Matias D. Cattaneo and Rocio Titiunik},
	Title = {Robust Nonparametric Confidence Intervals for Regression-Discontinuity Designs},
	Journal = {Econometrica}, Number = {6}, Volume = {82}, Pages = {2295--2326}, Year = {2014}
}

@article{Calonico-Cattaneo-Titiunik_2015_JASA,
  author  = {Calonico, Sebastian and Matias D. Cattaneo and Rocio Titiunik},
  title   = {Optimal Data-Driven Regression Discontinuity Plots},
  journal = {Journal of the American Statistical Association}, volume={110}, number={512}, pages={1753--1769}, year={2015}
}

@Article{Calonico-Cattaneo-Farrell_2018_JASA,
  Title                    = {On the Effect of Bias Estimation on Coverage Accuracy in Nonparametric Inference},
  Author                   = {Calonico, Sebastian and Matias D. Cattaneo and Max H. Farrell},
  Journal                  = {Journal of the American Statistical Association},
  Year                     = {2018},
  Number                   = {522},
  Pages                    = {767--779},
  Volume                   = {113}
}

@article{Calonico-Cattaneo-Farrell-Titiunik_2019_RESTAT,
  author  = {Calonico, Sebastian and Matias D. Cattaneo and Max H. Farrell and Rocio Titiunik},
  title   = {Regression Discontinuity Designs using Covariates},
  journal = {Review of Economics and Statistics}, volume={101}, number={3}, pages={442--451}, year={2019}
}

@article{Calonico-Cattaneo-Farrell_2020_ECTJ,
	author  = {Calonico, Sebastian and Matias D. Cattaneo and Max H. Farrell},
	title   = {Optimal Bandwidth Choice for Robust Bias Corrected Inference in Regression Discontinuity Designs},
	journal = {Econometrics Journal}, volume={23}, number={2}, pages={192--210}, year={2020}
}

@article{Calonico-Cattaneo-Farrell_2022_Bernoulli,
	author  = {Calonico, Sebastian and Matias D. Cattaneo and Max H. Farrell},
	title   = {Coverage Error Optimal Confidence Intervals for Local Polynomial Regression},
	journal = {Bernoulli}, volume={28}, number={4}, pages={2998--3022}, year={2022}
}

@article{Calonico-Cattaneo-Farrell-Palomba-Titiunik_2026_wp,
  author  = {Calonico, Sebastian and Matias D. Cattaneo and Max H. Farrell and Palomba, Filippo and Rocio Titiunik},
  title   = {Treatment Effect Heterogeneity in Regression Discontinuity Designs},
  journal = {arXiv preprint arXiv:2503.13696}, year={2026}
}

@article{Card-Lee-Pei-Weber_2015_ECMA,
	author  = {Card, David and David S. Lee and Zhuan Pei and Andrea Weber},
	title   = {Inference on Causal Effects in a Generalized Regression Kink Design},
	journal = {Econometrica}, volume={83}, number={6}, pages={2453--2483}, year={2015}
}

@Article{Cattaneo-Titiunik_2022_ARE,
  Title                    = {Regression Discontinuity Designs},
  Author                   = {Cattaneo, Matias D. and Rocio Titiunik},
  Journal                  = {Annual Review of Economics},
  Year                     = {2022},
  Pages                    = {821--851},
  Volume                   = {14}
}

@article{Cattaneo-Yu_2025_AOS,
    title={Strong Approximations for Empirical Processes Indexed by Lipschitz Functions},
    author={Cattaneo, Matias D. and Yu, Ruiqi (Rae)},
    journal={Annals of Statistics}, volume={53}, number={3}, pages={1203--1229}, year={2025}
}

@book{Cattaneo-Idrobo-Titiunik_2020_CUP,
	author    = {Cattaneo, Matias D. and Nicol\'{a}s Idrobo and Rocio Titiunik},
	title     = {A Practical Introduction to Regression Discontinuity Designs: Foundations},
	publisher = {Cambridge University Press}, year={2020}
}

@article{Cattaneo-Chandak-Jansson-Ma_2024_Bernoulli,
    title={Boundary Adaptive Local Polynomial Conditional Density Estimators},
    author={Cattaneo, Matias D. and Chandak, Rajita and Jansson, Michael and Ma, Xinwei},
    journal={Bernoulli}, volume={30}, number={4}, pages={3193--3223}, year={2024}
}

@article{Cattaneo-Crump-Farrell-Feng_2024_AER,
	author  = {Cattaneo, Matias D. and Richard K. Crump and Max H. Farrell and Yingjie Feng},
	title   = {On Binscatter},
	journal = {American Economic Review}, volume={114}, number={5}, pages={1488--1514}, year={2024}
}

@article{Cattaneo-Crump-Farrell-Feng_2026_RESTAT,
	author  = {Cattaneo, Matias D. and Richard K. Crump and Max H. Farrell and Yingjie Feng},
	title   = {Nonlinear Binscatter Methods},
	journal = {arXiv:2407.15276}, year={2026}
}

@article{Cattaneo-Feng-Underwood_2024_JASA,
	author  = {Cattaneo, Matias D. and Yingjie Feng and William G. Underwood},
	title   = {Uniform Inference for Kernel Density Estimators with Dyadic Data},
	journal = {Journal of the American Statistical Association}, volume={119}, number={524}, pages={2695--2708}, year={2024}
}

@book{Cattaneo-Idrobo-Titiunik_2024_CUP,
	author    = {Cattaneo, Matias D. and Nicol\'{a}s Idrobo and Rocio Titiunik},
	title     = {A Practical Introduction to Regression Discontinuity Designs: Extensions},
	publisher = {Cambridge University Press}, year={2024}
}

@article{Cattaneo-Keele-Titiunik-VazquezBare_2016_JOP,
  author  = {Cattaneo, Matias D. and Luke Keele and Rocio Titiunik and Gonzalo Vazquez-Bare},
  title   = {Interpreting Regression Discontinuity Designs with Multiple Cutoffs},
  journal = {Journal of Politics}, volume={78}, number={4}, pages={1229--1248}, year={2016}
}

@article{Cattaneo-Keele-Titiunik-VazquezBare_2021_JASA,
  author  = {Cattaneo, Matias D. and Luke Keele and Rocio Titiunik and Gonzalo Vazquez-Bare},
  title   = {Extrapolating Treatment Effects in Multi-Cutoff Regression Discontinuity Designs},
  journal = {Journal of the American Statistical Association}, volume={116}, number={536}, pages={1941--1952}, year={2021}
}

@article{Cattaneo-Titiunik-Yu_2026_BDD-Pooling,
	author  = {Cattaneo, Matias D. and Titiunik, Rocio and Yu, Ruiqi (Rae)},
	title   = {Estimation and Inference in Boundary Discontinuity Designs: Pooling-Based Methods},
	journal = {Working paper}, year={2026}
}

@article{Cattaneo-Titiunik-Yu_2026_JOE,
	author  = {Cattaneo, Matias D. and Titiunik, Rocio and Yu, Ruiqi (Rae)},
	title   = {Estimation and Inference in Boundary Discontinuity Designs: Distance-Based Methods},
	journal = {Working paper}, year={2026}
}

@article{Cattaneo-Titiunik-Yu_2025_rd2d,
	author  = {Cattaneo, Matias D. and Titiunik, Rocio and Yu, Ruiqi (Rae)},
	title   = {\texttt{rd2d}: Causal Inference in Boundary Discontinuity Designs},
	journal = {Working paper}, year={2025}
}

@incollection{Cattaneo-Titiunik-Yu_2026_BookCh,
	author    = {Cattaneo, Matias D. and Titiunik, Rocio and Yu, Ruiqi (Rae)},
	title     = {Boundary Discontinuity Designs: Theory and Practice},
	booktitle = {Invited book chapter for the 2025 Econometric Society World Congress},
    volume    = {1}, chapter = {2}, publisher={Cambridge University Press}, year={2026}
}

@article{Chen-Gao_2025_arXiv,
  title={Thin Sets Are Not Equally Thin: Minimax Learning of Submanifold Integrals},
  author={Chen, Xiaohong and Gao, Wayne Yuan},
  journal={arXiv preprint arXiv:2507.12673},
  year={2026}
}

@article{Chernozhukov-Chetverikov-Kato_2014a_AoS,
    title={Anti-Concentration and Honest, Adaptive Confidence Bands},
    author={Chernozhukov, Victor and Chetverikov, Denis and Kato, Kengo},
    journal={Annals of Statistics}, volume={42}, number={5}, pages={1787--1818}, year={2014}
}

@article{Chernozhukov-Chetverikov-Kato_2014b_AoS,
    title={Gaussian Approximation of Suprema of Empirical Processes},
    author={Chernozhukov, Victor and Chetverikov, Denis and Kato, Kengo},
    journal={Annals of Statistics}, volume={42}, number={4}, pages={1564--1597}, year={2014}
}

@article{Chernozhukov-Chetverikov-Kato-Koike_2022_AoS,
  title={Improved central limit theorem and bootstrap approximations in high dimensions},
  author={Chernozhukov, Victor and Chetverikov, Denis and Kato, Kengo and Koike, Yuta},
  journal={Annals of Statistics},
  volume={50},
  number={5},
  pages={2562--2586},
  year={2022}
}

@article{choi2023complier,
  title={Complier and monotonicity for Fuzzy Multi-score Regression Discontinuity with partial effects},
  author={Choi, Jin-young and Lee, Myoung-jae},
  journal={Economics Letters},
  volume={228},
  pages={111169},
  year={2023},
  publisher={Elsevier}
}

@article{DeMagalhaes-etal_2025_PA,
  title={When Can We Trust Regression Discontinuity Design Estimates from Close Elections? Evidence from Experimental Benchmarks},
  author={De Magalh{\~a}es, Leandro and Hangartner, Dominik and Hirvonen, Salomo and Meril{\"a}inen, Jaakko and Ruiz, Nelson A and Tukiainen, Janne},
  journal={Political Analysis},
  year={2025}
}

@article{Diaz-Zubizarreta_2023_AOAS,
  title={Complex Discontinuity Designs Using Covariates for Policy Impact Evaluation},
  author={Diaz, Juan D and Zubizarreta, Jose R},
  journal={Annals of Applied Statistics}, volume={17}, number={1}, pages={67--88}, year={2023}
}

@book{dudley2014uniform,
  title={Uniform central limit theorems},
  author={Dudley, Richard M},
  volume={142},
  year={2014},
  publisher={Cambridge university press}
}

@book{federer2014geometric,
  title={Geometric measure theory},
  author={Federer, Herbert},
  year={2014},
  publisher={Springer}
}

@book{folland2002advanced,
  title={Advanced Calculus},
  author={Folland, G.B.},
  series={Featured Titles for Advanced Calculus Series},
  year={2002},
  publisher={Prentice Hall}
}

@InCollection{Galiani-McEwan-Quistorff_2017_AIE,
  Title = {External and Internal Validity of a Geographic Quasi-Experiment Embedded in a Cluster-Randomized Experiment},
  Author = {Sebastian Galiani and Patrick J. McEwan and Brian Quistorff},
  Booktitle = {Regression Discontinuity Designs: Theory and Applications (Advances in Econometrics, volume 38)},
  Publisher = {Emerald Group Publishing},
  Year = {2017},
  Editor = {Cattaneo, Matias D. and Juan Carlos Escanciano},
  Pages = {195--236}
}

@book{Gine-Nickl_2016_Book,
title	={Mathematical Foundations of Infinite-dimensional Statistical Models},
author	={Gin{\'e}, Evarist and Nickl, Richard},
year	={2016},
publisher={Cambridge University Press},
address	={New York}
}

@article{Hahn-Todd-vanderKlaauw_2001_ECMA,
	author  = {Hahn, Jinyong and Petra Todd and Wilbert van der Klaauw},
	title   = {Identification and Estimation of Treatment Effects with a Regression-Discontinuity Design},
	journal = {Econometrica}, volume={69}, number={1}, pages={201--209}, year={2001}
}

@book{Hardle-etal_2004_Book,
  title     = {Nonparametric and Semiparametric Models},
  author={H{\"a}rdle, Wolfgang and M{\"u}ller, Marlene and Sperlich, Stefan and Werwatz, Axel},
  year={2004},
  publisher = {Springer},
  address   = {Heidelberg}
}

@book{Hernan-Robins_2020_Book,
  title={Causal Inference: What If},
  author={Hernán, Miguel A. and Robins, James M.},
  year={2020},
  publisher={Boca Raton: Chapman \& Hall/CRC}
}

@article{Hyytinen-Tukiainen-etal2018_QE,
  title={When does regression discontinuity design work? Evidence from random election outcomes},
  author={Hyytinen, Ari and Meril{\"a}inen, Jaakko and Saarimaa, Tuukka and Toivanen, Otto and Tukiainen, Janne},
  journal={Quantitative Economics},
  volume={9},
  number={2},
  pages={1019--1051},
  year={2018}
}

@article{jiang2026extrapolating,
  title={Extrapolating Treatment Effects in Multiple-Score Regression Discontinuity Designs},
  author={Jiang, Weiwei and Zhu, Rong JB},
  journal={Journal of Educational and Behavioral Statistics},
  pages={10769986261426399},
  year={2026},
  publisher={SAGE Publications Sage CA: Los Angeles, CA}
}

@article{Keele-Titiunik_2016_PSRM,
  title={Natural experiments based on geography},
  author={Keele, Luke and Titiunik, Rocio},
  journal={Political Science Research and Methods},
  volume={4},
  number={1},
  pages={65--95},
  year={2016},
  publisher={Cambridge University Press}
}

@article{Keele-Titiunik_2015_PA,
	author  = {Keele, Luke J. and Rocio Titiunik},
	title   = {Geographic Boundaries as Regression Discontinuities},
	journal = {Political Analysis}, volume={23}, number={1}, pages={127--155}, year={2015}
}

@Article{Keele-Titiunik-Zubizarreta_2015_JRSSA,
  Title                    = {Enhancing a Geographic Regression Discontinuity Design Through Matching to Estimate the Effect of Ballot Initiatives on Voter Turnout},
  Author                   = {Keele, Luke J. and Rocio Titiunik and Jose Zubizarreta},
  Journal                  = {Journal of the Royal Statistical Society: Series A},
  Year                     = {2015},
  Number                   = {1},
  Pages                    = {223--239},
  Volume                   = {178}
}

@InCollection{Keele-etal_2017_AIE,
  Title = {An Overview of Geographically Discontinuous Treatment Assignments with an Application to Children’s Health Insurance},
  Author = {Keele, Luke J. and Scott Lorch and Molly Passarella and Dylan Small and Rocio Titiunik},
  Booktitle = {Regression Discontinuity Designs: Theory and Applications (Advances in Econometrics, volume 38)},
  Publisher = {Emerald Group Publishing},
  Year = {2017},
  Editor = {Cattaneo, Matias D. and Juan Carlos Escanciano},
  Pages = {147--194}
}

@article{Korting-Lieberman-Matsudaira-Pei-Shen_2023_QJE,
  title={Visual Inference and Graphical Representation in Regression Discontinuity Designs},
  author={Korting, Christina and Lieberman, Carl and Matsudaira, Jordan and Pei, Zhuan and Shen, Yi},
  journal={Quarterly Journal of Economics},
  volume={138}, number={3}, pages={1977--2019}, year={2023}
}

@article{LondonoVelezRodriguezSanchez_2020_AEJ,
	title={Upstream and downstream impacts of college merit-based financial aid for low-income students: Ser Pilo Paga in Colombia},
	author={Londo{\~n}o-V{\'e}lez, Juliana and Rodr{\'\i}guez, Catherine and S{\'a}nchez, Fabio},
	journal={American Economic Journal: Economic Policy},
	volume={12},
	number={2},
	pages={193--227},
	year={2020}
}

@article{mackinnon2023cluster,
  title={Cluster-robust inference: A guide to empirical practice},
  author={MacKinnon, James G and Nielsen, Morten {\O}rregaard and Webb, Matthew D},
  journal={Journal of Econometrics},
  volume={232},
  number={2},
  pages={272--299},
  year={2023}
}

@Article{Papay-Willett-Murnane_2011_JoE,
  Title   = {Extending the regression-discontinuity approach to multiple assignment variables},
  Author  = {Papay, John P and Willett, John B and Murnane, Richard J},
  Journal = {Journal of Econometrics}, Volume = {161}, Number = {2}, Pages = {203--207}, Year = {2011}
}

@Article{Reardon-Robinson_2012_JREE,
  Title                    = {Regression Discontinuity Designs with Multiple Rating-Score Variables},
  Author                   = {Reardon, Sean F and Robinson, Joseph P},
  Journal                  = {Journal of Research on Educational Effectiveness},
  Year                     = {2012},
  Number                   = {1},
  Pages                    = {83--104},
  Volume                   = {5}
}

@article{Rischard-Branson-Miratrix-Bornn_2021_JASA,
  title={Do School Districts Affect NYC House Prices? Identifying Border Differences using a Bayesian Nonparametric Approach to Geographic Regression Discontinuity Designs},
  author={Rischard, Maxime and Branson, Zach and Miratrix, Luke and Bornn, Luke},
  journal={Journal of the American Statistical Association},
  volume={116},
  number={534},
  pages={619--631},
  year={2021}
}

@article{Shimizu_2025_JoE,
  title = {Nonparametric Regression under Cluster Sampling},
  author = {Shimizu, Yuya},
  year = 2025,
  journal = {Journal of Econometrics},
  volume = {252},
  pages = {106102}
}

@book{simon1984lectures,
  title={Lectures on geometric measure theory},
  author={Simon, Leon and others},
  year={1984},
  publisher={Centre for Mathematical Analysis, Australian National University Canberra}
}

@article{schwarz2025effect,
  title={Effect Identification and Unit Categorization in the Multi-Score Regression Discontinuity Design with Application to LED Manufacturing},
  author={Schwarz, Philipp Alexander and Schacht, Oliver and Klaassen, Sven and Oberpriller, Johannes and Spindler, Martin},
  journal={arXiv preprint arXiv:2508.15692},
  year={2025}
}

@book{tsybakov2008introduction,
  title={Introduction to Nonparametric Estimation},
  author={Tsybakov, A.B.},
  year={2008},
  publisher={Springer}
}

@book{van-der-Vaart-Wellner_1996_Book,
title	={Weak Convergence and Empirical Processes},
author	={{van der Vaart}, Aad W. and Wellner, Jon A.},
publisher={Springer}, 
year	={1996},
doi={10.1007/978-1-4757-2545-2}
}

@article{Wong-Steiner-Cook_2013_JEBS,
  author={Wong, Vivian C and Steiner, Peter M and Cook, Thomas D},
  title={Analyzing regression-discontinuity designs with multiple assignment variables: A comparative study of four estimation methods},
  journal={Journal of Educational and Behavioral Statistics}, volume={38}, number={2}, pages={107--141}, year={2013}
}
\bibliographystyle{plainnat}


\end{document}


\title{Estimation and Inference in Boundary Discontinuity Designs: Location-Based Methods\\ Supplemental Appendix\bigskip}
\author{Matias D. Cattaneo\thanks{Department of Operations Research and Financial Engineering, Princeton University.} \and
	    Rocio Titiunik\thanks{Department of Politics, Princeton University.} \and
	    Ruiqi (Rae) Yu\thanks{Department of Operations Research and Financial Engineering, Princeton University.}
	    }
\maketitle

\begin{abstract}
    This supplemental appendix presents more general theoretical results encompassing those reported in the paper, their theoretical proofs, and other technical results. It also provides a discussion of implementation details, and additional numerical results.
\end{abstract}

\textit{Keywords}: regression discontinuity, treatment effects estimation, causal inference.

\thispagestyle{empty}
\clearpage

\onehalfspacing
\setcounter{page}{1}
\pagestyle{plain}

\setcounter{tocdepth}{2}
\setcounter{secnumdepth}{4}
\tableofcontents

\clearpage

\section{Introduction}

This supplemental appendix considers four generalizations of the setup in the paper.
\begin{enumerate}
    \item The score $\bX_i$ is $d$-dimensional with $d\geq2$ and support $\mathcal{X}\subseteq\mathbb{R}^d$.
    \item The boundary region $\B$ is a substantively more general low-dimensional manifold with ``effective dimension'' $d-1$, and hence not necessarily piecewise linear with finitely many line segments.
    \item Derivative estimation and inference for all parameters of interest.
    \item Different bandwidths $\bh = (h_1, \ldots, h_d)^{\top}$ are allowed for each dimension of the score variable.
\end{enumerate}
In addition, this supplement presents several other technical and methodological results not covered in the main text due to space limitations.

The paper presents results for the special case $d=2$, $\B$ piecewise linear with finitely many line segments, $h=h_1=h_2$, and $\bnu=\mathbf{0}$. The assumptions in the main paper are more specialized primitive conditions; after imposing these restrictions on the supplemental results, the mapping between the paper and this supplemental appendix is as follows.
\begin{itemize}
    \item Theorem \ref{sa-thm: Convergence Rates: Sharp BATEC} $\implies$ Theorem 1.
    \item Theorem \ref{sa-thm: MSE Expansions: Sharp BATEC} $\implies$ Theorem 2.
    \item Theorems \ref{sa-thm: Confidence Intervals: Sharp BATEC} and \ref{sa-thm: Confidence Bands: Sharp BATEC -- Suprema} $\implies$ Theorem 3.
    \item Theorem \ref{sa-thm: MSE Expansion: Sharp WBATE} $\implies$ Theorem 4.
    \item Theorem \ref{sa-thm: Asymptotic Normality: Sharp WBATE} $\implies$ Theorem 5.
    \item Theorem \ref{sa-thm: Confidence Intervals: Sharp LBATE} $\implies$ Theorem 6.
    \item Theorem \ref{sa-thm: Convergence Rates: Fuzzy BATEC} $\implies$ Theorem 7.
    \item Theorem \ref{sa-thm: MSE Expansions: Fuzzy BATEC} $\implies$ Theorem 8.
    \item Theorems \ref{sa-thm: Confidence Intervals: Fuzzy BATEC} and \ref{sa-thm: Confidence Bands: Fuzzy BATEC -- suprema} $\implies$ Theorem 9.
    \item Theorem \ref{sa-thm: MSE Expansion: Fuzzy WBATE} $\implies$ Theorem 10.
    \item Theorem \ref{sa-thm: Asymptotic Normality: Fuzzy WBATE} $\implies$ Theorem 11.
    \item Theorem \ref{sa-thm: Confidence Intervals: Fuzzy LBATE} $\implies$ Theorem 12.
\end{itemize}

The generalizations considered in this supplemental appendix are conceptually straightforward, but technically more involved. Following the structure of the paper, we first consider sharp designs. Section \ref{sa-sec: Imperfect Compliance} generalizes the results to fuzzy designs. Section \ref{sa-sec: Gaussian Strong Approximation} presents a novel Gaussian strong approximation result that may be of independent interest.

\subsection{Notation and Definitions}

For textbook references on empirical process, see \cite{van-der-Vaart-Wellner_1996_Book}, \cite{dudley2014uniform}, and \cite{Gine-Nickl_2016_Book}. For textbook references on geometric measure theory, see \cite{simon1984lectures}, \cite{federer2014geometric}, and \cite{folland2002advanced}.

\begin{enumerate}[label=(\roman*)]
    \item \textit{Multi-index Notations}. For a multi-index $\bnu = (\nu_1, \ldots, \nu_d) \in \mathbb{N}_0^d$, let $|\bnu| = \sum_{l = 1}^d \nu_l$, $\bnu! = \nu_1!\cdots \nu_d!$, and $\bu^{\bnu} = \prod_{l=1}^d u_l^{\nu_l}$ for $\bu=(u_1,\ldots,u_d)\in\mathbb{R}^d$. For any integer $p\geq0$, let $\br_p(\bu) = (\bu^{\boldsymbol{\alpha}}: |\boldsymbol{\alpha}|\leq p)^\top \in \mathbb{R}^{\mathfrak{p}_p}$ collect all monomials of total degree at most $p$ in a fixed order, where $\mathfrak{p}_p = \frac{(d + p)!}{d! p !}$. Define $\be_{\bnu}$ to be the $\mathfrak{p}_p$-dimensional unit vector such that $\be_{\bnu}^{\top} \br_p(\bu) = \bu^{\bnu}$ for all $\bu \in \reals^d$ and $|\bnu|\leq p$. In addition, let $f^{(\bnu)}(\bx) = \partial^{\nu_1}\cdots\partial^{\nu_d} f(\bx)$.

    \item \textit{Norms}. For a vector $\ba \in \reals^k$, $\norm{\ba} = (\sum_{i = 1}^k \ba_i^2)^{1/2}$, $\infnorm{\ba} = \max_{1 \leq i \leq k}|\ba_i|$. For a matrix $\bA \in \reals^{m \times n}$, $\norm{\bA}_p = \sup_{\norm{\bx}_p = 1} \norm{\bA\bx}_p$, $p \in \mathbb{N} \cup \{\infty\}$, $\lambda_{\min}(\bA)$ denotes its minimum eigenvalue, and $|\bA|=|\det(\bA)|$ denotes the absolute value of its determinant. For a function $f$ on a metric space $(S, d)$, $\infnorm{f} = \sup_{s \in S} |f(s)|$. For a probability measure $Q$ on $(\mathcal{S}, \mathscr{S})$ and $p \geq 1$, define $\norm{f}_{Q,p} = (\int_{\mathcal{S}} |f|^p \dif Q)^{1/p}$, and $Q(f) = \int f \dif Q$. For a set $E \subseteq \reals^d$, denote by $\mathfrak{m}(E)$ the Lebesgue measure of $E$.

    \item \textit{Empirical Process}. We use standard empirical process notations: $\En[g(\bv_i)] = \frac{1}{n} \sum_{i = 1}^n g(\bv_i)$ and $\gn[g(\bv_i)] = \frac{1}{\sqrt{n}} \sum_{i = 1}^n (g(\bv_i) - \E[g(\bv_i)])$. Let $(\mathcal{S},d)$ be a semi-metric space. The covering number $N(\mathcal{S}, d, \varepsilon)$ is the minimal number of balls $B_s(\varepsilon) =\{t: d(t,s) < \varepsilon\}$ needed to cover $\mathcal{S}$. A \emph{$\P$-Brownian bridge} is a mean-zero Gaussian random function $\mathbb{G}_{\P}(f), f \in L_2(\X, \P)$ with covariance $\E[\mathbb{G}_{\P}(f)\mathbb{G}_{\P}(g)] = \P(fg) - \P(f)\P(g)$, for $f,g \in L_2(\X,\P)$. A class $\mathcal{F} \subseteq L_2(\X, \P)$ is \emph{$\P$-pregaussian} if there is a version of $\P$-Brownian bridge $\mathbb{G}_{\P}$ such that $\mathbb{G}_{\P} \in C(\mathcal{F}; \rho_{\P})$ almost surely, where $\rho_{\P}$ is the semi-metric on $L_2(\X,\P)$ is defined by $\rho_{\P}(f, g) = (\|f - g\|_{\P,2}^2 - (\int f \dif\P - \int g \dif\P)^2)^{1/2}$, for $f, g \in L_2(\X,\P)$.

    \item \textit{Geometric Measure Theory}. For a set $A \subseteq \X$, the \emph{De Giorgi perimeter of $A$ related to $\X$} is $\mathcal{L}(A) = \ttTV_{\{\Indicator_{A}\},\X}$. For $d \in \mathbb{N}$ and $0 \leq m \leq d$, the $m$-dimensional Hausdorff (outer) measure is $\Haus^m(A) = \lim_{\delta \downarrow 0}\Haus^m_{\delta}(A)$, $A \subseteq \reals^d$, where for each $\delta > 0$, $\Haus^m_{\delta}(A)$ is defined by taking $\Haus^m_{\delta}(\emptyset) = 0$, and for any non-empty $A \subseteq \reals^d$, $\Haus^m_{\delta}(A) = \frac{\pi^{m/2}}{\Gamma(m/2+1)} \inf \sum_{j = 1}^{\infty} (\diam(C_j)/2)^m$, and the infimum is taken over all countable collections $C_1, C_2, \cdots$ of subsets of $\reals^d$ such that $\diam(C_j) < \delta$ and $A \subseteq \cup_{j = 1}^{\infty}C_j$. Integration against $\Haus^m$ is defined via Carathéodory's Theorem following the classical measure-theoretic literature. The Hausdorff dimension $\dim_{\Haus}(A)$ of $A$ is defined by $\dim_{\Haus}(A) = \inf\{t \geq 0: \Haus^t(A) = 0\}$. A set $A \subseteq \reals^d$ is said to be $k$-rectifiable if $A$ is of Hausdorff dimension $k$, and there exist a countable collection $\{f_i\}$ of continuously differentiable maps $f_i: \reals^k \to \reals^d$ such that $\Haus^{k}(A \setminus \cup_{i = 0}^{\infty} f_i(\reals^k)) = 0$. $B$ is a \emph{rectifiable curve} if there exists a Lipschitz continuous function $\gamma:[0,1] \to \reals^d$ such that $B=\gamma([0,1])$. We define the curve length function of $B$ to be $\mathfrak{L}({B}) = \sup_{\pi \in \Pi} s(\pi, \gamma)$, where $\Pi = \left\{(t_0, t_1, \ldots, t_N): N \in \mathbb{N}, 0 \leq t_0 < t_1 < \ldots \leq t_N \leq 1\right\}$ and $s(\pi,\gamma) = \sum_{i = 0}^{N-1}\norm{\gamma(t_{i}) - \gamma(t_{i+1})}_2$ for $\pi = (t_0, t_1, \ldots, t_N)$.

    \item \textit{Bounds and Asymptotics}. For real sequences, we write $a_n = o(b_n)$ if $\limsup \frac{|a_n|}{|b_n|} = 0$, and $a_n \lesssim b_n$ if there exist constants $C$ and $N > 0$ such that $n > N$ implies $|a_n| \leq C |b_n|$. For sequences of random variables, we write $a_n = o_{\P}(b_n)$ if $\operatorname{plim}_{n \to \infty}\frac{|a_n|}{|b_n|} = 0$, and $|a_n| \lessp |b_n|$ if $\limsup_{M \to \infty} \limsup_{n \to \infty} \P[|\frac{a_n}{b_n}| \geq M] = 0$.

\end{enumerate}

Let $\Phi(\cdot)$ be the cumulative distribution function of a standard univariate Gaussian random variable.
For a bandwidth vector $\bh=(h_1,\ldots,h_d)^\top\in\mathbb{R}_{++}^d$, let $\bH=\diag(\bh)$, $\underline h=\min_{1\leq l\leq d}h_l$, $\bar{h}=\max_{1\leq l\leq d}h_l$, and $K_{\bH}(\bu)=|\bH|^{-1}K(\bH^{-1}\bu)$. All limits are taken such that $\bar{h}\to0$, $\bar{h}/\underline h\lesssim1$, and the bandwidths vanish as $n\to\infty$. Most of our results hold with fixed but small enough bandwidths. We do not make this distinction explicit to avoid overly-complex statements and technical discussions.

\section{Sharp Design Setup}

We partition $\mathcal{X}$ into two areas, $\A_t\subset\mathbb{R}^d$ with $t\in\{0,1\}$, which represent the control and treatment regions, respectively. That is, $\mathcal{X} = \A_0 \cup \A_1$, where $\A_0$ and $\A_1$ are two disjoint regions in $\mathbb{R}^d$. The observed outcome is $Y_i = \Indicator(\bX_i \in \A_0) Y_i(0) + \Indicator(\bX_i \in \A_1) Y_i(1)$. The ``boundary'' is $\B = \setbd(\A_0) \cap \setbd(\A_1) \subseteq\mathbb{R}^{d}$, where $\setbd(\cdot)$ denotes the topological boundary of a set, and has effective dimension $d-1$. We assume that $\B$ belongs to $\A_1$, that is, $\B \subseteq \A_1$ and $\B \cap \A_0 = \emptyset$.

\subsection{Assumptions}

Assumption 1 generalizes as follows.

\begin{assumption}[Data Generating Process]\label{sa-assump: Sharp DGP}
    Let $t\in\{0,1\}$.
    \begin{enumerate}[label=\normalfont(\roman*),noitemsep,leftmargin=*]

    \item $(Y_1(t), \bX_1^\top)^\top,\ldots, (Y_n(t), \bX_n^\top)^\top$ are independent and identically distributed random vectors.

    \item The distribution of $\bX_i$ has compact support $\X\subseteq\mathbb{R}^d$ with nonempty interior and a Lebesgue density $f_X(\bx)$ that is continuous and bounded away from zero on $\X$.

    \item $\mu_t(\bx) = \E[Y_i(t)| \bX_i = \bx]$ has a $(p+1)$-times continuously differentiable extension to an open neighborhood of $\X$.

    \item $\sigma^2_t(\bx) = \Var[Y_i(t)|\bX_i = \bx]$ is bounded away from zero and continuous on $\X$.

    \item $\sup_{\bx \in \X}\E[|Y_i(t)|^{2+v}|\bX_i = \bx] < \infty$ for some $v \geq 2$.
    \end{enumerate}
\end{assumption}

Assumption 2 generalizes as follows.

\begin{assumption}[Boundary and Kernel]\label{sa-assump: Boundary and Kernel} $ $
    \begin{enumerate}[label=\normalfont(\roman*)]
    \item $\B\subset\setint(\X)$, and $\B$ is compact $(d-1)$-rectifiable, with $\Haus^{d-1}(\B)$ positive and finite.
    \item $K: \reals^d \to [0,\infty)$ is compactly supported and Lipschitz continuous, or $K(\bu) = \Indicator(\bu \in [-1,1]^d)$.
    \item For each $t \in \{0,1\}$, there exists a Lebesgue-measurable set $U \subseteq \reals^d$ with $0<\leb(U)<\infty$, such that $K(\bu) \geq \kappa > 0$ for all $\bu \in U$, $\lambda_{\min} (\int_U \br_p(\bz) \br_p(\bz)^{\top} \dif \bz) > 0$, and
    \begin{align*}
        \liminf_{\bar{h} \downarrow 0}\inf_{\bx \in \B} \int_{U} K(\bu) \Indicator(\bx + \bH \bu \in \A_t) \dif \bu \gtrsim 1,
    \end{align*}
    uniformly over diagonal bandwidth matrices satisfying $\bar{h}/\underline h\lesssim1$.
    \end{enumerate}
\end{assumption}

For $d = 2$, Assumption~\ref{sa-assump: Boundary and Kernel}(i) holds whenever $\B$ is a compact rectifiable curve with positive and finite length. For the special case in the main paper, where $\B$ is piecewise linear with finitely many line segments and $K$ is positive in a neighborhood of the origin, Assumption~\ref{sa-assump: Boundary and Kernel}(iii) also holds: locally, each side of the boundary contains a half-plane or polygonal cone of positive aperture, and the finite number of line segments gives a uniform lower bound over $\B$. More generally, Assumption~\ref{sa-assump: Boundary and Kernel}(iii) is a local nondegeneracy condition linking the geometry of $\B$ to the support and shape of the kernel $K$. The density $f_X(\bx)$ does not appear explicitly in this condition because Assumption~\ref{sa-assump: Sharp DGP}(ii) already requires it to be bounded away from zero on $\X$.

More precisely, Assumption~\ref{sa-assump: Boundary and Kernel}(iii) rules out boundary-kernel configurations in which, after local rescaling by $\bH$, one side of the boundary receives too little kernel-weighted mass to guarantee the local polynomial regression is well-defined uniformly over $\B$. Thus, the condition ensures that the local design is sufficiently rich on each side of the boundary. Lemma~\ref{sa-lem: Invertibility} shows that Assumptions~\ref{sa-assump: Sharp DGP}(ii) and \ref{sa-assump: Boundary and Kernel} imply that the population Gram matrix is full rank for $\bar h$ small enough, uniformly in $\bx\in\B$. This guarantees that the treatment effect estimators are well-defined in large samples under the conditions imposed. Assumption~\ref{sa-assump: Boundary and Kernel}(iii) is stated uniformly over $\B$ in anticipation of our uniform estimation and inference results, but it can be relaxed to hold only pointwise in $\bx\in\B$ for pointwise results.

Assumption 3 generalizes as follows.

\begin{assumption}[Weight Function and Boundary]\label{sa-assump: Weight Function and Boundary}
    Let $w:\B \to (0,\infty)$ satisfy $\sup_{\bx\in\B} w(\bx) < \infty$, $\inf_{\bx\in\B} w(\bx) >0$, and $\int_{\B} w(\bx) \dif\Haus^{d-1}(\bx) < \infty$.
\end{assumption}

\subsection{Parameters}

The multidimensional generalizations of the three causal parameters studied in the paper are:
\begin{enumerate}
    \item \emph{Boundary Average Treatment Effect Curve} (BATEC):
    \begin{align*}
        \tau(\bx) = \E[Y_i(1) - Y_i(0)|\bX_i = \bx], \qquad \bx\in\B.
    \end{align*}

    \item \emph{Weighted Boundary Average Treatment Effect} (WBATE):
    \begin{align*}
        \tau_{\mathtt{WBATE}} = \frac{\int_{\B} \tau(\bx) w(\bx) \dif\Haus^{d-1}(\bx)}
                                     {\int_{\B} w(\bx) \dif\Haus^{d-1}(\bx)}.
    \end{align*}

    \item \emph{Largest Boundary Average Treatment Effect} (LBATE):
    \begin{align*}
        \tau_{\mathtt{LBATE}} = \sup_{\bx\in\B} \tau(\bx).
    \end{align*}

\end{enumerate}
More generally, this supplemental appendix considers the derivatives of the BATEC parameter
\begin{align*}
    \tau^{(\bnu)}(\bx) = \mu^{(\bnu)}_1(\bx) - \mu^{(\bnu)}_0(\bx), \qquad \bx \in \B,
\end{align*}
and the corresponding weighted and largest derivative parameters, respectively given by
\begin{align*}
    \tau_{\mathtt{WBATE}}^{(\bnu)} = \frac{\int_{\B} \tau^{(\bnu)}(\bx) w(\bx) \dif\Haus^{d-1}(\bx)}
                                     {\int_{\B} w(\bx) \dif\Haus^{d-1}(\bx)}
    \qquad\text{and}\qquad
    \tau_{\mathtt{LBATE}}^{(\bnu)} = \sup_{\bx\in\B} \tau^{(\bnu)}(\bx),
\end{align*}
where $|\bnu| \leq p$.

\subsection{Estimators and Other Related Quantities}

The estimator of the (derivative of the) BATEC parameters
\begin{align*}
    \widehat{\tau}^{(\bnu)}(\bx)
    = \widehat{\mu}^{(\bnu)}_1(\bx) - \widehat{\mu}^{(\bnu)}_0(\bx)
    \qquad \bx\in\B,
\end{align*}
where $\widehat{\mu}_t^{(\bnu)}(\bx) = \bnu! \be_{\bnu}^\top \widehat{\bbeta}_t(\bx)$ for $t\in\{0,1\}$ with
\begin{align}\label{sa-eq: locpoly estimator}
    \widehat{\bbeta}_t(\bx)
    = \argmin_{\bbeta \in \mathbb{R}^{\mathfrak{p}_p}}
      \En \big[ \big(Y_i - \br_p(\bX_i - \bx)^{\top}\bbeta \big)^2 K_{\bH}(\bX_i - \bx)\Indicator(\bX_i \in \A_t) \big], \qquad \bx \in \B,
\end{align}
with $K_{\bH}(\bu) = |\bH|^{-1} K(\bH^{-1}\bu)$ for a $d$-variate kernel function $K(\cdot)$ and a diagonal bandwidth matrix $\bH = \diag(\bh)$ with $\bh = (h_1, \ldots, h_d)^\top\in\mathbb{R}^d_{++}$. Thus,
\begin{align*}
    |\bH| = \prod_{l = 1}^d h_l
    \qquad\text{and}\qquad
    \bH^{-1} = \diag(h_1^{-1}, \ldots, h_d^{-1}).
\end{align*}
Furthermore, if all bandwidths are proportional to each other, that is, $h_l = c_l h$ for fixed constants $c_l > 0$, $l=1,\ldots,d$, and common bandwidth $h > 0$, then $\bH = h \diag(\bc)$ with $\bc=(c_1,\ldots,c_d)^\top$, and hence
\begin{align*}
    |\bH| = h^d \prod_{l = 1}^d c_l,
    \qquad\text{and}\qquad
    \bH^{-1} = h^{-1} \diag(c_1^{-1}, \ldots, c_d^{-1}).
\end{align*}
Therefore, $|\bH|\asymp h^d$ and $\bH^{-1} \asymp h^{-1} \bI_d$.

Under the assumptions imposed,
\begin{align*}
    \widehat{\bbeta}_{t}(\bx)
    = \bS^{-1} \widehat{\bGamma}_{t,\bx}^{-1} \En \big[\br_p(\bH^{-1}(\bX_i-\bx)) K_{\bH}(\bX_i - \bx) Y_i \Indicator(\bX_i \in \A_t) \big],
    \qquad
    \bS = \diag(\br_p(\bh)),
\end{align*}
and
\begin{align*}
    \widehat{\bGamma}_{t,\bx} = \En \big[ \br_p(\bH^{-1}(\bX_i-\bx)) \br_p(\bH^{-1}(\bX_i-\bx))^{\top} K_{\bH}(\bX_i - \bx) \Indicator(\bX_i \in \A_t)\big],
\end{align*}
where its population analog is
\begin{align*}
    \Gammax = \E \big[\br_p(\bH^{-1}(\bX_i-\bx)) \br_p(\bH^{-1}(\bX_i-\bx))^{\top} K_{\bH}(\bX_i - \bx) \Indicator(\bX_i \in \A_t) \big].
\end{align*}
For use in the proofs, let $\bbeta_t(\bx)$ denote the population Taylor coefficient vector satisfying
\begin{align*}
    \bbeta_t(\bx)^\top\br_p(\bu)
    = \sum_{|\bomega|\leq p}\frac{\mu_t^{(\bomega)}(\bx)}{\bomega!}\bu^{\bomega},
    \qquad \bu\in\reals^d.
\end{align*}
Note that $\norm{\be_{\bnu}^\top \bS^{-1}}_2 = \norm{\be_{\bnu}^\top \bS^{-1}}_{\infty} = \bh^{-\bnu}$. In addition, define
\begin{align*}
    \bQ_{t,\bx} = \En \big[\br_p(\bH^{-1}(\bX_i-\bx)) K_{\bH}(\bX_i - \bx) \Indicator(\bX_i \in \A_t) \varepsilon_i\big],
\end{align*}
where $\varepsilon_i = Y_i - [\Indicator(\bX_i \in \A_0) \mu_0(\bX_i) + \Indicator(\bX_i \in \A_1) \mu_1(\bX_i)] = Y_i - \E[Y_i|\bX_i]$.

For $\bx_1, \bx_2 \in \B$ and $t \in \{0,1\}$, we introduce the following population quantities
\begin{align*}
    \Omega_{\bx_1,\bx_2}^{(\bnu)}
    = \Omega_{0,\bx_1,\bx_2}^{(\bnu)} + \Omega_{1,\bx_1,\bx_2}^{(\bnu)},
    \qquad
    \Omega_{t,\bx_1,\bx_2}^{(\bnu)}
    = \frac{(\bnu!)^2}{n |\bH| \bh^{2\bnu}} \be_{\bnu}^{\top}  \bGamma_{t,\bx_1}^{-1} \bSigma_{t,\bx_1,\bx_2} \bGamma_{t,\bx_2}^{-1} \be_{\bnu},
\end{align*}
\begin{align*}
    \bSigma_{t, \bx_1,\bx_2}
    = |\bH| \E \big[\br_p(\bH^{-1}(\bX_i-\bx_1)) \br_p(\bH^{-1}(\bX_i-\bx_2))^{\top} K_{\bH}(\bX_i - \bx_1) K_{\bH}(\bX_i - \bx_2) \sigma_t^2(\bX_i) \Indicator(\bX_i \in \A_t)\big],
\end{align*}
and their sample analogs
\begin{align*}
    \widehat{\Omega}_{\bx_1,\bx_2}^{(\bnu)}
    = \widehat{\Omega}_{0,\bx_1,\bx_2}^{(\bnu)} + \widehat{\Omega}_{1,\bx_1, \bx_2}^{(\bnu)},
    \qquad
    \widehat{\Omega}_{t,\bx_1,\bx_2}^{(\bnu)}
    = \frac{(\bnu!)^2}{n |\bH| \bh^{2\bnu}} \be_{\bnu}^{\top} \widehat{\bGamma}_{t,\bx_1}^{-1} \widehat{\bSigma}_{t,\bx_1,\bx_2} \widehat{\bGamma}_{t,\bx_2}^{-1} \be_{\bnu},
\end{align*}
\begin{align*}
    \widehat{\bSigma}_{t,\bx_1,\bx_2}
    = |\bH| \En \big[\br_p(\bH^{-1}(\bX_i-\bx_1)) \br_p(\bH^{-1}(\bX_i-\bx_2))^{\top} K_{\bH}(\bX_i - \bx_1) K_{\bH}(\bX_i - \bx_2) \widehat{\varepsilon}_i(\bx_1) \widehat{\varepsilon}_i(\bx_2) \Indicator(\bX_i \in \A_t)\big],
\end{align*}
where $\widehat{\varepsilon}_{i}(\bx) = Y_i - \br_p(\bX_i - \bx)^\top[\Indicator(\bX_i \in \A_0) \widehat{\bbeta}_0(\bx) + \Indicator(\bX_i \in \A_1) \widehat{\bbeta}_1(\bx)]$.

Finally, we introduce more notation for the leading bias and variance of the estimator of the (derivative of the) BATEC parameter. For $\bx \in \B$ and $t \in \{0,1\}$, define
\begin{align*}
    B_{t,\bx}^{(\bnu)}(\bh)
    = \bnu!\be_{\bnu}^{\top} \bGamma_{t,\bx}^{-1} \sum_{|\bomega| = p + 1} \frac{\mu_t^{(\bomega)}(\bx)}{\bomega!}\bh^{\bomega-\bnu}
      \E \big[\br_p(\bH^{-1}(\bX_i-\bx)) (\bH^{-1}(\bX_i-\bx))^{\bomega} K_{\bH}(\bX_i - \bx) \Indicator(\bX_i \in \A_t)\big],
\end{align*}
where $B_{\bx}^{(\bnu)}(\bh) = B_{1,\bx}^{(\bnu)}(\bh) - B_{0,\bx}^{(\bnu)}(\bh)$. When $\bh = \bc h$ for a fixed vector $\bc$ and $h$ is a common bandwidth, $B_{\bx}^{(\bnu)}(\bh)=h^{p+1-|\bnu|}\widetilde B_{\bx}^{(\bnu)}(\bc)$ for some $\widetilde B_{\bx}^{(\bnu)}(\bc)$. Also, define
\begin{align*}
    V_{\bx}^{(\bnu)}  = V_{0,\bx}^{(\bnu)} + V_{1,\bx}^{(\bnu)},
\end{align*}
with
\begin{align*}
    V_{t,\bx}^{(\bnu)}
    = (\bnu!)^2\be_{\bnu}^{\top} \bGamma_{t,\bx}^{-1} \bSigma_{t,\bx,\bx} \bGamma_{t,\bx}^{-1}\be_{\bnu}
    = n |\bH| \bh^{2\bnu} \Omega_{t,\bx,\bx}^{(\bnu)}.
\end{align*}
The corresponding empirical analogs of these leading terms, leaving the unknown derivatives explicit, are
\begin{align*}
    \widehat{B}_{t,\bx}^{(\bnu)}(\bh)
    = \bnu!\be_{\bnu}^{\top} \widehat{\bGamma}_{t,\bx}^{-1} \sum_{|\bomega| = p + 1} \frac{\mu_t^{(\bomega)}(\bx)}{\bomega!}\bh^{\bomega-\bnu}
      \En \big[\br_p(\bH^{-1}(\bX_i-\bx)) (\bH^{-1}(\bX_i-\bx))^{\bomega} K_{\bH}(\bX_i - \bx) \Indicator(\bX_i \in \A_t)\big],
\end{align*}
and $\widehat{B}_{\bx}^{(\bnu)}(\bh) = \widehat{B}_{1,\bx}^{(\bnu)}(\bh) - \widehat{B}_{0,\bx}^{(\bnu)}(\bh)$. Also, $\widehat{V}_{\bx}^{(\bnu)} = \widehat{V}_{0,\bx}^{(\bnu)} + \widehat{V}_{1,\bx}^{(\bnu)}$ with $\widehat{V}_{t,\bx}^{(\bnu)} = n |\bH| \bh^{2\bnu} \widehat{\Omega}_{t,\bx,\bx}^{(\bnu)}$.

\subsection{Preliminary Lemmas}

Let $\bX = (\bX_1^{\top}, \cdots, \bX_n^{\top})$, and recall that $t \in \{0,1\}$.

\begin{lem}[Invertibility]\label{sa-lem: Invertibility}
    Suppose Assumptions \ref{sa-assump: Sharp DGP}(i,ii) and \ref{sa-assump: Boundary and Kernel} hold. Then,
    \begin{align*}
        \liminf_{\bar{h} \to 0} \inf_{\bx \in \B}\lambda_{\min}(\bGamma_{t,\bx}) > 0.
    \end{align*}
\end{lem}

\begin{lem}[Gram]\label{sa-lem: Gram Matrix}
    Suppose Assumptions \ref{sa-assump: Sharp DGP}(i,ii) and \ref{sa-assump: Boundary and Kernel} hold. If $\frac{\log(1/\bar{h})}{n |\bH|} = o(1)$, then
    \begin{align*}
        \sup_{\bx \in \B} \big\|\hGamma - \Gammax\big\|  \lessp \sqrt{\frac{\log(1/\bar{h})}{n |\bH|}},
        \qquad
        \sup_{\bx \in \B} \big\|\hGamma^{-1} - \Gammax^{-1}\big\| \lessp \sqrt{\frac{\log(1/\bar{h})}{n |\bH|}}.
    \end{align*}
    If, in addition, $\bar{h} = o(1)$, then $\inf_{\bx \in \B}\lambda_{\min}(\hGamma) \gtrsim_{\P} 1$, $\sup_{\bx \in \B}\|\hGamma\|\lessp 1$, and $\sup_{\bx \in \B}\|\hGamma^{-1}\|\lessp 1$.
\end{lem}

\begin{lem}[Linear Approximation]\label{sa-lem: Linear Approximation}
    Suppose Assumptions \ref{sa-assump: Sharp DGP}(i,ii,iv,v) and \ref{sa-assump: Boundary and Kernel} hold. If $\frac{\log(1/\bar{h})}{n |\bH|} = o(1)$, then
    \begin{align*}
        \sup_{\bx \in \B} \big\|\bQ_{t,\bx} \big\| \lessp \sqrt{\frac{\log(1/\bar{h})}{n |\bH|}} + \frac{\log(1/\bar{h})}{n^{\frac{1+v}{2+v}}|\bH|}.
    \end{align*}
    If, in addition,  $\bar{h} = o(1)$, then
    \begin{align*}
        \sup_{\bx \in \B} \big|\widehat{\mu}_t^{(\bnu)}(\bx)   - \E\big[\widehat{\mu}_t^{(\bnu)}(\bx)\big|\bX\big] - \bnu!\be_{\bnu}^{\top}\bS^{-1}\bGamma_{t,\bx}^{-1}\bQ_{t,\bx}\big| \lessp \bh^{-\bnu} \sqrt{\frac{\log(1/\bar{h})}{n |\bH|}} \bigg(\sqrt{\frac{\log(1/\bar{h})}{n |\bH|}} + \frac{\log(1/\bar{h})}{n^{\frac{1+v}{2+v}}|\bH|} \bigg).
    \end{align*}
\end{lem}

\begin{lem}[Covariance]\label{sa-lem: covariance}
    Suppose Assumptions \ref{sa-assump: Sharp DGP} and \ref{sa-assump: Boundary and Kernel} hold. If $\frac{\log(1/\bar{h})}{n^{\frac{v}{2+v}} |\bH|} = o(1)$ and $\bar{h}=o(1)$, then
    \begin{align*}
        \sup_{\bx_1,\bx_2 \in \B} \big\|\widehat{\bSigma}_{t, \bx_1,\bx_2} - \bSigma_{t, \bx_1,\bx_2}\big\|
        \lessp \sqrt{\frac{\log(1/\bar{h})}{n |\bH|}} + \frac{\log(1/\bar{h})}{n^{\frac{v}{2+v}}|\bH|} + \bar{h}^{p+1},
    \end{align*}
    \begin{align*}
        \sup_{\bx_1,\bx_2 \in \B} \big|\widehat{\Omega}_{\bx_1,\bx_2}^{(\bnu)} - \Omega_{\bx_1,\bx_2}^{(\bnu)}\big|
        \lessp (n |\bH|\bh^{2\bnu})^{-1}\bigg(\sqrt{\frac{\log(1/\bar{h})}{n |\bH|}} + \frac{\log(1/\bar{h})}{n^{\frac{v}{2+v}}|\bH|}+ \bar{h}^{p+1}\bigg),
    \end{align*}
    and
    \begin{align*}
        \sup_{\bx \in \B} \big|(\widehat{\Omega}_{\bx,\bx}^{(\bnu)})^{-1/2} - (\Omega_{\bx,\bx}^{(\bnu)})^{-1/2} \big|
        \lessp \sqrt{n |\bH|\bh^{2\bnu}}\bigg(\sqrt{\frac{\log(1/\bar{h})}{n |\bH|}} + \frac{\log(1/\bar{h})}{n^{\frac{v}{2+v}}|\bH|} + \bar{h}^{p+1}\bigg).
    \end{align*}
\end{lem}

\begin{lem}[Bias]\label{sa-lem: bias}
    Suppose Assumptions \ref{sa-assump: Sharp DGP}(i,ii,iii) and \ref{sa-assump: Boundary and Kernel} hold. If $\frac{\log(1/\bar{h})}{n |\bH|} = o(1)$ and $\bar{h} = o(1)$, then
    \begin{align*}
        \sup_{\bx \in \B} \big|\E[\widehat{\mu}_t^{(\bnu)}(\bx)|\bX] - \mu_t^{(\bnu)}(\bx)\big| \lessp \bar{h}^{p+1}\bh^{-\bnu},
    \end{align*}
    implying
    \begin{align*}
        \sup_{\bx \in \B} \big|\E[\widehat{\mu}_t^{(\bnu)}(\bx)|\bX] - \mu_t^{(\bnu)}(\bx) - \widehat{B}_{t,\bx}^{(\bnu)}(\bh)\big| = o_{\P}(\bar{h}^{p+1}\bh^{-\bnu}),
    \end{align*}
    with $\sup_{\bx \in \B} \big|\widehat{B}_{t,\bx}^{(\bnu)}(\bh) - B_{t,\bx}^{(\bnu)}(\bh) \big| \lessp \bar{h}^{p+1}\bh^{-\bnu}\sqrt{\frac{\log(1/\bar{h})}{n |\bH|}}$, and hence $\sup_{\bx \in \B}|B_{t,\bx}^{(\bnu)}(\bh)| \lesssim \bar{h}^{p+1}\bh^{-\bnu}$.
\end{lem}

\section{Boundary Average Treatment Effect Curve}

\subsection{Point Estimation and MSE Expansions}

\begin{thm}[Convergence Rates: Sharp BATEC]\label{sa-thm: Convergence Rates: Sharp BATEC}
    Suppose Assumptions \ref{sa-assump: Sharp DGP} and \ref{sa-assump: Boundary and Kernel} hold. If $\log(1/\bar{h})/(n |\bH|) = o(1)$ and $\bar{h} = o(1)$, then
    \begin{align*}
        \big|\widehat{\tau}^{(\bnu)}(\bx) - \tau^{(\bnu)}(\bx) \big|
        \lessp \bh^{-\bnu} \Big(\frac{1}{\sqrt{n |\bH|}} + \frac{1}{n^{\frac{1+v}{2+v}}|\bH|} + \bar{h}^{p+1}\Big)
    \end{align*}
    for $\bx \in \B$, and
    \begin{align*}
        \sup_{\bx \in \B} \big|\widehat{\tau}^{(\bnu)}(\bx) - \tau^{(\bnu)}(\bx)\big|
        \lessp \bh^{-\bnu} \Big(\sqrt{\frac{\log(1/\bar{h})}{ n |\bH|}} + \frac{\log(1/\bar{h})}{n^{\frac{1+v}{2+v}}|\bH|} + \bar{h}^{p+1}\Big).
    \end{align*}
\end{thm}

The conditional mean-squared error (MSE) is
\begin{align*}
    & \operatorname{MSE}^{(\bnu)}(\bx) = \E \big[(\widehat{\tau}^{(\bnu)}(\bx) - \tau^{(\bnu)}(\bx))^2 \big| \bX \big]
\end{align*}
for $\bx\in \B$, and the conditional integrated MSE (IMSE) is
\begin{align*}
    & \operatorname{IMSE}^{(\bnu)} = \int_{\B}  \operatorname{MSE}^{(\bnu)}(\bx) w(\bx) \dif\Haus^{d-1}(\bx),
\end{align*}
where $w(\bx)$ satisfies Assumption \ref{sa-assump: Weight Function and Boundary}.

\begin{thm}[MSE Expansions: Sharp BATEC]\label{sa-thm: MSE Expansions: Sharp BATEC}
    Suppose Assumptions \ref{sa-assump: Sharp DGP}, \ref{sa-assump: Boundary and Kernel} and \ref{sa-assump: Weight Function and Boundary} hold. If $\log(1/\bar{h})/(n^{\frac{v}{2+v}} |\bH|) = o(1)$ and $\bar{h} = o(1)$, then
    \begin{align*}
        \operatorname{MSE}^{(\bnu)}(\bx)
        = \big(B_{\bx}^{(\bnu)}(\bh)\big)^2
          + \frac{1}{n|\bH|\bh^{2\bnu}} V_{\bx}^{(\bnu)}
          + o_{\P} \big(\bar{h}^{2p+2}\bh^{-2\bnu} + (n|\bH|\bh^{2\bnu})^{-1}\big)
    \end{align*}
    for $\bx \in \B$, and
    \begin{align*}
        \operatorname{IMSE}^{(\bnu)}
        = \int_{\B} \big[ \big(B_{\bx}^{(\bnu)}(\bh)\big)^2
           + \frac{1}{n|\bH|\bh^{2\bnu}} V_{\bx}^{(\bnu)} \big] w(\bx) \dif\Haus^{d-1}(\bx)
           + o_{\P} \big(\bar{h}^{2p+2}\bh^{-2\bnu} + (n|\bH|\bh^{2\bnu})^{-1}\big).
    \end{align*}
\end{thm}

Theorem \ref{sa-thm: MSE Expansions: Sharp BATEC} can be used to develop feasible bandwidth selectors for diagonal bandwidth matrices by minimizing the displayed criterion over $\bh\in\mathbb{R}_{++}^d$. If $\bh = \bc h$ for a fixed vector $\bc$ and a common bandwidth $h$, the problem reduces to scalar bandwidth selection. If $\widetilde B_{\bx}^{(\bnu)}(\bc)\neq0$, the asymptotic MSE-optimal common scale is
\begin{align*}
    h_{\operatorname{MSE},\bnu}(\bx)
    = \left(\frac{(d + 2|\bnu|) V_{\bx}^{(\bnu)}}
                 {(2p+2 - 2 |\bnu|) (\widetilde B_{\bx}^{(\bnu)}(\bc))^2} \frac{1}{n}\right)^{\frac{1}{2p+d+2}}
\end{align*}
for $\bx \in \B$. Similarly, if $\int_{\B} (\widetilde B_{\bx}^{(\bnu)}(\bc))^2 w(\bx)\dif\Haus^{d-1}(\bx) \neq 0$, the asymptotic IMSE-optimal common scale is
\begin{align*}
    h_{\operatorname{IMSE},\bnu}
    = \left(\frac{(d + 2 |\bnu|) \int_{\B} V_{\bx}^{(\bnu)} w(\bx)\dif\Haus^{d-1}(\bx)}
                 {(2p+2 - 2|\bnu|) \int_{\B} (\widetilde B_{\bx}^{(\bnu)}(\bc))^2 w(\bx) \dif\Haus^{d-1}(\bx)} \frac{1}{n}\right)^{\frac{1}{2p+d+2}}.
\end{align*}

In practice, the unknown bias and variance quantities can be replaced with (consistent) estimators thereof. For example, the unknown functions $\mu_t^{(\bomega)}(\bx)$ can be estimated using higher-order local polynomial estimators in $\widehat{B}_{\bx}^{(\bnu)}(\bh)$, and
$\widehat{V}_{\bx}^{(\bnu)} = \widehat{V}_{0,\bx}^{(\bnu)} + \widehat{V}_{1,\bx}^{(\bnu)}$
with
$\widehat{V}_{t,\bx}^{(\bnu)} = (\bnu!)^2\be_{\bnu}^{\top} \widehat{\bGamma}_{t,\bx}^{-1} \widehat{\bSigma}_{t,\bx,\bx} \widehat{\bGamma}_{t,\bx}^{-1}\be_{\bnu}$, which corresponds to a standard variance estimator (which is also used for asymptotic inference as discussed below).

Finally, some pointwise statements admit weaker side conditions when the empirical-process classes are restricted to a fixed boundary point. We state the stronger uniform side conditions throughout this section to simplify the exposition.

\subsection{Distributional Approximation and Inference}

Let $\bV = ((Y_1, \bX_1^{\top}), \cdots, (Y_n, \bX_n^{\top}))^\top$, and recall that $t \in \{0,1\}$. For $|\bnu| \leq p$, define the feasible $t$-statistic
\begin{align*}
    \widehat{\Tstat}^{(\bnu)}(\bx)
    = \frac{\widehat{\tau}^{(\bnu)}(\bx) - \tau^{(\bnu)}(\bx)}{\sqrt{\widehat{\Omega}_{\bx,\bx}^{(\bnu)}}},
    \qquad \bx \in \B.
\end{align*}
The associated $100(1-\alpha)\%$ confidence interval estimator is
\begin{align*}
    \widehat{\CI}^{(\bnu)}_{\alpha}(\bx)
    = \bigg[\;\widehat{\tau}^{(\bnu)}(\bx) - \q_{\alpha} \sqrt{\widehat{\Omega}_{\bx,\bx}^{(\bnu)}}
            \; , \;
              \widehat{\tau}^{(\bnu)}(\bx) + \q_{\alpha} \sqrt{\widehat{\Omega}_{\bx,\bx}^{(\bnu)}}\;\bigg],
\end{align*}
where $\q_{\alpha}$ denotes an appropriate quantile depending on the desired confidence level $\alpha\in(0,1)$, and coverage objective (pointwise versus uniform over $\B$).

\begin{thm}[Confidence Intervals: Sharp BATEC]\label{sa-thm: Confidence Intervals: Sharp BATEC}
    Suppose Assumptions \ref{sa-assump: Sharp DGP} and \ref{sa-assump: Boundary and Kernel} hold. If $n |\bH| \to \infty$ and $n |\bH| \bar{h}^{2(p+1)} =o(1)$, then
    \begin{align*}
        \sup_{u \in \reals} \big|\P \big(\widehat{\Tstat}^{(\bnu)}(\bx) \leq u \big) - \Phi(u)\big| = o(1),
        \qquad \bx \in \B,
    \end{align*}
    and
    \begin{align*}
        \P \big(\tau^{(\bnu)}(\bx) \in \widehat{\CI}_{\alpha}^{(\bnu)}(\bx) \big) = 1 - \alpha + o(1),
        \qquad \bx \in \B,
    \end{align*}
    provided that $\q_{\alpha} = \inf \{c > 0: \P( |\widehat{Z}| \geq c | \bV) \leq \alpha \}$ with $\widehat{Z}|\bV \sim \mathsf{Normal}(0, 1)$.
\end{thm}

For uniform inference, we rely on a new strong approximation result established in Section \ref{sa-sec: Gaussian Strong Approximation}. First, we simplify the statistic $\widehat{\Tstat}^{(\bnu)}$, which is not a sum of independent random variables. Let
\begin{align*}
    \overline{\Tstat}^{(\bnu)}(\bx)
    = \E_n \Big[(\Omega_{\bx,\bx}^{(\bnu)})^{-1/2} \bnu!\be_{\bnu}^{\top} \bS^{-1} \big[\Indicator(\bX_i \in \A_1) \Gammap^{-1} - \Indicator(\bX_i \in \A_0) \Gamman^{-1}\big] \br_p(\bH^{-1}(\bX_i - \bx)) K_{\bH}(\bX_i - \bx)  \varepsilon_i \Big],
\end{align*}
where recall that $\varepsilon_i = Y_i - \sum_{t \in \{0,1\}}\Indicator(\bX_i \in \A_t) \mu_t(\bX_i) = Y_i -  \E[Y_i | \bX_i]$.

\begin{thm}[Stochastic Linearization: Sharp BATEC]\label{sa-thm: Stochastic Linearization: Sharp BATEC}
    Suppose Assumptions \ref{sa-assump: Sharp DGP} and \ref{sa-assump: Boundary and Kernel} hold. If $\log(1/\bar{h})/(n^{v/(2+v)}|\bH|) = o(1)$ and $\bar{h} = o(1)$, then
    \begin{align*}
        \sup_{\bx \in \B} \big|\widehat{\Tstat}^{(\bnu)}(\bx) - \overline{\Tstat}^{(\bnu)}(\bx) \big|
        \lessp \bar{h}^{p+1}\sqrt{n |\bH|} + \sqrt{\log(1/\bar{h})} \Big(\sqrt{\frac{\log(1/\bar{h})}{n |\bH|}} + \frac{\log(1/\bar{h})}{n^{\frac{v}{2+v}}|\bH|}\Big).
    \end{align*}
\end{thm}

We can now exploit the linear structure of $(\overline{\Tstat}^{(\bnu)}(\bx): \bx \in \B)$, that is, an average of i.n.i.d.\ random vectors. Define the following functions indexed by $\bx \in \B$:
\begin{align*}
    g_{\bx}(\bu) = \Indicator(\bu \in \A_1)  \K^{(\bnu)}_1(\bu;\bx) - \Indicator(\bu \in \A_0) \K^{(\bnu)}_0(\bu;\bx),
    \qquad \bu \in \X,
\end{align*}
and
\begin{align*}
    \K^{(\bnu)}_t(\bu;\bx) = n^{-1/2} (\Omega_{\bx,\bx}^{(\bnu)})^{-1/2}\bnu!\be_{\bnu}^{\top} \bS^{-1} \bGamma_{t,\bx}^{-1} \br_p(\bH^{-1}(\bu - \bx)) K_{\bH}(\bu - \bx), \qquad \bu \in \X,\quad t \in \{0,1\}.
\end{align*}
Define the associated class of functions $\G = \{g_{\bx}: \bx \in \B\}$ and $\R = \{\operatorname{Id}\}$, where $\operatorname{Id}(x) = x$, for all $x \in \reals$. Then, the \emph{residual-based empirical process} is
\begin{align*}
    R_n(g,r) = n^{-1/2}\sum_{i = 1}^n \Big[g(\bX_i)r(Y_i) - g(\bX_i) \E[r(Y_i)|\bX_i] \Big], \qquad g \in \G, r \in \R,
\end{align*}
and therefore
\begin{align*}
    \overline{\Tstat}^{(\bnu)}(\bx) = R_n(g_{\bx},\operatorname{Id}), \qquad \bx \in \B.
\end{align*}

Theorem~\ref{sa-thm: Residual Empirical Process Strong Approximation} gives a new Gaussian strong approximation that can be applied to our current setup. Specifically, our new theorem allows polynomial moment bounds on the conditional distribution of $Y_i|\bX_i$.

\begin{thm}[Gaussian Strong Approximation: $\overline{\Tstat}^{(\bnu)}$]\label{sa-thm: Gaussian Strong Approximation: Sharp Tstat}
    Suppose Assumptions \ref{sa-assump: Sharp DGP} and \ref{sa-assump: Boundary and Kernel} hold, and $\X=\prod_{l=1}^d[a_l,b_l]$, with $-\infty<a_l<b_l<\infty$ for $l=1,\ldots,d$. Suppose also that there exists a constant $C > 0$ such that for $t \in \{0,1\}$ and for any $\bx \in \B$, the De Giorgi perimeter of the set $E_{t,\bx} = \{\by \in \A_t: \bH^{-1}(\by - \bx) \in \supp(K)\}$ satisfies  $\mathcal{L}(E_{t,\bx}) \leq C \bar{h}^{d-1}$. For the associated classes $\G$ and $\R=\{\operatorname{Id}\}$ defined above, suppose the singleton regularity condition in part (2) of Theorem~\ref{sa-thm: Residual Empirical Process Strong Approximation} holds. If $\liminf_{n \to \infty} \log \bar{h}/\log n > - \infty$ and $n |\bH| \to \infty$, then (on a possibly enlarged probability space) there exists a mean-zero Gaussian process $Z^{(\bnu)}$ indexed by $\B$ with almost surely continuous sample path such that
    \begin{align*}
        \E\Big[\sup_{\bx \in \B} \big|\overline{\Tstat}^{(\bnu)}(\bx) - Z^{(\bnu)}(\bx) \big| \Big]
        \lesssim (\log n)^{\frac{3}{2}} \bigg(\frac{1}{n |\bH|}\bigg)^{\frac{1}{2d+2}\cdot \frac{v}{v + 2}} + \log (n) \left(\frac{1}{n^{\frac{v}{2+v}}|\bH|}\right)^{\frac{1}{2}},
    \end{align*}
    where $\lesssim$ is up to a universal constant, and $Z^{(\bnu)}$ has the same covariance structure as $\overline{\Tstat}^{(\bnu)}$; that is, $\Cov[\overline{\Tstat}^{(\bnu)}(\bx_1), \overline{\Tstat}^{(\bnu)}(\bx_2)] = \Cov[Z^{(\bnu)}(\bx_1), Z^{(\bnu)}(\bx_2)]$ for all $\bx_1, \bx_2 \in \B$.
\end{thm}

The rectangular support condition in Theorem~\ref{sa-thm: Gaussian Strong Approximation: Sharp Tstat} is imposed only for the process-level strong approximation. It permits an affine rescaling of the score support to $[0,1]^d$; combined with the bounded and bounded-away-from-zero density in Assumption~\ref{sa-assump: Sharp DGP}(ii), $\P_X$ can then be used as the surrogate measure and the Rosenblatt transform provides the normalizing transformation with bounded constants. See \citet[Section 3.1]{Cattaneo-Yu_2025_AOS} for details. This normalization simplifies the verification of that stronger coupling and is not used in the pointwise, aggregate, or supremum-based confidence-band results. Because the local-polynomial estimators use only shrinking neighborhoods of $\B\subset\setint(\X)$, the condition is not practically restrictive for the process-level result in settings where the relevant score support can be represented by a compact rectangle after rescaling.

The De Giorgi perimeter condition in Theorem~\ref{sa-thm: Gaussian Strong Approximation: Sharp Tstat} is used only to verify the total variation requirement of the residual empirical process strong approximation theorem (Theorem~\ref{sa-thm: Residual Empirical Process Strong Approximation}) for the class of boundary-indexed local-polynomial weighting functions. Even when the kernel is smooth, the factors $\Indicator(\bu\in\A_t)$ introduce discontinuities along the assignment boundary. The localized perimeter bound controls the total variation of these discontinuities uniformly over $\bx\in\B$, because the relevant jumps occur on the sets $E_{t,\bx}=\{\by \in \A_t: \bH^{-1}(\by-\bx)\in\supp(K)\}$. The separate singleton regularity condition concerns the truncated conditional means introduced by the polynomial-moment truncation step. Informally, these additional technical restrictions avoid an overly ``wiggly'' assignment boundary and irregular truncation-induced conditional means.

The following lemma shows that the simpler boundary assumptions imposed in the paper are sufficient to verify the De Giorgi perimeter condition for standard compactly supported kernels.

\begin{lem}[Localized Perimeter: Piecewise-Linear Boundary]\label{sa-lem: localized perimeter piecewise linear}
    Suppose $d=2$, $\bH=h\mathbf{I}_2$, $\B$ is piecewise linear with finitely many line segments, $\B$ lies in the interior of $\X$, and $\supp(K)$ is bounded. If either $K(\bu)=\Indicator(\bu\in[-1,1]^2)$ or $\supp(K)$ has finite De Giorgi perimeter, then, for $t\in\{0,1\}$,
    \begin{align*}
        \sup_{\bx\in\B}\mathcal{L}\big(\{\by\in\A_t:h^{-1}(\by-\bx)\in \supp(K)\}\big)
        \lesssim h.
    \end{align*}
\end{lem}

Theorem \ref{sa-thm: Gaussian Strong Approximation: Sharp Tstat} can be used to construct confidence bands for $(\tau^{(\bnu)}(\bx):\bx\in\B)$. Let $(\widehat{Z}^{(\bnu)}(\bx):\bx \in \B)$ be a (conditionally on $\bV$) mean-zero Gaussian process with feasible (conditional) covariance function
\begin{align*}
    \Cov \big[\widehat{Z}^{(\bnu)}(\bx_1),\widehat{Z}^{(\bnu)}(\bx_2) \big|\bV \big]
    = \frac{\widehat{\Omega}_{\bx_1,\bx_2}^{(\bnu)}}
           {\sqrt{\widehat{\Omega}_{\bx_1,\bx_1}^{(\bnu)} \widehat{\Omega}_{\bx_2,\bx_2}^{(\bnu)}}},
    \qquad \bx_1, \bx_2 \in \B.
\end{align*}

\begin{thm}[Confidence Bands: Sharp BATEC; process approximation]\label{sa-thm: Confidence Bands: Sharp BATEC -- process}
    Suppose the assumptions and conditions in Theorem \ref{sa-thm: Gaussian Strong Approximation: Sharp Tstat} hold. If $\liminf_{n \to \infty} \log \bar{h}/\log n > - \infty$, $(\log n)^{3}/(n^{v/(2+v)}|\bH|) = o(1)$ and $n |\bH| \bar{h}^{2(p+1)}\log n = o(1)$, then
    \begin{align*}
        \sup_{u \in \reals} \Big|\P \Big(\sup_{\bx \in \B} \big|\widehat{\Tstat}^{(\bnu)}(\bx)\big| \leq u \Big)
                            - \P \Big(\sup_{\bx \in \B} \big|\widehat{Z}^{(\bnu)}(\bx)\big| \leq u \Big| \bV \Big) \Big| = o_{\P}(1)
    \end{align*}
    and
    \begin{align*}
        \P\Big[\tau^{(\bnu)}(\bx) \in \widehat{\CI}_{\alpha}^{(\bnu)}(\bx), \text{ for all } \bx \in \B \Big] = 1 - \alpha + o(1),
    \end{align*}
    provided that $\q_{\alpha} = \inf \big\{c > 0: \P \big(\sup_{\bx \in \B} \big|\widehat{Z}^{(\bnu)}(\bx)\big|\geq c \big| \bV \big) \leq \alpha \big\}$.
\end{thm}

Theorem~\ref{sa-thm: Gaussian Strong Approximation: Sharp Tstat} gives a strong approximation of the entire process $\overline{\Tstat}^{(\bnu)}$ by a Gaussian process $Z^{(\bnu)}$. This process-level coupling is stronger than what is needed for pointwise normal approximation, and also stronger than the supremum approximation needed for confidence bands. Indeed, as demonstrated in the following theorem, approximating only the supremum of the process using the results in \cite{Chernozhukov-Chetverikov-Kato_2014b_AoS} removes the need for the De Giorgi perimeter condition, and allows for a wider class of boundaries $\B$.

\begin{thm}[Gaussian Approximation for $\sup_{\bx \in \B} |\overline{\Tstat}^{(\bnu)}(\bx)|$]\label{sa-thm: Gaussian Approximation Suprema -- Sharp}
Suppose Assumptions \ref{sa-assump: Sharp DGP} and \ref{sa-assump: Boundary and Kernel} hold. Then for $n$ large enough, on a possibly enriched probability space, there exists a Gaussian process $Z^{(\bnu)}(\bx)$, $\bx \in \B$, with almost sure continuous trajectories such that
\begin{align*}
    \Cov[Z^{(\bnu)}(\bx_1), Z^{(\bnu)}(\bx_2)]
    =
    \Cov[\overline{\Tstat}^{(\bnu)}(\bx_1), \overline{\Tstat}^{(\bnu)}(\bx_2)],
    \qquad \bx_1,\bx_2\in\B,
\end{align*}
and
\begin{align*}
    \P\Big[ \big|\sup_{\bx \in \B} |\overline{\Tstat}^{(\bnu)}(\bx)| - \sup_{\bx \in \B} |Z^{(\bnu)}(\bx)|\big| > \frac{\log n}{(n |\bH|)^{1/6}} + \frac{(\log n)^{5/4}}{(n |\bH|)^{1/4}} + \frac{(\log n)^{3/2}}{(n^{\frac{v}{2+v}} |\bH|)^{1/2}}\Big] \lesssim \frac{1}{\log n} + \frac{\log n}{n}.
\end{align*}
\end{thm}

Using this result, we obtain the following theorem for confidence bands that does not require the De Giorgi perimeter condition.

\begin{thm}[Confidence Bands: Sharp BATEC; suprema approximation]\label{sa-thm: Confidence Bands: Sharp BATEC -- Suprema}
    Suppose Assumptions \ref{sa-assump: Sharp DGP} and \ref{sa-assump: Boundary and Kernel} hold. If $\liminf_{n \to \infty} \log \bar{h}/\log n > -\infty$, $(\log n)^9/(n^{v/(2+v)}|\bH|)=o(1)$, and $\bar{h}^{2(p+1)}n|\bH|\log n=o(1)$, then
    \begin{align*}
        \sup_{u \in \reals} \Big|\P \Big(\sup_{\bx \in \B} \big|\widehat{\Tstat}^{(\bnu)}(\bx)\big| \leq u \Big)
                            - \P \Big(\sup_{\bx \in \B} \big|\widehat{Z}^{(\bnu)}(\bx)\big| \leq u \Big| \bV \Big) \Big| = o_{\P}(1)
    \end{align*}
    and
    \begin{align*}
        \P\Big[\tau^{(\bnu)}(\bx) \in \widehat{\CI}_{\alpha}^{(\bnu)}(\bx), \text{ for all } \bx \in \B \Big] = 1 - \alpha + o(1),
    \end{align*}
    provided that $\q_{\alpha} = \inf \big\{c > 0: \P \big(\sup_{\bx \in \B} \big|\widehat{Z}^{(\bnu)}(\bx)\big|\geq c \big| \bV \big) \leq \alpha \big\}$.
\end{thm}

\subsection{Covariate Adjustment}

Following \cite{Calonico-Cattaneo-Farrell-Titiunik_2019_RESTAT} and \cite{Calonico-Cattaneo-Farrell-Palomba-Titiunik_2026_wp}, we can incorporate covariate adjustment into our framework. To conserve space, we only give a brief description of the basic idea. Suppose that $\bZ_1,\ldots,\bZ_n$ are pre-intervention covariates of dimension $d_Z\geq1$.

For efficiency improvements, the covariate-adjusted BATEC estimator is
\begin{align*}
    \widetilde{\tau}(\bx) = \be_{\mathfrak{p}_p+1}^{\top}\widetilde{\bgamma}(\bx)
\end{align*}
with
\begin{align*}
    \widetilde{\bgamma}(\bx)
    = \argmin_{\bgamma \in \mathbb{R}^{2\mathfrak{p}_p+d_Z}}
      \En \big[ \big(Y_i - \widetilde{\br}_p(\bX_i-\bx,T_i,\bZ_i)^\top \bgamma \big)^2
                K_h(\bX_i - \bx) \big],
\end{align*}
where $T_i=\Indicator(\bX_i \in \A_1)$, and
\begin{align*}
    \widetilde{\br}_p(\bX_i-\bx,T_i,\bZ_i) = [\br_p(\bX_i - \bx)^{\top},T_i\cdot\br_p(\bX_i - \bx)^{\top},\bZ_i^\top]^\top
\end{align*}
contains the full polynomial basis function for each treatment group but the pre-intervention covariates are not interacted with the treatment indicator.

For heterogeneity analysis, the covariate-adjusted BATEC estimator is
\begin{align*}
    \check{\kappa}(\bx,\bz)
    = \be_{\mathfrak{p}_p+1}^{\top}\check{\bgamma}(\bx)
    + \bz^\top
      \begin{bmatrix}
        \mathbf{0}_{d_Z \times (2+d_Z)\mathfrak{p}_p}
        & \mathbf{I}_{d_Z} & \mathbf{0}_{d_Z \times (\mathfrak{p}_p-1)d_Z}
      \end{bmatrix}
      \check{\bgamma}(\bx)
\end{align*}
with
\begin{align*}
    \check{\bgamma}(\bx)
    = \argmin_{\bgamma \in \mathbb{R}^{2\mathfrak{p}_p(d_Z+1)}}
      \En \big[ \big(Y_i - \check{\br}_p(\bX_i-\bx,T_i,\bZ_i)^\top \bgamma \big)^2
                K_h(\bX_i - \bx) \big],
\end{align*}
where $\mathbf{I}_d$ denotes the $(d\times d)$ identity matrix, $\mathbf{0}_{d_1 \times d_2}$ denotes the $(d_1 \times d_2)$ matrix of zeros, $\bz$ takes values on the support of $\bZ_i$, and
\begin{align*}
    \check{\br}_p(\bX_i-\bx,T_i,\bZ_i)
     = [\br_p(\bX_i - \bx)^{\top},T_i\cdot\br_p(\bX_i - \bx)^{\top}, (\br_p(\bX_i - \bx)\otimes \bZ_i)^\top, T_i \cdot (\br_p(\bX_i - \bx)\otimes \bZ_i)^\top]^\top
\end{align*} contains the full interaction between the polynomial basis function, the treatment assignment indicator, and the pre-intervention covariates.

Employing the theoretical results in the supplemental appendix, it is not difficult to develop valid pointwise and uniform estimation and inference results for these covariate-adjusted BATEC estimators, transformations thereof, and their fuzzy counterparts. We plan to formalize these results in future research.

\section{Weighted Boundary Average Treatment Effect}

Without loss of generality, we set $\int_{\B} w(\bb) \dif\Haus^{d-1} (\bb) = 1$. For $|\bnu|\leq p$, the sharp (derivative) WBATE parameter is
\begin{align*}
    \tau_{\mathtt{WBATE}}^{(\bnu)} = \int_{\B} \tau^{(\bnu)}(\bb) w(\bb) \dif\Haus^{d-1}(\bb),
\end{align*}
where the weight function satisfies Assumption \ref{sa-assump: Weight Function and Boundary}. The original WBATE parameter is the special case $\bnu=\mathbf{0}$.

The sharp (derivative) WBATE estimator is
\begin{align*}
    \widehat{\tau}_{\mathtt{WBATE}}^{(\bnu)} = \int_\B \widehat{\tau}^{(\bnu)}(\bb) w(\bb) \dif\Haus^{d-1}(\bb),
\end{align*}

Our first lemma in this section studies the conditional bias of $\tau_{\mathtt{WBATE}}^{(\bnu)}$. Let
\begin{align*}
    B_{\mathtt{WBATE}}^{(\bnu)}(\bh) = B_{1,\mathtt{WBATE}}^{(\bnu)}(\bh) - B_{0,\mathtt{WBATE}}^{(\bnu)}(\bh),
    \qquad
    B_{t,\mathtt{WBATE}}^{(\bnu)}(\bh) = \int_{\B} B^{(\bnu)}_{t,\bb}(\bh) w(\bb) \dif\Haus^{d-1}(\bb),
\end{align*}
for $t\in\{0,1\}$.

\begin{lem}[Bias: Sharp WBATE]\label{sa-lem: Bias: Sharp WBATE}
    Suppose Assumption~\ref{sa-assump: Sharp DGP}(i)-(iii), \ref{sa-assump: Boundary and Kernel} and \ref{sa-assump: Weight Function and Boundary} hold. If $\frac{\log(1/\bar{h})}{n |\bH|} = o(1)$ and $\bar{h} = o(1)$, then
    \begin{align*}
        \E[\widehat{\tau}_{\mathtt{WBATE}}^{(\bnu)}|\bX] - \tau_{\mathtt{WBATE}}^{(\bnu)} = B_{\mathtt{WBATE}}^{(\bnu)}(\bh) + o_{\P}(\bar{h}^{p+1}\bh^{-\bnu}).
    \end{align*}
    If, in addition, $\bh = \bc h$ for a fixed vector $\bc$ and common bandwidth $h$, then $B_{\mathtt{WBATE}}^{(\bnu)}(\bh) = h^{p+1-|\bnu|}\widetilde{B}_{\mathtt{WBATE}}^{(\bnu)}(\bc) + o(h^{p+1-|\bnu|})$.
\end{lem}

The next lemma studies the conditional variance of $\tau_{\mathtt{WBATE}}^{(\bnu)}$, and a plug-in estimator thereof. Let
\begin{align*}
    \Omega_{\mathtt{WBATE}}^{(\bnu)} = \Omega_{1,\mathtt{WBATE}}^{(\bnu)} + \Omega_{0,\mathtt{WBATE}}^{(\bnu)},
    \qquad
    \Omega_{t,\mathtt{WBATE}}^{(\bnu)} = \int_{\B} \int_{\B} \Omega^{(\bnu)}_{t,\bb_1,\bb_2} w(\bb_1) w(\bb_2) \dif\Haus^{d-1}(\bb_1) \dif\Haus^{d-1}(\bb_2)
\end{align*}
and impose the aggregate variance nondegeneracy condition
\begin{align}
    \Omega_{\mathtt{WBATE}}^{(\bnu)}
    \gtrsim \frac{\bar{h}^{d-1}}{n|\bH|\bh^{2\bnu}}.
    \label{sa-eq: sharp WBATE aggregate nondegeneracy}
\end{align}
This scalar condition rules out cancellation of the local covariance kernel after integrating the weighted boundary process; it is automatically distinct from the pointwise variance lower bound when derivative aggregates are considered.
For the aggregate variance and inference results below, we also use the local boundary-measure regularity condition
\begin{align}
    \sup_{\bx\in\B}\Haus^{d-1}(\B\cap B(\bx,r)) \leq C_{\B}r^{d-1},
    \qquad 0<r\leq r_0,
    \label{sa-eq: local boundary measure regularity}
\end{align}
for some constants $C_{\B}<\infty$ and $r_0>0$. In the proofs below, write $\nu_{\B,k}$ for the $k$-fold product measure $(\Haus^{d-1})^k$ on $\B^k$.
Also let
\begin{align*}
    \widehat{\Omega}_{\mathtt{WBATE}}^{(\bnu)} = \widehat{\Omega}_{1,\mathtt{WBATE}}^{(\bnu)} + \widehat{\Omega}_{0,\mathtt{WBATE}}^{(\bnu)},
    \qquad
    \widehat{\Omega}_{t,\mathtt{WBATE}}^{(\bnu)} = \int_{\B} \int_{\B} \widehat{\Omega}^{(\bnu)}_{t,\bb_1, \bb_2} w(\bb_1) w(\bb_2) \dif\Haus^{d-1}(\bb_1) \dif\Haus^{d-1}(\bb_2),
\end{align*}
for $t\in\{0,1\}$.

\begin{lem}[Variance: Sharp WBATE]\label{sa-lem: Variance: Sharp WBATE}
    Suppose Assumptions \ref{sa-assump: Sharp DGP}, \ref{sa-assump: Boundary and Kernel} and \ref{sa-assump: Weight Function and Boundary} hold, and \eqref{sa-eq: sharp WBATE aggregate nondegeneracy} and \eqref{sa-eq: local boundary measure regularity} hold. If $\frac{\log(1/\bar{h})}{n |\bH|} = o(1)$ and $\bar{h} = o(1)$, then
    \begin{align*}
        \Var[\widehat{\tau}_{\mathtt{WBATE}}^{(\bnu)}|\bX]
        = \Omega_{\mathtt{WBATE}}^{(\bnu)} + o_{\P}\Big(\frac{\bar{h}^{d-1}}{n|\bH|\bh^{2\bnu}}\Big),
    \end{align*}
    where
    \begin{align*}
        \frac{\bar{h}^{d-1}}{n|\bH|\bh^{2\bnu}} \lesssim \Omega_{\mathtt{WBATE}}^{(\bnu)} \lesssim \frac{\bar{h}^{d-1}}{n|\bH|\bh^{2\bnu}}.
    \end{align*}

    Furthermore,
    \begin{align*}
        \Var[\widehat{\tau}_{\mathtt{WBATE}}^{(\bnu)}|\bX] & = \widehat{\Omega}_{\mathtt{WBATE}}^{(\bnu)} + o_{\P}\Big(\frac{\bar{h}^{d-1}}{n|\bH|\bh^{2\bnu}}\Big).
    \end{align*}
\end{lem}

\begin{thm}[MSE Expansion: Sharp WBATE]\label{sa-thm: MSE Expansion: Sharp WBATE}
    Suppose Assumptions \ref{sa-assump: Sharp DGP}, \ref{sa-assump: Boundary and Kernel} and \ref{sa-assump: Weight Function and Boundary} hold, and \eqref{sa-eq: sharp WBATE aggregate nondegeneracy} and \eqref{sa-eq: local boundary measure regularity} hold. If $\log(1/\bar{h})/(n^{v/(2+v)}|\bH|) = o(1)$ and $\bar{h} = o(1)$, then
    \begin{align*}
        \E [(\widehat{\tau}_{\mathtt{WBATE}}^{(\bnu)} - \tau_{\mathtt{WBATE}}^{(\bnu)})^2|\bX]
        = \Omega_{\mathtt{WBATE}}^{(\bnu)} + \big(B_{\mathtt{WBATE}}^{(\bnu)}(\bh)\big)^2
          + o_{\P}\Big(\frac{\bar{h}^{d-1}}{n|\bH|\bh^{2\bnu}}\Big) + o_{\P}(\bar{h}^{2p+2}\bh^{-2\bnu}).
    \end{align*}
\end{thm}

MSE-optimal diagonal bandwidth selection follows directly from Theorem \ref{sa-thm: MSE Expansion: Sharp WBATE}; scalar bandwidth selectors are obtained as the proportional-bandwidth special case $h_j=c_jh$.

For inference, we consider the feasible $t$-statistics
\begin{align*}
    \widehat{\Tstat}_{\mathtt{WBATE}}^{(\bnu)} = \frac{\widehat{\tau}_{\mathtt{WBATE}}^{(\bnu)} - \tau_{\mathtt{WBATE}}^{(\bnu)}}{\sqrt{\widehat{\Omega}_{\mathtt{WBATE}}^{(\bnu)}}}.
\end{align*}

\begin{thm}[Asymptotic Normality: Sharp WBATE]\label{sa-thm: Asymptotic Normality: Sharp WBATE}
    Suppose Assumptions \ref{sa-assump: Sharp DGP}, \ref{sa-assump: Boundary and Kernel} and \ref{sa-assump: Weight Function and Boundary} hold, and \eqref{sa-eq: sharp WBATE aggregate nondegeneracy} and \eqref{sa-eq: local boundary measure regularity} hold. If $\log(1/\bar{h})/(n^{v/(2+v)}|\bH|) = o(1)$ and $n|\bH|\bar{h}^{2p+2}/\bar{h}^{d-1} = o(1)$, then
    \begin{align*}
        \sup_{u \in \reals} \big|\P (\widehat{\Tstat}_{\mathtt{WBATE}}^{(\bnu)} \leq u )  - \Phi(u) \big| = o(1).
    \end{align*}
\end{thm}

Condition~\eqref{sa-eq: local boundary measure regularity} controls the amount of boundary measure in bandwidth-sized neighborhoods, which is what gives the $\bar h^{d-1}$ effective integration factor in aggregate variances. The piecewise-linear boundary and scalar bandwidth assumptions used in the main paper imply this condition.

\section{Largest Boundary Average Treatment Effect}

For $|\bnu|\leq p$, we consider the sharp (derivative) LBATE estimand and estimator
\begin{align*}
    \tau_{\mathtt{LBATE}}^{(\bnu)} = \sup_{\bb \in \B} \tau^{(\bnu)}(\bb)
    \qquad\text{and}\qquad
    \widehat{\tau}_{\mathtt{LBATE}}^{(\bnu)} = \sup_{\bb \in \B} \widehat{\tau}^{(\bnu)}(\bb).
\end{align*}
The original LBATE parameter is the special case $\bnu=\mathbf{0}$.

\begin{thm}[Convergence Rate: Sharp LBATE]\label{sa-thm: Convergence Rate: Sharp LBATE}
    Suppose Assumptions \ref{sa-assump: Sharp DGP} and \ref{sa-assump: Boundary and Kernel} hold. If $\log(1/\bar{h})/(n |\bH|) = o(1)$ and $\bar{h} = o(1)$, then
    \begin{align*}
        \big|\widehat{\tau}_{\mathtt{LBATE}}^{(\bnu)} - \tau_{\mathtt{LBATE}}^{(\bnu)} \big|
        \lessp \bh^{-\bnu}\Big(\sqrt{\frac{\log(1/\bar{h})}{ n |\bH|}} + \frac{\log(1/\bar{h})}{n^{\frac{1+v}{2+v}}|\bH|} + \bar{h}^{p+1}\Big).
    \end{align*}
\end{thm}

Recall from Theorem \ref{sa-thm: Confidence Bands: Sharp BATEC -- Suprema} that $(\widehat{Z}^{(\bnu)}(\bx):\bx \in \B)$ is a (conditionally on $\bV$) mean-zero Gaussian process with feasible (conditional) covariance function
\begin{align*}
    \Cov \Big[\widehat{Z}^{(\bnu)}(\bx_1),\widehat{Z}^{(\bnu)}(\bx_2) \Big|\bV \Big]
    = \frac{\widehat{\Omega}_{\bx_1,\bx_2}^{(\bnu)}}
           {\sqrt{\widehat{\Omega}_{\bx_1,\bx_1}^{(\bnu)} \widehat{\Omega}_{\bx_2,\bx_2}^{(\bnu)}}},
    \qquad \bx_1, \bx_2 \in \B.
\end{align*}
Consider the confidence interval
\begin{align*}
    \widehat{\CI}_{\alpha,\mathtt{LBATE}}^{(\bnu)} = \bigg[\sup_{\bb \in \B} \Big(\widehat{\tau}^{(\bnu)}(\bb)  - \q_{\alpha} \sqrt{\widehat{\Omega}_{\bb,\bb}^{(\bnu)}}\Big), \; \sup_{\bb \in \B} \Big(\widehat{\tau}^{(\bnu)}(\bb) + \q_{\alpha} \sqrt{\widehat{\Omega}_{\bb,\bb}^{(\bnu)}}\Big)\bigg],
\end{align*}
where $\q_{\alpha} = \inf \big\{c > 0: \P \big(\sup_{\bx \in \B} \big|\widehat{Z}^{(\bnu)}(\bx)\big|\geq c \big| \bV \big) \leq \alpha \big\}$.

\begin{thm}[Confidence Interval: Sharp LBATE]\label{sa-thm: Confidence Intervals: Sharp LBATE}
    Suppose the assumptions and conditions in Theorem \ref{sa-thm: Confidence Bands: Sharp BATEC -- Suprema} hold. Then
    \begin{align*}
            \P\Big[\tau_{\mathtt{LBATE}}^{(\bnu)} \in \widehat{\CI}_{\alpha,\mathtt{LBATE}}^{(\bnu)} \Big] \geq 1 - \alpha + o(1).
    \end{align*}
\end{thm}

\section{Imperfect Compliance}\label{sa-sec: Imperfect Compliance}

We extend our previous results to settings with imperfect treatment compliance, where treatment assignment and treatment status may not be equal for some units. See, e.g., \cite{Hernan-Robins_2020_Book} for a textbook introduction.

We generalize our notation and assumptions to explicitly account for imperfect treatment compliance. Let
\begin{align*}
    W_i = \Indicator(\bX_i \in \A_0) \cdot W_i(0) + \Indicator(\bX_i \in \A_1) \cdot W_i(1)
\end{align*}
be the observed treatment status, where $W_i(t)$ denotes the potential treatment status under treatment assignment $t\in\{0,1\}$ for each unit. Accordingly, the observed outcome now is
\begin{align*}
    Y_i = \Indicator(\bX_i \in \A_0) \cdot Y_i(0,W_i(0)) + \Indicator(\bX_i \in \A_1) \cdot Y_i(1,W_i(1)),
\end{align*}
where the potential outcomes are now functions of two arguments: $Y_i(t,r)$ denotes the potential outcome for unit $i$ when this unit is assigned to treatment $t\in\{0,1\}$ and takes treatment status $r\in\{0,1\}$.

The usual (fuzzy) estimands and estimators in settings with imperfect treatment compliance are, respectively,
\begin{align*}
    \zeta(\bx) = \frac{\tau_{Y}(\bx)}{\tau_{W}(\bx)}
    \qquad\text{and}\qquad
    \widehat{\zeta}(\bx) = \frac{\widehat{\tau}_{Y}(\bx)}{\widehat{\tau}_{W}(\bx)},
\end{align*}
where, for each $\bx \in \B$,
\begin{align*}
    \tau_{Y}(\bx) = \E[Y_i(1,W_i(1)) - Y_i(0,W_i(0)) | \bX_i = \bx]
    \qquad\text{and}\qquad
    \tau_{W}(\bx) = \E[W_i(1) - W_i(0) | \bX_i = \bx],
\end{align*}
and
\begin{align*}
    \widehat{\tau}_{Y}(\bx) = \be_{\mathbf{0}}^{\top}\widehat{\bbeta}_{Y,1}(\bx) - \be_{\mathbf{0}}^{\top}\widehat{\bbeta}_{Y,0}(\bx)
    \qquad\text{and}\qquad
    \widehat{\tau}_{W}(\bx) = \be_{\mathbf{0}}^{\top}\widehat{\bbeta}_{W,1}(\bx) - \be_{\mathbf{0}}^{\top}\widehat{\bbeta}_{W,0}(\bx)
\end{align*}
with $\widehat{\bbeta}_{A,t}(\bx)$ denoting the local polynomial fit \eqref{sa-eq: locpoly estimator} when using outcome variable $A \in \{Y,W\}$ on side $t\in\{0,1\}$.

To recycle the notation and assumptions from the sharp BD design setup, we redefine $Y_i(t) = Y_i(t,W_i(t))$ for each $t\in\{0,1\}$, which enables us to reuse Assumption \ref{sa-assump: Sharp DGP} in the fuzzy BD design setting. We also impose the following analogous assumption for the reduced-form potential outcome/status pairs.

\begin{assumption}[Fuzzy DGP]\label{sa-assump: Fuzzy DGP}
    Let $t\in\{0,1\}$.
    \begin{enumerate}[label=\normalfont(\roman*),noitemsep,leftmargin=*]

    \item $(Y_1(t), W_1(t), \bX_1^\top)^\top,\ldots, (Y_n(t), W_n(t), \bX_n^\top)^\top$ are independent and identically distributed random vectors.

    \item $\mu_{W,t}(\bx) = \E[W_i(t)| \bX_i = \bx]$ has a $(p+1)$-times continuously differentiable extension to an open neighborhood of $\X$.

    \item $\bUpsilon_{t}(\bx) = \Cov[(Y_i(t), W_i(t))^{\top}|\bX_i = \bx]$ has entries that are continuous on $\X$.

    \end{enumerate}
\end{assumption}

More generally, for $|\bnu|\leq p$, we consider Wald-type ratios of derivative discontinuities of the form
\begin{align*}
    \zeta^{(\bnu)}(\bx) = \frac{\tau^{(\bnu)}_{Y}(\bx)}{\tau^{(\bnu)}_{W}(\bx)}
    \qquad\text{and}\qquad
    \widehat{\zeta}^{(\bnu)}(\bx) = \frac{\widehat{\tau}^{(\bnu)}_{Y}(\bx)}{\widehat{\tau}^{(\bnu)}_{W}(\bx)}.
\end{align*}
For $\bnu=\mathbf{0}$ this is the usual fuzzy Wald curve. For $|\bnu|>0$, it is a ratio of derivative jumps, not the derivative of the level Wald ratio. Here, for each $\bx \in \B$,
\begin{align*}
    \tau^{(\bnu)}_{Y}(\bx) = \mu^{(\bnu)}_{Y,1}(\bx) - \mu^{(\bnu)}_{Y,0}(\bx)
    \qquad\text{and}\qquad
    \tau^{(\bnu)}_{W}(\bx) = \mu^{(\bnu)}_{W,1}(\bx) - \mu^{(\bnu)}_{W,0}(\bx),
\end{align*}
and
\begin{align*}
    \widehat{\tau}^{(\bnu)}_{Y}(\bx)
    = \bnu!\be_{\bnu}^{\top}\widehat{\bbeta}_{Y,1}(\bx) - \bnu!\be_{\bnu}^{\top}\widehat{\bbeta}_{Y,0}(\bx)
    \qquad\text{and}\qquad
    \widehat{\tau}_{W}^{(\bnu)}(\bx)
    = \bnu!\be_{\bnu}^{\top}\widehat{\bbeta}_{W,1}(\bx) - \bnu!\be_{\bnu}^{\top}\widehat{\bbeta}_{W,0}(\bx)
\end{align*}
with $\widehat{\bbeta}_{A,t}(\bx)$ denoting the local polynomial fit \eqref{sa-eq: locpoly estimator} when using outcome variable $A \in \{Y,W\}$ on side $t\in\{0,1\}$. That is, under the assumptions imposed,
\begin{align*}
    \widehat{\bbeta}_{A,t}(\bx)
     = \bS^{-1} \widehat{\bGamma}_{t,\bx}^{-1} \En \big[\br_p(\bH^{-1}(\bX_i-\bx)) K_{\bH}(\bX_i - \bx) A_i \Indicator(\bX_i \in \A_t) \big],
    \qquad A\in\{Y,W\},
\end{align*}
corresponds to the local polynomial regression fit when using either $Y_i$ or $W_i$ as outcome. We also define
\begin{align*}
    \bQ_{A,t,\bx} = \En \big[\br_p(\bH^{-1}(\bX_i - \bx)) K_{\bH}(\bX_i - \bx) \Indicator(\bX_i \in \A_t) \varepsilon_{A,i}\big],
\end{align*}
where $\varepsilon_{A,i} = A_i - [\Indicator(\bX_i \in \A_0) \mu_{A,0}(\bX_i) + \Indicator(\bX_i \in \A_1) \mu_{A,1}(\bX_i)] = A_i - \E[A_i|\bX_i]$, where $\mu_{A,t}(\bX_i) = \E[A_i(t)|\bX_i]$ and $A\in\{Y,W\}$. Let $\bbeta_{A,t}(\bx)$ denote the corresponding population Taylor coefficient vector satisfying
\begin{align*}
    \bbeta_{A,t}(\bx)^\top\br_p(\bu)
    = \sum_{|\bomega|\leq p}\frac{\mu_{A,t}^{(\bomega)}(\bx)}{\bomega!}\bu^{\bomega},
    \qquad \bu\in\reals^d.
\end{align*}
Finally, for $A_1,A_2\in\{Y,W\}$, $\bx_1, \bx_2 \in \B$ and $t \in \{0,1\}$, we introduce the following population quantities
\begin{align*}
    \Omega_{A_1,A_2,\bx_1,\bx_2}^{(\bnu)}
    = \Omega_{A_1,A_2,0,\bx_1,\bx_2}^{(\bnu)} + \Omega_{A_1,A_2,1,\bx_1,\bx_2}^{(\bnu)},
    \qquad
    \Omega_{A_1,A_2,t,\bx_1,\bx_2}^{(\bnu)}
    = \frac{(\bnu!)^2}{n |\bH|\bh^{2\bnu}} \be_{\bnu}^{\top}  \bGamma_{t,\bx_1}^{-1} \bSigma_{A_1,A_2,t,\bx_1,\bx_2} \bGamma_{t,\bx_2}^{-1} \be_{\bnu},
\end{align*}
where
\begin{align*}
    \bSigma_{A_1,A_2,t, \bx_1,\bx_2}
    &= |\bH| \E \big[\br_p(\bH^{-1}(\bX_i-\bx_1)) \br_p(\bH^{-1}(\bX_i-\bx_2))^{\top}\\
    &\hspace{1in} \cdot K_{\bH}(\bX_i - \bx_1) K_{\bH}(\bX_i - \bx_2) \sigma_{A_1,A_2,t}^2(\bX_i) \Indicator(\bX_i \in \A_t)\big],
\end{align*}
with $\sigma_{A_1,A_2,t}^2(\bX_i) = \Cov(A_1(t),A_2(t)|\bX_i)$, and their sample analogs
\begin{align*}
    \widehat{\Omega}_{A_1,A_2,\bx_1,\bx_2}^{(\bnu)}
    = \widehat{\Omega}_{A_1,A_2,0,\bx_1,\bx_2}^{(\bnu)} + \widehat{\Omega}_{A_1,A_2,1,\bx_1, \bx_2}^{(\bnu)},
    \qquad
    \widehat{\Omega}_{A_1,A_2,t,\bx_1,\bx_2}^{(\bnu)}
    = \frac{(\bnu!)^2}{n |\bH|\bh^{2\bnu}} \be_{\bnu}^{\top} \widehat{\bGamma}_{t,\bx_1}^{-1} \widehat{\bSigma}_{A_1,A_2,t,\bx_1,\bx_2} \widehat{\bGamma}_{t,\bx_2}^{-1} \be_{\bnu},
\end{align*}
where
\begin{align*}
    \widehat{\bSigma}_{A_1,A_2,t,\bx_1,\bx_2}
    &= |\bH| \En \big[\br_p(\bH^{-1}(\bX_i-\bx_1)) \br_p(\bH^{-1}(\bX_i-\bx_2))^{\top}\\
    &\hspace{1in} \cdot K_{\bH}(\bX_i - \bx_1) K_{\bH}(\bX_i - \bx_2) \widehat{\varepsilon}_{A_1,i}(\bx_1) \widehat{\varepsilon}_{A_2,i}(\bx_2) \Indicator(\bX_i \in \A_t)\big],
\end{align*}
where $\widehat{\varepsilon}_{A,i}(\bx) = A_i - \br_p(\bH^{-1}(\bX_i - \bx))^\top\bS[\Indicator(\bX_i \in \A_0) \widehat{\bbeta}_{A,0}(\bx) + \Indicator(\bX_i \in \A_1) \widehat{\bbeta}_{A,1}(\bx)]$, with $A\in\{Y,W\}$.

\subsection{Second-order Linearization}\label{sec: Second-Order Linearization: Fuzzy BATEC}

The results in this section will follow from the following exact second-order linearization:
\begin{align*}
    \widehat{\zeta}^{(\bnu)}(\bx) - \zeta^{(\bnu)}(\bx)
    &= \mathfrak{w}(\bx)^\top \mathfrak{T}(\bx) + \mathfrak{R}_{\mathtt{F}}(\bx)
\end{align*}
with
\begin{align*}
    \mathfrak{w}(\bx)
    &= \Big[ \frac{1}{\tau^{(\bnu)}_W(\bx)}, -\frac{\tau^{(\bnu)}_Y(\bx)}{\tau^{(\bnu)}_W(\bx)^{2}} \Big]^\top
    \qquad\text{and}\qquad
    \mathfrak{T}(\bx)
    = \Big[ \widehat{\tau}^{(\bnu)}_Y(\bx) - \tau^{(\bnu)}_Y(\bx), \widehat{\tau}^{(\bnu)}_W(\bx) - \tau^{(\bnu)}_W(\bx) \Big]^\top,
\end{align*}
and second-order error
\begin{align*}
    \mathfrak{R}_{\mathtt{F}}(\bx)
    &= \frac{\tau^{(\bnu)}_Y(\bx)}{\tau^{(\bnu)}_W(\bx)^{2}\widehat{\tau}^{(\bnu)}_W(\bx)}
       \big(\widehat{\tau}^{(\bnu)}_W(\bx) - \tau^{(\bnu)}_W(\bx)\big)^2
     - \frac{1}{\tau^{(\bnu)}_W(\bx)\widehat{\tau}^{(\bnu)}_W(\bx)}
       \big(\widehat{\tau}^{(\bnu)}_Y(\bx) - \tau^{(\bnu)}_Y(\bx) \big)
       \big(\widehat{\tau}^{(\bnu)}_W(\bx) - \tau^{(\bnu)}_W(\bx) \big).
\end{align*}

For example, under the assumptions imposed, if $\inf_{\bx\in\B} |\tau^{(\bnu)}_W(\bx)| > 0$, $\log(1/\bar{h})/(n^{\frac{1+v}{2+v}}|\bH|)=o(1)$, $\bar{h}=o(1)$, and
\begin{align*}
    \bh^{-\bnu}\bigg(\sqrt{\frac{\log(1/\bar{h})}{n|\bH|}}+\frac{\log(1/\bar{h})}{n^{\frac{1+v}{2+v}}|\bH|}+\bar{h}^{p+1}\bigg)=o(1),
\end{align*}
then two applications of Theorem \ref{sa-thm: Convergence Rates: Sharp BATEC} (taking as outcomes $Y$ and $W$) imply
\begin{align}\label{sa-eq: second order term -- fuzzy}
    \nonumber \sup_{\bx\in\B} |\mathfrak{R}_{\mathtt{F}}(\bx)|
    &\lesssim_\P \sup_{\bx\in\B} |\widehat{\tau}^{(\bnu)}_W(\bx) - \tau^{(\bnu)}_W(\bx)|^2
                + \sup_{\bx\in\B} |\widehat{\tau}^{(\bnu)}_Y(\bx) - \tau^{(\bnu)}_Y(\bx)|
                  \sup_{\bx\in\B} |\widehat{\tau}^{(\bnu)}_W(\bx) - \tau^{(\bnu)}_W(\bx)|\\
    &\lesssim_\P \bh^{-2\bnu} \Big(\sqrt{\frac{\log(1/\bar{h})}{ n |\bH|}} + \frac{\log(1/\bar{h})}{n^{\frac{1+v}{2+v}}|\bH|} + \bar{h}^{p+1}\Big)^2,
\end{align}
thereby demonstrating that the second-order term $\mathfrak{R}_{\mathtt{F}}(\bx)$ is (uniformly) higher-order relative to the leading term $\mathfrak{w}(\bx)^\top \mathfrak{T}(\bx)$.

\subsection{Fuzzy BATEC}

\subsubsection{Point Estimation and AMSE Expansions}

\begin{thm}[Convergence Rates: Fuzzy BATEC]\label{sa-thm: Convergence Rates: Fuzzy BATEC}
    Suppose Assumptions \ref{sa-assump: Sharp DGP}, \ref{sa-assump: Boundary and Kernel}, and \ref{sa-assump: Fuzzy DGP} hold, and $\inf_{\bx\in\B} |\tau^{(\bnu)}_W(\bx)| > 0$. If $\log(1/\bar{h})/(n^{\frac{1+v}{2+v}} |\bH|) = o(1)$, $\bar{h} = o(1)$, and
    \begin{align*}
        \bh^{-\bnu}\bigg(\sqrt{\frac{\log(1/\bar{h})}{n|\bH|}}+\frac{\log(1/\bar{h})}{n^{\frac{1+v}{2+v}}|\bH|}+\bar{h}^{p+1}\bigg)=o(1),
    \end{align*}
    then
    \begin{align*}
        \big| \widehat{\zeta}^{(\bnu)}(\bx) - \zeta^{(\bnu)}(\bx) \big|
        \lessp \bh^{-\bnu} \Big(\frac{1}{\sqrt{n |\bH|}} + \frac{1}{n^{\frac{1+v}{2+v}}|\bH|} + \bar{h}^{p+1}\Big)
    \end{align*}
    for $\bx \in \B$, and
    \begin{align*}
        \sup_{\bx \in \B} \big| \widehat{\zeta}^{(\bnu)}(\bx) - \zeta^{(\bnu)}(\bx) \big|
        \lessp \bh^{-\bnu} \Big(\sqrt{\frac{\log(1/\bar{h})}{ n |\bH|}} + \frac{\log(1/\bar{h})}{n^{\frac{1+v}{2+v}}|\bH|} + \bar{h}^{p+1}\Big).
    \end{align*}
\end{thm}

The approximate (conditional) mean-squared error (AMSE) is
\begin{align*}
    \operatorname{AMSE}_{\mathtt{F}}^{(\bnu)}(\bx)
    = \E \big[(\mathfrak{w}(\bx)^\top \mathfrak{T}(\bx))^2 \big| \bX \big]
    = \mathfrak{w}(\bx)^\top \E \big[\mathfrak{T}(\bx) \mathfrak{T}(\bx)^\top \big| \bX \big] \mathfrak{w}(\bx)
\end{align*}
for $\bx\in \B$, and the approximate (conditional) integrated MSE (AIMSE) is
\begin{align*}
    & \operatorname{AIMSE}_{\mathtt{F}}^{(\bnu)} = \int_{\B}  \operatorname{AMSE}_{\mathtt{F}}^{(\bnu)}(\bx) w(\bx) \dif\Haus^{d-1}(\bx),
\end{align*}
where $w(\bx)$ satisfies Assumption \ref{sa-assump: Weight Function and Boundary}. These constructions disregard the higher-order term $\mathfrak{R}_{\mathtt{F}}(\bx)$.

To state the AMSE expansions with imperfect compliance, we generalize the notation introduced previously for the leading bias and variance. For the bias, letting $A\in\{Y,W\}$, we define
\begin{align*}
    B_{A,\bx}^{(\bnu)}(\bh) = B_{A,1,\bx}^{(\bnu)}(\bh) - B_{A,0,\bx}^{(\bnu)}(\bh),
\end{align*}
\begin{align*}
    B_{A,t,\bx}^{(\bnu)}(\bh)
    = \bnu!\be_{\bnu}^{\top} \bGamma_{t,\bx}^{-1} \sum_{|\bomega| = p + 1} \frac{\mu_{A,t}^{(\bomega)}(\bx)}{\bomega!}\bh^{\bomega-\bnu}
      \E \big[\br_p(\bH^{-1}(\bX_i-\bx)) (\bH^{-1}(\bX_i - \bx))^{\bomega} K_{\bH}(\bX_i - \bx) \Indicator(\bX_i\in\A_t) \big],
\end{align*}
where recall that $\mu_{A,t}(\bx) = \E[A_i(t)|\bX_i=\bx]$. Therefore, the leading bias takes the form:
\begin{align}\label{sa-eq: leading bias fuzzy}
    B_{\mathtt{F},\bx}^{(\bnu)}(\bh)
    = \mathfrak{w}(\bx)^\top \bB_{\bx}^{(\bnu)}(\bh),
    \qquad
    \bB_{\bx}^{(\bnu)}(\bh) = \big( B_{Y,\bx}^{(\bnu)}(\bh), B_{W,\bx}^{(\bnu)}(\bh) \big)^\top.
\end{align}
When $\bh = \bc h$ for a fixed vector $\bc$ and $h$ is a common bandwidth, $B_{\mathtt{F},\bx}^{(\bnu)}(\bh)=h^{p+1-|\bnu|}\widetilde B_{\mathtt{F},\bx}^{(\bnu)}(\bc)$ for some $\widetilde B_{\mathtt{F},\bx}^{(\bnu)}(\bc)$.

For the variance, letting $A_1,A_2\in\{Y,W\}$, we define
\begin{align*}
    V_{A_1,A_2,\bx}^{(\bnu)}  = V_{A_1,A_2,0,\bx}^{(\bnu)} + V_{A_1,A_2,1,\bx}^{(\bnu)},
\end{align*}
where
\begin{align*}
    V_{A_1,A_2,t,\bx}^{(\bnu)}
    = (\bnu!)^2\be_{\bnu}^{\top} \bGamma_{t,\bx}^{-1} \bSigma_{A_1,A_2,t,\bx,\bx} \bGamma_{t,\bx}^{-1}\be_{\bnu}
    = n|\bH|\bh^{2\bnu} \Omega_{A_1,A_2,t,\bx,\bx}^{(\bnu)}.
\end{align*}
Therefore, the leading variance takes the form:
\begin{align*}
    V_{\mathtt{F},\bx}^{(\bnu)}
    = \mathfrak{w}(\bx)^\top \mathbf{C}_{\bx}^{(\bnu)} \mathfrak{w}(\bx),
    \qquad
    \mathbf{C}_{\bx}^{(\bnu)}
    = \begin{pmatrix}
        V_{Y,Y,\bx}^{(\bnu)} & V_{Y,W,\bx}^{(\bnu)} \\
        V_{W,Y,\bx}^{(\bnu)} & V_{W,W,\bx}^{(\bnu)}
      \end{pmatrix}.
\end{align*}
For inference, we impose $\inf_{\bx\in\B}V_{\mathtt{F},\bx}^{(\bnu)}>0$ directly. This scalar nondegeneracy condition applies to the linearized fuzzy Wald score and does not require the reduced-form conditional covariance matrix of $(Y_i(t),W_i(t))$ to be uniformly positive definite.

\begin{thm}[AMSE Expansions: Fuzzy BATEC]\label{sa-thm: MSE Expansions: Fuzzy BATEC}
    Suppose Assumptions \ref{sa-assump: Sharp DGP}, \ref{sa-assump: Boundary and Kernel}, \ref{sa-assump: Weight Function and Boundary} and \ref{sa-assump: Fuzzy DGP} hold, and $\inf_{\bx\in\B}|\tau_W^{(\bnu)}(\bx)|>0$. If $\log(1/\bar{h})/(n^{\frac{v}{2+v}} |\bH|) = o(1)$ and $\bar{h} = o(1)$, then
    \begin{align*}
        \operatorname{AMSE}_{\mathtt{F}}^{(\bnu)}(\bx)
        = \big(B_{\mathtt{F},\bx}^{(\bnu)}(\bh)\big)^2
          + \frac{1}{n|\bH|\bh^{2\bnu}} V_{\mathtt{F},\bx}^{(\bnu)}
          + o_{\P} \big(\bar{h}^{2p+2}\bh^{-2\bnu} + (n|\bH|\bh^{2\bnu})^{-1}\big)
    \end{align*}
    for $\bx \in \B$, and
    \begin{align*}
        \operatorname{AIMSE}_{\mathtt{F}}^{(\bnu)}
        = \int_{\B} \Big[ \big(B_{\mathtt{F},\bx}^{(\bnu)}(\bh)\big)^2
           + \frac{1}{n|\bH|\bh^{2\bnu}} V_{\mathtt{F},\bx}^{(\bnu)} \Big] w(\bx) \dif\Haus^{d-1}(\bx)
           + o_{\P} \big(\bar{h}^{2p+2}\bh^{-2\bnu} + (n|\bH|\bh^{2\bnu})^{-1}\big).
    \end{align*}
\end{thm}

Theorem \ref{sa-thm: MSE Expansions: Fuzzy BATEC} can be used to develop feasible bandwidth selectors for diagonal bandwidth matrices by minimizing the displayed criterion over $\bh\in\mathbb{R}_{++}^d$. If $\bh=\bc h$ for a fixed vector $\bc$ and a common bandwidth $h$, the problem reduces to scalar bandwidth selection. If $\widetilde B_{\mathtt{F},\bx}^{(\bnu)}(\bc) \neq 0$, the approximate MSE-optimal common scale is
\begin{align*}
    h_{\mathtt{F},\operatorname{AMSE},\bnu}(\bx)
    = \left(\frac{(d + 2|\bnu|) V_{\mathtt{F},\bx}^{(\bnu)}}
                 {(2p+2 - 2 |\bnu|) (\widetilde B_{\mathtt{F},\bx}^{(\bnu)}(\bc))^2} \frac{1}{n}\right)^{\frac{1}{2p+d+2}}
\end{align*}
for $\bx \in \B$. Similarly, if $\int_{\B} (\widetilde B_{\mathtt{F},\bx}^{(\bnu)}(\bc))^2 w(\bx)\dif\Haus^{d-1}(\bx) \neq 0$, the approximate IMSE-optimal common scale is
\begin{align*}
    h_{\mathtt{F},\operatorname{AIMSE},\bnu}
    = \left(\frac{(d + 2 |\bnu|) \int_{\B} V_{\mathtt{F},\bx}^{(\bnu)} w(\bx)\dif\Haus^{d-1}(\bx)}
                 {(2p+2 - 2|\bnu|) \int_{\B} (\widetilde B_{\mathtt{F},\bx}^{(\bnu)}(\bc))^2 w(\bx) \dif\Haus^{d-1}(\bx)} \frac{1}{n}\right)^{\frac{1}{2p+d+2}}.
\end{align*}

In practice, the unknown bias and variance quantities can be replaced with (consistent) estimators thereof, as discussed before.

\subsubsection{Distributional Approximation and Inference}

Let $\bV = ((Y_1,W_1,\bX_1^{\top}), \cdots, (Y_n,W_n,\bX_n^{\top}))^{\top}$, and recall that $t \in \{0,1\}$. For $|\bnu| \leq p$, define the feasible $t$-statistic
\begin{align*}
    \widehat{\Tstat}_\mathtt{F}^{(\bnu)}(\bx)
    = \frac{\widehat{\zeta}^{(\bnu)}(\bx) - \zeta^{(\bnu)}(\bx)}{\sqrt{\widehat{\Omega}_{\mathtt{F},\bx,\bx}^{(\bnu)}}},
    \qquad \bx \in \B,
\end{align*}
where we define
\begin{align*}
    \widehat{\mathfrak{w}}(\bx)
    &= \Big[ \frac{1}{\widehat{\tau}^{(\bnu)}_W(\bx)}, -\frac{\widehat{\tau}^{(\bnu)}_Y(\bx)}{\big(\widehat{\tau}^{(\bnu)}_W(\bx)\big)^2} \Big]^\top,
\end{align*}
and
\begin{align*}
    \widehat{\Omega}_{\mathtt{F},\bx_1,\bx_2}^{(\bnu)}
    = \widehat{\mathfrak{w}}(\bx_1)^\top \widehat{\bTheta}_{\ttF, \bx_1, \bx_2}^{(\bnu)} \widehat{\mathfrak{w}}(\bx_2),
    \qquad
    \widehat{\bTheta}_{\ttF, \bx_1, \bx_2}^{(\bnu)}
    = \begin{pmatrix}
        \widehat{\Omega}_{Y,Y,\bx_1, \bx_2}^{(\bnu)} & \widehat{\Omega}_{Y,W,\bx_1, \bx_2}^{(\bnu)} \\
        \widehat{\Omega}_{W,Y,\bx_1, \bx_2}^{(\bnu)} & \widehat{\Omega}_{W,W,\bx_1, \bx_2}^{(\bnu)}
      \end{pmatrix}, \qquad \bx_1, \bx_2 \in \B.
\end{align*}
We also define the population version of the denominator by
\begin{align*}
    \Omega_{\mathtt{F},\bx_1, \bx_2}^{(\bnu)}
    = \mathfrak{w}(\bx_1)^\top \bTheta_{\ttF, \bx_1, \bx_2}^{(\bnu)} \mathfrak{w}(\bx_2),
    \qquad
    \bTheta_{\ttF, \bx_1, \bx_2}^{(\bnu)}
    = \begin{pmatrix}
        \Omega_{Y,Y,\bx_1, \bx_2}^{(\bnu)} & \Omega_{Y,W,\bx_1, \bx_2}^{(\bnu)} \\
        \Omega_{W,Y,\bx_1, \bx_2}^{(\bnu)} & \Omega_{W,W,\bx_1, \bx_2}^{(\bnu)}
      \end{pmatrix}, \qquad \bx_1, \bx_2 \in \B.
\end{align*}

\begin{lem}[Covariance: Fuzzy BATEC]\label{sa-lem: Covariance: Fuzzy BATEC}
    Suppose Assumptions \ref{sa-assump: Sharp DGP}, \ref{sa-assump: Boundary and Kernel} and \ref{sa-assump: Fuzzy DGP} hold, $\inf_{\bx\in\B}|\tau_W^{(\bnu)}(\bx)|>0$, and $\inf_{\bx\in\B}V_{\mathtt{F},\bx}^{(\bnu)}>0$. If $\frac{\log(1/\bar{h})}{n^{\frac{v}{2+v}} |\bH|} = o(1)$, $\bar{h}=o(1)$, and
    \begin{align*}
        \bh^{-\bnu}\bigg(\sqrt{\frac{\log(1/\bar{h})}{n |\bH|}}+\frac{\log(1/\bar{h})}{n^{\frac{v}{2+v}}|\bH|}+\bar{h}^{p+1}\bigg)=o(1),
    \end{align*}
    then
    \begin{gather*}
        \sup_{\bx_1,\bx_2 \in \B} \big|\widehat{\Omega}_{\ttF, \bx_1,\bx_2}^{(\bnu)} - \Omega_{\ttF, \bx_1,\bx_2}^{(\bnu)}\big|
        \lessp \frac{\bh^{-\bnu}}{n |\bH|\bh^{2\bnu}}\bigg(\sqrt{\frac{\log(1/\bar{h})}{n |\bH|}} + \frac{\log(1/\bar{h})}{n^{\frac{v}{2+v}}|\bH|}+ \bar{h}^{p+1}\bigg), \\
        \sup_{\bx \in \B} \big|(\widehat{\Omega}_{\ttF, \bx,\bx}^{(\bnu)})^{-1/2} - (\Omega_{\ttF, \bx,\bx}^{(\bnu)})^{-1/2} \big|
        \lessp \bh^{-\bnu}\sqrt{n |\bH|\bh^{2\bnu}}\bigg(\sqrt{\frac{\log(1/\bar{h})}{n |\bH|}} + \frac{\log(1/\bar{h})}{n^{\frac{v}{2+v}}|\bH|} + \bar{h}^{p+1}\bigg).
    \end{gather*}
\end{lem}

The associated $100(1-\alpha)\%$ confidence interval estimator is
\begin{align*}
    \widehat{\CI}^{(\bnu)}_{\mathtt{F},\alpha}(\bx)
    = \bigg[\;\widehat{\zeta}^{(\bnu)}(\bx) - \q_{\alpha} \sqrt{\widehat{\Omega}_{\mathtt{F},\bx,\bx}^{(\bnu)}}
            \; , \;
              \widehat{\zeta}^{(\bnu)}(\bx) + \q_{\alpha} \sqrt{\widehat{\Omega}_{\mathtt{F},\bx,\bx}^{(\bnu)}}\;\bigg],
\end{align*}
where $\q_{\alpha}$ denotes an appropriate quantile depending on the desired confidence level $\alpha\in(0,1)$, and coverage objective (pointwise versus uniform over $\B$).

The following theorem establishes pointwise asymptotic normality and validity of confidence intervals.

\begin{thm}[Confidence Intervals: Fuzzy BATEC]\label{sa-thm: Confidence Intervals: Fuzzy BATEC}
    For a fixed boundary point $\bx\in\B$, suppose Assumptions \ref{sa-assump: Sharp DGP}, \ref{sa-assump: Boundary and Kernel} and \ref{sa-assump: Fuzzy DGP} hold, $|\tau^{(\bnu)}_W(\bx)| > 0$, and $V_{\mathtt{F},\bx}^{(\bnu)}>0$. If $n |\bH| \to \infty$, $n|\bH|\bh^{2\bnu}\to\infty$, and $n |\bH| \bar{h}^{2(p+1)} \to 0$, then
    \begin{align*}
        \sup_{u \in \reals} \Big|\P \big(\widehat{\Tstat}_\mathtt{F}^{(\bnu)}(\bx) \leq u \big) - \Phi(u)\Big| = o(1),
    \end{align*}
    and
    \begin{align*}
        \P \big(\zeta^{(\bnu)}(\bx) \in \widehat{\CI}_{\mathtt{F},\alpha}^{(\bnu)}(\bx) \big) = 1 - \alpha + o(1),
    \end{align*}
    provided that $\q_{\alpha} = \inf \{c > 0: \P( |\widehat{Z}| \geq c | \bV) \leq \alpha \}$ with $\widehat{Z}|\bV \sim \mathsf{Normal}(0, 1)$.
\end{thm}

For uniform inference, we rely on the strong approximation result established in Section~\ref{sa-sec: Gaussian Strong Approximation}. Since $\widehat{\Tstat}^{(\bnu)}_{\mathtt{F}}$ is not directly a sum of independent terms, we introduce the linearized statistic
\begin{align*}
    \overline{\Tstat}_{\mathtt{F}}^{(\bnu)}(\bx)
    &=
    \E_n\Big[
    (\Omega_{\mathtt{F},\bx,\bx}^{(\bnu)})^{-1/2}
    \bnu!\be_{\bnu}^{\top}\bS^{-1}
    \big[\Indicator(\bX_i\in\A_1)\bGamma_{1,\bx}^{-1}-\Indicator(\bX_i\in\A_0)\bGamma_{0,\bx}^{-1}\big]
    \br_p(\bH^{-1}(\bX_i-\bx))\\
    &\hspace{1in} \cdot
    K_{\bH}(\bX_i-\bx) \mathfrak{w}(\bx)^\top \boldsymbol{\varepsilon}_{i}
    \Big],
\end{align*}
where $\boldsymbol{\varepsilon}_i = (\varepsilon_{Y,i}, \varepsilon_{W,i})^\top = (Y_i - \E[Y_i|\bX_i], W_i - \E[W_i|\bX_i])^\top$ is the vector of regression errors.

The following theorem quantifies the approximation error of the stochastic linearization:

\begin{thm}[Stochastic Linearization: Fuzzy BATEC]\label{sa-thm: Stochastic Linearization: Fuzzy BATEC}
    Suppose Assumptions \ref{sa-assump: Sharp DGP}, \ref{sa-assump: Boundary and Kernel} and \ref{sa-assump: Fuzzy DGP} hold, $\inf_{\bx \in \B}|\tau^{(\bnu)}_W(\bx)| > 0$, and $\inf_{\bx\in\B}V_{\mathtt{F},\bx}^{(\bnu)}>0$. If $\log(1/\bar{h})/(n^{v/(2+v)}|\bH|) = o(1)$, $\bar{h} = o(1)$, and
    \begin{align*}
        \bh^{-\bnu}\bigg(\sqrt{\frac{\log(1/\bar{h})}{n |\bH|}}+\frac{\log(1/\bar{h})}{n^{\frac{v}{2+v}}|\bH|}+\bar{h}^{p+1}\bigg)=o(1),
    \end{align*}
    then
    \begin{align*}
        \sup_{\bx \in \B} \Big|\widehat{\Tstat}^{(\bnu)}_{\mathtt{F}}(\bx) - \overline{\Tstat}^{(\bnu)}_{\mathtt{F}}(\bx) \Big|
        \lessp \bar{h}^{p+1}\sqrt{n |\bH|} + \bh^{-\bnu}\; \sqrt{\log(1/\bar{h})} \Big(\sqrt{\frac{\log(1/\bar{h})}{n |\bH|}} +  \frac{\log(1/\bar{h})}{n^{\frac{v}{2+v}}|\bH|} + \bar{h}^{p+1}\Big).
    \end{align*}
\end{thm}

Next, we present a coupling result for the supremum of the fuzzy linearized T-statistics. Unlike Theorem~\ref{sa-thm: Gaussian Strong Approximation: Sharp Tstat}, this result uses a Gaussian approximation for suprema and therefore requires only VC-type entropy and moment bounds, not the total-variation/perimeter condition used for the process-level coupling.

\begin{thm}[Gaussian Approximation for $\sup_{\bx \in \B} |\overline{\Tstat}^{(\bnu)}_{\mathtt{F}}(\bx)|$]\label{sa-thm: Gaussian Approximation Suprema -- Fuzzy}
Suppose Assumptions \ref{sa-assump: Sharp DGP}, \ref{sa-assump: Boundary and Kernel} and \ref{sa-assump: Fuzzy DGP} hold, $\inf_{\bx \in \B}|\tau^{(\bnu)}_W(\bx)| > 0$, and $\inf_{\bx\in\B}V_{\mathtt{F},\bx}^{(\bnu)}>0$. Then for $n$ large enough, on a possibly enriched probability space, there exists a Gaussian process $Z^{(\bnu)}_{\mathtt{F}}(\bx), \bx \in \B$ with almost sure continuous trajectories such that
\begin{align*}
    \Cov[Z^{(\bnu)}_{\mathtt{F}}(\bx_1), Z^{(\bnu)}_{\mathtt{F}}(\bx_2)] =
    \Cov[\overline{\Tstat}^{(\bnu)}_{\mathtt{F}}(\bx_1), \overline{\Tstat}^{(\bnu)}_{\mathtt{F}}(\bx_2)], \qquad \bx_1, \bx_2 \in \B,
\end{align*}
and
\begin{align*}
    \P\Big[ \big|\sup_{\bx \in \B} |\overline{\Tstat}^{(\bnu)}_{\mathtt{F}}(\bx)| - \sup_{\bx \in \B} |Z^{(\bnu)}_{\mathtt{F}}(\bx)|\big| > \frac{\log n}{(n |\bH|)^{1/6}} + \frac{(\log n)^{5/4}}{(n |\bH|)^{1/4}} + \frac{(\log n)^{3/2}}{(n^{\frac{v}{2+v}} |\bH|)^{1/2}}\Big] \lesssim \frac{1}{\log n} + \frac{\log n}{n}.
\end{align*}
\end{thm}

To construct a feasible confidence band for $(\zeta^{(\bnu)}(\bx): \bx \in \B)$, let $(\widehat{Z}^{(\bnu)}_{\mathtt{F}}(\bx): \bx \in \B)$ be a mean-zero Gaussian process conditional on $\bV$. Its feasible conditional covariance function is defined as
\begin{equation*}
    \Cov \left[ \widehat{Z}^{(\bnu)}_{\mathtt{F}}(\bx_1), \widehat{Z}^{(\bnu)}_{\mathtt{F}}(\bx_2) \mid \bV \right]
    = \frac{\widehat{\Omega}_{\mathtt{F}, \bx_1, \bx_2}^{(\bnu)}}{\sqrt{\widehat{\Omega}_{\mathtt{F}, \bx_1, \bx_1}^{(\bnu)} \widehat{\Omega}_{\mathtt{F}, \bx_2, \bx_2}^{(\bnu)}}},
    \quad \bx_1, \bx_2 \in \B.
\end{equation*}

\begin{thm}[Confidence Bands: Fuzzy BATEC; suprema approximation]\label{sa-thm: Confidence Bands: Fuzzy BATEC -- suprema}
    Suppose Assumptions \ref{sa-assump: Sharp DGP}, \ref{sa-assump: Boundary and Kernel}, and \ref{sa-assump: Fuzzy DGP} hold, $\inf_{\bx \in \B} |\tau^{(\bnu)}_W(\bx)| > 0$, and $\inf_{\bx\in\B}V_{\mathtt{F},\bx}^{(\bnu)}>0$. If $\liminf_{n \to \infty} \log \bar{h}/\log n > -\infty$, $(\log n)^9/(n^{v/(2+v)} |\bH|) = o(1)$, $\bar{h}^{p+1} \sqrt{n |\bH|\log n} = o(1)$, and
    \begin{align*}
        &\bh^{-\bnu}\sqrt{\log(1/\bar{h})}
        \left(\sqrt{\log(1/\bar{h})/(n|\bH|)}+
        \log(1/\bar{h})/(n^{v/(2+v)}|\bH|)+\bar{h}^{p+1}\right)=o(1),\\
        &\log^2(n)\bh^{-\bnu}
        \left(
        \sqrt{\frac{\log n}{n|\bH|}}
        +\frac{\log n}{n^{v/(2+v)}|\bH|}
        +\bar h^{p+1}
        \right)=o(1),
    \end{align*}
    then
    \begin{equation*}
        \sup_{u \in \reals} \left| \P \left( \sup_{\bx \in \B} \big| \widehat{\Tstat}_{\mathtt{F}}^{(\bnu)}(\bx) \big| \leq u \right) - \P \left( \sup_{\bx \in \B} \big| \widehat{Z}_{\mathtt{F}}^{(\bnu)}(\bx) \big| \leq u \;\middle|\; \bV \right) \right| = o_{\P}(1)
    \end{equation*}
    and
    \begin{equation*}
        \P \left( \zeta^{(\bnu)}(\bx) \in \widehat{\CI}_{\mathtt{F},\alpha}^{(\bnu)}(\bx) \text{ for all } \bx \in \B \right) = 1 - \alpha + o(1),
    \end{equation*}
    provided that $\q_{\alpha} = \inf \left\{ c > 0 : \P \left( \sup_{\bx \in \B} \big| \widehat{Z}_{\mathtt{F}}^{(\bnu)}(\bx) \big| \geq c \;\middle|\; \bV \right) \leq \alpha \right\}$.
\end{thm}

\subsection{Fuzzy WBATE}

We now consider WBATE summaries of these fuzzy ratios of derivative discontinuities. For $|\bnu|\le p$, the parameter is
\begin{align*}
    \zeta_{\mathtt{WBATE}}^{(\bnu)} = \int_{\B} \zeta^{(\bnu)}(\bb) w(\bb) \dif\Haus^{d-1} (\bb),
\end{align*}
where the weight function $w(\cdot)$ satisfies Assumption \ref{sa-assump: Weight Function and Boundary}. Without loss of generality, we assume that $\int_{\B} w(\bb) \dif\Haus^{d-1} (\bb) = 1$. The corresponding plug-in estimator is
\begin{align*}
    \widehat{\zeta}_{\mathtt{WBATE}}^{(\bnu)} = \int_\B \widehat{\zeta}^{(\bnu)}(\bb) w(\bb) \dif\Haus^{d-1}(\bb).
\end{align*}
The original fuzzy WBATE parameter is the special case $\bnu=\mathbf{0}$.

Our AMSE results for the fuzzy WBATE estimator leverage the leading linear component from the second-order linearization established in Section \ref{sec: Second-Order Linearization: Fuzzy BATEC},
\begin{align*}
    \check{\zeta}_{\mathtt{WBATE}}^{(\bnu)}
    = \int_{\B}\mathfrak{w}(\bb)^\top\mathfrak{T}(\bb) w(\bb) \dif\Haus^{d-1}(\bb).
\end{align*}
The corresponding approximate MSE is
\begin{align*}
    \operatorname{AMSE}_{\mathtt{F,WBATE}}^{(\bnu)}
    = \E\big[\big(\check{\zeta}_{\mathtt{WBATE}}^{(\bnu)}\big)^2\big|\bX\big].
\end{align*}
The leading bias is defined as the integral of the pointwise leading bias of the fuzzy BATEC estimator:
\begin{align*}
    B_{\mathtt{F,WBATE}}^{(\bnu)}(\bh) = \int_{\B} B_{\mathtt{F},\bb}^{(\bnu)}(\bh) \; w(\bb) \dif\Haus^{d-1}(\bb),
\end{align*}
where $B_{\mathtt{F},\bb}^{(\bnu)}(\bh)$ is given in Equation~\eqref{sa-eq: leading bias fuzzy}.

\begin{lem}[Bias: Fuzzy WBATE Linearization]\label{sa-lem: Fuzzy Integral: Bias}
    Suppose Assumptions \ref{sa-assump: Sharp DGP}, \ref{sa-assump: Boundary and Kernel}, \ref{sa-assump: Weight Function and Boundary} and \ref{sa-assump: Fuzzy DGP} hold, and $\inf_{\bx\in\B}|\tau_W^{(\bnu)}(\bx)|>0$. If $\frac{\log(1/\bar{h})}{n^{\frac{1+v}{2+v}}|\bH|} = o(1)$ and $\bar{h} = o(1)$, then
    \begin{align*}
        \E[\check{\zeta}_{\mathtt{WBATE}}^{(\bnu)}|\bX] = B_{\mathtt{F,WBATE}}^{(\bnu)}(\bh) + o_{\P}(\bar{h}^{p+1}\bh^{-\bnu}).
    \end{align*}
    If, in addition, $\bh=\bc h$ for a fixed vector $\bc$ and common bandwidth $h$, then $B_{\mathtt{F,WBATE}}^{(\bnu)}(\bh)=h^{p+1-|\bnu|}\widetilde B_{\mathtt{F,WBATE}}^{(\bnu)}(\bc)$.
\end{lem}

The conditional variance of this linearized fuzzy WBATE component involves the integration of the cross-covariance structure of the reduced-form processes. Let
\begin{align*}
    \Omega_{\mathtt{F,WBATE}}^{(\bnu)} = \int_{\B} \int_{\B} \Omega_{\mathtt{F},\bb_1, \bb_2}^{(\bnu)} w(\bb_1) w(\bb_2) \dif\Haus^{d-1}(\bb_1) \dif\Haus^{d-1}(\bb_2),
\end{align*}
and impose the aggregate variance nondegeneracy condition
\begin{align}
    \Omega_{\mathtt{F,WBATE}}^{(\bnu)}
    \gtrsim \frac{\bar{h}^{d-1}}{n|\bH|\bh^{2\bnu}}.
    \label{sa-eq: fuzzy WBATE aggregate nondegeneracy}
\end{align}
This condition rules out cancellation after weighting and integrating the linearized fuzzy Wald process.
Let its plug-in estimator be
\begin{align*}
    \widehat{\Omega}_{\mathtt{F,WBATE}}^{(\bnu)} = \int_{\B} \int_{\B} \widehat{\Omega}_{\mathtt{F},\bb_1, \bb_2}^{(\bnu)} w(\bb_1) w(\bb_2) \dif\Haus^{d-1}(\bb_1) \dif\Haus^{d-1}(\bb_2).
\end{align*}

\begin{lem}[Variance: Fuzzy WBATE Linearization]\label{sa-lem: Fuzzy Integral: Variance}
    Suppose the assumptions and conditions of Lemma \ref{sa-lem: Covariance: Fuzzy BATEC} and Assumption \ref{sa-assump: Weight Function and Boundary} hold, and \eqref{sa-eq: fuzzy WBATE aggregate nondegeneracy} and \eqref{sa-eq: local boundary measure regularity} hold. Then
    \begin{align*}
        \Var[\check{\zeta}_{\mathtt{WBATE}}^{(\bnu)}|\bX]
        = \Omega_{\mathtt{F,WBATE}}^{(\bnu)} + o_{\P}\Big(\frac{\bar{h}^{d-1}}{n|\bH|\bh^{2\bnu}}\Big),
    \end{align*}
    where
    \begin{align*}
        \frac{\bar{h}^{d-1}}{n|\bH|\bh^{2\bnu}}
         \lesssim \Omega_{\mathtt{F,WBATE}}^{(\bnu)}
         \lesssim \frac{\bar{h}^{d-1}}{n|\bH|\bh^{2\bnu}}.
    \end{align*}

    In addition,
    \begin{align*}
        \Var[\check{\zeta}_{\mathtt{WBATE}}^{(\bnu)}|\bX]
        & = \widehat{\Omega}_{\mathtt{F,WBATE}}^{(\bnu)} + o_{\P}\Big(\frac{\bar{h}^{d-1}}{n|\bH|\bh^{2\bnu}}\Big).
    \end{align*}
\end{lem}

\begin{thm}[AMSE Expansion: Fuzzy WBATE]\label{sa-thm: MSE Expansion: Fuzzy WBATE}
    Suppose the assumptions of Lemma \ref{sa-lem: Fuzzy Integral: Variance} hold, including \eqref{sa-eq: local boundary measure regularity}. If $\log(1/\bar{h})/(n^{v/(2+v)}|\bH|) = o(1)$ and $\bar{h} = o(1)$, then
    \begin{align*}
        \operatorname{AMSE}_{\mathtt{F,WBATE}}^{(\bnu)}
        = \Omega_{\mathtt{F,WBATE}}^{(\bnu)} + \big(B_{\mathtt{F,WBATE}}^{(\bnu)}(\bh)\big)^2
        + o_{\P}\Big(\frac{\bar{h}^{d-1}}{n|\bH|\bh^{2\bnu}} + \bar{h}^{2p+2}\bh^{-2\bnu}\Big).
    \end{align*}
\end{thm}

For inference, we define the feasible $t$-statistic for the aggregate effect:
\begin{align*}
    \widehat{\Tstat}_{\mathtt{F,WBATE}}^{(\bnu)} = \frac{\widehat{\zeta}_{\mathtt{WBATE}}^{(\bnu)} - \zeta_{\mathtt{WBATE}}^{(\bnu)}}{\sqrt{\widehat{\Omega}_{\mathtt{F,WBATE}}^{(\bnu)}}}.
\end{align*}

\begin{thm}[Asymptotic Normality: Fuzzy WBATE]\label{sa-thm: Asymptotic Normality: Fuzzy WBATE}
    Suppose Assumptions \ref{sa-assump: Sharp DGP}, \ref{sa-assump: Boundary and Kernel}, \ref{sa-assump: Weight Function and Boundary} and \ref{sa-assump: Fuzzy DGP} hold, $\inf_{\bx\in\B}|\tau_W^{(\bnu)}(\bx)|>0$, $\inf_{\bx\in\B}V_{\mathtt{F},\bx}^{(\bnu)}>0$, and \eqref{sa-eq: fuzzy WBATE aggregate nondegeneracy} and \eqref{sa-eq: local boundary measure regularity} hold. If $\log(1/\bar{h})/(n^{v/(2+v)}|\bH|) = o(1)$, $n|\bH|\bar{h}^{2p+2}/\bar{h}^{d-1} = o(1)$, and
    \begin{align*}
        \frac{\bh^{-2\bnu}\log^2(1/\bar{h})}{n|\bH|\bar{h}^{d-1}}=o(1),
        \qquad
        \frac{\bh^{-2\bnu}\log^4(1/\bar{h})}{n^{\frac{4v}{2+v}-1}|\bH|^3\bar{h}^{d-1}}=o(1),
    \end{align*}
    then
    \begin{align*}
        \sup_{u \in \reals} \big|\P (\widehat{\Tstat}_{\mathtt{F,WBATE}}^{(\bnu)} \leq u )  - \Phi(u) \big| = o(1).
    \end{align*}
\end{thm}

The local boundary-measure condition in \eqref{sa-eq: local boundary measure regularity} is used here for the same reason as in the sharp WBATE results: it controls the effective size of the boundary pairs that can have nonzero local-kernel covariance. The more specialized piecewise-linear boundary conditions in the main paper imply this condition.

\subsection{Fuzzy LBATE}

For $|\bnu|\leq p$, we consider the LBATE summary of the fuzzy ratio of derivative discontinuities, with estimand and estimator
\begin{align*}
    \zeta_{\mathtt{LBATE}}^{(\bnu)} = \sup_{\bb \in \B} \zeta^{(\bnu)}(\bb)
    \qquad\text{and}\qquad
    \widehat{\zeta}_{\mathtt{LBATE}}^{(\bnu)} = \sup_{\bb \in \B} \widehat{\zeta}^{(\bnu)}(\bb).
\end{align*}
The original fuzzy LBATE parameter is the special case $\bnu=\mathbf{0}$.

\begin{thm}[Convergence Rate: Fuzzy LBATE]\label{sa-thm: Convergence Rate: Fuzzy LBATE}
    Suppose the assumptions and conditions of Theorem~\ref{sa-thm: Convergence Rates: Fuzzy BATEC} hold. Then
    \begin{align*}
        \big|\widehat{\zeta}_{\mathtt{LBATE}}^{(\bnu)} - \zeta_{\mathtt{LBATE}}^{(\bnu)} \big|
        \lessp \bh^{-\bnu}\Big(\sqrt{\frac{\log(1/\bar{h})}{ n |\bH|}} + \frac{\log(1/\bar{h})}{n^{\frac{1+v}{2+v}}|\bH|} + \bar{h}^{p+1}\Big).
    \end{align*}
\end{thm}

Recall from Theorem~\ref{sa-thm: Confidence Bands: Fuzzy BATEC -- suprema} that $(\widehat{Z}_{\mathtt{F}}^{(\bnu)}(\bx):\bx \in \B)$ is a (conditionally on $\bV$) mean-zero Gaussian process with feasible (conditional) covariance function
\begin{align*}
    \Cov \Big[\widehat{Z}_{\mathtt{F}}^{(\bnu)}(\bx_1),\widehat{Z}_{\mathtt{F}}^{(\bnu)}(\bx_2) \Big|\bV \Big]
    = \frac{\widehat{\Omega}_{\mathtt{F},\bx_1,\bx_2}^{(\bnu)}}
           {\sqrt{\widehat{\Omega}_{\mathtt{F},\bx_1,\bx_1}^{(\bnu)} \widehat{\Omega}_{\mathtt{F},\bx_2,\bx_2}^{(\bnu)}}},
    \qquad \bx_1, \bx_2 \in \B.
\end{align*}
Consider the confidence interval
\begin{align*}
    \widehat{\CI}_{\alpha,\mathtt{F,LBATE}}^{(\bnu)}
    =
    \bigg[\sup_{\bb \in \B} \Big(\widehat{\zeta}^{(\bnu)}(\bb)  - \q_{\alpha} \sqrt{\widehat{\Omega}_{\mathtt{F},\bb,\bb}^{(\bnu)}}\Big), \;
    \sup_{\bb \in \B} \Big(\widehat{\zeta}^{(\bnu)}(\bb) + \q_{\alpha} \sqrt{\widehat{\Omega}_{\mathtt{F},\bb,\bb}^{(\bnu)}}\Big)\bigg],
\end{align*}
where $\q_{\alpha} = \inf \big\{c > 0: \P \big(\sup_{\bx \in \B} \big|\widehat{Z}_{\mathtt{F}}^{(\bnu)}(\bx)\big|\geq c \big| \bV \big) \leq \alpha \big\}$.

\begin{thm}[Confidence Interval: Fuzzy LBATE]\label{sa-thm: Confidence Intervals: Fuzzy LBATE}
    Suppose the assumptions and conditions in Theorem~\ref{sa-thm: Confidence Bands: Fuzzy BATEC -- suprema} hold. Then
    \begin{align*}
            \P\Big[\zeta_{\mathtt{LBATE}}^{(\bnu)} \in \widehat{\CI}_{\alpha,\mathtt{F,LBATE}}^{(\bnu)} \Big] \geq 1 - \alpha + o(1).
    \end{align*}
\end{thm}

\section{Gaussian Strong Approximation}\label{sa-sec: Gaussian Strong Approximation}

We present a Gaussian strong approximation theorem, which is the key technical tool behind Theorem~\ref{sa-thm: Gaussian Strong Approximation: Sharp Tstat}. The theorem builds on and generalizes the results in \cite{Cattaneo-Yu_2025_AOS}. Consider the \emph{residual-based empirical process} given by
\begin{align*}
    R_n[g,r] = \frac{1}{\sqrt{n}} \sum_{i = 1}^n \Big[g(\bx_i) r(y_i) - \E[g(\bx_i) r(y_i)|\bx_i] \Big], \qquad g \in \G, r \in \R,
\end{align*}
where $\G$ and $\R$ are classes of functions satisfying certain regularity conditions.

\subsection{Definitions for Function Spaces}

Let $\F$ be a class of measurable functions from a probability space $(\reals^q, \B(\reals^q), \P)$ to $\reals$. We introduce several definitions that capture properties of $\F$.

\begin{enumerate}[label=(\roman*)]
    \item  $\F$ is pointwise measurable if it contains a countable subset $\G$ such that for any $f \in \F$, there exists a sequence $(g_m:m\geq1) \subseteq \G$ such that $\lim_{m \to \infty} g_m(\bu) = f(\bu)$ for all $\bu \in \reals^q$.
    \item  Let $\supp(\F) = \cup_{f \in \F}\supp(f)$. A probability measure $\bbQ_\F$ on $(\reals^q,\B(\reals^q))$ is a surrogate measure for $\P$ with respect to $\F$ if
    \begin{enumerate}[label=(\roman*)]
        \item $\bbQ_\F$ agrees with $\P$ on $\supp(\P) \cap \supp(\F)$.
        \item $\bbQ_\F(\supp(\F) \setminus \supp(\P)) = 0$.
    \end{enumerate}
    Let $\mathcal{Q}_\F=\supp(\bbQ_\F)$.
    \item For $q=1$ and an interval $\mathcal{I}\subseteq\reals$, the pointwise total variation of $\F$ over $\mathcal{I}$ is
    \begin{align*}
        \mathtt{pTV}_{\F,\mathcal{I}} = \sup_{f \in \F} \sup_{P\geq1}\sup_{\calP_P \in \mathcal{I}} \sum_{i = 1}^{P-1}|f(a_{i+1}) - f(a_i)|,
    \end{align*}
    where $\calP_P=\{(a_1,\dots,a_P):a_1 \leq \cdots \leq a_P\}$ denotes the collection of all partitions of $\mathcal{I}$.
    \item For a non-empty $\mathcal{C} \subseteq \reals^q$, the total variation of $\F$ over $\mathcal{C}$ is
    \begin{align*}
        \ttTV_{\F, \mathcal{C}} = \inf_{\mathcal{U} \in \mathcal{O}(\mathcal{C})}\sup_{f \in \F} \sup_{\phi \in \mathscr{D}_{q}(\mathcal{U})} \int_{\reals^q} f(\bu)\operatorname{div}(\phi)(\bu) \dif \bu / \infnorm{\norm{\phi}_2},
    \end{align*}
    where $\mathcal{O}(\mathcal{C})$ denotes the collection of all open sets that contains $\mathcal{C}$, and $\mathscr{D}_{q}(\mathcal{U})$ denotes the space of infinitely differentiable functions from $\reals^q$ to $\reals^q$ with compact support contained in $\mathcal{U}$.
    \item For a non-empty $\mathcal{C} \subseteq \reals^q$, the local total variation constant of $\F$ over $\mathcal{C}$, is a positive number $\ttK_{\F,\mathcal{C}}$ such that for any cube $\mathcal{D} \subseteq \reals^q$ with edges of length $\ell$ parallel to the coordinate axes,
    \begin{align*}
        \ttTV_{\F, \mathcal{D} \cap \mathcal{C}}  \leq \ttK_{\F, \mathcal{C}} \ell^{q-1}.
    \end{align*}
    \item For a non-empty $\mathcal{C} \subseteq \reals^q$, the envelopes of $\F$ over $\mathcal{C}$ are
    \begin{align*}
        \ttM_{\F,\C} = \sup_{\bu \in \C }M_{\F,\C}(\bu),
        \qquad M_{\F,\C}(\bu) = \sup_{f \in \F}|f(\bu)|,
        \qquad \bu \in \C.
    \end{align*}
    \item  For a non-empty $\mathcal{C} \subseteq \reals^q$, the Lipschitz constant of $\F$ over $\C$ is
    \begin{align*}
        \ttL_{\F,\C} = \sup_{f \in \F}\sup_{\bu_1, \bu_2 \in \C} \frac{|f(\bu_1) - f(\bu_2)|}{\|\bu_1 - \bu_2\|_\infty}.
    \end{align*}
    \item For a non-empty $\mathcal{C} \subseteq \reals^q$, the $L_1$ bound of $\F$ over $\C$ is
    \begin{align*}
        \mathtt{E}_{\F,\C} = \sup_{f \in \F} \int_{\C} |f| \dif\P.
    \end{align*}
    \item   For a non-empty $\mathcal{C} \subseteq \reals^q$, the uniform covering number of $\F$ with envelope $M_{\F,\C}$ over $\C$ is
    \begin{align*}
        \ttN_{\F,\C}(\delta,M_{\F,\C}) = \sup_{\mu} N(\F,\norm{\cdot}_{\mu,2},\delta \norm{M_{\F,\C}}_{\mu,2}),
        \qquad \delta \in (0, \infty),
    \end{align*}
    where the supremum is taken over all finite discrete measures on $(\C, \B(\C))$. We assume that $M_{\F,\C}(\bu)$ is finite for every $\bu \in \C$.
    \item  For a non-empty $\mathcal{C} \subseteq \reals^q$, the uniform entropy integral of $\F$ with envelope $M_{\F,\C}$ over $\C$ is
    \begin{align*}
        J_\C(\delta, \F, M_{\F,\C}) = \int_0^{\delta} \sqrt{1 + \log \ttN_{\F,\C}(\varepsilon,M_{\F,\C})} \dif \varepsilon,
    \end{align*}
    where it is assumed that $M_{\F,\C}(\bu)$ is finite for every $\bu \in \C$.
    \item For a non-empty $\mathcal{C} \subseteq \reals^q$, $\F$ is a VC-type class with envelope $M_{\F,\C}$ over $\C$ if (i) $M_{\F,\C}$ is measurable and $M_{\F,\C}(\bu)$ is finite for every $\bu \in \C$, and (ii) there exist $\ttc_{\F,\C}>0$ and $\mathtt{d}_{\F,\C}>0$ such that
    \begin{align*}
        \ttN_{\F,\C}(\varepsilon,M_{\F,\C}) \leq \ttc_{\F,\C} \varepsilon^{-\mathtt{d}_{\F,\C}},
        \qquad \varepsilon\in(0,1).
    \end{align*}

\end{enumerate}

If a surrogate measure $\bbQ_\F$ for $\P$ with respect to $\F$ has been assumed, and it is clear from the context, we drop the dependence on $\C = \mathcal{Q}_{\F}$ for all quantities in the previous definitions. That is, to save notation, we set $\ttTV_{\F}=\ttTV_{\F,\mathcal{Q}_{\F}}$, $\ttK_{\F}=\ttK_{\F,\mathcal{Q}_{\F}}$, $\ttM_{\F}=\ttM_{\F,\mathcal{Q}_{\F}}$, $M_{\F}(\bu)=M_{\F,\mathcal{Q}_{\F}}(\bu)$, $\ttL_{\F}=\ttL_{\F,\mathcal{Q}_{\F}}$, and so on, whenever there is no confusion.

\subsection{Residual-based Empirical Process}

The following theorem generalizes \citet[Theorem 2]{Cattaneo-Yu_2025_AOS} by requiring only bounded polynomial moments for $y_i$ conditional on $\bx_i$.

\begin{thm}[Strong Approximation for Residual-based Empirical Processes]\label{sa-thm: Residual Empirical Process Strong Approximation}
    Suppose $(\bz_i=(\bx_i, y_i): 1 \leq i \leq n)$ are i.i.d.\ random vectors taking values in $(\reals^{d+1}, \B(\reals^{d+1}))$ with common law $\P_Z$, where $\bx_i$ has distribution $\P_X$ supported on $\X\subseteq\reals^d$, $y_i$ has distribution $\P_Y$ supported on $\Y\subseteq\reals$, $\sup_{\bx \in \X}\E[|y_i|^{2 + v}|\bx_i = \bx] \leq C_y<\infty$ for some $v > 0$, and the following conditions hold:

    \begin{enumerate}[label=\emph{(\roman*)}]
        \item $\G$ is a real-valued pointwise measurable class of functions on $(\reals^d, \B(\reals^d), \P_X)$.
        \item There exists a surrogate measure $\bbQ_\G$ for $\P_X$ with respect to $\G$ such that $\bbQ_\G = \mathfrak{m} \circ \phi_\G$, where the \textit{normalizing transformation} $\phi_{\G}: \Q_\G \to [0,1]^d$ is a diffeomorphism.
        \item $\G$ is a VC-type class with envelope $\ttM_{\G}$ over $\Q_\G$ with $\ttc_{\G} \geq e$ and $\mathtt{d}_{\G} \geq 1$.
        \item $\R$ is a real-valued pointwise measurable class of functions on $(\reals, \mathsf{Borel}(\reals),\P_Y)$.
        \item $\R$ is a VC-type class with envelope $M_{\R,\Y}$ over $\Y$ with $\ttc_{\R,\Y}\geq e$ and $\mathtt{d}_{\R,\Y}\geq 1$, where $M_{\R,\Y}(y) + \mathtt{pTV}_{\R,(-|y|,|y|)} \leq \mathtt{v} (1 + |y|)$ for all $y \in \Y$, for some $\mathtt{v}>0$.
        \item There exists a constant $\mathtt{k}$ such that $|\log_2 \mathtt{E}_{\G}| + |\log_2 \ttTV| + |\log_2 \ttM_{\G}| \leq \mathtt{k} \log_2 n$, where the constant $\ttTV = \max \{\ttTV_{\G}, \ttTV_{\G \times \mathscr{U}_{\R},\Q_\G}\}$ with $\mathscr{U}_{\R} = \{\theta(\cdot,r, \tau) : r \in \R, \tau \in (0,\infty]\}$, and $\theta(\bx,r,\tau) = \E[r(y_i) \Indicator(|y_i| \leq \tau)|\bx_i = \bx]$ for $\bx \in \X$.
    \end{enumerate}
    Define the residual-based empirical process
    \begin{align*}
        R_n(g,r) = \frac{1}{\sqrt{n}} \sum_{i = 1}^n g(\bx_i)(r(y_i) - \E[r(y_i)|\bx_i]), \qquad g \in \G, r \in \R.
    \end{align*}
    Then (1) on a possibly enlarged probability space, there exists a sequence of mean-zero Gaussian processes $(Z_n^R(g,r): g\in\G, r \in \R)$ with almost sure continuous trajectories such that:
    \begin{itemize}
        \item $\E[R_n(g_1, r_1) R_n(g_2, r_2)] = \E[Z^R_n(g_1, r_1) Z^R_n(g_2, r_2)]$ for all $(g_1, r_1), (g_2, r_2) \in \G \times \R$, and
        \item $\E\big[\norm{R_n - Z_n^R}_{\G \times \R}\big] \leq C \mathtt{v} \ttd \log(\ttc n) \rho_n$,
    \end{itemize}
    with
    \begin{align*}
        \rho_n = \sqrt{\ttd \log (\ttc n)} \; \mathtt{r}_n^{\frac{v}{v +2}}(\sqrt{\ttM_{\G}\ttE_{\G}})^{\frac{2}{v+2}} + \ttM_{\G} n^{-\frac{v/2}{2+v}} +  \frac{\ttM_{\G}}{\sqrt{n}}  \Big(\frac{\sqrt{\ttM_{\G} \ttE_{\G}}}{\mathtt{r}_n}\Big)^{\frac{2}{v+2}},
    \end{align*}
    where $C$ is a positive constant that depends only on $v$ and $C_y$, $\ttc = \ttc_{\G} \ttc_{\R,\Y} \ttk$, $\ttd = \ttd_{\G} + \ttd_{\R,\Y} + \ttk$, and
    \begin{gather*}
        \mathtt{r}_n = \min\Big\{\frac{(\mathtt{c}_1^d \ttM_{\G}^{d+1} \ttTV^d \mathtt{E}_{\G})^{1/(2d+2)} }{n^{1/(2d+2)}}, \frac{(\mathtt{c}_1^{d/2} \mathtt{c}_2^{d/2} \ttM_{\G} \ttTV^{d/2} \mathtt{E}_{\G} \ttL^{d/2})^{1/(d+2)}}{n^{1/(d+2)}} \Big\}, \\
        \ttc_1 = d \sup_{\bx \in \mathcal{Q}_{\G}} \prod_{j = 1}^{d-1} \sigma_j(\nabla \phi_{\G}(\bx)), \qquad
        \ttc_2 = \sup_{\bx \in \mathcal{Q}_{\G}} \frac{1}{\sigma_{d}(\nabla \phi_{\G}(\bx))}, \qquad
        \ttL = \max\{\ttL_{\G}, \ttL_{\G \times \mathscr{U}_{\R},\Q_{\G}}\};
    \end{gather*}
    and (2) if $\R$ is a singleton and the truncated conditional-mean class $\mathscr{U}_{\R}$ satisfies
    \[
        \ttTV_{\G\times\mathscr{U}_{\R},\Q_{\G}}\lesssim \ttTV_{\G\times\V_{\R},\Q_{\G}},
        \qquad
        \ttL_{\G\times\mathscr{U}_{\R},\Q_{\G}}\lesssim \ttL_{\G\times\V_{\R},\Q_{\G}},
    \]
    then we can replace $\ttTV$ and $\ttL$ in the previous conditions and statements by $\ttTV_{\text{sing}} = \max\{\ttTV_{\G}, \ttTV_{\G \times \V_{\R}, \Q_{\G}}\}$, and $\ttL_{\text{sing}} = \max\{\ttL_{\G}, \ttL_{\G \times \V_{\R},\Q_{\G}}\}$, respectively, with $\V_{\R} = \{\theta(\cdot,r) : r \in \R\}$, and $\theta(\bx,r) = \E[r(y_i) |\bx_i = \bx]$ for $\bx \in \X$.
\end{thm}

\begin{remark}
The class $\mathscr{U}_{\mathcal R}$ comprises truncated conditional means at all truncation levels. Its Lipschitz and total-variation constants can be bounded, for example, if $f(y \mid \bx)$ is Lipschitz in $\bx$ uniformly over $(\bx,y)$ in the support of $(\bx_i,y_i)$. When $\mathcal R$ is a singleton, regularity of the untruncated conditional mean class $\V_{\mathcal R}$ is enough whenever the truncated conditional means inherit the same bounds, as stated in part (2) of the theorem; this is automatic under bounded support for the residual component and holds under standard smooth conditional-density conditions.
\end{remark}

\subsection{Proof of Theorem~\ref{sa-thm: Residual Empirical Process Strong Approximation}}\label{sa-sec: proof of residual empirical process strong approximation}

We use a truncation argument. Let $\kappa_n \geq 1$ be the level of truncation. For each $r \in \R$, define
\begin{align*}
    \tilde{r}(y) = r(y) \Indicator(|y| \leq \kappa_n), \qquad y \in \reals,
\end{align*}
and define the class $\tilde{\R} = \{\tilde{r}: r \in \R\}$. Suppose $Z_n^R$ is a mean-zero Gaussian process indexed by $\G \times \R \cup \G \times \tilde{\R}$, whose construction is given below. Then
\begin{align*}
    R_n(g,r) - Z_n^R(g,r)
    & = \big[R_n(g,\tilde{r}) - Z_n^R(g,\tilde{r})\big] + \big[R_n(g,r) - R_n(g,\tilde{r})\big] + \big[Z_n^R(g,\tilde{r}) - Z_n^R(g,r)\big].
\end{align*}

\subsubsection*{Part 1: Strong approximation for truncated residual empirical process.}

The truncation class $\tilde{\R}$ is pointwise measurable whenever $\R$ is, and it is VC-type with envelope $M_{\tilde{\R},\Y} = M_{\R,\Y} \Indicator(|\cdot| \leq \kappa_n)$ over $\Y$ and constants $\ttc_{\R,\Y}$ and $\ttd_{\R,\Y}$ up to universal factors. Moreover, the envelope and pointwise total variation satisfy
\begin{align*}
    M_{\tilde{\R},\Y}(y) + \mathtt{pTV}_{\tilde{\R},(-|y|,|y|)} \lesssim \ttv \kappa_n, \qquad y \in \Y,
\end{align*}
because truncating $r$ only adds jumps at $\pm\kappa_n$. Therefore \citet[Theorem 2]{Cattaneo-Yu_2025_AOS}, applied with $\alpha=0$ and with the bounded-outcome constant proportional to $\ttv\kappa_n$, implies on a possibly enlarged probability space that there exists a sequence of mean-zero Gaussian processes $(Z_n^R(g,r): (g,r)\in \G\times \tilde{\R})$ with almost sure continuous trajectories on $(\G \times \tilde{\R}, \rho_{\P})$ such that $\E[R_n(g_1, r_1) R_n(g_2, r_2)] = \E[Z^R_n(g_1, r_1) Z^R_n(g_2, r_2)]$ for all $(g_1, r_1), (g_2, r_2) \in \G \times \tilde{\R}$, and, after integrating the tail bound in that theorem,
\begin{align*}
    & \E[\lVert R_n(g,\tilde{r}) - Z_n^R(g,\tilde{r}) \rVert_{\G \times \R}] \\
    & \leq C_1 \ttv \kappa_n \bigg(\sqrt{d} \min\Big\{\frac{(\ttc_1^d \ttM_{\G}^{d+1} \ttTV^d \mathtt{E}_{\G})^{\frac{1}{2d+2}} }{n^{1/(2d+2)}},
                     \frac{(\ttc_1^{\frac{d}{2}} \ttc_2^{\frac{d}{2}}\ttM_{\G} \ttTV^{\frac{d}{2}} \mathtt{E}_{\G} \ttL^{\frac{d}{2}})^{\frac{1}{d+2}}}{n^{1/(d+2)}} \Big\} (\ttd\log (\ttc n))^{3/2}
        + \frac{\ttd\log (\ttc n)}{\sqrt{n}} \ttM_{\G} \bigg) \\
    & = C_1 \ttv \kappa_n \bigg(\sqrt{d} \mathtt{r}_n (\ttd\log (\ttc n))^{\frac{3}{2}}
        + \frac{\ttd\log (\ttc n)}{\sqrt{n}} \ttM_{\G}\bigg),
\end{align*}
where $C_1$ is a positive universal constant. The total-variation constant $\ttTV$ used in the theorem controls the truncated class because
\begin{align*}
    \V_{\tilde{\R}}
    = \{\theta(\cdot,r,\kappa_n): r\in\R\}
    \subseteq \mathscr{U}_{\R}.
\end{align*}
Thus $\ttTV = \max \{\ttTV_{\G}, \ttTV_{\G \times \mathscr{U}_{\R},\Q_\G}\}$ is an upper bound for $\max \{\ttTV_{\G}, \ttTV_{\G \times \V_{\tilde{\R}},\Q_\G}\}$. The same argument applies to $\ttL$.

In the special case that $\R = \{r_{\ast}\}$ is a singleton, take $\tilde{y}_i = r_{\ast}(y_i) \Indicator(|y_i| \leq \kappa_n)/(\ttv \kappa_n)$. Then $\E[\exp(|\tilde{y}_i|)] \leq 2$, and $\tilde{y}_i$ is supported on $\tilde{\Y} = [-1,1]$. Moreover,
\begin{align*}
    \frac{1}{\ttv \kappa_n}R_n(g, \tilde{r}_{\ast}) = \frac{1}{\sqrt{n}} \sum_{i = 1}^n g(\bx_i) (\tilde{y}_i - \E[\tilde{y}_i|\bx_i]), \qquad g \in \G.
\end{align*}
The right-hand side is a residual empirical process based on the sample $(\bx_i, \tilde{y}_i)$, $1 \leq i \leq n$, indexed by $\G \times \{\operatorname{Id}\}$, where $\operatorname{Id}: \reals \to \reals$ is the identity function. Under the singleton domination condition in part (2) of the theorem, applying \citet[Theorem 2]{Cattaneo-Yu_2025_AOS} with $\alpha = 0$ to this empirical process gives the same upper bound with $\ttTV$ and $\ttL$ replaced by $\ttTV_{\text{sing}}$ and $\ttL_{\text{sing}}$.

\subsubsection*{Part 2: Truncation error for the empirical process --- $\lVert R_n(g,r) - R_n(g, \widetilde{r})\rVert_{\G \times \R}$}

Consider the class of differences due to truncation, $\Delta \R = \{r - \tilde{r}: r \in \R\}$. The assumptions imply $\G \times \Delta \R$ is VC-type in the sense that for all $0 <\varepsilon < 1$,
\begin{align*}
    \sup_{\Q} N(\G \times \Delta \R, \norm{\cdot}_{\Q,2}, \varepsilon \lVert \ttM_{\G} (M_{\R,\Y} - M_{\tilde{\R},\Y}) \rVert_{\Q,2}) \leq \ttc_{\G} \ttc_{\R,\Y} (\varepsilon^2/4)^{-\ttd_{\G} - \ttd_{\R,\Y}},
\end{align*}
where the supremum is over all finite discrete measures on $\reals^{d+1}$, and $M_{\tilde{\R},\Y}(y) = M_{\R,\Y}(y) \Indicator(|y| \leq \kappa_n)$. The function $\ttM_{\G} (M_{\R,\Y} - M_{\tilde{\R},\Y})$ is an envelope for $\G \times \Delta \R$, since all functions in $\Delta\R$ vanish on $[-\kappa_n,\kappa_n]$. Denote $\bX = (\bx_i)_{1 \leq i \leq n}$. The linear envelope condition on $\R$ and the conditional $(2+v)$ moment bound on $y_i$ imply
\begin{align*}
    \E \Big[\max_{1 \leq i \leq n}\ttM_{\G}^2 (M_{\R,\Y}(y_i) - M_{\tilde{\R},\Y}(y_i))^2\Big|\bX\Big]^{\frac{1}{2}}
    & \lesssim \ttM_{\G} \E \Big[ \big( \max_{1 \leq i \leq n}M_{\R,\Y}(y_i)\big)^2\Big| \bX \Big]^{\frac{1}{2}} \lesssim  \ttM_{\G} n^{\frac{1}{2+v}}, \\
    \sup_{(g,r)  \in \G \times \R} \E[g(\bx_i)^2 r(y_i)^2 \Indicator(|y_i| \geq \kappa_n)]^{\frac{1}{2}}
    & \lesssim \sup_{(g,r)  \in \G \times \R} \E \left[g(\bx_i)^2 \E[|r(y_i)|^{2+v}|\bx_i]^{\frac{2}{2+v}} \P(|y_i| \geq \kappa_n| \bx_i)^{\frac{v}{2+v}}\right]^{\frac{1}{2}} \\
    & \lesssim \sqrt{\ttM_{\G} \ttE_{\G} \kappa_n^{-v}}.
\end{align*}
By Jensen's inequality, we also have
\begin{align*}
    & \E \Big[\max_{1 \leq i \leq n}\ttM_{\G}^2 (\E[M_{\R,\Y}(y_i) - M_{\tilde{\R},\Y}(y_i)|\bx_i])^2\Big|\bX\Big]^{\frac{1}{2}} \lesssim \ttM_{\G} n^{\frac{1}{2+v}}, \\
    & \sup_{(g,r)  \in \G \times \R} \E[g(\bx_i)^2 \E[r(y_i) - \tilde{r}(y_i)|\bx_i]^2]^{\frac{1}{2}} \lesssim \sqrt{\ttM_{\G} \ttE_{\G} \kappa_n^{-v}}, \\
    & \E[\ttM_{\G}^2(M_{\R,\Y}(y_i) - M_{\tilde{\R},\Y}(y_i))^2]^{1/2} \lesssim \ttM_{\G} \kappa_n^{-v/2}.
\end{align*}
Denote $A =  (\ttc_{\G} \ttc_{\R,\Y})^{\frac{1}{2 \ttd_{\G} + 2 \ttd_{\R,\Y}}}/4$ and $D = 2 \ttd_{\G} + 2 \ttd_{\R,\Y}$. \citet[Corollary 5.1]{Chernozhukov-Chetverikov-Kato_2014b_AoS} gives
\begin{align*}
    \E \left[\lVert R_n(g, r) - R_n(g,\widetilde{r})\rVert_{\G \times \R}\right] &
    \lesssim \E \left[\sup_{g \in \G} \sup_{h \in \Delta \R} \frac{1}{\sqrt{n}} \sum_{i = 1}^n g(\bx_i)(h(y_i) - \E[h(y_i)|\bx_i])\right]\\
    & \lesssim \sqrt{D \ttM_{\G} \ttE_{\G}\kappa_n^{-v} \log (A \sqrt{\ttM_{\G}/\ttE_{\G}})} + \frac{D \ttM_{\G} n^{\frac{1}{2+v}}}{\sqrt{n}} \log (A \sqrt{\ttM_{\G}/\ttE_{\G}})\\
    & \lesssim \sqrt{D \log (A \sqrt{\ttM_{\G}/\ttE_{\G}})} \sqrt{\ttM_{\G} \ttE_{\G}} \kappa_n^{-v/2} + \frac{D \log (A \sqrt{\ttM_{\G}/\ttE_{\G}}) \ttM_{\G}}{\sqrt{n^{\frac{v}{2+v}}}}.
\end{align*}

\subsubsection*{Part 3: Truncation error for the Gaussian process --- $\lVert Z_n^R(g,r) - Z_n^R(g,\tilde{r}) \rVert_{\G \times \R}$}

The class $\G \times \tilde{\R} \cup \G \times \R$ is VC-type with respect to envelope function $2 \ttM_{\G} \ttM_{\R,\Y}$ in the sense that for all $0 <\varepsilon < 1$,
\begin{align*}
    \sup_{Q} N(\G \times \R \cup \G \times \tilde{\R}, \norm{\cdot}_{Q,2}, 2 \varepsilon \lVert \ttM_{\G} \ttM_{\R,\Y} \rVert_{Q,2}) \leq \ttc_{\G} \ttc_{\R} (\varepsilon^2/4)^{-\ttd_{\G}-\ttd_{\R}},
\end{align*}
where the supremum is over all finite discrete measures on $\reals^{d+1}$. Here $\ttc_{\R}$ and $\ttd_{\R}$ are VC-type constants for $\R\cup\tilde{\R}$; truncation by the fixed indicator $\Indicator(|y|\leq\kappa_n)$ preserves the entropy exponent, so these constants can be chosen with $\ttc_{\R}\geq \ttc_{\R,\Y}$ and $\ttd_{\R}=\ttd_{\R,\Y}$ up to universal constants. Hence $\G \times \tilde{\R} \cup \G \times \R$ is pre-Gaussian, and on some probability space there exists a mean-zero Gaussian process $\bar{Z}_n^R$ indexed by $\F = \G \times \tilde{\R} \cup \G \times \R$ with the same covariance structure as $R_n$ and almost surely continuous paths with respect to the metric $\rho$, given by
\begin{align*}
    \rho((g_1,r_1),(g_2,r_2)) =  \E[(\bar{Z}_n^R(g_1, r_1) - \bar{Z}_n^R(g_2, r_2))^2]^{\frac{1}{2}} = \E[(R_n(g_1, r_1) - R_n(g_2, r_2))^2]^{\frac{1}{2}},
\end{align*}
for $(g_1, r_1), (g_2, r_2) \in \F$. Recall the definition of $\Delta\R$ in Part 2. The increment diameter between each original function and its truncated counterpart satisfies
\begin{align*}
    \sigma \equiv \sup_{g \in \G, r \in \R}\rho((g,r),(g,\tilde r))
    \leq \sup_{g \in \G, h \in \Delta\R} \E[g(\bx_i)^2 h(y_i)^2]^{1/2}
    \lesssim \sqrt{\ttM_{\G} \ttE_{\G} \kappa_n^{-v}}.
\end{align*}
The entropy bound above also implies, for all $0 < \varepsilon < 1$,
\begin{align*}
    N(\G \times \R \cup \G \times \tilde{\R}, \rho, 2\varepsilon\lVert \ttM_{\G} M_{\R,\Y}\rVert_{\P,2})
    \leq \ttc_{\G} \ttc_{\R,\Y}(\varepsilon^2/4)^{-\ttd_{\G}-\ttd_{\R,\Y}}.
\end{align*}
Denote $A =  (\ttc_{\G} \ttc_{\R,\Y})^{\frac{1}{2 \ttd_{\G} + 2 \ttd_{\R,\Y}}}/4$ and $D = 2 \ttd_{\G} + 2 \ttd_{\R,\Y}$. By \citet[Corollary 2.2.8]{van-der-Vaart-Wellner_1996_Book}, choosing any $(g_0, r_0) \in \G \times \R$,
\begin{align*}
    \E\bigg[\lVert \bar{Z}_n^R(g,r) - \bar{Z}_n^R(g,\tilde{r}) \rVert_{\G \times \R}\bigg] & \lesssim \E \left[\left|\bar{Z}_n^R(g_0, r_0) - \bar{Z}_n^R(g_0, \tilde{r}_0)\right|\right] + \int_{0}^{\sigma} \sqrt{\log\Big( \ttc_{\G} \ttc_{\R}\big(\frac{\ttM_{\G}}{\varepsilon}\big)^{\ttd_{\G} + \ttd_{\R}}\Big)} \dif \varepsilon \\
    & \leq \sqrt{D \log (A \sqrt{\ttM_{\G}/\ttE_{\G}})} \sqrt{\ttM_{\G} \ttE_{\G}} \kappa_n^{-v/2} \\
    & \lesssim \sqrt{(\ttd_{\G} + \ttd_{\R,\Y}) \log(\ttc_{\G} \ttc_{\R,\Y} \ttk n)} \sqrt{\ttM_{\G} \ttE_{\G}} \kappa_n^{-v/2}.
\end{align*}
Since the restriction of $(\bar{Z}_n^R(g,r): (g,r)\in\F)$ to $\G\times\tilde{\R}$ has the same distribution as the Gaussian process constructed in Part 1, the Vorob'ev--Berkes--Philipp theorem \citep[Theorem 1.31]{dudley2014uniform} allows us to construct $\bar{Z}_n^R$ on the same probability space as $(\bx_i,y_i)_{1 \leq i \leq n}$ and $Z_n^R$, with $\bar{Z}_n^R$ and $Z_n^R$ coinciding on $\G \times \tilde{\R}$. By an abuse of notation, call this extended Gaussian process $Z_n^R$.

\subsubsection*{Part 4: Putting Together}

Combining the previous three parts, choose $\kappa_n$ so that
\begin{align*}
    \mathtt{r}_n \kappa_n \asymp \sqrt{\ttM_{\G} \ttE_{\G}} \kappa_n^{-v/2},
    \qquad\text{that is}\qquad
    \kappa_n \asymp \left(\frac{\sqrt{\ttM_{\G}\ttE_{\G}}}{\mathtt{r}_n}\right)^{\frac{2}{v+2}}.
\end{align*}
Then
\begin{align*}
    \E \big[\lVert R_n - Z_n^R \rVert_{\G \times \R} \big] & \lesssim \ttv(\ttd\log (\ttc n))^{3/2} \mathtt{r}_n^{\frac{v}{v +2}}(\sqrt{\ttM_{\G}\ttE_{\G}})^{\frac{2}{v+2}} + \ttv\ttd\log(\ttc n) \ttM_{\G} n^{-\frac{v/2}{2+v}} \\
    & \qquad +  \ttv\ttd\log(\ttc n) \ttM_{\G} n^{-1/2} \Big(\frac{\sqrt{\ttM_{\G} \ttE_{\G}}}{\mathtt{r}_n}\Big)^{\frac{2}{v+2}} \\
    & \lesssim \ttv \ttd\log(\ttc n)\rho_n,
\end{align*}
where $\ttd = \ttd_{\G} + \ttd_{\R,\Y} + \ttk$, and $\ttc = \ttc_{\G} \ttc_{\R,\Y} \ttk$. This proves the stated bound.
\qed

\section{Proofs}\label{sa-sec: Proofs}

\subsection{Proof of Lemma~\ref{sa-lem: Invertibility}}

Because $\B$ is compact and $\B\subset\setint(\X)$, compact support of $K$ implies that $\bx+\bH\supp(K)\subseteq\X$ uniformly over $\bx\in\B$ for all sufficiently small $\bar h$. Assumption~\ref{sa-assump: Sharp DGP}(ii) and uniform continuity of $f_X$ on the compact set $\X$ imply
\begin{align*}
    \bGamma_{t,\bx} & = \E \Big[\br_p(\bH^{-1}(\bX_i-\bx)) \br_p(\bH^{-1}(\bX_i-\bx))^{\top} K_{\bH}(\bX_i - \bx) \Indicator(\bX_i \in \A_t) \Big] \\
    & = \int_{\A_t} \br_p(\bH^{-1}(\bu - \bx)) \br_p(\bH^{-1}(\bu - \bx))^{\top} K_{\bH}(\bu - \bx) f_X(\bu) \dif \bu \\
    & = f_X(\bx)\int_{\A_t} \br_p(\bH^{-1}(\bu - \bx)) \br_p(\bH^{-1}(\bu - \bx))^{\top} K_{\bH}(\bu - \bx) \dif \bu + o(1),
\end{align*}
where in the last line we have used $\sup_{\bu\in\bx+\bH\supp(K)}|f_X(\bu)-f_X(\bx)|=o(1)$ uniformly over $\bx\in\B$ and $\int_{\A_t} (\bH^{-1}(\bu - \bx))^{\bv} K_{\bH}(\bu - \bx) \dif \bu = O(1)$ for any multi-index $\bv$ from a standard change of variable argument.

\begin{center}
\textbf{I. Polynomial Representation of Minimum Eigenvalue}
\end{center}
For each bandwidth matrix $\bH$, define
\begin{align*}
    \bA_{t,\bx}(\bH) = \int_{\A_t} \br_p(\bH^{-1}(\bu - \bx)) \br_p(\bH^{-1}(\bu - \bx))^{\top} K_{\bH}(\bu - \bx) \dif \bu.
\end{align*}
The argument below establishes a uniform positive lower bound for $\lambda_{\min}(\bA_{t,\bx}(\bH))$ for all sufficiently small $\bar{h}$; no pointwise limit of $\bA_{t,\bx}(\bH)$ is needed.
A change of variable gives
\begin{align*}
    \bA_{t,\bx}(\bH) = \int \br_p(\bz) \br_p(\bz)^{\top} K(\bz) \Indicator(\bx + \bH \bz \in \A_t) \dif \bz.
\end{align*}
Let $\ba \in \reals^{\mathfrak{p}_p}$, where $\mathfrak{p}_p = \frac{(d + p)!}{d! p !}$.
Then the equivalent representation of minimum eigenvalue gives
\begin{align}\label{sa-eq: min eigenvalue}
    \nonumber \lambda_{\min}(\bA_{t,\bx}(\bH))
    & = \min_{\norm{\ba} = 1} \int (\ba^{\top} \br_p(\bz))^2 K(\bz) \Indicator(\bx + \bH \bz \in \A_t) \dif \bz \\
    & \geq \kappa \min_{\norm{\ba} = 1} \int_{U} (\ba^{\top} \br_p(\bz))^2  \Indicator(\bx + \bH \bz \in \A_t) \dif \bz,
\end{align}
where in the last line we have used $K(\bu) \geq \kappa$ for all $\bu \in U$.

\begin{center}
\textbf{II. Mass Retaining Ratio in Treatment/Control Region}
\end{center}

Denote $E_{\bH}(\bx,t) = \{\bz \in U: \bx + \bH \bz \in \A_t \}$. Assumption~\ref{sa-assump: Boundary and Kernel} (ii) implies there is some upper bound $\Lambda > 0$ of $K(\cdot)$. Hence for $c_0 = 1/2 \; \liminf_{\bar{h} \downarrow 0}\inf_{\bx \in \B} \int_{U} K(\bu) \Indicator(\bx + \bH \bu \in \A_t) \dif \bu$, we have
\begin{align*}
    \Lambda \leb(E_{\bH}(\bx,t))
    & \geq \int_{U} K(\bu) \Indicator(\bx + \bH \bu \in \A_t) \dif \bu
    \geq c_0
\end{align*}
for small enough $\bar{h}$, which implies
\begin{align}\label{sa-eq: mass relation}
      \leb(E_{\bH}(\bx,t))  \geq \alpha \leb(U), \qquad \alpha = \frac{c_0}{\Lambda \leb(U)}.
\end{align}

\begin{center}
\textbf{III. $L_2$ Integral of Polynomials in Full Versus Treatment/Control Regions}
\end{center}

Consider $S = \{f \in \mathcal{P}_{p}: \int_U f(\bu)^2 \dif \bu = 1\}$, where $\mathcal{P}_{p}$ is the collection of all polynomials in $\bu$ with total degree at most $p$. Let $(\phi_j, 1 \leq j \leq \mathfrak{p}_p)$ be an orthonormal basis of $(\mathcal{P}_{p}, \lVert \cdot \rVert_{L_2})$. Then $T(\mathbf{a}) = \sum_{j = 1}^{\mathfrak{p}_p} a_j \phi_j$ is an isometry. Since $T(S) = \{\mathbf{a} \in \reals^{\mathfrak{p}_p}: \norm{\ba} = 1\}$ is compact, $S$ is also compact in $(\mathcal{P}_{p}, \lVert \cdot \rVert_{L_2})$. Since $\mathcal{P}_{p}$ is $\mathfrak{p}_p$-dimensional, equivalence of norms implies that $S$ is also compact in $(\mathcal{P}_{p}, \lVert \cdot \rVert_{L_{\infty}})$. Now consider
\begin{align*}
    \Phi_q(\varepsilon) = \leb(\{\bu \in U: |q(\bu)| < \varepsilon \}), \qquad q \in S, \varepsilon > 0,
\end{align*}
and
\begin{align*}
    \psi(q) = \sup \Big\{\varepsilon > 0: \Phi_q(\varepsilon) \leq \frac{\alpha}{2} \leb(U) \Big\}.
\end{align*}
Since $\int_U q^2 = 1$ and $q$ is polynomial, $\lim_{\varepsilon \downarrow 0} \Phi_q(\varepsilon) = 0$ and $\Phi_q(\lVert q\rVert_{\infty}) = \leb(U)$. Continuity and Lipschitzness of $q \in S$ imply $\psi(q) > 0$ for all $q \in S$.

Next, we want to show $\psi$ is lower-semicontinuous on $(\mathcal{P}_{p}, \lVert \cdot \rVert_{L_{\infty}})$. Suppose $q_n \to q$ uniformly on $U$. For every $\varepsilon_0 \in (0, \psi(q))$, there exists $\eta > 0$ such that $\Phi_q(\varepsilon_0) \leq \frac{\alpha}{2} \leb(U) - \eta$. Continuity of polynomials and the fact that level sets of nonzero polynomials have zero Lebesgue measure imply $\Indicator_{\{|q_n| < \varepsilon_0\}}(\cdot) \to \Indicator_{\{|q| < \varepsilon_0\}}(\cdot)$ almost surely. By Dominated Convergence Theorem, $\Phi_{q_n}(\varepsilon_0) \to \Phi_q(\varepsilon_0)$. Hence for large enough $n$, $\Phi_{q_n}(\varepsilon_0) \leq \frac{\alpha}{2}\leb(U)$, which implies $\varepsilon_0 \leq \psi(q_n)$. This implies $\liminf_{n \to \infty} \psi(q_n) \geq \varepsilon_0$. Since $\varepsilon_0$ is arbitrary in $(0, \psi(q))$, we have $\liminf_{n \to \infty} \psi(q_n) \geq \psi(q)$.

Compactness of $S$ and lower-semicontinuity of $\psi$ implies $\psi$ attains its minimum on $S$. Since $\psi(q) > 0$ for all $q \in S$, we know $\varepsilon_* = \inf_{q \in S} \psi(q) > 0$. Then for every $q \in S$,
\begin{align*}
    \int_{E_{\bH}(\bx,t)} q^2
    & \geq \varepsilon_*^2 \; \leb \Big(E_{\bH}(\bx,t) \setminus \{|q| \leq \varepsilon_*\} \Big) \\
    & \geq \varepsilon_*^2 \; \Big(\leb(E_{\bH}(\bx,t)) - \leb(\{|q| \leq \varepsilon_*\}) \Big) \\
    & \geq \varepsilon_*^2 \; \frac{\alpha}{2} \leb(U).
\end{align*}
Scaling $q$ from $S$ gives
\begin{align}\label{sa-eq: integral relation}
    \int_{E_{\bH}(\bx,t)} q^2 \geq \varepsilon_*^2 \; \frac{\alpha}{2} \int_{U} q^2, \qquad q \in \mathcal{P}_{p}.
\end{align}

\begin{center}
\textbf{IV. Lower Bound of Minimum Eigenvalue}
\end{center}

Equations~\eqref{sa-eq: min eigenvalue}, \eqref{sa-eq: mass relation} and \eqref{sa-eq: integral relation} together give for small enough $\bar{h}$,
\begin{align*}
    \inf_{\bx \in \B}\lambda_{\min}(\bA_{t,\bx}(\bH))
    & \geq \kappa \inf_{\bx \in \B}\min_{\norm{\ba} = 1} \int_{E_{\bH}(\bx,t)} (\ba^{\top} \br_p(\bz))^2   \dif \bz, \\
    & \geq \kappa \varepsilon_*^2 \; \frac{\alpha}{2}  \min_{\norm{\ba} = 1} \int_{U} (\ba^{\top} \br_p(\bz))^2 \dif \bz \\
    & \geq \kappa \varepsilon_*^2 \; \frac{\alpha}{2}  \lambda_{\min} \Big(\int_U \br_p(\bz) \br_p(\bz)^{\top} \dif \bz \Big),
\end{align*}
which implies $\liminf_{\bar{h} \to 0} \inf_{\bx \in \B}\lambda_{\min}(\bA_{t,\bx}(\bH)) > 0$. Since Assumption~\ref{sa-assump: Sharp DGP}(ii) gives $\inf_{\bx\in\X}f_X(\bx)>0$ and the preceding expansion of $\bGamma_{t,\bx}$ is uniform over $\bx\in\B$, it follows that
\begin{align*}
    \liminf_{\bar h\to0}\inf_{\bx\in\B}\lambda_{\min}(\bGamma_{t,\bx})
    \geq
    \inf_{\bx\in\X}f_X(\bx)\,
    \liminf_{\bar h\to0}\inf_{\bx\in\B}\lambda_{\min}(\bA_{t,\bx}(\bH))
    >0.
\end{align*}

\subsection{Proof of Lemma~\ref{sa-lem: Gram Matrix}}

Since $\widehat{\bGamma}_{t,\bx}$ is a finite dimensional matrix, it suffices to show the stated rate of convergence for each entry. Let $\bv$ be a multi-index such that $|\bv| \leq 2 p$. Define
\begin{align*}
    g_n(\xi,\bx) = (\bH^{-1}(\xi - \bx))^{\bv} K_{\bH}(\xi - \bx)\Indicator(\xi \in \A_t), \qquad \xi \in \mathcal{X}, \bx \in \B.
\end{align*}
and $\F = \{g_n(\cdot, \bx): \bx \in \B\}$. We will show $\F$ is a VC-type class. In order to do this, we study the following quantities.

\medskip\textit{Constant Envelope Function}. We assume $K$ is continuous and has compact support, or $K = \Indicator(\cdot \in [-1,1]^d)$. Hence there exists a constant $C_1$ such that for all $l \in \mathcal{F}$ and all $\xi \in \X$, $|l(\xi)| \leq C_1 |\bH|^{-1} = F$.

\medskip\textit{Diameter of $\F$ in $L_2$}. $ \sup_{l \in \mathcal{F}} \norm{l}_{\P,2}
= \sup_{\bx \in \B}(\int_{\bH^{-1}(\A_t - \bx)} |\bH|^{-1}\by^{2\bv}K(\by)^2 f_X(\bx + \bH \by) \dif \by)^{1/2} \leq C_2 |\bH|^{-1/2}$ for some constant $C_2$. We can take $C_1$ large enough so that $\sigma = C_2 |\bH|^{-1/2} \leq F = C_1 |\bH|^{-1}$.

\medskip\textit{Ratio}. For some constant $C_3$, $\delta = \frac{\sigma}{F}  = C_3 \sqrt{|\bH|}$.

\medskip\textit{Covering Numbers}. Case 1: $K$ is Lipschitz. Let $\bx,\bx' \in \B$. Then, for generic evaluation points $\bx = (x_1,\ldots,x_d)^\top$ and $\bx' = (x_1',\ldots,x_d')^\top$,
\begin{align*}
    \sup_{\xi \in \X} \left|g_n(\xi,\bx) - g_n(\xi, \px) \right|
    & \leq  \left|(\bH^{-1}(\xi-\bx))^{\bv}  - (\bH^{-1}(\xi-\px))^{\bv} \right| K_{\bH}(\xi - \bx) \\
    & \qquad + |(\bH^{-1}(\xi-\px))^{\bv}| \Big|K_{\bH}(\xi - \bx) - K_{\bH}(\xi - \px) \Big| \\
    & \lesssim |\bH|^{-1}\underline h^{-1} \infnorm{\bx - \px},
\end{align*}
since we have assumed that $K$ has compact support and is Lipschitz continuous. Hence, for any $\varepsilon \in (0,1]$ and for any finitely supported measure $Q$ and metric $\norm{\cdot}_{Q,2}$ based on $L_2(Q)$,
\begin{align*}
    N(\F, \norm{\cdot}_{Q,2}, \varepsilon \norm{F}_{Q,2})
    \leq N(\X, \infnorm{\cdot}, \varepsilon \norm{F}_{Q,2} |\bH|\underline h)
    \stackrel{(i)}{\lesssim} \left(\frac{\diam(\X)}{ \varepsilon \norm{F}_{Q,2} |\bH|\underline h}\right)^d
    \lesssim \left(\frac{\diam(\X)}{\varepsilon \underline h}\right)^d,
\end{align*}
where in ($i$) we used compactness of $\X$ and the fact that $\varepsilon \norm{F}_{Q,2}|\bH|\underline h \lesssim \varepsilon \underline h \lesssim 1$. Hence, $\F$ forms a VC-type class, and taking $A_1 = \diam(\X)/\underline h$ and $A_2 = d$, $\sup_Q N(\F, \norm{\cdot}_{Q,2}, \varepsilon \norm{F}_{Q,2}) \lesssim (A_1/ \varepsilon)^{A_2}$, $\varepsilon \in (0,1]$, and where the supremum is over all finite discrete measures.

Case 2: $K = \Indicator(\cdot \in [-1,1]^d)$. Consider
\begin{align*}
    m_n(\xi, \bx)
    = (\bH^{-1}(\xi - \bx))^{\bv} |\bH|^{-1} \Indicator(\xi \in \A_t),
    \qquad \xi, \bx \in \mathcal{X},
\end{align*}
$\M = \{m_n(\cdot, \bx): \bx \in \B\}$ and the constant envelope function $M = C_4 |\bH|^{-1}\underline h^{-|\bv|}$, for some constant $C_4$ only depending on diameter of $\X$. The same argument as before shows that for any discrete measure $Q$, we have
\begin{align*}
    N(\M, \norm{\cdot}_{Q,2}, \varepsilon \norm{M}_{Q,2})
    & \leq N(\X, \infnorm{\cdot}, \varepsilon \norm{M}_{Q,2} |\bH|\underline h^{|\bv|+1})
    \lesssim \Big(\frac{\diam(\X)}{\varepsilon \norm{M}_{Q,2} |\bH|\underline h^{|\bv|+1}}\Big)^d
    \lesssim \Big(\frac{\diam(\X)}{\varepsilon \underline h}\Big)^d.
\end{align*}
The class $\G = \{\Indicator(\bH^{-1}(\cdot - \bx) \in [-1,1]^d): \bx \in \B\}$ has VC dimension no greater than $2d$ \cite[Example 2.6.1]{van-der-Vaart-Wellner_1996_Book}, and by \citet[Theorem 2.6.4]{van-der-Vaart-Wellner_1996_Book}, for any discrete measure $Q$, $N(\G, \norm{\cdot}_{Q,2}, \varepsilon) \leq 2d (4 e)^{2d} \varepsilon^{-4d}$, $0 < \varepsilon \leq 1$. The class $\F$ is contained in the product class $\M\cdot\G$. Since the envelope ratio satisfies $\norm{F}_{Q,2}/\norm{M}_{Q,2}\gtrsim\underline h^{|\bv|}$ uniformly in $Q$, the standard product-covering inequality gives, for a constant $c>0$,
\begin{align*}
    N(\F, \norm{\cdot}_{Q,2}, \varepsilon \norm{F}_{Q,2})
    &\leq
    N(\M, \norm{\cdot}_{Q,2}, c\varepsilon \underline h^{|\bv|} \norm{M}_{Q,2})
    N(\G, \norm{\cdot}_{Q,2}, c\varepsilon \underline h^{|\bv|})\\
    &\lesssim
    \left(\frac{\diam(\X)}{\varepsilon \underline h^{|\bv|+1}}\right)^d
    (\varepsilon\underline h^{|\bv|})^{-4d}
    \lesssim
    \varepsilon^{-5d}\underline h^{-d(5|\bv|+1)}.
\end{align*}
Hence, taking $A_1 = C\underline h^{-(|\bv|+1)}$ and $A_2 = 5d$ for a large enough constant $C$, $\sup_Q N(\F, \norm{\cdot}_{Q,2}, \varepsilon \norm{F}_{Q,2}) \lesssim (A_1/ \varepsilon)^{A_2}$, $\varepsilon \in (0,1]$, where the supremum is over all finite discrete measures.

\medskip\textit{Maximal Inequality}. By Corollary 5.1 in \cite{Chernozhukov-Chetverikov-Kato_2014b_AoS} for the empirical process on class $\F$,
\begin{align*}
     \E \Big[\sup_{\bx \in \B} \big|\En [g_n(\bX_i, \bx)]  - \E[g_n(\bX_i, \bx)]\big| \Big]
    & \lesssim \frac{\sigma}{\sqrt{n}}\sqrt{A_2\log(A_1/\delta)} + \frac{\pnorm{F} A_2\log(A_1/\delta)}{n} \\
    & \lesssim \sqrt{\frac{\log(1/\bar{h})}{n |\bH|}} + \frac{\log(1/\bar{h})}{n |\bH|},
\end{align*}
where $A_1, A_2, \sigma, F, \delta$ are all given previously. Assuming $\frac{\log(1/\bar{h})}{n |\bH|} \to 0$, we conclude that $\sup_{\bx \in \B} \big\|\widehat{\bGamma}_{t,\bx} - \bGamma_{t,\bx}\big\| \lessp \sqrt{\frac{\log(1/\bar{h})}{n |\bH|}}$. By Weyl's Theorem, $\sup_{\bx \in \B}|\lambda_{\min}(\widehat{\bGamma}_{t,\bx}) - \lambda_{\min}(\bGamma_{t,\bx})| \leq \sup_{\bx \in \B} \big\|\widehat{\bGamma}_{t,\bx} - \bGamma_{t,\bx}\big\| \lessp \sqrt{\frac{\log(1/\bar{h})}{n |\bH| }}$. Lemma~\ref{sa-lem: Invertibility} gives $\lambda_{\min}(\bGamma_{t,\bx}) \gtrsim 1$ uniformly over $\bx\in\B$. Therefore, $\inf_{\bx \in \B}\lambda_{\min}(\widehat{\bGamma}_{t,\bx}) \gtrsim_{\P} 1$, $\sup_{\bx \in \B} \big\|\widehat{\bGamma}_{t,\bx}\big\| \lessp 1$, and $\sup_{\bx \in \B} \big\|\widehat{\bGamma}_{t,\bx}^{-1}\big\| \lessp 1$. Hence $\sup_{\bx \in \B} \big\|\widehat{\bGamma}_{t,\bx}^{-1} - \bGamma_{t,\bx}^{-1}\big\| \leq \sup_{\bx \in \B} \big\|\bGamma_{t,\bx}^{-1}\big\| \big\|\bGamma_{t,\bx} -\widehat{\bGamma}_{t,\bx}\big\| \big\|\widehat{\bGamma}_{t,\bx}^{-1}\big\| \lessp \sqrt{\frac{\log(1/\bar{h})}{n |\bH|}}$.

\subsection{Proof of Lemma~\ref{sa-lem: Linear Approximation}}

The proof is similar to the proof of Lemma~\ref{sa-lem: Gram Matrix}. Let $\bv$ be a multi-index such that $ 0 \leq |\bv| \leq p$. Let
\begin{align*}
    g_n(\xi,\bx) = (\bH^{-1}(\xi - \bx))^{\bv} K_{\bH}(\xi - \bx)\Indicator(\xi\in \A_t), \qquad \xi , \bx \in \X.
\end{align*}
Define the class of functions $\F = \{(\xi, u) \in \X \times \reals \mapsto g_n(\xi,\bx)u: \bx \in \B\}$.

\medskip\textit{Envelope Function}. Since $K$ is continuous on its compact support, there exists a constant $C_1 > 0$ such that $|g_n(\xi,\bx)u| \leq C_1 |\bH|^{-1} |u|$, for $\xi, \bx \in \X$ and $ u \in \reals$. We define the envelope function $F(\xi, u) = C_1 |\bH|^{-1}|u|$, for $\xi \in \X$ and $ u \in \reals$. Moreover, by Assumption~\ref{sa-assump: Sharp DGP}(v), let $M = \max_{1 \leq i \leq n} F(\bX_i, \varepsilon_i)$, then
\begin{align*}
    \E[M^2]^{1/2}
    \lesssim |\bH|^{-1} \E \big[\max_{1 \leq i \leq n} |\varepsilon_i|^2 \big]^{1/2}
    \lesssim |\bH|^{-1} \E \big[\max_{1 \leq i \leq n} |\varepsilon_i|^{2+v}\big]^{1/(2+v)}
    \lesssim n^{1/(2+v)} |\bH|^{-1}.
\end{align*}

\medskip\textit{Diameter of $\F$ in $L_2$}. Recall we denote $\varepsilon_i = Y_i - \E[Y_i|\bX_i]$, then
\begin{align*}
    \sup_{l \in \mathcal{F}} \E[l(\bX_i, \varepsilon_i)^2]^{1/2}
    \leq \sup_{\xi \in \X} \E[\varepsilon_i^2 | \bX_i = \xi]^{1/2} \sup_{\xi \in \X} \E[g_n(\bX_i, \xi)^2]^{1/2} \leq C_3 |\bH|^{-1/2}
    = \sigma.
\end{align*}

\medskip\textit{Ratio}. We set $\delta = \frac{\sigma}{\norm{F}_{\P,2}} \lesssim |\bH|^{1/2}$.

\medskip\textit{Covering Numbers}. Case 1: $K$ is Lipschitz. Let $\mathbb{Q}$ be a finite distribution on $(\X \times \reals, \B(\X) \otimes \mathsf{Borel}(\reals))$. Let $\bx, \bx' \in \X$. In the proof of Lemma~\ref{sa-lem: Gram Matrix}, we showed that $\sup_{\xi \in \mathcal{X}} \sup_{\bx, \bx' \in \X } \frac{|g_n(\xi,\bx) - g_n(\xi, \bx')|}{\infnorm{\bx - \bx'}} \lesssim |\bH|^{-1}\underline h^{-1}$. Hence,
\begin{align*}
    \| g_n(\bX_i, \bx) \varepsilon_i - g_n(\bX_i, \bx')\varepsilon_i \|_{\mathbb{Q},2}
    \leq \infnorm{g_n(\cdot, \bx) - g_n(\cdot,\bx')} \norm{\varepsilon_i}_{\mathbb{Q},2}
    \lesssim \underline h^{-1} \norm{F}_{\mathbb{Q},2} \infnorm{\bx- \px}.
\end{align*}
It follows that $\sup_{Q} N(\F, \lVert \cdot \rVert_{Q,2}, \epsilon \| F \|_{Q,2}) \lesssim (\frac{\diam(\X)}{\epsilon \underline h})^d$, where $\sup$ is over all finite probability distributions on $(\X \times \reals, \B(\X) \otimes \mathsf{Borel}(\reals))$. Letting $A_1 = \frac{\diam(\X)}{\underline h}$ and $ A_2 = d$, we conclude that
\begin{align*}
    \sup_Q N(\F, \norm{\cdot}_{Q,2}, \epsilon \norm{F}_{Q,2}) \lesssim (A_1/\epsilon)^{A_2}, \qquad \epsilon \in (0,1].
\end{align*}

Case 2: $K$ is the uniform kernel. Let
\begin{align*}
    m_n(\xi, \bx) =  (\bH^{-1}(\xi - \bx))^{\bv} |\bH|^{-1} \Indicator(\xi \in \A_t), \qquad \xi, \bx \in \mathcal{X},
\end{align*}
with $\M = \{(\xi,u) \in \X \times \reals \to m_n(\xi,\bx)u: \bx \in \B\}$ and envelope function $M(\xi,u) = C_1 |\bH|^{-1}\underline h^{-|\bv|} |u|$, for a positive constant $C_1$ depending only on $K$. By similar arguments as Case 1 and the proof of Lemma~\ref{sa-lem: Gram Matrix}, it follows that $\sup_Q N(\M, \norm{\cdot}_{Q,2}, \varepsilon \|M\|_{Q,2}) \lesssim \big(\frac{\diam(\X)}{\varepsilon \underline h}\big)^d$, where the supremum is taken over all finite discrete measures. Taking $\G = \{\Indicator(\bH^{-1}(\cdot - \bx) \in [-1,1]^d): \bx \in \B\}$, the proof of Lemma~\ref{sa-lem: Gram Matrix} shows that
\begin{align*}
    \sup_Q N(\G, \norm{\cdot}_{Q,2}, \varepsilon) \leq 2d (4 e)^{2d} \varepsilon^{-4d}, \qquad \varepsilon \in (0,1],
\end{align*}
where the supremum is taken over all finite discrete measures. The class $\F$ is contained in $\M\cdot\G$. Since $\norm{F}_{Q,2}/\norm{M}_{Q,2}\gtrsim\underline h^{|\bv|}$, the product-covering inequality gives, for a constant $c>0$,
\begin{align*}
    N(\F, \norm{\cdot}_{Q,2}, \varepsilon \norm{F}_{Q,2})
    &\leq
    N(\M, \norm{\cdot}_{Q,2}, c\varepsilon\underline h^{|\bv|}\norm{M}_{Q,2})
    N(\G, \norm{\cdot}_{Q,2}, c\varepsilon\underline h^{|\bv|})\\
    &\lesssim
    \varepsilon^{-5d}\underline h^{-d(5|\bv|+1)}.
\end{align*}
Taking $A_1 = C\underline h^{-(|\bv|+1)}$ and $A_2 = 5d$ for a large enough constant $C$, we have
\begin{align*}
    \sup_Q N(\F, \norm{\cdot}_{Q,2}, \varepsilon \norm{F}_{Q,2}) \lesssim (A_1/ \varepsilon)^{A_2}, \qquad \varepsilon \in (0,1],
\end{align*}
the supremum is over all finite discrete measures.

\medskip\textit{Maximal Inequality}. By Corollary 5.1 in \cite{Chernozhukov-Chetverikov-Kato_2014b_AoS},
\begin{align*}
    \E \Big[\sup_{\bx \in \B} \big|\En [g_n(\bX_i,\bx)\varepsilon_i] \big| \Big]
    & \lesssim \frac{\sigma}{\sqrt{n}}\sqrt{A_2\log(A_1/\delta)} + \frac{\pnorm{M} A_2\log(A_1/\delta)}{n}\\
    & \lesssim \sqrt{\frac{\log(1/\bar{h})}{n |\bH|}} +\frac{\log(1/\bar{h})}{n^{\frac{1+v}{2+v}} |\bH|}.
\end{align*}
Since $\bQ_{t,\bx}$ is finite-dimensional, entry-wise convergence implies convergence in norm with the same rate. Hence, $\sup_{\bx \in \B} \big\|\bQ_{t,\bx}\big\| \lessp \sqrt{\frac{\log(1/\bar{h})}{n |\bH|}} + \frac{\log(1/\bar{h})}{n^{\frac{1+v}{2+v}}|\bH|}$. By Lemma~\ref{sa-lem: Gram Matrix},
\begin{align*}
    \sup_{\bx \in \B} \big|\widehat{\mu}_t^{(\bnu)}(\bx) - \E [\widehat{\mu}_t^{(\bnu)}(\bx) | \bX ]
                           - \bnu!\be_{\bnu}^{\top} \bS^{-1} \bGamma_{t,\bx}^{-1} \bQ_{t,\bx} \big|
    &= \sup_{\bx \in \B} \big| \bnu!\be_{\bnu}^{\top} \bS^{-1} \big(\widehat{\bGamma}_{t,\bx}^{-1} - \bGamma_{t,\bx}^{-1} \big) \bQ_{t,\bx}\big| \\
    & \lessp \bh^{-\bnu} \sqrt{\frac{\log(1/\bar{h})}{n |\bH|}} \Big(\sqrt{\frac{\log(1/\bar{h})}{n |\bH|}} + \frac{\log(1/\bar{h})}{n^{\frac{1+v}{2+v}}|\bH|} \Big),
\end{align*}
and
\begin{align*}
    \sup_{\bx \in \B} \big|\widehat{\mu}_t^{(\bnu)}(\bx) - \E[\widehat{\mu}_t^{(\bnu)}(\bx) | \bX ] \big|
    \lessp \bh^{-\bnu} \Big(\sqrt{\frac{\log(1/\bar{h})}{n |\bH|}} + \frac{\log(1/\bar{h})}{n^{\frac{1+v}{2+v}}|\bH|}\Big),
\end{align*}
which completes the proof.
\qed

\subsection{Proof of Lemma~\ref{sa-lem: covariance}}

Let $\eta_i(\bx) = \sum_{t \in \{0,1\}} \Indicator(\bX_i \in \A_t) (\mu_t(\bX_i) - \widehat{\bbeta}_t(\bx)^\top \br_p(\bX_i - \bx))$. Then, for all $\bx, \by \in \B$, the difference between the estimated and true variance matrices is
\begin{align*}
    \widehat{\bSigma}_{t,\bx,\by} - \bSigma_{t,\bx,\by} = \bM_{1,\bx,\by} + \bM_{2,\bx,\by} + \bM_{3,\bx,\by} + \bM_{4,\bx,\by}
\end{align*}
where
\begin{align*}
    \bM_{1,\bx,\by}
    &= \En \Big[\br_p(\bH^{-1}(\bX_i - \bx)) \br_p(\bH^{-1}(\bX_i - \by))^{\top} |\bH| K_{\bH}(\bX_i - \bx)K_{\bH}(\bX_i - \by)\eta_i(\bx) \eta_i(\by) \Indicator(\bX_i \in  \A_t)\Big],\\
    \bM_{2,\bx,\by}
    &= \En \Big[\br_p(\bH^{-1}(\bX_i - \bx)) \br_p(\bH^{-1}(\bX_i - \by))^{\top} |\bH| K_{\bH}(\bX_i - \bx)K_{\bH}(\bX_i - \by) (\eta_i(\bx) + \eta_i(\by)) \varepsilon_i \Indicator(\bX_i \in \A_t)\Big],\\
    \bM_{3,\bx,\by}
    &= \En \Big[\br_p(\bH^{-1}(\bX_i - \bx))\br_p(\bH^{-1}(\bX_i - \by))^{\top}|\bH|K_{\bH}(\bX_i - \bx)K_{\bH}(\bX_i - \by)(\varepsilon_i^2 - \sigma_t(\bX_i)^2)\Indicator(\bX_i \in \A_t)\Big], \\
    \bM_{4,\bx,\by}
    &= \En \Big[\br_p(\bH^{-1}(\bX_i - \bx))\br_p(\bH^{-1}(\bX_i - \by))^{\top}|\bH|K_{\bH}(\bX_i - \bx)K_{\bH}(\bX_i - \by)\sigma_t(\bX_i)^2\Indicator(\bX_i \in  \A_t)\Big]\\
    &\qquad - \E \Big[\br_p(\bH^{-1}(\bX_i - \bx))\br_p(\bH^{-1}(\bX_i - \by))^{\top}|\bH|K_{\bH}(\bX_i - \bx)K_{\bH}(\bX_i - \by)\sigma_t(\bX_i)^2 \Indicator(\bX_i \in  \A_t)\Big].
\end{align*}

For $\bu$ and $\bv$ multi-indices, let $g_n(\bX_i; \bx, \by) = |\bH|K_{\bH}(\bX_i - \bx)K_{\bH}(\bX_i - \by)(\bH^{-1}(\bX_i - \bx))^{\bu}(\bH^{-1}(\bX_i - \by))^{\bv} \Indicator(\bX_i \in  \A_t)$. Set
\begin{align*}
    \mathfrak{R}_{\mathtt{EP}} = \sqrt{\frac{\log (1/\bar{h})}{n |\bH|}} + \frac{\log (1/\bar{h})}{n^{\frac{1+v}{2+v}}|\bH|}.
\end{align*}

First, we present a bound on $\max_{1 \leq i \leq n} |\eta_i(\bx)| \Indicator(\bH^{-1}(\bX_i - \bx) \in \supp(K))$. By Lemma~\ref{sa-lem: bias} and Lemma~\ref{sa-lem: Linear Approximation}, for any multi-index $\boldsymbol{\alpha}$ such that $|\boldsymbol{\alpha}| \leq p$,
\begin{align*}
    \sup_{\bx \in \B}|\be_{\boldsymbol{\alpha}}^\top (\widehat{\bbeta}_t(\bx) - \bbeta_t(\bx))| \lessp \bh^{-\boldsymbol{\alpha}} (\bar{h}^{p+1} + \mathfrak{R}_{\mathtt{EP}}).
\end{align*}
Since $K$ is compactly supported, we have
\begin{align*}
     \max_{1 \leq i \leq n} \Big|\sum_{t \in \{0,1\}}\Indicator(\bX_i \in \A_t) (\widehat{\bbeta}_t(\bx) - \bbeta_t(\bx))^\top \br_p(\bX_i - \bx) \Indicator(\bH^{-1}(\bX_i - \bx) \in \supp(K))\Big|
     \lessp \bar{h}^{p+1} + \mathfrak{R}_{\mathtt{EP}}.
\end{align*}
Since $\mu_t$ is $p+1$ times continuously differentiable,
\begin{align*}
    \max_{1 \leq i \leq n} \Big|\sum_{t \in \{0,1\}}\Indicator(\bX_i \in \A_t) (\mu_t(\bX_i)
    - \bbeta_t(\bx)^\top \br_p(\bX_i - \bx)) \Indicator(\bH^{-1}(\bX_i - \bx) \in \supp(K))\Big|
    \lesssim \bar{h}^{p+1}.
\end{align*}
It follows that
\begin{align*}
    \sup_{\bx \in \B}\max_{1 \leq i \leq n} |\eta_i(\bx)| \Indicator(\bH^{-1}(\bX_i - \bx) \in \supp(K))
    \lessp \bar{h}^{p+1} + \mathfrak{R}_{\mathtt{EP}}.
\end{align*}

\medskip\textit{Term $\bM_{1,\bx,\by}$}. From the proof for Lemma~\ref{sa-lem: Gram Matrix}, applied to the absolute-value class generated by $g_n$, $\sup_{\bx,\by \in \B}\En[ |g_n(\bX_i; \bx,\by)|]\lessp 1$. Thus,
\begin{align*}
    &\sup_{\bx, \by \in \B} \big|\En[g_n(\bX_i; \bx, \by) \eta_i(\bx) \eta_i(\by)] \big|\\
    & \leq
      \Big(\sup_{\bx \in \B}\max_{1 \leq i \leq n} |\eta_i(\bx)| \Indicator(\bH^{-1}(\bX_i - \bx) \in \supp(K))\Big)^2
      \sup_{\bx, \by \in \B}\En[|g_n(\bX_i; \bx, \by)|] \\
    & \lessp (\bar{h}^{p+1} + \mathfrak{R}_{\mathtt{EP}})^2,
\end{align*}
where we have used the preceding local-support bound on $\eta_i(\bx)$. Finite dimensionality of $\bM_{1,\bx,\by}$ then implies
\begin{align*}
    \sup_{\bx, \by \in \B}\norm{\bM_{1,\bx,\by}} \lessp (\bar{h}^{p+1} + \mathfrak{R}_{\mathtt{EP}})^2.
\end{align*}

\medskip\textit{Term $\bM_{2,\bx,\by}$}. The same maximal-inequality argument used for Lemma~\ref{sa-lem: Linear Approximation}, applied to the absolute-value class $|g_n(\cdot;\bx,\by)u|$, gives $\sup_{\bx,\by \in \B}\En[|g_n(\bX_i; \bx, \by)\varepsilon_i|]\lessp 1$. Thus,
\begin{align*}
    &\sup_{\bx, \by \in \B} \big|\En[g_n(\bX_i; \bx, \by)(\eta_i(\bx) + \eta_i(\by)) \varepsilon_i] \big|\\
    & \leq
      2\sup_{\bx \in \B}\max_{1 \leq i \leq n} |\eta_i(\bx)|\Indicator(\bH^{-1}(\bX_i-\bx)\in\supp(K))
      \sup_{\bx, \by \in \B} \En[|g_n(\bX_i; \bx, \by) \varepsilon_i|]\\
    &\lessp \bar{h}^{p+1} + \mathfrak{R}_{\mathtt{EP}},
\end{align*}
which implies that
\begin{align*}
    \sup_{\bx, \by \in \B}\norm{\bM_{2,\bx,\by}} \lessp \bar{h}^{p+1} + \mathfrak{R}_{\mathtt{EP}}.
\end{align*}

\medskip\textit{Term $\bM_{3,\bx,\by}$}. Define $l_n(\cdot, \cdot;\bx,\by): \X \times \reals \to \reals$ as
\begin{align*}
    l_n(\xi, \varepsilon;\bx, \by)
    = |\bH|K_{\bH}(\xi - \bx)K_{\bH}(\xi - \by)(\bH^{-1}(\xi - \bx))^{\bu}(\bH^{-1}(\xi - \by))^{\bv} \Indicator(\xi \in \A_t) (\varepsilon^2 - \sigma_t^2(\xi)),
\end{align*}
and consider the function class $\mathcal{L} = \{l_n(\cdot,\cdot;\bx,\by): \bx,\by \in \B\}$. Let $L: \X \times \reals \to \reals$ be $L(\xi, \varepsilon) = c|\bH|^{-1}|\varepsilon^2 - \sigma_t^2(\xi)|$ with $c = \sup_{\bx, \by \in \B} \big|(\bH^{-1}(\xi - \bx))^{\bu}(\bH^{-1}(\xi - \by))^{\bv}K(\bH^{-1}(\xi - \bx)) K(\bH^{-1}(\xi - \by))\big|$. By similar argument as in the proof for Lemma~\ref{sa-lem: Linear Approximation}, we can show $\mathcal{L}$ is a VC-type class such that $\E [l_n(\bX_i,\varepsilon_i; \bx, \by)] = 0$, for all $\bx,\by \in \B$,
\begin{align*}
    \sup_{\bx,\by \in \B}\E[l_n(\bX_i,\varepsilon_i;\bx,\by)^2]^{\frac{1}{2}}
    \lesssim \sup_{\bx, \by \in \B} \E[g_n(\bX_i; \bx, \by)^2]^{\frac{1}{2}}
    \sup_{\xi \in \X}\E[(\varepsilon_i^2-\sigma_t^2(\xi))^2|\bX_i = \xi]^{1/2}
    \lesssim |\bH|^{-1/2}
\end{align*}
because Assumption~\ref{sa-assump: Sharp DGP}(v), with $v\ge2$, gives a uniformly bounded conditional fourth moment for $\varepsilon_i$.
and
\begin{align*}
    \E\big[\max_{1 \leq i \leq n} L(\bX_i,\varepsilon_i)^2\big]^{\frac{1}{2}}
    \lesssim |\bH|^{-1} \E\big[\max_{1 \leq i \leq n}\varepsilon_i^4\big]^{1/2}
    \lesssim |\bH|^{-1} \E\big[\max_{1 \leq i \leq n} |\varepsilon_i|^{2+v}\big]^{\frac{2}{2+v}}
    \lesssim |\bH|^{-1}n^{\frac{2}{2+v}}.
\end{align*}
Applying Corollary 5.1 in \cite{Chernozhukov-Chetverikov-Kato_2014b_AoS}, we obtain
\begin{align*}
    \sup_{\bx, \by \in \B} \big|\En[l_n(\bX_i,\varepsilon_i;\bx,\by)] \big|
    \lessp \sqrt{\frac{\log (1/\bar{h})}{n |\bH|}} + \frac{\log (1/\bar{h})}{n^{\frac{v}{2+v}}|\bH|}
\end{align*}
and
\begin{align*}
    \sup_{\bx, \by \in \B} \norm{\bM_{3,\bx,\by}}
    \lessp \sqrt{\frac{\log (1/\bar{h})}{n |\bH|}} + \frac{\log (1/\bar{h})}{n^{\frac{v}{2+v}}|\bH|}.
\end{align*}

\medskip\textit{Term $\bM_{4,\bx,\by}$}. Notice that $\{g_n(\cdot; \bx, \by)\sigma_t^2(\cdot): \bx, \by \in \B\}$ is a VC-type class with constant envelope function $C |\bH|^{-1}$ for some positive constant $C$, where
\begin{align*}
    \sup_{\bx, \by \in \B} \sup_{\xi \in \X}|g_n(\xi; \bx, \by)\sigma_t^2(\xi)| \lesssim |\bH|^{-1}
\end{align*}
and
\begin{align*}
    \sup_{\bx, \by \in \B} \E[g_n(\bX_i; \bx, \by)^2 \sigma_t(\bX_i)^2]^{\frac{1}{2}} \lesssim |\bH|^{-1/2}.
\end{align*}
Then, similar to the proof of $\bM_{1,\bx,\by}$, we conclude that
\begin{align*}
     \sup_{\bx, \by \in \B} \big|\En[g_n(\bX_i; \bx, \by)\sigma_t^2(\bX_i)] -\E[g_n(\bX_i; \bx, \by)\sigma_t^2(\bX_i)] \big|
     \lessp \sqrt{\frac{\log (1/\bar{h})}{n |\bH|}}
\end{align*}
and
\begin{align*}
     \sup_{\bx, \by \in \B} \norm{\bM_{4,\bx,\by}}
     \lessp \sqrt{\frac{\log (1/\bar{h})}{n |\bH|}}.
\end{align*}

\medskip\textit{Final result}. Combining the upper bounds of the four terms,
\begin{align*}
    \sup_{\bx,\by \in \B} \big\|\widehat{\bSigma}_{t, \bx,\by} - \bSigma_{t, \bx,\by} \big\|
    \lessp \bar{h}^{p+1} + \sqrt{\frac{\log (1/\bar{h})}{n |\bH|}} + \frac{\log (1/\bar{h})}{n^{\frac{v}{2+v}}|\bH|},
\end{align*}
which implies $\sup_{\bx, \by \in \B} \|\widehat{\bSigma}_{t,\bx,\by}\| \lessp 1$. It follows that, for each $t\in\{0,1\}$,
\begin{align*}
    \sup_{\bx, \by \in \B}  |\widehat{\Omega}_{t,\bx,\by}^{(\bnu)} - \Omega_{t,\bx,\by}^{(\bnu)}|
    & \leq  \frac{1}{n |\bH|\bh^{2\bnu}}  \Big( \sup_{\bx, \by \in \B} \big\|\widehat{\bGamma}_{t,\bx}^{-1} - \bGamma_{t,\bx}^{-1} \big\| \big\|\widehat{\bSigma}_{t,\bx,\by}\big\| \big\|\widehat{\bGamma}_{t,\by}^{-1}\big\|\\
    &\qquad +  \sup_{\bx, \by \in \B} \big\|\bGamma_{t,\bx}^{-1}\big\|  \big\|\widehat{\bSigma}_{t,\bx,\by} - \bSigma_{t,\bx,\by}\big\| \big\|\widehat{\bGamma}_{t,\by}^{-1}\big\|\\
    &\qquad +  \sup_{\bx, \by \in \B} \big\|\bGamma_{t,\bx}^{-1}\big\| \big\|\bSigma_{t,\bx,\by}\big\| \big\|\widehat{\bGamma}_{t,\by}^{-1} - \bGamma_{t,\by}^{-1}\big\| \Big) \\
    & \lesssim_\P \frac{1}{n |\bH|\bh^{2\bnu}} \Big(\bar{h}^{p+1} + \sqrt{\frac{\log (1/\bar{h})}{n |\bH|}} + \frac{\log (1/\bar{h})}{n^{\frac{v}{2+v}}|\bH|}\Big).
\end{align*}
Summing the last display over $t\in\{0,1\}$ gives the asserted bound for $\widehat{\Omega}_{\bx,\by}^{(\bnu)} - \Omega_{\bx,\by}^{(\bnu)}$. By Assumption~\ref{sa-assump: Sharp DGP}(iv) and Assumption~\ref{sa-assump: Boundary and Kernel}(ii), $\inf_{\bx \in \B}\Omega_{\bx,\bx}^{(\bnu)} \gtrsim (n |\bH|\bh^{2\bnu})^{-1}$. Therefore, $\inf_{\bx \in \B}\widehat{\Omega}_{\bx,\bx}^{(\bnu)} \gtrsim_{\P} (n |\bH|\bh^{2\bnu})^{-1}$. Furthermore,
\begin{align*}
    & \sup_{\bx \in \B} \Big|\sqrt{\widehat{\Omega}_{\bx,\bx}^{(\bnu)}} - \sqrt{\Omega_{\bx,\bx}^{(\bnu)}}\Big|
    \lessp \sup_{\bx \in \B}\sqrt{n |\bH|\bh^{2\bnu}} \Big|\widehat{\Omega}_{\bx,\bx}^{(\bnu)} - \Omega_{\bx,\bx}^{(\bnu)}\Big|
    \lessp \frac{1}{\sqrt{n |\bH|\bh^{2\bnu}}} \Big(\bar{h}^{p+1} + \sqrt{\frac{\log (1/\bar{h})}{n |\bH|}} + \frac{\log (1/\bar{h})}{n^{\frac{v}{2+v}}|\bH|}\Big)
\end{align*}
and
\begin{align*}
    \sup_{\bx \in \B} \left|\frac{\bh^{-\bnu}}{\sqrt{\widehat{\Omega}_{\bx,\bx}^{(\bnu)}}} - \frac{\bh^{-\bnu}}{\sqrt{\Omega_{\bx,\bx}^{(\bnu)}}}\right|
    = \bh^{-\bnu}\sup_{\bx \in \B} \left|\frac{\sqrt{\widehat{\Omega}_{\bx,\bx}^{(\bnu)}} - \sqrt{\Omega_{\bx,\bx}^{(\bnu)}}}{\sqrt{\widehat{\Omega}_{\bx,\bx}^{(\bnu)} \Omega_{\bx,\bx}^{(\bnu)}}} \right|
    \lessp \sqrt{n |\bH|} \Big(\bar{h}^{p+1} + \sqrt{\frac{\log (1/\bar{h})}{n |\bH|}} + \frac{\log (1/\bar{h})}{n^{\frac{v}{2+v}}|\bH|}\Big),
\end{align*}
which completes the proof.
\qed

\subsection{Proof of Lemma~\ref{sa-lem: bias}}

Define
\begin{align*}
    \boldsymbol{\chi}_{t, \bx}
    = \En \Big[\br_p(\bH^{-1}(\bX_i-\bx)) K_{\bH}(\bX_i - \bx)\Indicator(\bX_i \in \A_t)\fr_t(\bX_i;\bx) \Big],
    \quad
    \fr_t(\xi;\bx) = \mu_t(\xi) - \sum_{0 \leq |\bomega| \leq p}\frac{\mu^{(\bomega)}_t(\bx)}{\bomega!}(\xi - \bx)^{\bomega}.
\end{align*}
Since $\mu_t$ is $(p+1)$-times continuously differentiable, Taylor's theorem and compact support of $K$ imply
\begin{align*}
    \sup_{\bx\in\B}\max_{1\leq i\leq n}
    \big|\fr_t(\bX_i;\bx)\big|\Indicator\{K_{\bH}(\bX_i-\bx)\neq0\}
    \lesssim \bar{h}^{p+1}.
\end{align*}
Since $\frac{\log(1/\bar{h})}{n|\bH|} = o(1)$, the same argument as the proof of Lemma~\ref{sa-lem: Gram Matrix} shows
\begin{align*}
    \sup_{\bx\in\B}
    \En \Big[\big\lVert \br_p(\bH^{-1}(\bX_i-\bx)) \big\rVert K_{\bH}(\bX_i - \bx)\Indicator(\bX_i \in \A_t)\Big] \lessp 1.
\end{align*}
Therefore, $\sup_{\bx\in\B}\norm{\boldsymbol{\chi}_{t,\bx}}\lessp \bar{h}^{p+1}$.
It then follows from Lemma~\ref{sa-lem: Gram Matrix} that
\begin{align*}
    \sup_{\bx \in \B} \big|\E[\widehat{\mu}_t^{(\bnu)}(\bx)|\bX] - \mu_t^{(\bnu)}(\bx)\big|
    = \sup_{\bx \in \B} \big|\bnu!\be_{\bnu}^{\top} \bS^{-1} \widehat{\bGamma}_{t,\bx}^{-1} \boldsymbol{\chi}_{t,\bx}\big|
    \lessp \bar{h}^{p + 1}\bh^{-\bnu}.
\end{align*}

Now also assume that $\bar{h} = o(1)$. By Assumption~\ref{sa-assump: Sharp DGP}(iii), for all $\bx \in \B$ and $\xi \in \X$,
\begin{align*}
    \fr_t(\xi;\bx)
    = \sum_{|\bv|=p+1}\frac{\mu_t^{(\bv)}(\bx)}{\bv!}(\xi-\bx)^{\bv}
      + \sum_{|\bv|=p+1} r_{t,\bv}(\xi;\bx)(\xi-\bx)^{\bv},
\end{align*}
where
\begin{align*}
    \max_{|\bv|=p+1}\sup_{\bx\in\B,\xi\in\X}
    |r_{t,\bv}(\xi;\bx)|\Indicator\{K_{\bH}(\xi-\bx)\neq0\}=o(1).
\end{align*}
Letting
\begin{align*}
    \widetilde{\boldsymbol{\chi}}_{t,\bx}
    = \En \Big[\br_p(\bH^{-1}(\bX_i-\bx)) K_{\bH}(\bX_i - \bx)\Indicator(\bX_i \in \A_t) ( \sum_{|\bv| = p+1} \frac{\mu_t^{(\bv)}(\bx)}{\bv!}(\bX_i - \bx)^{\bv})\Big],
\end{align*}
we conclude that
\begin{align*}
   \sup_{\bx \in \B} \big\| \boldsymbol{\chi}_{t,\bx} - \widetilde{\boldsymbol{\chi}}_{t,\bx} \big\|
   &\lesssim M_n \sup_{\bx \in \B} \Big\| \En \Big[\br_p(\bH^{-1}(\bX_i-\bx)) K_{\bH}(\bX_i - \bx) \Indicator(\bX_i \in \A_t) \Big(\sum_{|\bv| = p+1} \frac{\left|\bX_i - \bx\right|^{\bv}}{\bv!} \Big) \Big] \Big\|\\
   &= o_{\P}(\bar{h}^{p+1}),
\end{align*}
where the last equality employs the same arguments as in the proof of Lemma~\ref{sa-lem: Gram Matrix}. Hence,
\begin{align*}
    \sup_{\bx \in \B} \Big|\E[\widehat{\mu}_t^{(\bnu)}(\bx)|\bX] - \mu_t^{(\bnu)}(\bx) - \widehat{B}_{t,\bx}^{(\bnu)}(\bh)\Big|
    &= \sup_{\bx \in \B} \Big|\bnu!\be_{\bnu}^{\top} \bS^{-1} \widehat{\bGamma}_{t,\bx}^{-1} \boldsymbol{\chi}_{t,\bx} - \bnu!\be_{\bnu}^{\top} \bS^{-1} \widehat{\bGamma}_{t,\bx}^{-1} \widetilde{\boldsymbol{\chi}}_{t,\bx}\Big|\\
    &= o_{\P}(\bar{h}^{p+1}\bh^{-\bnu}).
\end{align*}
Using Lemma~\ref{sa-lem: Gram Matrix} and the maximal inequality as in the proof of Lemma~\ref{sa-lem: Gram Matrix}, we conclude that
\begin{align*}
    \max_{t \in \{0,1\}} \sup_{\bx \in \B} \big|\widehat{B}_{t,\bx}^{(\bnu)}(\bh) - B_{t,\bx}^{(\bnu)}(\bh)\big|
    \lesssim_{\P} \bar{h}^{p+1}\bh^{-\bnu}\sqrt{\frac{\log(1/\bar{h})}{n |\bH|}}.
\end{align*}
Since $\max_{t \in \{0,1\}}\sup_{\bx \in \B}|B_{t,\bx}^{(\bnu)}(\bh)| \lesssim \bar{h}^{p+1}\bh^{-\bnu}$, it follows that $\max_{t \in \{0,1\}} \sup_{\bx \in \B}|\widehat{B}_{t,\bx}^{(\bnu)}(\bh)| \lessp \bar{h}^{p+1}\bh^{-\bnu}$.
\qed

\subsection{Proof of Theorem~\ref{sa-thm: Convergence Rates: Sharp BATEC}}

For a fixed $\bx\in\B$, the pointwise versions of Lemma~\ref{sa-lem: Linear Approximation} and Lemma~\ref{sa-lem: bias} decompose
$\widehat{\mu}^{(\bnu)}_t(\bx)-\mu_t^{(\bnu)}(\bx)$ into a stochastic component and a conditional bias component of the stated pointwise orders. Applying this decomposition to both $t=0$ and $t=1$ and using the triangle inequality gives the pointwise rate for $\widehat{\tau}^{(\bnu)}(\bx)-\tau^{(\bnu)}(\bx)$. Taking suprema over $\bx\in\B$ and using the displayed uniform bounds in the two lemmas gives the uniform rate.
\qed

\subsection{Proof of Theorem~\ref{sa-thm: MSE Expansions: Sharp BATEC}}

For the conditional bias, by Lemma~\ref{sa-lem: bias},
\begin{align*}
    & \sup_{\bx \in \B} \big|\E \big[\widehat{\tau}^{(\bnu)}(\bx) - \tau^{(\bnu)}(\bx) \big| \bX \big]^2 - (B_{\bx}^{(\bnu)}(\bh))^2 \big| \\
    & \leq \sup_{\bx \in \B} \big|\E \big[\widehat{\tau}^{(\bnu)}(\bx) - \tau^{(\bnu)}(\bx) \big| \bX \big] - B_{\bx}^{(\bnu)}(\bh) \big| \cdot \sup_{\bx \in \B} \big|\E \big[\widehat{\tau}^{(\bnu)}(\bx) - \tau^{(\bnu)}(\bx) \big| \bX \big] + B_{\bx}^{(\bnu)}(\bh) \big| \\
    & = o_{\P}(\bar{h}^{2p+2}\bh^{-2\bnu}).
\end{align*}

For the conditional variance, by Lemma~\ref{sa-lem: covariance},
\begin{align*}
    \sup_{\bx \in \B} \big|\Var\big[\widehat{\tau}^{(\bnu)}(\bx) \big|\bX \big] -  (n|\bH|\bh^{2\bnu})^{-1} V_{\bx}^{(\bnu)} \big|
    = o_{\P}((n|\bH|\bh^{2\bnu})^{-1}).
\end{align*}

The pointwise MSE expansion follows directly. For the IMSE expansion, notice that
\begin{align*}
    & \Big|\operatorname{IMSE}^{(\bnu)} - \int_{\B} \big[(B_{\bx}^{(\bnu)}(\bh))^2 + (n|\bH|\bh^{2\bnu})^{-1} V_{\bx}^{(\bnu)}\big] w(\bx) \dif\Haus^{d-1}(\bx)\Big|\\
    & \leq \int_{\B} |w(\bx)| \dif\Haus^{d-1}(\bx) \cdot \sup_{\bx \in \B} \big|\operatorname{MSE}^{(\bnu)}(\bx) -  (B_{\bx}^{(\bnu)}(\bh))^2 - (n|\bH|\bh^{2\bnu})^{-1} V_{\bx}^{(\bnu)}\big|\\
    & = o_{\P} \big(\bar{h}^{2p+2}\bh^{-2\bnu} + (n|\bH|\bh^{2\bnu})^{-1} \big),
\end{align*}
which completes the proof.
\qed

\subsection{Proof of Theorem~\ref{sa-thm: Confidence Intervals: Sharp BATEC}}

We have $\overline{\Tstat}^{(\bnu)}(\bx) = \sum_{i = 1}^n \chi_i$ with
\begin{align*}
    \chi_i = \sum_{t \in \{0,1\}} s_t n^{-1} (\Omega_{\bx,\bx}^{(\bnu)})^{-1/2} \bnu!\be_{\bnu}^{\top} \bS^{-1} \bGamma_{t,\bx}^{-1} \br_p(\bH^{-1}(\bX_i - \bx)) K_{\bH}(\bX_i - \bx) \Indicator(\bX_i \in \A_t) \varepsilon_i,
\end{align*}
where $s_1=1$, $s_0=-1$, $\E[\chi_i] = 0$, and $\Var[\chi_i] = n^{-1}$. By the Berry-Esseen Theorem,
\begin{align*}
    \sup_{u \in \reals} \Big|\P \big(\overline{\Tstat}^{(\bnu)}(\bx) \leq u\big) - \Phi(u) \Big|
    \lesssim B_n^{-1} \sum_{i = 1}^n \E[|\chi_i|^3],
\end{align*}
where $B_n  = \sum_{i = 1}^n \Var[\chi_i] = 1$. Moreover,
\begin{align*}
    \sum_{i = 1}^n \E[|\chi_i|^3]
    & =  n^{-3} (\Omega_{\bx,\bx}^{(\bnu)})^{-3/2} \sum_{i = 1}^n \E \Big[\Big|\sum_{t \in \{0,1\}} s_t\bnu!\be_{\bnu}^{\top} \bS^{-1} \bGamma_{t,\bx}^{-1} \br_p(\bH^{-1}(\bX_i - \bx)) K_{\bH}(\bX_i - \bx) \Indicator(\bX_i \in \A_t) \varepsilon_i \Big|^3 \Big]\\
    & \lesssim n^{-3} (\Omega_{\bx,\bx}^{(\bnu)})^{-3/2} \sum_{i = 1}^n \E \Big[\Big|\sum_{t \in \{0,1\}}s_t\bnu!\be_{\bnu}^{\top} \bS^{-1} \bGamma_{t,\bx}^{-1} \br_p(\bH^{-1}(\bX_i - \bx)) K_{\bH}(\bX_i - \bx) \Indicator(\bX_i \in \A_t)\Big|^3 \Big]\\
    & \lesssim n^{-2} \bh^{-\bnu}|\bH|^{-1} (\Omega_{\bx,\bx}^{(\bnu)})^{-3/2} \E \Big[\Big|\sum_{t \in \{0,1\}}s_t\bnu!\be_{\bnu}^{\top} \bS^{-1} \bGamma_{t,\bx}^{-1} \br_p(\bH^{-1}(\bX_i - \bx)) K_{\bH}(\bX_i - \bx) \Indicator(\bX_i \in \A_t)\Big|^2 \Big]\\
    & \lesssim n^{-1} \bh^{-\bnu}|\bH|^{-1}  (\Omega_{\bx,\bx}^{(\bnu)})^{-1/2}\\
    & \lesssim (n |\bH|)^{-1/2},
\end{align*}
where the second line uses Assumption~\ref{sa-assump: Sharp DGP}(v): since $v\ge2$, the conditional third moment of $\varepsilon_i$ is uniformly bounded, so conditioning on $\bX_i$ absorbs the factor $|\varepsilon_i|^3$ into a constant. The third line uses
\begin{align*}
    \bigg|\sum_{t \in \{0,1\}}s_t\bnu!\be_{\bnu}^{\top} \bS^{-1} \bGamma_{t,\bx}^{-1} \br_p(\bH^{-1}(\bX_i - \bx)) K_{\bH}(\bX_i - \bx) \Indicator(\bX_i \in \A_t)\bigg| \lesssim \bh^{-\bnu}|\bH|^{-1}
\end{align*}
and the fourth line uses the definition of $\Omega_{\bx,\bx}^{(\bnu)}$.

Finally, although Lemma~\ref{sa-lem: Gram Matrix} through Lemma~\ref{sa-lem: covariance} establish uniform convergence over $\bx \in \B$, the corresponding pointwise results for a fixed $\bx$ follow by restricting the function classes in those proofs to singletons indexed by $\bx$. In particular, when handling the variance terms in Lemma~\ref{sa-lem: covariance} (specifically $\bM_{3,\bx,\by}$), we bound terms involving $\varepsilon_i^2$ using either the Fuk--Nagaev inequality (under $2+v$ moments) or Markov's inequality. Thus, for each fixed $\bx\in\B$,
\begin{align*}
    \|\widehat{\bGamma}_{t,\bx}^{-1}-\bGamma_{t,\bx}^{-1}\|
    &=O_{\P}\big((n|\bH|)^{-1/2}\big),\\
    \|\widehat{\bSigma}_{t,\bx,\bx}-\bSigma_{t,\bx,\bx}\|
    &=O_{\P}\big((n|\bH|)^{-1/2}+\bar h^{p+1}\big),
\end{align*}
and therefore
\begin{align*}
    \left|\frac{\widehat{\Omega}_{\bx,\bx}^{(\bnu)}}{\Omega_{\bx,\bx}^{(\bnu)}}-1\right|
    &=O_{\P}\big((n|\bH|)^{-1/2}+\bar h^{p+1}\big),\\
    \left|(\widehat{\Omega}_{\bx,\bx}^{(\bnu)})^{-1/2}
          -(\Omega_{\bx,\bx}^{(\bnu)})^{-1/2}\right|
    &\lessp \sqrt{n|\bH|\bh^{2\bnu}}
      \big((n|\bH|)^{-1/2}+\bar h^{p+1}\big).
\end{align*}
Together with
\begin{align*}
    \big|\bnu!\be_{\bnu}^{\top}\bS^{-1}\bGamma_{t,\bx}^{-1}\bQ_{t,\bx}\big|
    =O_{\P}\big(\bh^{-\bnu}(n|\bH|)^{-1/2}\big)
\end{align*}
and the bias bound
\begin{align*}
    \big|\E[\widehat{\tau}^{(\bnu)}(\bx)|\bX]-\tau^{(\bnu)}(\bx)\big|
    =O_{\P}(\bar h^{p+1}\bh^{-\bnu}),
\end{align*}
this yields
\begin{align}\label{sa-eq: lin error 2d}
\Big|\widehat{\Tstat}^{(\bnu)}(\bx) - \overline{\Tstat}^{(\bnu)}(\bx)\Big|
\lessp
\bar{h}^{p+1}\sqrt{n |\bH|} + \frac{1}{\sqrt{n |\bH|}},
\end{align}
provided that $\bar{h} \to 0$ and $n |\bH| \to \infty$.

The final results follow by weak convergence to a Gaussian distribution, and properties of the distribution function.
\qed

\subsection{Proof of Theorem~\ref{sa-thm: Stochastic Linearization: Sharp BATEC}}

For all $\bx \in \B$, we have $\widehat{\Tstat}^{(\bnu)}(\bx) = \overline{\Tstat}^{(\bnu)}(\bx) + G_1^{(\bnu)}(\bx) +  G_2^{(\bnu)}(\bx)$, where
\begin{align*}
  G_1^{(\bnu)}(\bx)
  = \Big(\E\big[\widehat{\tau}^{(\bnu)}(\bx) \big| \bX \big] - \tau^{(\bnu)}(\bx)\Big) (\widehat{\Omega}_{\bx,\bx}^{(\bnu)})^{-1/2},
\end{align*}
and
\begin{align*}
  G_2^{(\bnu)}(\bx)
  = \bnu!\be_{\bnu}^{\top}\bS^{-1} \Big[\big(\hGammap^{-1} \bQ_{1, \bx}- \hGamman^{-1} \bQ_{0, \bx}\big) (\widehat{\Omega}_{\bx,\bx}^{(\bnu)})^{-\frac{1}{2}} - \big(\Gammap^{-1} \bQ_{1, \bx}- \Gamman^{-1} \bQ_{0, \bx}\big)(\Omega_{\bx,\bx}^{(\bnu)})^{-\frac{1}{2}}\Big].
\end{align*}
By Lemma~\ref{sa-lem: bias} and Lemma~\ref{sa-lem: covariance},
\begin{align*}
    \sup_{\bx \in \B} \big|G_1^{(\bnu)}(\bx)\big| \lessp \bar{h}^{p+1}\bh^{-\bnu} (n |\bH|\bh^{2\bnu})^{1/2} \lesssim \bar{h}^{p+1}\sqrt{n |\bH|}.
\end{align*}
By Lemma~\ref{sa-lem: Gram Matrix}, Lemma~\ref{sa-lem: Linear Approximation} and Lemma~\ref{sa-lem: covariance},
\begin{align*}
    \sup_{\bx \in \B} \Big|\bnu!\be_{\bnu}^{\top} \bS^{-1} \big[\widehat{\bGamma}_{t,\bx}^{-1} - \bGamma_{t,\bx}^{-1}\big] \bQ_{t,\bx} (\widehat{\Omega}_{\bx,\bx}^{(\bnu)})^{-1/2}\Big|
    \lessp \sqrt{\log(1/\bar{h})} \bigg(\sqrt{\frac{\log(1/\bar{h})}{n |\bH|}} + \frac{\log(1/\bar{h})}{n^{\frac{1 + v}{2+v}}|\bH|}\bigg)
\end{align*}
and
\begin{align*}
    & \sup_{\bx \in \B} \Big|\bnu!\be_{\bnu}^{\top} \bS^{-1} \bGamma_{t,\bx}^{-1} \bQ_{t,\bx} \Big[(\widehat{\Omega}_{\bx,\bx}^{(\bnu)})^{-1/2} - (\Omega_{\bx,\bx}^{(\bnu)})^{-1/2} \Big] \Big| \\
    & \lessp \bh^{-\bnu} \cdot \bigg(\sqrt{\frac{\log(1/\bar{h})}{n |\bH|}}  + \frac{\log(1/\bar{h})}{n^{\frac{1+v}{2+v}}|\bH|} \bigg) \cdot \sqrt{n|\bH|\bh^{2\bnu}} \bigg(\sqrt{\frac{\log(1/\bar{h})}{n |\bH|}}  + \frac{\log(1/\bar{h})}{n^{\frac{v}{2+v}}|\bH|} + \bar{h}^{p+1}\bigg) \\
    & \lesssim \frac{\log(1/\bar{h})}{\sqrt{n |\bH|}} + \frac{(\log(1/\bar{h}))^{3/2}}{n^{\frac{v}{2+v}}|\bH|}
      + \bar{h}^{p+1}\sqrt{n|\bH|}\bigg(\sqrt{\frac{\log(1/\bar{h})}{n |\bH|}}  + \frac{\log(1/\bar{h})}{n^{\frac{1+v}{2+v}}|\bH|} \bigg) \\
    & \lesssim \frac{\log(1/\bar{h})}{\sqrt{n |\bH|}} + \frac{(\log(1/\bar{h}))^{3/2}}{n^{\frac{v}{2+v}}|\bH|} + o(\bar{h}^{p+1}\sqrt{n |\bH|}).
\end{align*}
The last term is absorbed by the bound for $G_1^{(\bnu)}$, and the two displayed logarithmic terms give the second component of the rate in the theorem.
The result now follows from combining the bounds above.
\qed

\subsection{Proof of Theorem~\ref{sa-thm: Gaussian Strong Approximation: Sharp Tstat}}

We verify the high-level conditions of Theorem~\ref{sa-thm: Residual Empirical Process Strong Approximation}. We will employ the following technical lemma.

\begin{lem}[VC Class to VC2 Class]\label{sa-lem: VC Class to VC2 Class}
    Assume $\F$ is a VC class on a measure space $(\X, \mathcal{B})$: there exists an envelope function $F$ and positive constants $c(\F), d(\F)$ such that for all $\varepsilon\in(0,1)$,
    \begin{align*}
        \sup_{Q}N(\mathcal{F}, \norm{\cdot}_{Q,1}, \varepsilon \norm{F}_{Q,1}) \leq c(\F)\varepsilon^{-d(\F)},
    \end{align*}
    where the supremum is taken over all finite discrete measures. Then, $\F$ is also VC2 class: for all $\varepsilon\in(0,1)$,
    \begin{align*}
        \sup_{Q} N(\mathcal{F}, \norm{\cdot}_{Q,2}, \varepsilon \norm{F}_{Q,2}) \leq c(\F) (\varepsilon^2/2)^{-d(\mathcal{F})},
    \end{align*}
    where the supremum is taken over all finite discrete measures.
\end{lem}

\noindent\textbf{Proof of Lemma~\ref{sa-lem: VC Class to VC2 Class}}. Let $Q$ be a finite discrete probability measure. Let $f,g \in \mathcal{F}$. Then, $\int |f - g|^2 \dif Q \leq 2 \int |f - g| |F| \dif Q$. Define another probability measure $\tilde{Q}(c_k) = F(c_k) Q(c_k) / \norm{F}_{Q,1}$ on the support of $Q$, denoted by $\{c_1, \ldots, c_k, \ldots\}$. Then,
\begin{align*}
    \int |f - g|^2 \dif Q
    \leq 2 \norm{F}_{Q,1} \int |f - g| \dif \tilde{Q}
    \leq 2 \norm{F}_{Q,1} \norm{f - g}_{\tilde{Q},1}.
\end{align*}
Hence, if we take an $\varepsilon^2/2$-net in $(\mathcal{F}, \norm{\cdot}_{\tilde{Q},1})$ with cardinality no greater than $c(\mathcal{F})(\varepsilon^2/2)^{- d (\mathcal{F})}$, then for any $f \in \mathcal{F}$, there exists a $g \in \mathcal{F}$ such that $\norm{f - g}_{\tilde{Q},1} \leq \varepsilon^2/2 \norm{F}_{\tilde{Q},1}$, and hence
\begin{align*}
    \norm{f - g}_{Q,2}^2 \leq 2 \varepsilon^2/2 \norm{F}_{Q,1} \norm{F}_{\tilde{Q},1} \leq \varepsilon^2 \norm{F}_{Q,2}^2,
\end{align*}
which gives the result.
\qed

By the rectangular-support condition imposed in Theorem~\ref{sa-thm: Gaussian Strong Approximation: Sharp Tstat}, after a componentwise affine rescaling we can take $\X=[0,1]^d$. Since $f_X$ is continuous and bounded away from zero on this compact rectangle, \citet[Section 3.1]{Cattaneo-Yu_2025_AOS} implies that $\Q_{\F_t}=\P_X$ is a valid surrogate measure for $\P_X$ with respect to $\F_t$, and that the Rosenblatt transform $\phi_{\F_t}=T_{\P_X}$ is a valid normalizing transformation. The constants $\ttc_1$ and $\ttc_2$ from Theorem~\ref{sa-thm: Residual Empirical Process Strong Approximation} are therefore finite and uniformly bounded by constants depending only on the density bounds and the support geometry.

Consider first the class of functions $\F_t = \{\K_t^{(\bnu)}(\cdot; \bx): \bx \in \B\}$, for $t \in \{0,1\}$.

\medskip\textit{Envelope Function}. By Lemma~\ref{sa-lem: Invertibility}, Assumption~\ref{sa-assump: Sharp DGP}(iv), and Assumption~\ref{sa-assump: Boundary and Kernel}, uniformly in $\bx \in \B$,
\begin{align*}
    \Omega_{\bx,\bx}^{(\bnu)} \asymp (n|\bH|\bh^{2\bnu})^{-1}.
\end{align*}
Together with the compact support of $K$ and $\|\bS^{-1}\be_{\bnu}\|\asymp \bh^{-\bnu}$, this gives
\begin{align*}
    \sup_{\bx \in \B} \sup_{\xi \in \X} \big|\K_t^{(\bnu)}(\xi; \bx)\big|
    \lesssim n^{-1/2} (n|\bH|\bh^{2\bnu})^{1/2} \bh^{-\bnu} |\bH|^{-1}
    \lesssim |\bH|^{-1/2}.
\end{align*}
Hence, there exists a constant $C_1 > 0$ such that $\ttM_{\F_t} = C_1|\bH|^{-1/2}$ is a constant envelope function.

\medskip\textit{$L_1$ Bound}. Since $\K_t^{(\bnu)}(\cdot;\bx)$ is supported on $\bx+\bH\supp(K)$, whose Lebesgue measure is of order $|\bH|$,
\begin{align*}
    \ttE_{\F_t} = \sup_{\bx \in \B} \E[|\K_t^{(\bnu)}(\bX_i; \bx)|] \lesssim |\bH|^{1/2}.
\end{align*}

\medskip\textit{Uniform Variation}. Case 1: $K$ is Lipschitz. By Assumption \ref{sa-assump: Sharp DGP}(iv) and Assumption \ref{sa-assump: Boundary and Kernel},
\begin{align*}
    \ttL_{\F_t}
    = \sup_{\bx \in \B}\sup_{\xi, \xi' \in \X} \frac{|\K_t^{(\bnu)}(\xi; \bx) - \K_t^{(\bnu)}(\xi'; \bx)|}{\infnorm{\xi - \xi'}}
    \lesssim |\bH|^{-1/2}\underline h^{-1}.
\end{align*}
Each entry of $\bGamma_{t,\bx}$ and $\bSigma_{t,\bx,\bx}$ is a smooth local integral with argument $\bH^{-1}(\xi-\bx)$ and is therefore $\underline h^{-1}$-Lipschitz in $\bx$. Similarly,
\begin{align*}
    \big|\Omega_{t,\bx,\bx}^{(\bnu)} - \Omega_{t,\px,\px}^{(\bnu)}\big|
    \lesssim (n|\bH|\bh^{2\bnu}\underline h)^{-1}\infnorm{\bx-\px},
\end{align*}
which, using $\Omega_{t,\bx,\bx}^{(\bnu)}\asymp(n|\bH|\bh^{2\bnu})^{-1}$, implies
\begin{align*}
    \big|(\Omega_{t,\bx,\bx}^{(\bnu)})^{-1/2} - (\Omega_{t,\px,\px}^{(\bnu)})^{-1/2}\big|
    \lesssim \underline h^{-1}(n|\bH|\bh^{2\bnu})^{1/2}\infnorm{\bx-\px}.
\end{align*}
It follows that the class is Lipschitz with respect to the point of evaluation:
\begin{align*}
    \mathtt{l}_{\F_t}
    = \sup_{\xi \in \X}\sup_{\bx, \px \in \B} \frac{\left|\K_t^{(\bnu)}(\xi; \bx) - \K_t^{(\bnu)}(\xi; \px) \right|}{\infnorm{\bx - \px}}
    \lesssim |\bH|^{-1/2}\underline h^{-1}.
\end{align*}
Since $\K_t^{(\bnu)}(\cdot;\bx)$ is supported on $\bx+\bH\supp(K)$, we have
\begin{align*}
    \ttTV_{\F_t}
    \lesssim |\bH| \, |\bH|^{-1/2}\underline h^{-1}
    \lesssim |\bH|^{1/2}\underline h^{-1}.
\end{align*}

Case 2: $K = \Indicator(\cdot \in [-1,1]^d)$. Define the smooth part
\begin{align*}
    \tilde{\K}^{(\bnu)}_t(\bu;\bx)
    = n^{-1/2} (\Omega_{\bx,\bx}^{(\bnu)})^{-1/2}\bnu!\be_{\bnu}^{\top} \bS^{-1} \bGamma_{t,\bx}^{-1} \br_p(\bH^{-1}(\bu - \bx)) |\bH|^{-1},
    \qquad \bu \in \X,\quad t \in \{0,1\}.
\end{align*}
Then, $\K_t^{(\bnu)}(\bu;\bx) = \tilde{\K}_t^{(\bnu)}(\bu;\bx)\Indicator(\bH^{-1}(\bu-\bx)\in[-1,1]^d)$, and the preceding argument gives $\ttTV_{\tilde{\F}_t}\lesssim |\bH|^{1/2}\underline h^{-1}$. For $\mathscr{L}=\{\Indicator(\bH^{-1}(\cdot-\bx)\in[-1,1]^d):\bx\in\B\}$, the boundary of each rectangle has perimeter $O(|\bH|\underline h^{-1})$, so the product rule yields
\begin{align*}
    \ttTV_{\F_t}
    \leq \ttTV_{\tilde{\F}_t}\ttM_{\mathscr{L}} + \ttM_{\tilde{\F}_t}\ttTV_{\mathscr{L}}
    \lesssim |\bH|^{1/2}\underline h^{-1}.
\end{align*}

\medskip\textit{VC-type Class}. Case 1: $K$ is Lipschitz. We apply \citet[Lemma 7]{Cattaneo-Chandak-Jansson-Ma_2024_Bernoulli} to the rescaled maps
\begin{align*}
    f_{\bx}(\cdot)
    = \frac{\bnu!}{\sqrt{n \Omega_{\bx,\bx}^{(\bnu)}}}\be_{\bnu}^{\top} \bS^{-1} \bGamma_{t,\bx}^{-1} \br_p(\cdot)K(\cdot),
    \qquad \bx \in \B,
\end{align*}
indexed by $\mathcal{H}=\{f_{\bx}(\bH^{-1}(\cdot-\bx)):\bx\in\B\}$. The boundedness and compact support conditions are unchanged; the Lipschitz constant in the index is now of order $\underline h^{-1}$. Therefore, for $0<\varepsilon\leq1$,
\begin{align*}
    \sup_{Q} N\big(\F_t,\norm{\cdot}_{Q,2},(2c+1)^{d+1}\varepsilon\ttM_{\F_t}\big)
    \leq \bc'2^{2d+4}\varepsilon^{-2d-4}+1,
\end{align*}
where the supremum is taken over all finite discrete measures.

\emph{Case 2: Suppose $K=\Indicator(\cdot\in[-1,1]^d)$.} The smooth class $\tilde{\F}_t$ satisfies the same entropy bound. The indicator class $\mathscr{L}$ is a VC class of axis-aligned rectangles with side lengths $2h_j$, so
\begin{align*}
    \sup_Q N(\mathscr{L},\norm{\cdot}_{Q,2},\varepsilon)\leq C\varepsilon^{-4d},\qquad 0<\varepsilon\leq1.
\end{align*}
The product entropy bound then gives, for a constant $c>0$,
\begin{align*}
    \sup_Q N(\F_t,\norm{\cdot}_{Q,2},\varepsilon C_1\ttM_{\tilde{\F}_t})
    &\leq
    \sup_Q N(\tilde{\F}_t,\norm{\cdot}_{Q,2},c\varepsilon\ttM_{\tilde{\F}_t})
    \sup_Q N(\mathscr{L},\norm{\cdot}_{Q,2},c\varepsilon)\\
    &\leq C_2\varepsilon^{-(6d+4)},
\end{align*}
for constants depending only on $d$.

\bigskip
Consider next the class of functions $\G = \{g_{\bx}: \bx \in \B\}$, where $g_{\bx}(\bu)=\Indicator(\bu\in\A_1)\K_1^{(\bnu)}(\bu;\bx)-\Indicator(\bu\in\A_0)\K_0^{(\bnu)}(\bu;\bx)$. We have immediately that $\ttM_{\G}\lesssim|\bH|^{-1/2}$, $\ttE_{\G}\lesssim|\bH|^{1/2}$, and
\begin{align*}
    \sup_Q N(\G,\norm{\cdot}_{Q,2},\varepsilon(2c+1)^{d+1}\ttM_{\G})\leq 2\bc'\varepsilon^{-4d-4}+2.
\end{align*}

\medskip\textit{Total Variation}. Observe that $\Indicator(\bu\in\A_t)\K_t^{(\bnu)}(\bu;\bx)\neq0$ implies $\bu\in E_{t,\bx}=\{\by\in\A_t:\bH^{-1}(\by-\bx)\in\supp(K)\}$. By assumption, $\mathcal{L}(E_{t,\bx})\leq C\bar{h}^{d-1}$. Using the product rule and the bandwidth comparability condition,
\begin{align*}
    \ttTV_{\G}
    &\leq \sup_{\bx\in\B}\sum_{t\in\{0,1\}}\ttTV_{\{\K_t^{(\bnu)}(\cdot;\bx)\}}+\ttM_{\F_t}\ttTV_{\{\Indicator_{E_{t,\bx}}\}}\\
    &\lesssim |\bH|^{1/2}\underline h^{-1}+|\bH|^{-1/2}\bar{h}^{d-1}
    \lesssim |\bH|^{1/2}\underline h^{-1}.
\end{align*}

The imposed singleton regularity condition verifies the remaining truncated conditional-mean requirement in Theorem~\ref{sa-thm: Residual Empirical Process Strong Approximation}. We have therefore verified all high-level sufficient conditions, which immediately give the result.
\qed

\subsection{Proof of Lemma~\ref{sa-lem: localized perimeter piecewise linear}}

Let $\mathcal{N}_{\bx,h}=\bx+h\,\supp(K)$. Since $\supp(K)$ is bounded, there is a constant $R<\infty$ such that $\mathcal{N}_{\bx,h}\subseteq B(\bx,Rh)$ for all $\bx\in\B$ and all $h>0$. Since $\B$ is a finite union of line segments, say $\B=\cup_{j=1}^{J}L_j$, uniformly over $\bx\in\B$,
\begin{align*}
    \Haus^1(\B\cap \mathcal{N}_{\bx,h})
    \leq \Haus^1(\B\cap B(\bx,Rh))
    \leq \sum_{j=1}^{J}\Haus^1(L_j\cap B(\bx,Rh))
    \leq 2JRh
    \lesssim h.
\end{align*}
Because $\B$ lies in the interior of $\X$, for all small enough $h$ the window $\mathcal{N}_{\bx,h}$ does not intersect $\setbd(\X)$ uniformly over $\bx\in\B$. Write $\mathcal{L}(\A_t;\mathcal{N}_{\bx,h})$ for the perimeter of $\A_t$ relative to $\mathcal{N}_{\bx,h}$. The relative boundary of $\A_t$ inside $\mathcal{N}_{\bx,h}$ is contained in $\B\cap \mathcal{N}_{\bx,h}$, up to finitely many endpoints, and therefore
\begin{align*}
    \mathcal{L}(\A_t;\mathcal{N}_{\bx,h})\lesssim \Haus^1(\B\cap \mathcal{N}_{\bx,h})\lesssim h.
\end{align*}

If $K(\bu)=\Indicator(\bu\in[-1,1]^2)$, then $\mathcal{N}_{\bx,h}$ is a square with perimeter $8h$. More generally, if $\supp(K)$ has finite De Giorgi perimeter, the scaling property of perimeter gives $\mathcal{L}(\mathcal{N}_{\bx,h})=h\mathcal{L}(\supp(K))\lesssim h$. By the standard subadditivity/product rule for finite-perimeter sets,
\begin{align*}
    \mathcal{L}(\A_t\cap \mathcal{N}_{\bx,h})
    \leq \mathcal{L}(\A_t;\mathcal{N}_{\bx,h})+\mathcal{L}(\mathcal{N}_{\bx,h})
    \lesssim h,
\end{align*}
uniformly over $\bx\in\B$ and $t\in\{0,1\}$. Since $d=2$ and $\bar h=h$, this is the claimed localized perimeter bound. \qed

\subsection{Proof of Theorem~\ref{sa-thm: Confidence Bands: Sharp BATEC -- process}}

The proof is divided into three theorem-specific technical lemmas. We first record a generic comparison lemma that will be used to translate a uniform process approximation into a Kolmogorov distance bound for suprema.

\begin{lem}[Generic Suprema Comparison]\label{sa-lem: generic suprema comparison}
Let $\mathbb{T}_j=(\mathbb{T}_j(\bx):\bx\in\B)$, $j=1,2$, be real-valued processes defined on a common probability space. Write
\[
    M_j=\sup_{\bx\in\B}|\mathbb{T}_j(\bx)|,\quad j=1,2,
    \qquad
    \Delta_{12}=\sup_{\bx\in\B}|\mathbb{T}_1(\bx)-\mathbb{T}_2(\bx)|.
\]
For any $\eta>0$,
\begin{align*}
    \sup_{u\in\reals}\left|\P(M_1\le u)-\P(M_2\le u)\right|
    \le
    \mathsf{AC}_2(\eta)
    +\P(\Delta_{12}>\eta),
\end{align*}
where
\[
    \mathsf{AC}_2(\eta)
    =
    \sup_{u\in\reals}\P(|M_2-u|\le \eta).
\]
If, in addition, $\mathbb{T}_2$ is a separable mean-zero Gaussian process and the Gaussian anti-concentration bound applies to $M_2$, then
\begin{align*}
    \sup_{u\in\reals}\left|\P(M_1\le u)-\P(M_2\le u)\right|
    \le
    4\eta\{\E[M_2]+1\}
    +\P(\Delta_{12}>\eta).
\end{align*}
Consequently, if $\E[\Delta_{12}]\le r_n$, then the last probability can be further bounded by $r_n/\eta$.
\end{lem}

\noindent\textbf{Proof of Lemma~\ref{sa-lem: generic suprema comparison}.}
On the event $\{\Delta_{12}\le\eta\}$, the inequalities
\[
    M_1\le u \implies M_2\le u+\eta,
    \qquad
    M_2\le u-\eta \implies M_1\le u
\]
hold for every $u\in\reals$. Therefore,
\begin{align*}
    \P(M_1\le u)-\P(M_2\le u)
    &\le \P(u<M_2\le u+\eta)
       +\P(\Delta_{12}>\eta),\\
    \P(M_2\le u)-\P(M_1\le u)
    &\le \P(u-\eta<M_2\le u)
       +\P(\Delta_{12}>\eta).
\end{align*}
Taking absolute values and then the supremum over $u$ gives the first result. The second display follows from the Gaussian anti-concentration inequality of \citet[Theorem 2.1]{Chernozhukov-Chetverikov-Kato_2014a_AoS}, applied to the symmetrized Gaussian process whose supremum is $M_2$. The final claim follows from Markov's inequality.
\qed

For the theorem-specific lemmas below, define the strong-approximation rate
\[
    \mathfrak{R}_{\mathtt{sa}}
    =
    (\log n)^{\frac{3}{2}}\Big(\frac{1}{n |\bH|}\Big)^{\frac{1}{2d+2}\cdot \frac{v}{v+2}}
    + \log(n)\sqrt{\frac{1}{n^{\frac{v}{2+v}}|\bH|}},
\]
the stochastic-linearization remainder
\[
    \mathfrak{R}_{\mathtt{lin}}
    =
    \sqrt{\log(1/\bar{h})}
    \Big(\sqrt{\frac{\log(1/\bar{h})}{n|\bH|}}
    + \frac{\log(1/\bar{h})}{n^{\frac{v}{2+v}}|\bH|} \Big)
    + \bar{h}^{p+1} \sqrt{n|\bH|},
\]
and the covariance-estimation rate
\[
    \mathfrak{R}_{\mathtt{cov}}
    =
    \sqrt{\frac{\log n}{n|\bH|}}
    + \frac{\log n}{n^{\frac{v}{2+v}}|\bH|}.
\]
Also set the covariance-comparison rate $\mathfrak{R}_{\mathtt{cc}}=\mathfrak{R}_{\mathtt{cov}}+\bar h^{p+1}$.

\begin{lem}[KS Distance Between $\overline{\Tstat}^{(\bnu)}$ and $Z^{(\bnu)}$]\label{sa-lem: infeasible gaussian to bahadur representation}
Suppose the conditions of Theorem~\ref{sa-thm: Confidence Bands: Sharp BATEC -- process} hold. Then,
\begin{align*}
    \sup_{u \in \reals} \Big|\P \Big(\sup_{\bx \in \B} \big|\overline{\Tstat}^{(\bnu)}(\bx)\big| \leq u\Big)
     - \P \Big(\sup_{\bx \in \B} \big|Z^{(\bnu)}(\bx)\big| \leq u\Big)\Big|
    \lesssim (\log n)^{1/4}\mathfrak{R}_{\mathtt{sa}}^{1/2}.
\end{align*}
\end{lem}

\noindent\textbf{Proof of Lemma~\ref{sa-lem: infeasible gaussian to bahadur representation}.} Let $a_n$ be a positive sequence to be determined below. Applying Lemma~\ref{sa-lem: generic suprema comparison} with $\mathbb{T}_1=\overline{\Tstat}^{(\bnu)}$, $\mathbb{T}_2=Z^{(\bnu)}$, and $\eta=a_n$, and then using the coupling tail bound in Theorem~\ref{sa-thm: Gaussian Strong Approximation: Sharp Tstat}, gives
\begin{align*}
    &\sup_{u\in\reals}\Big|
    \P \Big(\sup_{\bx \in \B} \big|\overline{\Tstat}^{(\bnu)}(\bx)\big| \leq u\Big)
    -\P \Big(\sup_{\bx \in \B} \big|Z^{(\bnu)}(\bx)\big| \leq u\Big)
    \Big|\\
    &\qquad\le
    4a_n\left(\E\left[\sup_{\bx\in\B}|Z^{(\bnu)}(\bx)|\right]+1\right)
    + \frac{C\mathfrak{R}_{\mathtt{sa}}}{a_n}.
\end{align*}

For the next displays, write $\K(\bu,\bx)=g_{\bx}(\bu)$ and $\sigma^2(\bu)=\Var(Y_i|\bX_i=\bu)$. Notice that $Z^{(\bnu)}(\bx), \bx \in \B$ is a mean-zero Gaussian process satisfying
\begin{align*}
    \E \big[\big(Z^{(\bnu)}(\bx) - Z^{(\bnu)}(\by)\big)^2\big]^{\frac{1}{2}}
    &= \E \big[(\K(\bX_i, \bx) - \K(\bX_i, \by))^2 \sigma(\bX_i)^2\big]^{\frac{1}{2}}  \\
    &\leq C' l_{n,2} \|\bx - \by\|_{\infty},
\end{align*}
where $C'$ is a constant and $l_{n,2} \asymp \underline h^{-1}$. Hence the canonical metric
$$
\rho(\bx,\by)
:= \E \big[\big(Z^{(\bnu)}(\bx) - Z^{(\bnu)}(\by)\big)^2\big]^{1/2}
\lesssim \underline h^{-1}\|\bx-\by\|_{\infty}.
$$
Moreover,
$$
\sup_{\bx\in\B} \E \big[ Z^{(\bnu)}(\bx)^2 \big]
= \sup_{\bx\in\B} \E \big[\K(\bX_i,\bx)^2 \sigma^2(\bX_i)\big]
\lesssim 1.
$$

Let $N(\varepsilon,\B,\rho)$ denote the covering number of $\B$ under the metric $\rho$.
Since $\rho(\bx,\by)\lesssim \underline h^{-1}\|\bx-\by\|_{\infty}$ and $\B$ is compact in $\R^d$,
$$
N(\varepsilon,\B,\rho)
\lesssim
N(\underline h\varepsilon,\B,\|\cdot\|_{\infty})
\lesssim
\Big(\frac{1}{\underline h\varepsilon}\Big)^d,
\qquad 0<\varepsilon\le 1,
$$
and therefore
$$
\log N(\varepsilon,\B,\rho)
\lesssim
d\log\!\Big(\frac{1}{\underline h\varepsilon}\Big).
$$

Applying Corollary 2.2.8 in \cite{van-der-Vaart-Wellner_1996_Book} yields
\begin{align*}
\E\!\left[\sup_{\bx\in\B}\big|Z^{(\bnu)}(\bx)\big|\right]
&\lesssim
\int_0^{1}
\sqrt{\log N(\varepsilon,\B,\rho)}\dif \varepsilon  \\
&\lesssim
\int_0^{1}
\sqrt{\log\!\Big(\frac{1}{\underline h\varepsilon}\Big)}\dif \varepsilon
\lesssim
\sqrt{\log(1/\underline h)} \lesssim \sqrt{\log n}.
\end{align*}

Choosing $a_n \asymp (\mathfrak{R}_{\mathtt{sa}}/\sqrt{\log n})^{1/2}$, the result now follows.
\qed

\begin{lem}[KS Distance Between $\widehat{\Tstat}^{(\bnu)}$ and $\overline{\Tstat}^{(\bnu)}$]\label{sa-lem: t stats to bahadur}
Suppose the conditions in Theorem~\ref{sa-thm: Confidence Bands: Sharp BATEC -- process} hold. Then, for any multi-index $|\bnu| \leq p$,
\begin{align*}
    \sup_{u \in \reals} \Big| \P \Big(\sup_{\bx \in \B} \big|\widehat{\Tstat}^{(\bnu)}(\bx)\big| \leq u\Big)
    - \P \Big(\sup_{\bx \in \B} \big|\overline{\Tstat}^{(\bnu)}(\bx)\big| \leq u \Big)\Big| = o(1).
\end{align*}
\end{lem}

\noindent\textbf{Proof of Lemma~\ref{sa-lem: t stats to bahadur}.} Choose a deterministic sequence $a_n$ such that $\mathfrak{R}_{\mathtt{lin}}=o(a_n)$ and $a_n\sqrt{\log n}=o(1)$; such a sequence exists under the side conditions of Theorem~\ref{sa-thm: Confidence Bands: Sharp BATEC -- process}. Then,
\[
    \Delta_{\widehat T,\overline T}
    :=
    \sup_{\bx \in \B} \big|\overline{\Tstat}^{(\bnu)}(\bx) - \widehat{\Tstat}^{(\bnu)}(\bx)\big|
    =
    o_{\P}(a_n).
\]
Define
\[
    M_{\widehat T}=\sup_{\bx\in\B}|\widehat{\Tstat}^{(\bnu)}(\bx)|,
    \qquad
    M_{\overline T}=\sup_{\bx\in\B}|\overline{\Tstat}^{(\bnu)}(\bx)|,
    \qquad
    M_Z=\sup_{\bx\in\B}|Z^{(\bnu)}(\bx)|.
\]
Applying the first part of Lemma~\ref{sa-lem: generic suprema comparison} with $\mathbb{T}_1=\widehat{\Tstat}^{(\bnu)}$ and $\mathbb{T}_2=\overline{\Tstat}^{(\bnu)}$ gives
\[
    \sup_{u\in\reals}\big|\P(M_{\widehat T}\le u)-\P(M_{\overline T}\le u)\big|
    \le
    \sup_{u\in\reals}\P(|M_{\overline T}-u|\le a_n)
    +\P(\Delta_{\widehat T,\overline T}>a_n).
\]
The anti-concentration term for $M_{\overline T}$ can be bounded through the Gaussian approximation in Lemma~\ref{sa-lem: infeasible gaussian to bahadur representation}: for every $u\in\reals$,
\begin{align*}
    \P(|M_{\overline T}-u|\le a_n)
    &\le
    \P(|M_Z-u|\le a_n)
    +2\sup_{v\in\reals}\big|\P(M_{\overline T}\le v)-\P(M_Z\le v)\big| \\
    &\le
    4a_n\{\E[M_Z]+1\}
    +2\sup_{v\in\reals}\big|\P(M_{\overline T}\le v)-\P(M_Z\le v)\big|.
\end{align*}
By Lemma~\ref{sa-lem: infeasible gaussian to bahadur representation}, the last Kolmogorov distance is $O((\log n)^{1/4}\mathfrak{R}_{\mathtt{sa}}^{1/2})=o(1)$. From the proof of that lemma, $\E[M_Z]\lesssim\sqrt{\log n}$. Since $a_n\sqrt{\log n}=o(1)$ and $\P(\Delta_{\widehat T,\overline T}>a_n)=o(1)$, the result follows.
\qed

\begin{lem}[KS Distance Between $Z^{(\bnu)}$ and $\widehat{Z}^{(\bnu)}$]\label{sa-lem: feasible gaussian to infeasible gaussian}
Suppose the conditions for Theorem~\ref{sa-thm: Confidence Bands: Sharp BATEC -- process} hold. Then, for any multi-index $|\bnu| \leq p$,
\begin{align*}
    \sup_{u \in \reals} \Big|\P \Big(\sup_{\bx \in \B} \big|Z^{(\bnu)}(\bx)\big| \leq u \Big)
                             - \P \Big(\sup_{\bx \in \B} \big|\widehat{Z}^{(\bnu)}(\bx) \big| \leq u \big| \bV \Big) \Big|
    \lessp \log(n)\mathfrak{R}_{\mathtt{cc}}^{1/2}.
\end{align*}
\end{lem}

\noindent \textbf{Proof of Lemma~\ref{sa-lem: feasible gaussian to infeasible gaussian}.} In this proof, we suppress the superscript $(\bnu)$ on $\Omega$ and $\widehat{\Omega}$. First, using Lemma~\ref{sa-lem: covariance}, we provide an upper bound between covariance functions of the feasible Gaussian process and the infeasible Gaussian process. Letting $\bPi_{\bx, \by} = \Omega_{\bx, \by} / \sqrt{\Omega_{\bx,\bx} \Omega_{\by,\by}}$ and $\widehat{\bPi}_{\bx,\by} = \widehat{\Omega}_{\bx,\by} / \sqrt{\widehat{\Omega}_{\bx,\bx} \widehat{\Omega}_{\by,\by}}$,
\begin{align*}
    \sup_{\bx,\by \in \B} \big|\bPi_{\bx, \by} - \widehat{\bPi}_{\bx,\by}\big|
    = \sup_{\bx, \by \in \B}
      \Big|\frac{\Omega_{\bx, \by} - \widehat{\Omega}_{\bx, \by}}{\sqrt{\Omega_{\bx,\bx} \Omega_{\by,\by}}}
         + \frac{\widehat{\Omega}_{\bx, \by}}{\sqrt{\widehat{\Omega}_{\bx,\bx} \widehat{\Omega}_{\by,\by}}} \Big(\sqrt{\frac{\widehat{\Omega}_{\bx,\bx} \widehat{\Omega}_{\by,\by}}{\Omega_{\bx,\bx} \Omega_{\by,\by}}} - 1\Big)\Big|
\end{align*}
From Lemma~\ref{sa-lem: covariance} and the fact that $|\sqrt{x} - \sqrt{y}| \leq (x \wedge y)^{-1/2} |x - y|/2$ for $x, y > 0$,
\begin{align*}
    \sup_{\bx,\by \in \B}
    \frac{\big|\big(\widehat{\Omega}_{\bx,\bx} \widehat{\Omega}_{\by,\by}\big)^{1/2} - \big(\Omega_{\bx,\bx} \Omega_{\by,\by}\big)^{1/2}\big|}
         {\big(\Omega_{\bx,\bx} \Omega_{\by,\by}\big)^{1/2}}
    \lesssim \frac{\sup_{\bx, \by \in \B} \big|\widehat{\Omega}_{\bx,\bx} \widehat{\Omega}_{\by,\by}  - \Omega_{\bx,\bx} \Omega_{\by,\by}\big|}
                  {\inf_{\bx,\by\in\B} \widehat{\Omega}_{\bx,\bx} \widehat{\Omega}_{\by,\by} \wedge \inf_{\bx,\by\in\B} \Omega_{\bx,\bx} \Omega_{\by,\by}}
    \lessp \bar{h}^{p+1} + \mathfrak{R}_{\mathtt{cov}}
\end{align*}
and
\begin{align*}
    \sup_{\bx, \by \in \B} \frac{\big|\Omega_{\bx,\by} - \widehat{\Omega}_{\bx,\by}\big|}{\sqrt{\Omega_{\bx,\bx} \Omega_{\by,\by}}}
    \lessp \bar{h}^{p+1} + \mathfrak{R}_{\mathtt{cov}}.
\end{align*}
Therefore, $\sup_{\bx, \by \in \B} |\bPi_{\bx, \by} - \widehat{\bPi}_{\bx,\by}| \lessp \mathfrak{R}_{\mathtt{cc}}$. Then, we bound the KS distance between the maximum of $Z^{(\bnu)}$ and $\widehat{Z}^{(\bnu)}$ on a $\delta_n$-net of $\B$, denoted by $\B_{\delta_n}$: for all $\bx \in \B$, there exists $\bz \in \B_{\delta_n}$ such that $\infnorm{\bx - \bz} \leq \delta_n$. Since $\B$ is compact, we can assume $M := \operatorname{Card}\left(\B_{\delta_n}\right) \lesssim \delta_n^{-d}$. Denote $\bZ^{\delta_n}$ and $\widehat{\bZ}^{\delta_n}$ as the restrictions of $Z^{(\bnu)}$ and $\widehat{Z}^{(\bnu)}$ to $\B_{\delta_n}$, respectively. Then, by \citet[Theorem 2.1]{Chernozhukov-Chetverikov-Kato-Koike_2022_AoS},
\begin{align*}
    \sup_{\by \in \reals^{M}} \big|\P(\bZ^{\delta_n}\leq \by) - \P(\widehat{\bZ}^{\delta_n} \leq \by | \bV) \big|
    \lesssim \log(M) \sup_{\bx,\by \in \B} \big|\bPi_{\bx, \by} - \widehat{\bPi}_{\bx,\by}\big|^{\frac{1}{2}}
    \lessp \log(M) \mathfrak{R}_{\mathtt{cc}}^{\frac{1}{2}},
\end{align*}
and hence
\begin{align*}
    \sup_{x \in \reals} \big|\P(\infnorm{\bZ^{\delta_n}} \leq x ) - \P(\infnorm{\widehat{\bZ}^{\delta_n}} \leq x | \bV)\big|
    &\leq \sup_{x \in \reals} \big|\P(- x\mathbf{1} \leq \bZ^{\delta_n} \leq x \mathbf{1}) - \P(- x\mathbf{1} \leq \widehat{\bZ}^{\delta_n}  \leq x \mathbf{1} | \bV)\big| \\
    & \lessp  \log(M) \mathfrak{R}_{\mathtt{cc}}^{\frac{1}{2}} =: \mathfrak{R}_M.
\end{align*}

Finally, we extend the comparison from the finite net to the full boundary. Let $\delta_n=n^{-s}$ for a fixed constant $s>0$ chosen large enough that
\begin{align*}
    \sqrt{\log n}\,|\bH|^{-1/2}\underline h^{-1}\delta_n=o(\mathfrak{R}_{\mathtt{cc}}),
    \qquad
    \sqrt{\log n}\,\underline h^{-1}\delta_n=o(\mathfrak{R}_{\mathtt{cc}}).
\end{align*}
Such an $s$ exists under the polynomial lower bound on $\bar h$ and the bounded ratio $\bar h/\underline h$. Put $b_n=\mathfrak{R}_{\mathtt{cc}}^{1/2}$ and define the modulus probabilities
\begin{align*}
    \Psi_{\delta_n}(b_n)
    &= \P \Big(\sup_{\substack{\bx,\by \in \B \\ \infnorm{\bx - \by} \leq \delta_n}} \big|Z^{(\bnu)}(\bx) - Z^{(\bnu)}(\by)\big| \geq b_n \Big),\\
    \widehat{\Psi}_{\delta_n}(b_n)
    &= \P \Big(\sup_{\substack{\bx,\by \in \B \\ \infnorm{\bx - \by} \leq \delta_n}} \big|\widehat{Z}^{(\bnu)}(\bx) - \widehat{Z}^{(\bnu)}(\by)\big| \geq b_n \Big| \bV \Big).
\end{align*}
The usual chaining argument for the local-polynomial Gaussian metrics gives
\begin{align*}
    \E \Big[\sup_{\substack{\bx,\by \in \B \\ \infnorm{\bx - \by}\leq \delta_n}} \big|Z^{(\bnu)}(\bx) - Z^{(\bnu)}(\by)\big|\Big]
    &\lesssim \sqrt{\log n}\,\underline h^{-1}\delta_n,\\
    \E \Big[\sup_{\substack{\bx,\by \in \B \\ \infnorm{\bx - \by}\leq \delta_n}} \big|\widehat{Z}^{(\bnu)}(\bx) - \widehat{Z}^{(\bnu)}(\by)\big| \Big| \bV \Big]
    &\lessp \sqrt{\log n}\,|\bH|^{-1/2}\underline h^{-1}\delta_n.
\end{align*}
The second display follows from Lemma~\ref{sa-lem: covariance}, Lemma~\ref{sa-lem: Gram Matrix}, and the Lipschitz bound for the feasible local-polynomial weights; the first display follows from the corresponding population bound. Therefore Markov's inequality implies
\begin{align*}
    \Psi_{\delta_n}(b_n)+\widehat{\Psi}_{\delta_n}(b_n)=o_{\P}(\mathfrak{R}_{\mathtt{cc}}^{1/2}).
\end{align*}
Moreover, the same entropy calculation gives
\[
    \E\Big[\sup_{\bx\in\B}|Z^{(\bnu)}(\bx)|\Big]\lesssim\sqrt{\log n},
    \qquad
    \E\Big[\sup_{\bx\in\B}|\widehat Z^{(\bnu)}(\bx)|\Big|\bV\Big]\lessp\sqrt{\log n}.
\]
For every $t>0$, the net approximation, the finite-dimensional comparison, and Gaussian anti-concentration give
\begin{align*}
    &\P \Big( \sup_{\bx \in \B} \big|Z^{(\bnu)}(\bx)\big| \leq t \Big)\\
    &\leq \P \Big( \sup_{\bx \in \B} \big|\widehat{Z}^{(\bnu)}(\bx)\big| \leq t \Big| \bV \Big)
       +4b_n\Big(\E\Big[\sup_{\bx\in\B}|\widehat Z^{(\bnu)}(\bx)|\Big|\bV\Big]+1\Big)
       +\Psi_{\delta_n}(b_n)+\widehat\Psi_{\delta_n}(b_n)+\mathfrak R_M,
\end{align*}
and the same argument with the roles of $Z^{(\bnu)}$ and $\widehat Z^{(\bnu)}$ reversed gives the corresponding lower bound. Since $\log M\lesssim \log n$, $\mathfrak R_M\lessp \log(n)\mathfrak{R}_{\mathtt{cc}}^{1/2}$; the anti-concentration term is $O_{\P}(\sqrt{\log n}\mathfrak{R}_{\mathtt{cc}}^{1/2})$; and the two modulus probabilities are $o_{\P}(\mathfrak{R}_{\mathtt{cc}}^{1/2})$. Combining the upper and lower bounds gives the stated result.
\qed

\bigskip
The proof of Theorem \ref{sa-thm: Confidence Bands: Sharp BATEC -- process} now follows directly from Lemma~\ref{sa-lem: infeasible gaussian to bahadur representation}, Lemma~\ref{sa-lem: t stats to bahadur} and Lemma~\ref{sa-lem: feasible gaussian to infeasible gaussian}. Furthermore, by definition of $\widehat{\CI}_{\alpha}^{(\bnu)}(\bx)$,
\begin{align*}
    \P\Big[\tau^{(\bnu)}(\bx) \in \widehat{\CI}_{\alpha}^{(\bnu)}(\bx), \text{ for all } \bx \in \B \Big]
    & = \P \Big[\sup_{\bx \in \B} \big|\widehat{\Tstat}^{(\bnu)}(\bx)\big| \leq \q_{\alpha}\Big] \\
    & = \P \Big[\sup_{\bx \in \B} \big|\widehat{Z}^{(\bnu)}(\bx)\big|\leq \q_{\alpha}\Big] + o(1) \\
    & = \E \Big[\P \Big[\sup_{\bx \in \B} \big|\widehat{Z}^{(\bnu)}(\bx)\big| \leq \q_{\alpha}\Big| \bV \Big]\Big] + o(1) \\
    & = 1 - \alpha + o(1),
\end{align*}
which completes the proof of the theorem.
\qed

\subsection{Proof of Theorem~\ref{sa-thm: Gaussian Approximation Suprema -- Sharp}}

We employ the following result established by \citet[Corollary 2.2]{Chernozhukov-Chetverikov-Kato_2014b_AoS}.

\begin{lem}[Gaussian Approximation of Suprema]\label{sa-lem: suprema coupling lemma}
    Let $X_1, \dots, X_n$ be i.i.d.\ random variables in a measurable space $(S, \mathcal{S})$ with law $P$, where $n \ge 3$. Let $\mathcal{F}$ be a class of measurable functions $f: S \to \mathbb{R}$ that is $P$-centered, i.e., $Pf = 0$ for all $f \in \mathcal{F}$. Define the empirical process $\mathbb{G}_n$ indexed by $\mathcal{F}$ as $\mathbb{G}_n f = \frac{1}{\sqrt{n}} \sum_{i=1}^n f(X_i)$ for $f \in \mathcal{F}$. Suppose that:
    \begin{enumerate}
        \item The class $\mathcal{F}$ is pointwise measurable, that is, it contains a countable subset $\mathcal{G}$ such that for every $f \in \mathcal{F}$ there exists a sequence $g_m \in \mathcal{G}$ with $g_m(x) \to f(x)$ for every $x \in S$.
        \item There exists a measurable envelope $F$ such that $|f| \le F$ for all $f \in \mathcal{F}$, and constants $A \ge e$ and $\upsilon \ge 1$ such that $\sup_Q N(\mathcal{F}, e_Q, \epsilon \|F\|_{Q,2}) \le (A/\epsilon)^{\upsilon}$ for all $0 < \epsilon \le 1$, where the supremum is taken over all finite discrete probability measures on $(S, \mathcal{S})$.
        \item For some $b \ge \sigma > 0$ and $q \in [4, \infty]$, we have $\sup_{f \in \mathcal{F}} P|f|^k \le \sigma^2 b^{k-2}$ for $k = 2, 3, 4$ and $\|F\|_{P,q} \le b$.
    \end{enumerate}
    Let $Z = \sup_{f \in \mathcal{F}} \mathbb{G}_n f$. Then for every $\gamma \in (0, 1)$, there exists, on a possibly extended probability space, a tight Gaussian process $(\mathbb{G}_P f)_{f \in \mathcal{F}}$ with mean zero, covariance function $\E[\mathbb{G}_P(f)\mathbb{G}_P(g)] = P(fg)$, and almost surely uniformly $e_P$-continuous trajectories, such that for $\tilde{Z} = \sup_{f \in \mathcal{F}} \mathbb{G}_P f$:
    $$ P\left\{ |Z - \tilde{Z}| > \frac{b K_n}{\gamma^{1/q} n^{1/2-1/q}} + \frac{(b \sigma)^{1/2} K_n^{3/4}}{\gamma^{1/2} n^{1/4}} + \frac{(b \sigma^2 K_n^2)^{1/3}}{\gamma^{1/3} n^{1/6}} \right\} \le C \left( \gamma + \frac{\log n}{n} \right) $$
    where $K_n = \upsilon(\log n \vee \log(A b/\sigma))$, and $C > 0$ is a constant that depends only on $q$.
\end{lem}

Consider $\bz_i=(\bX_i,\varepsilon_i)$, $1\leq i\leq n$, where $\varepsilon_i=Y_i-\E[Y_i|\bX_i]$. For each $\bx\in\B$, let $g_{\bx}$ be the boundary-indexed local-polynomial weight function defined before Theorem~\ref{sa-thm: Gaussian Strong Approximation: Sharp Tstat}, and consider the class
\begin{align*}
    \F=\{(\bu,u)\in\reals^d\times\reals\mapsto g_{\bx}(\bu)u:\bx\in\B\},
    \qquad
    \F_{\pm}=\F\cup(-\F).
\end{align*}
Then
\begin{align*}
    \sup_{\bx\in\B}|\overline{\Tstat}^{(\bnu)}(\bx)|
    =
    \sup_{f\in\F_{\pm}}
    \frac{1}{\sqrt n}\sum_{i=1}^n\{f(\bz_i)-\E[f(\bz_i)]\}.
\end{align*}
We apply Lemma~\ref{sa-lem: suprema coupling lemma} to this empirical process. First, the fixed indicators $\Indicator(\bu\in\A_t)$ are absorbed into $g_{\bx}$ and do not increase $L_2(Q)$ covering numbers beyond a constant factor, because multiplication by a fixed indicator is a contraction in $L_2(Q)$. Therefore, using the same local-polynomial entropy calculation as in the proofs of Lemmas~\ref{sa-lem: Gram Matrix} and \ref{sa-lem: Linear Approximation}, but without the total-variation step, $\{g_{\bx}:\bx\in\B\}$ is pointwise measurable and has an envelope $\ttM_{\G}\lesssim|\bH|^{-1/2}$ such that, for constants $\bc,\bc'>0$ depending only on the compact set $\X$,
\begin{align*}
    \sup_Q N(\G,\norm{\cdot}_{Q,2},\varepsilon(2\bc+1)^{d+1}\ttM_{\G})
    \leq 2\bc'\varepsilon^{-4d-4}+2,
    \qquad 0<\varepsilon<1,
\end{align*}
where the supremum is taken over all finite discrete measures. Since the residual class is the singleton $\{\operatorname{Id}\}$, the class $\F_{\pm}$ satisfies the same VC-type entropy bound up to a universal constant; centering the functions changes only the envelope by a constant factor.

For the moment condition, take $q=2+v\geq4$ in Lemma~\ref{sa-lem: suprema coupling lemma}. Then, Assumption~\ref{sa-assump: Sharp DGP} implies $\sup_{\bx\in\X}\E[|\varepsilon_i|^{2+v}|\bX_i=\bx]<\infty$. The envelope $F(\bu,u)=\ttM_{\G}|u|$ satisfies $\|F\|_{\P,2+v}\lesssim|\bH|^{-1/2}$. The same local-polynomial moment calculation used in the proof of Lemma~\ref{sa-lem: Linear Approximation} gives, for $\sigma=1$ and $b=|\bH|^{-1/2}$, that $\sup_{f\in\F_{\pm}}\P|f|^k\leq\sigma^2b^{k-2}$ for $k=2,3,4$ and $\|F\|_{\P,2+v}\leq b$. Taking $\gamma=(\log n)^{-1}$ in Lemma~\ref{sa-lem: suprema coupling lemma} yields
\begin{align*}
    \P\left\{ \left|\sup_{\bx\in\B}|\overline{\Tstat}^{(\bnu)}(\bx)|-\sup_{\bx\in\B}|Z^{(\bnu)}(\bx)|\right|>
    \frac{\log n}{(n|\bH|)^{1/6}}
    +\frac{(\log n)^{5/4}}{(n|\bH|)^{1/4}}
    +\frac{(\log n)^{3/2}}{(n^{\frac{v}{2+v}}|\bH|)^{1/2}}
    \right\}
    \leq C\left(\frac{1}{\log n}+\frac{\log n}{n}\right),
\end{align*}
for a constant $C>0$. This proves the theorem.
\qed

\subsection{Proof of Theorem~\ref{sa-thm: Confidence Bands: Sharp BATEC -- Suprema}}
Define
\begin{equation*}
    \mathfrak{R}_{\mathtt{sup}}
    =
    \frac{\log n}{(n|\bH|)^{1/6}}
    +\frac{(\log n)^{5/4}}{(n|\bH|)^{1/4}}
    +\frac{(\log n)^{3/2}}{(n^{\frac{v}{2+v}}|\bH|)^{1/2}}.
\end{equation*}
In this proof, let $\mathfrak{R}_{\mathtt{lin}}$ denote the stochastic-linearization remainder defined at the beginning of the proof of Theorem~\ref{sa-thm: Confidence Bands: Sharp BATEC -- process}.
The entropy calculation in the proof of Theorem~\ref{sa-thm: Gaussian Approximation Suprema -- Sharp} gives
\[
    \E\left[\sup_{\bx\in\B}|Z^{(\bnu)}(\bx)|\right]\lesssim\sqrt{\log n}.
\]
The side condition $(\log n)^9/(n^{v/(2+v)}|\bH|)=o(1)$ implies $\mathfrak{R}_{\mathtt{sup}}\sqrt{\log n}=o(1)$. Let
\[
    M_{\overline T}=\sup_{\bx\in\B}|\overline{\Tstat}^{(\bnu)}(\bx)|,
    \qquad
    M_Z=\sup_{\bx\in\B}|Z^{(\bnu)}(\bx)|.
\]
Applying Lemma~\ref{sa-lem: generic suprema comparison} with $\mathbb{T}_1=M_{\overline T}$ and $\mathbb{T}_2=M_Z$, viewed as singleton-indexed processes, and $\eta=\mathfrak{R}_{\mathtt{sup}}$, gives
\[
    \sup_{u\in\reals}\left|\P(M_{\overline T}\le u)-\P(M_Z\le u)\right|
    \le
    \sup_{u\in\reals}\P(|M_Z-u|\le \mathfrak{R}_{\mathtt{sup}})
    +\P(|M_{\overline T}-M_Z|>\mathfrak{R}_{\mathtt{sup}}).
\]
The first term is $O(\mathfrak{R}_{\mathtt{sup}}\sqrt{\log n})=o(1)$ by Gaussian anti-concentration, and the second term is $o(1)$ by Theorem~\ref{sa-thm: Gaussian Approximation Suprema -- Sharp}. Therefore,
\begin{equation}\label{sa-eq: t-to-z sharp suprema}
    \sup_{u\in\reals}
    \left|
    \P\left(\sup_{\bx\in\B}|\overline{\Tstat}^{(\bnu)}(\bx)|\leq u\right)
    -
    \P\left(\sup_{\bx\in\B}|Z^{(\bnu)}(\bx)|\leq u\right)
    \right|
    =o(1).
\end{equation}

Next, by Theorem~\ref{sa-thm: Stochastic Linearization: Sharp BATEC}, the stochastic-linearization error is $O_{\P}(\mathfrak{R}_{\mathtt{lin}})$. The polynomial lower bound on $\bar h$, together with $(\log n)^9/(n^{v/(2+v)}|\bH|)=o(1)$ and $\bar h^{p+1}\sqrt{n|\bH|\log n}=o(1)$, implies $\mathfrak{R}_{\mathtt{lin}}\sqrt{\log n}=o(1)$. Choose a deterministic sequence $a_n$ such that $\mathfrak{R}_{\mathtt{lin}}=o(a_n)$ and $a_n\sqrt{\log n}=o(1)$. Applying Lemma~\ref{sa-lem: generic suprema comparison} with $\mathbb{T}_1=\widehat{\Tstat}^{(\bnu)}$, $\mathbb{T}_2=\overline{\Tstat}^{(\bnu)}$, and $\eta=a_n$, and then bounding the anti-concentration of $M_{\overline T}$ through Equation~\eqref{sa-eq: t-to-z sharp suprema}, gives
\begin{equation}\label{sa-eq: t-to-t sharp suprema}
    \sup_{u\in\reals}
    \left|
    \P\left(\sup_{\bx\in\B}|\widehat{\Tstat}^{(\bnu)}(\bx)|\leq u\right)
    -
    \P\left(\sup_{\bx\in\B}|\overline{\Tstat}^{(\bnu)}(\bx)|\leq u\right)
    \right|
    =o(1).
\end{equation}

Finally, set
\[
    \mathfrak{R}_{\mathtt{cc}}
    =
    \sqrt{\frac{\log n}{n|\bH|}}
    +\frac{\log n}{n^{\frac{v}{2+v}}|\bH|}
    +\bar h^{p+1}.
\]
This is the same covariance-comparison rate used in Lemma~\ref{sa-lem: feasible gaussian to infeasible gaussian}. The covariance-comparison and net-modulus arguments in that lemma do not use the De Giorgi perimeter condition. Under the present side conditions, they give
\begin{equation}\label{sa-eq: z-to-z sharp suprema}
    \sup_{u\in\reals}
    \left|
    \P\left(\sup_{\bx\in\B}|Z^{(\bnu)}(\bx)|\leq u\right)
    -
    \P\left(\sup_{\bx\in\B}|\widehat Z^{(\bnu)}(\bx)|\leq u\;\middle|\;\bV\right)
    \right|
    \lessp \log(n)\mathfrak{R}_{\mathtt{cc}}^{1/2}
    =o_{\P}(1).
\end{equation}
The first conclusion follows from the triangle inequality applied to Equations~\eqref{sa-eq: t-to-z sharp suprema}, \eqref{sa-eq: t-to-t sharp suprema}, and \eqref{sa-eq: z-to-z sharp suprema}.

For the coverage statement, by definition of $\widehat{\CI}_{\alpha}^{(\bnu)}(\bx)$ and the first conclusion,
\begin{align*}
    \P\Big[\tau^{(\bnu)}(\bx)\in\widehat{\CI}_{\alpha}^{(\bnu)}(\bx),\text{ for all }\bx\in\B\Big]
    &=
    \P\left(\sup_{\bx\in\B}|\widehat{\Tstat}^{(\bnu)}(\bx)|\leq\q_{\alpha}\right)\\
    &=
    \E\left[
    \P\left(\sup_{\bx\in\B}|\widehat Z^{(\bnu)}(\bx)|\leq\q_{\alpha}\;\middle|\;\bV\right)
    \right]+o(1)\\
    &=1-\alpha+o(1),
\end{align*}
which completes the proof.
\qed

\subsection{Proof of Lemma~\ref{sa-lem: Bias: Sharp WBATE}}

By Lemma~\ref{sa-lem: bias},
\begin{align*}
    \sup_{\bx\in\B}
    \big|\E[\widehat{\tau}^{(\bnu)}(\bx)|\bX]-\tau^{(\bnu)}(\bx)-B_{\bx}^{(\bnu)}(\bh)\big|
    = o_{\P}(\bar{h}^{p+1}\bh^{-\bnu}).
\end{align*}
Integrating this uniform expansion against $w(\bx)\dif\Haus^{d-1}(\bx)$ gives the displayed WBATE bias expansion because $\int_{\B}|w(\bx)|\dif\Haus^{d-1}(\bx)<\infty$. When $\bh=\bc h$, the pointwise leading bias is $h^{p+1-|\bnu|}\widetilde{B}_{\bx}^{(\bnu)}(\bc)$; integrating this leading term over $\B$ gives $\widetilde{B}_{\mathtt{WBATE}}^{(\bnu)}(\bc)$, and the integrated remainder remains $o(h^{p+1-|\bnu|})$ by the same integrability argument.
\qed

\subsection{Proof of Lemma~\ref{sa-lem: Variance: Sharp WBATE}}

Since $\Var[\widehat{\tau}_{\mathtt{WBATE}}^{(\bnu)}|\bX] = \Var[\int_{\B}\widehat{\mu}^{(\bnu)}_0(\bb) w(\bb) \dif\Haus^{d-1}(\bb)|\bX] + \Var[\int_{\B}\widehat{\mu}^{(\bnu)}_1(\bb) w(\bb) \dif\Haus^{d-1}(\bb)|\bX]$, it is enough to consider only one treatment assignment group $t \in \{0,1\}$. The conditional cross-covariance between the two side-specific integrals is zero because the side indicators are disjoint observation by observation, and observations are conditionally independent given $\bX$. In addition,
\begin{align*}
    \Var \Big[\int_{\B}\widehat{\mu}^{(\bnu)}_t(\bb) w(\bb) \dif\Haus^{d-1}(\bb) \Big| \bX \Big]
    = \int_{\B} \int_{\B} \Cov \big[\widehat{\mu}^{(\bnu)}_t(\bb_1), \widehat{\mu}^{(\bnu)}_t(\bb_2) \big| \bX \big] w(\bb_1) w(\bb_2) \dif\Haus^{d-1}(\bb_1) \dif\Haus^{d-1}(\bb_2)
\end{align*}
and
\begin{align*}
    \Omega_{t,\mathtt{WBATE}}^{(\bnu)}
    = \int_{\B} \int_{\B} \Omega_{t,\bb_1, \bb_2}^{(\bnu)} w(\bb_1) w(\bb_2) \dif\Haus^{d-1}(\bb_1) \dif\Haus^{d-1}(\bb_2).
\end{align*}

Proceeding as in the proof of Lemma \ref{sa-lem: covariance}, we have
\begin{align*}
    \sup_{\bb_1, \bb_2 \in \B} \big| \Cov[\widehat{\mu}^{(\bnu)}_t(\bb_1), \widehat{\mu}^{(\bnu)}_t(\bb_2)| \bX ] - \Omega_{t, \bb_1, \bb_2}^{(\bnu)} \big|
    \lessp \frac{1}{n|\bH|\bh^{2\bnu}}\sqrt{\frac{\log(1/\bar{h})}{n|\bH|}}.
\end{align*}

Since $K$ is supported on a compact set, let $R\in(0,\infty)$ denote the diameter of the support, and define the ``effective domain'' $\mathcal{E}(\bH) = \{(\bx, \by) \in \B \times \B: \lVert \bx - \by \rVert \leq \bar{h} R\}$. Let $\nu_{\B,2}=\Haus^{d-1}\times\Haus^{d-1}$ on $\B\times\B$. By \eqref{sa-eq: local boundary measure regularity}, $\nu_{\B,2}(\mathcal{E}(\bH)) \lesssim \bar{h}^{d-1}$. Therefore,
\begin{align*}
    &\Big| \Var \Big[\int_{\B}\widehat{\mu}^{(\bnu)}_t(\bb) w(\bb) \dif\Haus^{d-1}(\bb) \Big| \bX \Big] - \Omega_{t,\mathtt{WBATE}}^{(\bnu)} \Big|\\
    &= \Big| \int_{\B} \int_{\B} \big( \Cov[\widehat{\mu}^{(\bnu)}_t(\bb_1), \widehat{\mu}^{(\bnu)}_t(\bb_2)| \bX] - \Omega_{t,\bb_1, \bb_2}^{(\bnu)} \big) w(\bb_1) w(\bb_2) \dif\Haus^{d-1}(\bb_1) \dif\Haus^{d-1}(\bb_2)\Big|\\
    & \lesssim \sup_{\bb_1, \bb_2 \in \B} \big| \Cov[\widehat{\mu}^{(\bnu)}_t(\bb_1), \widehat{\mu}^{(\bnu)}_t(\bb_2)|\bX] -
    \Omega_{t, \bb_1, \bb_2}^{(\bnu)}\big| \int_{\B} \int_{\B} \Indicator((\bb_1, \bb_2) \in \mathcal{E}(\bH)) w(\bb_1) w(\bb_2) \dif\Haus^{d-1}(\bb_1) \dif\Haus^{d-1}(\bb_2) \\
    &= o_{\P}\Big(\frac{\bar{h}^{d-1}}{n|\bH|\bh^{2\bnu}}\Big),
\end{align*}
because $\frac{\log(1/\bar{h})}{n|\bH|} = o(1)$ and $\bar{h}=o(1)$. This proves the first claim. Next, for each $t\in\{0,1\}$,
\begin{align*}
    \Omega_{t,\mathtt{WBATE}}^{(\bnu)}
    & = \int_{\B} \int_{\B} \Omega_{t,\bb_1, \bb_2}^{(\bnu)} w(\bb_1) w(\bb_2) \dif\Haus^{d-1}(\bb_1) \dif\Haus^{d-1}(\bb_2)\\
    & \leq \sup_{\bb_1, \bb_2 \in \B} \big|\Omega_{t,\bb_1, \bb_2}^{(\bnu)}\big| \int_{\B} \int_{\B} \Indicator((\bb_1, \bb_2) \in \mathcal{E}(\bH)) w(\bb_1) w(\bb_2) \dif\Haus^{d-1}(\bb_1) \dif\Haus^{d-1}(\bb_2) \\
    & \lesssim (n|\bH|\bh^{2\bnu})^{-1} \nu_{\B,2}(\mathcal{E}(\bH)) \lesssim \frac{\bar{h}^{d-1}}{n|\bH|\bh^{2\bnu}},
\end{align*}
which verifies the upper bound after summing over $t$. The lower bound follows from the aggregate variance nondegeneracy condition \eqref{sa-eq: sharp WBATE aggregate nondegeneracy}. This condition is needed because the covariance kernel can contain negative regions after local-polynomial weighting, so pointwise variance lower bounds alone do not rule out cancellation in the weighted boundary integral.

The third and final claim of the lemma follows from Lemma~\ref{sa-lem: covariance} and the same analysis as above.
\qed

\subsection{Proof of Theorem~\ref{sa-thm: MSE Expansion: Sharp WBATE}}

Lemma~\ref{sa-lem: Bias: Sharp WBATE} gives
\begin{align*}
    \E[\widehat{\tau}_{\mathtt{WBATE}}^{(\bnu)}|\bX]-\tau_{\mathtt{WBATE}}^{(\bnu)}
    = B_{\mathtt{WBATE}}^{(\bnu)}(\bh)+o_{\P}(\bar{h}^{p+1}\bh^{-\bnu}),
\end{align*}
so the squared conditional bias equals $(B_{\mathtt{WBATE}}^{(\bnu)}(\bh))^2+o_{\P}(\bar{h}^{2p+2}\bh^{-2\bnu})$. Lemma~\ref{sa-lem: Variance: Sharp WBATE} gives the conditional variance expansion with remainder $o_{\P}(\bar{h}^{d-1}/(n|\bH|\bh^{2\bnu}))$. Adding the squared-bias and variance expansions gives the stated MSE expansion.
\qed

\subsection{Proof of Theorem~\ref{sa-thm: Asymptotic Normality: Sharp WBATE}}

Since $\widehat{\tau}_{\mathtt{WBATE}}^{(\bnu)} - \tau_{\mathtt{WBATE}}^{(\bnu)} = (\widehat{\mu}_{1,\mathtt{WBATE}}^{(\bnu)} - \mu_{1,\mathtt{WBATE}}^{(\bnu)}) - (\widehat{\mu}_{0,\mathtt{WBATE}}^{(\bnu)} - \mu_{0,\mathtt{WBATE}}^{(\bnu)})$, it is enough to start with only one treatment assignment group $t \in \{0,1\}$. The same argument applies to both side-specific components, and the two components combine by summing their variances because the treatment- and control-side kernel-weighted summands have disjoint side indicators observation by observation. Furthermore,
\begin{align*}
    \widehat{\mu}_{t,\mathtt{WBATE}}^{(\bnu)} - \mu_{t,\mathtt{WBATE}}^{(\bnu)}
    &= \int_{\B} (\widehat{\mu}^{(\bnu)}_{t}(\bb) - \mu^{(\bnu)}_t(\bb) ) w(\bb) \dif\Haus^{d-1}(\bb)\\
    &= \int_{\B} \bnu!\be_{\bnu}^{\top}\bS^{-1} \bGamma_{t,\bb}^{-1} \bQ_{t,\bb} w(\bb) \dif\Haus^{d-1}(\bb)\\
    &\qquad  + \int_{\B} \bnu!\be_{\bnu}^{\top}\bS^{-1}(\widehat{\bGamma}_{t,\bb}^{-1} - \bGamma_{t,\bb}^{-1}) \bQ_{t,\bb} w(\bb) \dif\Haus^{d-1}(\bb)
      + O_\P(\bar{h}^{p+1}\bh^{-\bnu})
\end{align*}
using Lemma~\ref{sa-lem: bias} to bound the approximation error.
The side condition $n|\bH|\bar{h}^{2p+2}/\bar{h}^{d-1}=o(1)$ implies that the bias term is $o_{\P}((\Omega_{\mathtt{WBATE}}^{(\bnu)})^{1/2})$.

For the second integral, let
\begin{align*}
    \overline{\bSigma}_{t,\bx_1,\bx_2}
    = \En \Big[\br_p(\bH^{-1}(\bX_i - \bx_1)) \br_p(\bH^{-1}(\bX_i - \bx_2))^{\top} |\bH| K_{\bH}(\bX_i - \bx_1) K_{\bH}(\bX_i - \bx_2) \sigma_t^2(\bX_i) \Indicator(\bX_i \in \A_t)\Big],
\end{align*}
and since $\overline{\bSigma}_{t,\bb_1, \bb_2} = 0$ if $\lVert \bb_1-\bb_2\rVert$ is larger than a constant multiple of $\bar{h}$, let
\[
    \mathcal{E}(\bH)=\{(\bb_1,\bb_2)\in\B\times\B:\lVert\bb_1-\bb_2\rVert\lesssim\bar h\}.
\]
By \eqref{sa-eq: local boundary measure regularity}, $\nu_{\B,2}(\mathcal{E}(\bH))\lesssim \bar h^{d-1}$. Then
\begin{align*}
    &\E \bigg[ \bigg(\int_{\B} \bnu!\be_{\bnu}^{\top}\bS^{-1}(\widehat{\bGamma}_{t,\bb}^{-1} - \bGamma_{t,\bb}^{-1}) \bQ_{t,\bb} w(\bb) \dif\Haus^{d-1}(\bb) \bigg)^2\bigg|\bX\bigg]\\
    &= \int_{\B} \int_{\B} (\bnu!)^2\be_{\bnu}^{\top}\bS^{-1}(\widehat{\bGamma}_{t,\bb_1}^{-1} - \bGamma_{t,\bb_1}^{-1}) (n|\bH|)^{-1} \overline{\bSigma}_{t,\bb_1, \bb_2} (\widehat{\bGamma}_{t,\bb_2}^{-1} - \bGamma_{t,\bb_2}^{-1}) \bS^{-1}\be_{\bnu} w(\bb_1) w(\bb_2) \dif\Haus^{d-1}(\bb_1) \dif\Haus^{d-1}(\bb_2),\\
    &\leq \bh^{-2\bnu}\sup_{\bb \in \B} \|\widehat{\bGamma}_{t,\bb}^{-1} - \bGamma_{t,\bb}^{-1}\|^2 \sup_{\bb_1, \bb_2 \in \B} \lVert \overline{\bSigma}_{t,\bb_1, \bb_2}  \rVert \sup_{\bb \in \B} |w(\bb)| (n|\bH|)^{-1} \nu_{\B,2}(\mathcal{E}(\bH)) \\
    &= o_\P\Big(\frac{\bar{h}^{d-1}}{n|\bH|\bh^{2\bnu}}\Big),
\end{align*}
and hence $\int_{\B} \bnu!\be_{\bnu}^{\top}\bS^{-1}(\widehat{\bGamma}_{t,\bb}^{-1} - \bGamma_{t,\bb}^{-1}) \bQ_{t,\bb} w(\bb) \dif\Haus^{d-1}(\bb) = o_\P\big((\bar{h}^{d-1}/(n|\bH|\bh^{2\bnu}))^{1/2}\big)$.

Next, combine the two side-specific expansions with signs $s_1=1$ and $s_0=-1$. Define the signed linearized statistic
\begin{align*}
    \overline{\Tstat}_{\mathtt{WBATE}}^{(\bnu)}
    =
    (\Omega_{\mathtt{WBATE}}^{(\bnu)})^{-1/2}
    \sum_{t\in\{0,1\}}s_t
    \int_{\B} \bnu!\be_{\bnu}^{\top}\bS^{-1}\bGamma_{t,\bb}^{-1} \bQ_{t,\bb} w(\bb) \dif\Haus^{d-1}(\bb).
\end{align*}
Using Lemma~\ref{sa-lem: Variance: Sharp WBATE} and the previous side-specific bounds,
\begin{align*}
    \widehat{\Tstat}_{\mathtt{WBATE}}^{(\bnu)} - \overline{\Tstat}_{\mathtt{WBATE}}^{(\bnu)}
    = \big((\widehat{\Omega}_{\mathtt{WBATE}}^{(\bnu)})^{-1/2} - (\Omega_{\mathtt{WBATE}}^{(\bnu)})^{-1/2}\big)
      \sum_{t\in\{0,1\}}s_t
      \int_{\B} \bnu!\be_{\bnu}^{\top}\bS^{-1}\bGamma_{t,\bb}^{-1} \bQ_{t,\bb} w(\bb) \dif\Haus^{d-1}(\bb) + o_{\P}(1)
    = o_{\P}(1),
\end{align*}

Finally, apply the Berry--Esseen lemma to this signed treatment-control statistic. Write $\overline{\Tstat}_{\mathtt{WBATE}}^{(\bnu)} = \sum_{i = 1}^n \chi_i$, where
\begin{align*}
    \chi_i
    &= n^{-1}(\Omega_{\mathtt{WBATE}}^{(\bnu)})^{-1/2}
       \sum_{t\in\{0,1\}}s_t
       \int_{\B} \bnu!\be_{\bnu}^{\top}\bS^{-1}\bGamma_{t,\bb}^{-1}
       \br_p(\bH^{-1}(\bX_i - \bb)) K_{\bH}(\bX_i - \bb) \\
    &\qquad\qquad\qquad\qquad\qquad\qquad\qquad
       \cdot \Indicator(\bX_i \in \A_t) \varepsilon_i w(\bb) \dif\Haus^{d-1}(\bb).
\end{align*}
This satisfies $\E[\chi_i] = 0$. The definition of $\Omega_{\mathtt{WBATE}}^{(\bnu)}$ implies that $\sum_{i = 1}^n\Var[\chi_i] = 1$. Since the indicators $\Indicator(\bX_i\in\A_t)$ are disjoint observation by observation, the Lyapunov bound is the same as the side-specific bound up to a fixed constant. Hence, it remains to bound
\begin{align*}
    &\sum_{i = 1}^n \E[|\chi_i|^3]\\
    &\lesssim n^{-3} (\Omega_{\mathtt{WBATE}}^{(\bnu)})^{-3/2} \sum_{i = 1}^n\sum_{t\in\{0,1\}} \E \bigg[\bigg| \int_{\B} \bnu!\be_{\bnu}^{\top}\bS^{-1}\bGamma_{t,\bb}^{-1} \br_p(\bH^{-1}(\bX_i - \bb)) K_{\bH}(\bX_i - \bb) \Indicator(\bX_i \in \A_t) \varepsilon_i w(\bb) \dif\Haus^{d-1}(\bb) \bigg|^3\bigg].
\end{align*}

Let $R$ denote the diameter of the (compact) support of $K$, and define $\mathcal{E}_3(\bH) = \{(\bb_1, \bb_2, \bb_3) \in \B^3: \lVert \bb_i - \bb_j\rVert \leq R\bar{h}, i,j = 1,2,3\}$. Let $\nu_{\B,3}$ denote the product measure $\Haus^{d-1}\times\Haus^{d-1}\times\Haus^{d-1}$ on $\B^3$. Iterating \eqref{sa-eq: local boundary measure regularity} gives $\nu_{\B,3}(\mathcal{E}_3(\bH)) \lesssim \bar{h}^{2d-2}$. Then,
\begin{align*}
    & \E \bigg[\bigg| \int_{\B} \bnu!\be_{\bnu}^{\top}\bS^{-1}\bGamma_{t,\bb}^{-1} \br_p(\bH^{-1}(\bX_i - \bb)) K_{\bH}(\bX_i - \bb) \Indicator(\bX_i \in \A_t) \varepsilon_i w(\bb) \dif\Haus^{d-1}(\bb) \bigg|^3\bigg] \\
    & \leq \E \bigg[\int_{\bb_1 \in \B} \int_{\bb_2 \in \B} \int_{\bb_3 \in \B} |G(\bb_1, \bb_2, \bb_3)| w(\bb_1) w(\bb_2) w(\bb_3) \dif\Haus^{d-1}(\bb_1) \dif\Haus^{d-1}(\bb_2) \dif\Haus^{d-1}(\bb_3)\bigg] \\
    & \lesssim \nu_{\B,3}(\mathcal{E}_3(\bH))  \sup_{\bb_1, \bb_2, \bb_3 \in \B} \E[|G(\bb_1, \bb_2, \bb_3)|],
\end{align*}
where $G(\bb_1, \bb_2, \bb_3) = g(\bX_i, \varepsilon_i, \bb_1) g(\bX_i, \varepsilon_i, \bb_2) g(\bX_i, \varepsilon_i, \bb_3)$ with
\begin{align*}
    g(\bX_i, \varepsilon_i, \bb)
    = \bnu!\be_{\bnu}^{\top}\bS^{-1}\bGamma_{t,\bb}^{-1} \br_p(\bH^{-1}(\bX_i - \bb)) K_{\bH}(\bX_i - \bb) \Indicator(\bX_i \in \A_t) \varepsilon_i.
\end{align*}

Proceeding as in the proof of Lemma~\ref{sa-lem: Gram Matrix} and Lemma~\ref{sa-lem: Variance: Sharp WBATE}, it can be shown that
\begin{align*}
    \sup_{\bb_1, \bb_2, \bb_3 \in \B} \E[|G(\bb_1, \bb_2, \bb_3)|] \lesssim \bh^{-3\bnu}|\bH|^{-2}
\end{align*}
provided that $\frac{\log(1/\bar{h})}{n|\bH|} = o(1)$. Therefore, together with the rate of $\Omega_{\mathtt{WBATE}}^{(\bnu)}$ from Lemma~\ref{sa-lem: Variance: Sharp WBATE}, we have $\sum_{i = 1}^n \E[|\chi_i|^3] \lesssim (\bar{h}^{d-1}/(n|\bH|))^{1/2} \asymp (n\bar{h})^{-1/2}$, and the result follows.
\qed

\subsection{Proof of Theorem~\ref{sa-thm: Convergence Rate: Sharp LBATE}}

Follows by Theorem \ref{sa-thm: Convergence Rates: Sharp BATEC} after noting that
\begin{align*}
    |\widehat{\tau}_{\mathtt{LBATE}}^{(\bnu)} - \tau_{\mathtt{LBATE}}^{(\bnu)}|
    \leq \sup_{\bx\in\B} |\widehat{\tau}^{(\bnu)}(\bx) - \tau^{(\bnu)}(\bx)|.
\end{align*}
\qed

\subsection{Proof of Theorem~\ref{sa-thm: Confidence Intervals: Sharp LBATE}}

Consider the event $E = \Big\{\sup_{\bb \in \B} \frac{|\widehat{\tau}^{(\bnu)}(\bb) - \tau^{(\bnu)}(\bb)|}{(\widehat{\Omega}_{\bb, \bb}^{(\bnu)})^{1/2}} \leq \q_{\alpha}\Big\}$. Theorem~\ref{sa-thm: Confidence Bands: Sharp BATEC -- Suprema} implies that $\P(E) = 1 - \alpha + o(1)$. On the event $E$, we also have
\begin{align*}
    \widehat{\tau}^{(\bnu)}(\bb) - \q_{\alpha}(\widehat{\Omega}_{\bb,\bb}^{(\bnu)})^{1/2} \leq \tau^{(\bnu)}(\bb)
    \leq \widehat{\tau}^{(\bnu)}(\bb) + \q_{\alpha}(\widehat{\Omega}_{\bb,\bb}^{(\bnu)})^{1/2}, \qquad \forall \; \bb \in \B,
\end{align*}
which implies
\begin{align*}
    \sup_{\bb \in \B}  \widehat{\tau}^{(\bnu)}(\bb) - \q_{\alpha}(\widehat{\Omega}_{\bb,\bb}^{(\bnu)})^{1/2}
    \leq \sup_{\bb \in \B} \tau^{(\bnu)}(\bb)
    \leq \sup_{\bb \in \B} \widehat{\tau}^{(\bnu)}(\bb) + \q_{\alpha}(\widehat{\Omega}_{\bb,\bb}^{(\bnu)})^{1/2}.
\end{align*}
The stated result then follows.
\qed

\subsection{Proof of Theorem~\ref{sa-thm: Convergence Rates: Fuzzy BATEC}}

Theorem \ref{sa-thm: Convergence Rates: Sharp BATEC} applies directly to the reduced-form estimator $\widehat{\tau}^{(\bnu)}_Y$. For the treatment-status estimator $\widehat{\tau}^{(\bnu)}_W$, the same convergence-rate proof applies with $W$ in place of $Y$: Assumption~\ref{sa-assump: Fuzzy DGP}(ii) gives the required smoothness of $\mu_{W,t}$, and $W_i(t)\in\{0,1\}$ gives the required moment bounds. The variance lower bound in Assumption~\ref{sa-assump: Sharp DGP}(iv) is not used for these consistency rates. The strengthened bandwidth condition makes the uniform reduced-form convergence rate vanish, so $\inf_{\bx\in\B}|\widehat{\tau}_W^{(\bnu)}(\bx)|$ is bounded away from zero with probability approaching one. Since $\inf_{\bx\in\B}|\tau_W^{(\bnu)}(\bx)|>0$, the displayed rates follow from the exact expansion
\begin{align*}
    \widehat{\zeta}^{(\bnu)}(\bx) - \zeta^{(\bnu)}(\bx)
    = \frac{\widehat{\tau}^{(\bnu)}_Y(\bx)-\tau^{(\bnu)}_Y(\bx)}{\tau_W^{(\bnu)}(\bx)}
      - \frac{\tau_Y^{(\bnu)}(\bx)}{\tau_W^{(\bnu)}(\bx)\widehat{\tau}_W^{(\bnu)}(\bx)}
        \big(\widehat{\tau}^{(\bnu)}_W(\bx)-\tau^{(\bnu)}_W(\bx)\big),
\end{align*}
and the same argument uniformly over $\bx\in\B$.
\qed


\subsection{Proof of Theorem~\ref{sa-thm: MSE Expansions: Fuzzy BATEC}}

\begin{lem}[Covariance for Fuzzy Designs]\label{sa-lem: covariance for fuzzy designs}
    Suppose Assumptions \ref{sa-assump: Sharp DGP}, \ref{sa-assump: Boundary and Kernel} and \ref{sa-assump: Fuzzy DGP} hold. If $\frac{\log(1/\bar{h})}{n^{\frac{v}{2+v}} |\bH|} = o(1)$ and $\bar{h}=o(1)$, then for $t = 0,1$, and $A_1, A_2 \in \{Y, W\}$, we have
    \begin{align*}
        \sup_{\bx_1,\bx_2 \in \B} \big\|\widehat{\bSigma}_{A_1,A_2,t,\bx_1,\bx_2} - \bSigma_{A_1,A_2,t,\bx_1,\bx_2}\big\|
        \lessp \sqrt{\frac{\log(1/\bar{h})}{n |\bH|}} + \frac{\log(1/\bar{h})}{n^{\frac{v}{2+v}}|\bH|} + \bar{h}^{p+1},
    \end{align*}
    \begin{align*}
        \sup_{\bx_1,\bx_2 \in \B} \big|\widehat{\Omega}_{A_1, A_2, \bx_1,\bx_2}^{(\bnu)} - \Omega_{A_1, A_2, \bx_1,\bx_2}^{(\bnu)}\big|
        \lessp (n |\bH|\bh^{2\bnu})^{-1}\bigg(\sqrt{\frac{\log(1/\bar{h})}{n |\bH|}} + \frac{\log(1/\bar{h})}{n^{\frac{v}{2+v}}|\bH|}+ \bar{h}^{p+1}\bigg),
    \end{align*}
\end{lem}

\begin{proof}
The argument parallels the proof of Lemma~\ref{sa-lem: covariance}.
The only additional moment input is for the treatment-status residuals and cross-products: $W_i$ is binary, hence $\varepsilon_{W,i}$ is bounded, while Assumption~\ref{sa-assump: Sharp DGP}(v) gives the required conditional fourth-moment control for $\varepsilon_{Y,i}$; mixed terms with $A_1,A_2\in\{Y,W\}$ then follow by Cauchy--Schwarz.
We begin with the decomposition
\begin{align*}
    \widehat{\varepsilon}_{A_1,i}\widehat{\varepsilon}_{A_2,i}
    = (\varepsilon_{A_1,i} + \widehat{\varepsilon}_{A_1,i}-\varepsilon_{A_1,i})
      (\varepsilon_{A_2,i} + \widehat{\varepsilon}_{A_2,i}-\varepsilon_{A_2,i}),
\end{align*}
which allows us to control the difference between estimated and population covariances through uniform bounds on
$\widehat{\varepsilon}_{A,i}-\varepsilon_{A,i}$ for $A\in\{Y,W\}$. Fix $A\in\{Y,W\}$. For each $i$,
\begin{align*}
    &|\widehat{\varepsilon}_{A,i}-\varepsilon_{A,i}|\\
    &\le
    \sup_{\bx\in\B}
    \Big|
        \sum_{s\in\{0,1\}}\Indicator(\bX_i\in\A_s)\mu_{A,s}(\bX_i)\\
    &\hspace{1in}
        - \br_p(\bH^{-1}(\bX_i-\bx))^\top\bS
        \big[
            \Indicator(\bX_i\in\A_0)\widehat{\bbeta}_{A,0}(\bx)
            + \Indicator(\bX_i\in\A_1)\widehat{\bbeta}_{A,1}(\bx)
        \big]
    \Big|
    \Indicator(K_{\bH}(\bX_i-\bx)\neq0).
\end{align*}
Taking the maximum over $i$ and decomposing the difference into approximation and estimation components yields
\begin{align*}
    & \max_{1\le i\le n}|\widehat{\varepsilon}_{A,i}-\varepsilon_{A,i}| \\
    &\le
    \max_{1\le i\le n}\sup_{\bx\in\B}
    \Big|
        \sum_{s\in\{0,1\}}\Indicator(\bX_i\in\A_s)\mu_{A,s}(\bX_i)\\
    &\hspace{1in}
        - \br_p(\bH^{-1}(\bX_i-\bx))^\top\bS
        \big[
            \Indicator(\bX_i\in\A_0)\widehat{\bbeta}_{A,0}(\bx)
            + \Indicator(\bX_i\in\A_1)\widehat{\bbeta}_{A,1}(\bx)
        \big]
    \Big|
    \Indicator(K_{\bH}(\bX_i-\bx)\neq 0) \\
    &\le
    \sup_{\bx\in\B}\sup_{\by\in\X}
    \Big|
        \sum_{s\in\{0,1\}}\Indicator(\by\in\A_s)\mu_{A,s}(\by)\\
    &\hspace{1in}
        - \br_p(\bH^{-1}(\by-\bx))^\top\bS
        \big[
            \Indicator(\by\in\A_0)\bbeta_{A,0}(\bx)
            + \Indicator(\by\in\A_1)\bbeta_{A,1}(\bx)
        \big]
    \Big|
    \Indicator(K_{\bH}(\by-\bx)\neq 0) \\
    &\quad
    + \max_{t=0,1}\sup_{\bx\in\B}\sup_{\by\in\X}
    \Big|
        \br_p(\bH^{-1}(\by-\bx))^\top\bS
        \big(\widehat{\bbeta}_{A,t}(\bx)-\bbeta_{A,t}(\bx)\big)
    \Big|
    \Indicator(K_{\bH}(\by-\bx)\neq 0).
\end{align*}

The first term is the usual local polynomial approximation error and is of order $\bar{h}^{p+1}$ uniformly over $\bx$.
The second term is controlled by the uniform convergence rate of
$\widehat{\bbeta}_{A,t}(\bx)$, giving
\begin{align*}
    \max_{1\le i\le n}
    |\widehat{\varepsilon}_{A,i}-\varepsilon_{A,i}|
    \lessp
    \bar{h}^{p+1}
    + \sqrt{\frac{\log(1/\bar{h})}{n |\bH|}}
    + \frac{\log(1/\bar{h})}{n^{\frac{1+v}{2+v}}|\bH|}.
\end{align*}

Substituting this bound into the covariance decomposition and proceeding exactly as in Lemma~\ref{sa-lem: covariance} yields the stated uniform bounds for
$\widehat{\bSigma}_{A_1,A_2,t,\bx_1,\bx_2}$,
$\widehat{\Omega}_{A_1,A_2,\bx_1,\bx_2}^{(\bnu)}$.
\end{proof}

\noindent\textbf{Completion of the AMSE proof.}
Applying Lemma~\ref{sa-lem: bias} with $A \in \{Y,W\}$ yields
\begin{align*}
    \sup_{\bx \in \B}
    \Big|
        \E[\widehat{\tau}_{A}^{(\bnu)}(\bx)\mid\bX]
        - \tau_{A}^{(\bnu)}(\bx)
        - B_{A,\bx}^{(\bnu)}(\bh)
    \Big|
    = o_{\P}\!\left(\bar{h}^{p+1}\bh^{-\bnu}\right).
\end{align*}
Premultiplying the reduced-form bias vector by $\mathfrak{w}(\bx)$ gives $B_{\mathtt{F},\bx}^{(\bnu)}(\bh)$ in \eqref{sa-eq: leading bias fuzzy}. Combining this bias expansion with Lemma~\ref{sa-lem: covariance for fuzzy designs} gives the AMSE expansion for the leading linear component, and integrating the same pointwise expansion over $\B$ establishes the AIMSE result. \qed

\subsection{Proof of Lemma~\ref{sa-lem: Covariance: Fuzzy BATEC}}
Let
\begin{align*}
    a_n = \sqrt{\frac{\log(1/\bar{h})}{n |\bH|}}
          + \frac{\log(1/\bar{h})}{n^{\frac{v}{2+v}}|\bH|}
          + \bar{h}^{p+1}.
\end{align*}
Lemma~\ref{sa-lem: covariance for fuzzy designs} gives
\begin{align*}
    \sup_{\bx_1,\bx_2\in\B}
    \big\|\widehat{\bTheta}_{\ttF,\bx_1,\bx_2}^{(\bnu)}
          -\bTheta_{\ttF,\bx_1,\bx_2}^{(\bnu)}\big\|
    \lessp (n|\bH|\bh^{2\bnu})^{-1} a_n .
\end{align*}
The reduced-form convergence rates, together with $\inf_{\bx\in\B}|\tau_W^{(\bnu)}(\bx)|>0$, imply
\begin{align*}
    \sup_{\bx\in\B}\|\widehat{\mathfrak{w}}(\bx)-\mathfrak{w}(\bx)\|
    \lessp \bh^{-\bnu}a_n.
\end{align*}
The displayed rate condition in Lemma~\ref{sa-lem: Covariance: Fuzzy BATEC} makes the right-hand side $o_{\P}(1)$, so the feasible weights are uniformly bounded with probability approaching one.
Because $\mathfrak{w}(\cdot)$ is bounded and
$\sup_{\bx_1,\bx_2\in\B}\|\bTheta_{\ttF,\bx_1,\bx_2}^{(\bnu)}\|
\lesssim (n|\bH|\bh^{2\bnu})^{-1}$, the feasible plug-in covariance satisfies
\begin{align*}
    \sup_{\bx_1,\bx_2 \in \B} \big|\widehat{\Omega}_{\ttF, \bx_1,\bx_2}^{(\bnu)} - \Omega_{\ttF, \bx_1,\bx_2}^{(\bnu)}\big|
    \lessp \frac{\bh^{-\bnu}}{n|\bH|\bh^{2\bnu}}a_n.
\end{align*}
Finally, $\Omega_{\ttF,\bx,\bx}^{(\bnu)}=V_{\mathtt{F},\bx}^{(\bnu)}/(n|\bH|\bh^{2\bnu})$, so $\inf_{\bx\in\B}V_{\mathtt{F},\bx}^{(\bnu)}>0$ gives the required lower bound on the population denominator. The inverse-square-root bound follows from the displayed covariance bound and a mean-value expansion of $x\mapsto x^{-1/2}$ on this bounded-away-from-zero neighborhood. \qed

\subsection{Proof of Theorem~\ref{sa-thm: Confidence Intervals: Fuzzy BATEC}}
For the fixed boundary point $\bx$ in the theorem, a similar argument as the proof for Theorem~\ref{sa-thm: Confidence Intervals: Sharp BATEC} gives the following stochastic linearization,
\begin{align*}
    \Big|\widehat{\Tstat}_{\mathtt{F}}^{(\bnu)}(\bx) - \overline{\Tstat}_{\mathtt{F}}^{(\bnu)}(\bx)\Big|
    \lessp
    \bar{h}^{p+1}\sqrt{n |\bH|} + \frac{1}{\sqrt{n |\bH|}},
\end{align*}
provided that $\bar{h} \to 0$ and $n |\bH| \to \infty$. To see this, the pointwise reduced-form rates give
\begin{align*}
    |\widehat{\tau}_{A}^{(\bnu)}(\bx)-\tau_A^{(\bnu)}(\bx)|
    =O_{\P}\big(\bh^{-\bnu}((n|\bH|)^{-1/2}+\bar h^{p+1})\big),
    \qquad A\in\{Y,W\}.
\end{align*}
Since $|\tau_W^{(\bnu)}(\bx)|>0$, this implies $|\widehat{\tau}_W^{(\bnu)}(\bx)|^{-1}=O_{\P}(1)$ and
\begin{align*}
    \|\widehat{\mathfrak w}(\bx)-\mathfrak w(\bx)\|
    =O_{\P}\big(\bh^{-\bnu}((n|\bH|)^{-1/2}+\bar h^{p+1})\big).
\end{align*}
The exact second-order expansion in Section~\ref{sec: Second-Order Linearization: Fuzzy BATEC} gives
\begin{align*}
    |\mathfrak R_{\mathtt F}(\bx)|
    =O_{\P}\big(\bh^{-2\bnu}((n|\bH|)^{-1/2}+\bar h^{p+1})^2\big),
\end{align*}
and hence
\begin{align*}
    \frac{|\mathfrak R_{\mathtt F}(\bx)|}{\sqrt{\Omega_{\mathtt F,\bx,\bx}^{(\bnu)}}}
    =O_{\P}\!\left(
        \bh^{-\bnu}\sqrt{n|\bH|}
        ((n|\bH|)^{-1/2}+\bar h^{p+1})^2
      \right)=o_{\P}(1)
\end{align*}
under the theorem's bandwidth conditions. Combining these bounds with the pointwise Gram and covariance bounds displayed in the proof of Theorem~\ref{sa-thm: Confidence Intervals: Sharp BATEC} yields the displayed linearization. Here
\begin{align*}
    \overline{\Tstat}_{\mathtt{F}}^{(\bnu)}(\bx) = \sum_{i=1}^n \chi_i,
\end{align*}
with $\mathfrak{w}(\bx) = (\mathfrak{w}_1(\bx), \mathfrak{w}_2(\bx))$ and
\begin{align*}
    \chi_i
    &= n^{-1} (\Omega_{\mathtt{F}, \bx,\bx}^{(\bnu)})^{-1/2}
    \bnu!\be_{\bnu}^{\top} \bS^{-1}
    \Big[\Indicator(\bX_i\in\A_1)\bGamma_{1,\bx}^{-1}-\Indicator(\bX_i\in\A_0)\bGamma_{0,\bx}^{-1}\Big]
    \br_p(\bH^{-1}(\bX_i - \bx))\\
    &\hspace{1in} \cdot K_{\bH}(\bX_i - \bx)
    \big(\mathfrak{w}_1(\bx) \varepsilon_{Y,i} + \mathfrak{w}_2(\bx) \varepsilon_{W,i}\big).
\end{align*}
In particular, we have $\E[\chi_i] = 0$ and $\Var\big[\sum_{i=1}^n \chi_i\big] = 1$. Since $\mathfrak w(\bx)$ is fixed at the point $\bx$, $\varepsilon_{W,i}$ is bounded, and Assumption~\ref{sa-assump: Sharp DGP}(v) gives a uniformly bounded conditional third moment for $\varepsilon_{Y,i}$,
\begin{align*}
    \sum_{i=1}^n\E[|\chi_i|^3]\lesssim (n|\bH|\bh^{2\bnu})^{-1/2}=o(1).
\end{align*}
Berry--Esseen gives the normal approximation for $\overline{\Tstat}_{\mathtt F}^{(\bnu)}(\bx)$, and the stochastic linearization transfers it to $\widehat{\Tstat}_{\mathtt F}^{(\bnu)}(\bx)$. \qed

\subsection{Proof of Theorem~\ref{sa-thm: Stochastic Linearization: Fuzzy BATEC}}
Use the exact second-order expansion in Section~\ref{sec: Second-Order Linearization: Fuzzy BATEC}. The leading linear component is a fixed linear combination, with weights $\mathfrak{w}(\bx)$, of the reduced-form stochastic expansions for $Y$ and $W$. Applying Theorem~\ref{sa-thm: Stochastic Linearization: Sharp BATEC} to these two reduced-form estimators and then using Lemma~\ref{sa-lem: Covariance: Fuzzy BATEC} to replace $\Omega_{\mathtt{F},\bx,\bx}^{(\bnu)}$ by $\widehat{\Omega}_{\mathtt{F},\bx,\bx}^{(\bnu)}$ gives
\begin{align*}
    &\sup_{\bx \in \B}
    \left|\frac{\mathfrak{w}(\bx)^\top\mathfrak{T}(\bx)}
    {\sqrt{\widehat{\Omega}_{\mathtt{F},\bx,\bx}^{(\bnu)}}}
    - \overline{\Tstat}^{(\bnu)}_{\mathtt{F}}(\bx)\right| \\
    &\qquad \lessp
    \bar{h}^{p+1}\sqrt{n |\bH|}
    + \bh^{-\bnu}\sqrt{\log(1/\bar{h})}
      \left(\sqrt{\frac{\log(1/\bar{h})}{n|\bH|}}
      +\frac{\log(1/\bar{h})}{n^{\frac{v}{2+v}}|\bH|}
      +\bar{h}^{p+1}\right).
\end{align*}
Equation~\eqref{sa-eq: second order term -- fuzzy} and the lower bound
$\inf_{\bx\in\B}\Omega_{\mathtt{F},\bx,\bx}^{(\bnu)}\gtrsim(n|\bH|\bh^{2\bnu})^{-1}$ from $\inf_{\bx\in\B}V_{\mathtt{F},\bx}^{(\bnu)}>0$ imply that the normalized second-order remainder is bounded by the same rate. Combining the two displays proves the theorem.
\qed

\subsection{Proof of Theorem~\ref{sa-thm: Gaussian Approximation Suprema -- Fuzzy}}
Consider $\bz_i = (\bx_i, \varepsilon_{Y,i}, \varepsilon_{W,i})$, $1 \leq i \leq n$. For each $\bx \in \B$, define
\begin{align*}
    g_{\ttF, \bx}(\bu) = \Indicator(\bu \in \A_1)  \K^{(\bnu)}_{\ttF,1}(\bu;\bx) - \Indicator(\bu \in \A_0) \K^{(\bnu)}_{\ttF,0}(\bu;\bx),
    \qquad \bu \in \X,
\end{align*}
where
\begin{align*}
    \K^{(\bnu)}_{\ttF, t}(\bu;\bx) = n^{-1/2} (\Omega_{\ttF, \bx,\bx}^{(\bnu)})^{-1/2}\bnu!\be_{\bnu}^{\top} \bS^{-1} \bGamma_{t,\bx}^{-1} \br_p(\bH^{-1}(\bu - \bx)) K_{\bH}(\bu - \bx), \qquad \bu \in \X,\quad t \in \{0,1\}.
\end{align*}
Consider the class of functions
\begin{align*}
    \F = \{(\bu, \varepsilon_Y, \varepsilon_W) \in \reals^{d} \times \reals \times \reals \mapsto g_{\ttF, \bx}(\bu) \; (\mathfrak{w}_1(\bx)\varepsilon_{Y}+\mathfrak{w}_2(\bx)\varepsilon_{W}): \bx \in \B\},
\end{align*}
and let $\F_{\pm}=\F\cup(-\F)$. Also define
\begin{gather*}
    \G_{\ttF} = \{g_{\ttF, \bx}: \bx \in \B \}, \qquad
    \R_{\ttF} = \{(\varepsilon_Y, \varepsilon_W) \mapsto \mathfrak{w}_1(\bx)\varepsilon_{Y}+\mathfrak{w}_2(\bx)\varepsilon_{W}: \bx \in \B\}.
\end{gather*}
Then we have
\begin{align*}
    \sup_{\bx \in \B} |\overline{\Tstat}^{(\bnu)}_{\mathtt{F}}(\bx)| = \sup_{f \in \F_{\pm}} \frac{1}{\sqrt{n}} \sum_{i = 1}^n \Big\{f(\bz_i) - \E[f(\bz_i)]\Big\}.
\end{align*}
We use Lemma~\ref{sa-lem: suprema coupling lemma} to approximate the distribution of the right hand side of the above equation. First, we verify the VC-type entropy and moment conditions. The fixed indicators $\Indicator(\bu\in\A_t)$ are absorbed into $\G_{\ttF}$ and do not increase $L_2(Q)$ covering numbers beyond a constant factor, because multiplication by a fixed indicator is a contraction in $L_2(Q)$. Therefore, using the same local-polynomial entropy calculation as in the proofs of Lemmas~\ref{sa-lem: Gram Matrix} and \ref{sa-lem: Linear Approximation}, but without the total-variation step, we obtain that: (i) $\G_{\ttF}$ is pointwise measurable; (ii) there exists a constant envelope function $\ttM_{\G_{\ttF}} \lesssim |\bH|^{-1/2}$ such that for all $0 < \varepsilon < 1$,
\begin{align*}
    \sup_{Q} N(\G_{\ttF}, \norm{\cdot}_{Q,2}, \varepsilon (2 \bc +1)^{d+1} \ttM_{\G_{\ttF}}) \leq 2 \bc' \varepsilon^{-4d-4} + 2,
\end{align*}
where the supremum is taken over all finite discrete measures and some constants $\bc$ and $\bc^{\prime}$ only depending on the compact set $\X$. Moreover, Assumption~\ref{sa-assump: Sharp DGP} and \ref{sa-assump: Fuzzy DGP} imply that $\R_{\ttF}$ is also pointwise measurable and for an envelope function $M_{\R_{\ttF}}(e_1, e_2) = |e_1| + |e_2|$, $(e_1, e_2) \in \reals^2$, we have for all $0 < \varepsilon < 1$,
\begin{align*}
    \sup_{Q} N(\R_{\ttF}, \norm{\cdot}_{Q,2}, \varepsilon \bc M_{\R_{\ttF}}) \leq \bc' \varepsilon^{-2d},
\end{align*}
where the supremum is taken over all finite discrete measures. The last entropy bound follows because Assumption~\ref{sa-assump: Sharp DGP}(iii), Assumption~\ref{sa-assump: Fuzzy DGP}(ii), and $|\bnu|\leq p$ imply that $\tau_Y^{(\bnu)}(\bx)$ and $\tau_W^{(\bnu)}(\bx)$ are continuously differentiable on a neighborhood of $\B$; combined with $\inf_{\bx\in\B}|\tau_W^{(\bnu)}(\bx)|>0$, the map $\bx\mapsto\mathfrak w(\bx)$ is bounded and Lipschitz on the compact set $\B$. Thus the coefficient set $\{\mathfrak w(\bx):\bx\in\B\}\subset\reals^2$ has polynomial covering numbers inherited from $\B$. Since $\F\subseteq \G_{\ttF}\cdot\R_{\ttF}$, the product-covering inequality gives, for a constant $c>0$,
\begin{align*}
    N(\F, \norm{\cdot}_{Q,2}, \varepsilon c \ttM_{\G_{\ttF}}M_{\R_{\ttF}})
    &\leq
    N(\G_{\ttF}, \norm{\cdot}_{Q,2}, c\varepsilon\ttM_{\G_{\ttF}})
    N(\R_{\ttF}, \norm{\cdot}_{Q,2}, c\varepsilon M_{\R_{\ttF}}).
\end{align*}
Hence $\F_{\pm}$ is also pointwise measurable and satisfies the same VC-type bound as $\F$ up to a universal constant: for all $0 < \varepsilon < 1$,
\begin{align*}
    \sup_{Q} N(\F_{\pm}, \norm{\cdot}_{Q,2}, \varepsilon \bc \; \ttM_{\G_{\ttF}} M_{\R_{\ttF}}) \leq 64 \bc'^2 \varepsilon^{-6d-4} + 4 \bc' \varepsilon^{-2d},
\end{align*}
where the supremum is taken over all finite discrete measures. The denominator condition $\inf_{\bx \in \B}|\tau_W^{(\bnu)}(\bx)|>0$, continuity, and compactness of $\B$ imply that the population Wald weights $\mathfrak{w}(\bx)$ are uniformly bounded. For the moment condition, take $q=2+v\geq4$ in Lemma~\ref{sa-lem: suprema coupling lemma}. Since $\varepsilon_{W,i}=W_i-\E[W_i|\bX_i]$ is bounded and Assumption~\ref{sa-assump: Sharp DGP} gives $\sup_{\bx \in \X}\E[|Y_i(t)|^{2+v}|\bX_i = \bx] < \infty$, the envelope
\[
    F(\bx, \varepsilon_Y, \varepsilon_W) = \ttM_{\G_{\ttF}} M_{\R_{\ttF}}(\varepsilon_Y, \varepsilon_W)
\]
satisfies $\|F\|_{\P,2+v}\lesssim |\bH|^{-1/2}$. Together with the local-polynomial moment calculation used in the proof of Lemma~\ref{sa-lem: Linear Approximation}, for
\begin{align*}
    \sigma = 1, \qquad b = |\bH|^{-1/2},
\end{align*}
we have $\sup_{f \in \F_{\pm}} \P|f|^k \le \sigma^2 b^{k-2}$ for $k = 2, 3, 4$ and $\|F\|_{\P,2+v} \le b$.
Taking $\gamma = (\log n)^{-1}$ in Lemma~\ref{sa-lem: suprema coupling lemma},
\begin{align*}
    \P\left\{ |\sup_{\bx \in \B} |\overline{\Tstat}^{(\bnu)}_{\mathtt{F}}(\bx)| - \sup_{\bx \in \B} |Z^{(\bnu)}_{\mathtt{F}}(\bx)|| > \frac{\log n}{(n |\bH|)^{1/6}} + \frac{(\log n)^{5/4}}{(n |\bH|)^{1/4}} + \frac{(\log n)^{3/2}}{(n^{\frac{v}{2+v}} |\bH|)^{1/2}}\right\} \le C \left( \frac{1}{\log n} + \frac{\log n}{n} \right).
\end{align*}
\qed

\begin{lem}[KS Distance Between $Z_{\mathtt{F}}^{(\bnu)}$ and $\widehat{Z}_{\mathtt{F}}^{(\bnu)}$]\label{sa-lem: feasible fuzzy gaussian to infeasible fuzzy gaussian}
Suppose the conditions for Theorem~\ref{sa-thm: Confidence Bands: Fuzzy BATEC -- suprema} hold. Then, for any multi-index $|\bnu|\leq p$,
\begin{align*}
    \sup_{u \in \reals}
    \left|
    \P \left( \sup_{\bx \in \B} \big| \widehat{Z}_{\mathtt{F}}^{(\bnu)}(\bx) \big| \leq u \;\middle|\; \bV \right)
    - \P \left( \sup_{\bx \in \B} \big| Z_{\mathtt{F}}^{(\bnu)}(\bx) \big| \leq u \right)
    \right|
    = o_{\P}(1).
\end{align*}
\end{lem}

\begin{proof}
Let
\begin{align*}
    \mathfrak{R}_{\mathtt{F,cc}}
    =
    \bh^{-\bnu}
    \left(
    \sqrt{\frac{\log n}{n|\bH|}}
    +\frac{\log n}{n^{v/(2+v)}|\bH|}
    +\bar h^{p+1}
    \right).
\end{align*}
For $\bx,\by\in\B$, define
\[
    \bPi_{\mathtt{F},\bx,\by}
    =\frac{\Omega_{\mathtt{F},\bx,\by}^{(\bnu)}}
    {\sqrt{\Omega_{\mathtt{F},\bx,\bx}^{(\bnu)}\Omega_{\mathtt{F},\by,\by}^{(\bnu)}}},
    \qquad
    \widehat{\bPi}_{\mathtt{F},\bx,\by}
    =\frac{\widehat{\Omega}_{\mathtt{F},\bx,\by}^{(\bnu)}}
    {\sqrt{\widehat{\Omega}_{\mathtt{F},\bx,\bx}^{(\bnu)}\widehat{\Omega}_{\mathtt{F},\by,\by}^{(\bnu)}}}.
\]
Lemma~\ref{sa-lem: Covariance: Fuzzy BATEC}, the lower bound
$\inf_{\bx\in\B}V_{\mathtt{F},\bx}^{(\bnu)}>0$, and the mean-value expansion of $x\mapsto x^{-1/2}$ give
\begin{align*}
    \sup_{\bx,\by\in\B}
    \big|\widehat{\bPi}_{\mathtt{F},\bx,\by}
    -\bPi_{\mathtt{F},\bx,\by}\big|
    \lessp \mathfrak{R}_{\mathtt{F,cc}}.
\end{align*}
The Gaussian processes $Z_{\mathtt{F}}^{(\bnu)}$ and $\widehat Z_{\mathtt{F}}^{(\bnu)}$ have the same boundary-indexed local-polynomial structure as the sharp Gaussian processes, with the scalar residual replaced by the linear combination $\mathfrak w(\bx)^\top\boldsymbol{\varepsilon}_i$, whose coefficients are uniformly bounded by the denominator nondegeneracy condition. The entropy and modulus calculations in Lemma~\ref{sa-lem: feasible gaussian to infeasible gaussian} therefore apply with the sharp covariance-comparison rate $\mathfrak{R}_{\mathtt{cc}}$ used there replaced by the present fuzzy rate $\mathfrak{R}_{\mathtt{F,cc}}$. In particular, on a $\delta_n=n^{-s}$ net with $s$ large enough,
\begin{align*}
    \sup_{u \in \reals}
    \left|
    \P \left( \sup_{\bx \in \B} \big| \widehat{Z}_{\mathtt{F}}^{(\bnu)}(\bx) \big| \leq u \;\middle|\; \bV \right)
    - \P \left( \sup_{\bx \in \B} \big| Z_{\mathtt{F}}^{(\bnu)}(\bx) \big| \leq u \right)
    \right|
    \lessp \log(n)\mathfrak{R}_{\mathtt{F,cc}}^{1/2}.
\end{align*}
The second displayed rate condition in Theorem~\ref{sa-thm: Confidence Bands: Fuzzy BATEC -- suprema} gives $\log^2(n)\mathfrak{R}_{\mathtt{F,cc}}=o(1)$, and hence the right-hand side is $o_{\P}(1)$.
\end{proof}

\subsection{Proof of Theorem~\ref{sa-thm: Confidence Bands: Fuzzy BATEC -- suprema}}

Define the convergence rates
\begin{equation*}
    \mathfrak{R}_{\mathtt{sup}}
    =
    \frac{\log n}{(n |\bH|)^{1/6}}
    + \frac{(\log n)^{5/4}}{(n |\bH|)^{1/4}}
    + \frac{(\log n)^{3/2}}{(n^{v/(2+v)} |\bH|)^{1/2}},
\end{equation*}
and
\[
    \mathfrak{R}_{\mathtt{F,lin}}
    =
    \bar{h}^{p+1}\sqrt{n |\bH|}
    + \bh^{-\bnu}\sqrt{\log(1/\bar{h})}
      \left(\sqrt{\frac{\log(1/\bar{h})}{n|\bH|}}
      +\frac{\log(1/\bar{h})}{n^{\frac{v}{2+v}}|\bH|}
      +\bar{h}^{p+1}\right).
\]
First, we establish the coupling between the intermediate processes. Let
\[
    M_{\overline T,\mathtt{F}}=\sup_{\bx\in\B}|\overline{\Tstat}_{\mathtt{F}}^{(\bnu)}(\bx)|,
    \qquad
    M_{Z,\mathtt{F}}=\sup_{\bx\in\B}|Z_{\mathtt{F}}^{(\bnu)}(\bx)|.
\]
Applying Lemma~\ref{sa-lem: generic suprema comparison} with $\mathbb{T}_1=M_{\overline T,\mathtt{F}}$ and $\mathbb{T}_2=M_{Z,\mathtt{F}}$, viewed as singleton-indexed processes, and $\eta=\mathfrak{R}_{\mathtt{sup}}$, gives
\[
    \sup_{u\in\reals}\left|\P(M_{\overline T,\mathtt{F}}\le u)-\P(M_{Z,\mathtt{F}}\le u)\right|
    \le
    \sup_{u\in\reals}\P(|M_{Z,\mathtt{F}}-u|\le \mathfrak{R}_{\mathtt{sup}})
    +\P(|M_{\overline T,\mathtt{F}}-M_{Z,\mathtt{F}}|>\mathfrak{R}_{\mathtt{sup}}).
\]
The first term is $O(\mathfrak{R}_{\mathtt{sup}}\sqrt{\log n})=o(1)$ by Gaussian anti-concentration and the entropy calculation in the proof of Theorem~\ref{sa-thm: Gaussian Approximation Suprema -- Fuzzy}. The second term is $o(1)$ by Theorem~\ref{sa-thm: Gaussian Approximation Suprema -- Fuzzy}. Therefore,
\begin{equation}\label{sa-eq: t-to-z fuzzy}
    \sup_{u \in \reals} \left| \P \left( \sup_{\bx \in \B} \big| \overline{\Tstat}_{\mathtt{F}}^{(\bnu)}(\bx) \big| \leq u \right) - \P \left( \sup_{\bx \in \B} \big| Z_{\mathtt{F}}^{(\bnu)}(\bx) \big| \leq u \right) \right| = o(1).
\end{equation}

Furthermore, Theorem~\ref{sa-thm: Stochastic Linearization: Fuzzy BATEC} gives
\[
    \sup_{\bx\in\B}
    |\widehat{\Tstat}_{\mathtt{F}}^{(\bnu)}(\bx)
    -\overline{\Tstat}_{\mathtt{F}}^{(\bnu)}(\bx)|
    =
    O_{\P}(\mathfrak{R}_{\mathtt{F,lin}}).
\]
The side conditions of the theorem imply $\mathfrak{R}_{\mathtt{F,lin}}\sqrt{\log n}=o(1)$. Choose a deterministic sequence $a_n$ such that $\mathfrak{R}_{\mathtt{F,lin}}=o(a_n)$ and $a_n\sqrt{\log n}=o(1)$. Applying Lemma~\ref{sa-lem: generic suprema comparison} with $\mathbb{T}_1=\widehat{\Tstat}_{\mathtt{F}}^{(\bnu)}$, $\mathbb{T}_2=\overline{\Tstat}_{\mathtt{F}}^{(\bnu)}$, and $\eta=a_n$, and then bounding the anti-concentration of $M_{\overline T,\mathtt{F}}$ through Equation~\eqref{sa-eq: t-to-z fuzzy}, gives
\begin{equation}\label{sa-eq: t-to-t fuzzy}
    \sup_{u \in \reals} \left| \P \left( \sup_{\bx \in \B} \big| \widehat{\Tstat}_{\mathtt{F}}^{(\bnu)}(\bx) \big| \leq u \right) - \P \left( \sup_{\bx \in \B} \big| \overline{\Tstat}_{\mathtt{F}}^{(\bnu)}(\bx) \big| \leq u \right) \right| = o(1),
\end{equation}
and Lemma~\ref{sa-lem: feasible fuzzy gaussian to infeasible fuzzy gaussian} gives
\begin{equation}\label{sa-eq: z-to-z fuzzy}
    \sup_{u \in \reals} \left| \P \left( \sup_{\bx \in \B} \big| \widehat{Z}_{\mathtt{F}}^{(\bnu)}(\bx) \big| \leq u \;\middle|\; \bV \right) - \P \left( \sup_{\bx \in \B} \big| Z_{\mathtt{F}}^{(\bnu)}(\bx) \big| \leq u \right) \right| = o_{\P}(1).
\end{equation}
The distributional approximation follows from the triangle inequality applied to Equations~\eqref{sa-eq: t-to-z fuzzy}, \eqref{sa-eq: t-to-t fuzzy}, and \eqref{sa-eq: z-to-z fuzzy}. For coverage of the feasible confidence band, let
\[
    E_{\mathtt{F},\alpha}
    =
    \left\{
    \zeta^{(\bnu)}(\bx)\in\widehat{\CI}_{\mathtt{F},\alpha}^{(\bnu)}(\bx)
    \text{ for all }\bx\in\B
    \right\}.
\]
By definition of the band, $E_{\mathtt{F},\alpha}=\{\sup_{\bx\in\B}|\widehat{\Tstat}_{\mathtt{F}}^{(\bnu)}(\bx)|\leq \q_\alpha\}$. Hence the preceding approximation and the conditional definition of $\q_\alpha$ give
\begin{align*}
    \P(E_{\mathtt{F},\alpha})
    &= \P\left(\sup_{\bx\in\B}|\widehat{\Tstat}_{\mathtt{F}}^{(\bnu)}(\bx)|\leq \q_\alpha\right)\\
    &= \E\left[
        \P\left(\sup_{\bx\in\B}|\widehat Z_{\mathtt{F}}^{(\bnu)}(\bx)|\leq \q_\alpha\;\middle|\;\bV\right)
       \right]+o(1)\\
    &= 1-\alpha+o(1),
\end{align*}
using continuity of the supremum distribution of the feasible Gaussian process.
\qed

\subsection{Proof of Lemma~\ref{sa-lem: Fuzzy Integral: Bias}}

A similar argument to the proof of Lemma~\ref{sa-lem: bias} shows that under the assumptions imposed,
\begin{align*}
    \sup_{\bx \in \B} \big|\E[\mathfrak{w}(\bx)^\top\mathfrak{T}(\bx)|\bX] - B_{\mathtt{F},\bx}^{(\bnu)}(\bh) \big| = o_{\P}(\bar{h}^{p+1}\bh^{-\bnu}).
\end{align*}
Because $\mathfrak{w}(\cdot)$ is bounded under the denominator nondegeneracy condition and the reduced-form bias expansion is uniform in $\bx$, the displayed remainder is uniform over $\B$. Integrating against $w(\bx)\dif\Haus^{d-1}(\bx)$ therefore preserves the order by $\int_{\B} |w(\bx)| \dif\Haus^{d-1}(\bx)<\infty$, and the integral of $B_{\mathtt{F},\bx}^{(\bnu)}(\bh)$ is $B_{\mathtt{F,WBATE}}^{(\bnu)}(\bh)$ by definition. Under the common-bandwidth restriction $\bh=\bc h$, the same integration of the pointwise leading term gives the stated $h^{p+1-|\bnu|}\widetilde B_{\mathtt{F,WBATE}}^{(\bnu)}(\bc)$ form.
\qed

\subsection{Proof of Lemma~\ref{sa-lem: Fuzzy Integral: Variance}}

The proof is the fuzzy analog of Lemma~\ref{sa-lem: Variance: Sharp WBATE}. By Fubini's theorem,
\begin{align*}
    \Var[\check{\zeta}_{\mathtt{WBATE}}^{(\bnu)}|\bX]
    = \int_{\B}\int_{\B}
      \Cov[\mathfrak{w}(\bb_1)^\top\mathfrak{T}(\bb_1),\mathfrak{w}(\bb_2)^\top\mathfrak{T}(\bb_2)|\bX]
      w(\bb_1)w(\bb_2)\dif\Haus^{d-1}(\bb_1)\dif\Haus^{d-1}(\bb_2).
\end{align*}
Lemma~\ref{sa-lem: Covariance: Fuzzy BATEC} replaces this covariance uniformly by $\Omega_{\mathtt{F},\bb_1,\bb_2}^{(\bnu)}$ with an integrated error of order $o_{\P}(\bar{h}^{d-1}/(n|\bH|\bh^{2\bnu}))$. Compact support of $K$ implies that $\Omega_{\mathtt{F},\bb_1,\bb_2}^{(\bnu)}=0$ unless $\|\bb_1-\bb_2\|\lesssim \bar{h}$; by \eqref{sa-eq: local boundary measure regularity}, the effective integration set has product Hausdorff measure $O(\bar{h}^{d-1})$. The pointwise covariance is of order $(n|\bH|\bh^{2\bnu})^{-1}$, which gives the stated upper bound. The lower bound follows from the aggregate fuzzy variance nondegeneracy condition \eqref{sa-eq: fuzzy WBATE aggregate nondegeneracy}, which rules out cancellation after applying the fuzzy Wald weights and integrating over $\B$. Replacing $\Omega_{\mathtt{F},\bb_1,\bb_2}^{(\bnu)}$ by $\widehat{\Omega}_{\mathtt{F},\bb_1,\bb_2}^{(\bnu)}$ uses Lemma~\ref{sa-lem: Covariance: Fuzzy BATEC} once more.
\qed

\subsection{Proof of Theorem~\ref{sa-thm: MSE Expansion: Fuzzy WBATE}}

By Lemma~\ref{sa-lem: Fuzzy Integral: Bias},
\begin{align*}
    \E[\check{\zeta}_{\mathtt{WBATE}}^{(\bnu)}|\bX]
    = B_{\mathtt{F,WBATE}}^{(\bnu)}(\bh)+o_{\P}(\bar{h}^{p+1}\bh^{-\bnu}).
\end{align*}
Squaring the bias expansion gives $\big(B_{\mathtt{F,WBATE}}^{(\bnu)}(\bh)\big)^2+o_{\P}(\bar{h}^{2p+2}\bh^{-2\bnu})$. Lemma~\ref{sa-lem: Fuzzy Integral: Variance} gives the variance term $\Omega_{\mathtt{F,WBATE}}^{(\bnu)}$ with remainder $o_{\P}(\bar{h}^{d-1}/(n|\bH|\bh^{2\bnu}))$. Adding these two components gives the displayed AMSE expansion.
\qed

\subsection{Proof of Theorem~\ref{sa-thm: Asymptotic Normality: Fuzzy WBATE}}

The proof follows the same Lyapunov argument as Theorem~\ref{sa-thm: Asymptotic Normality: Sharp WBATE}, applied to the linearized fuzzy statistic obtained by integrating $\overline{\Tstat}^{(\bnu)}_{\mathtt{F}}(\bb)$ against $w(\bb)$. Lemma~\ref{sa-lem: Fuzzy Integral: Variance} gives the variance rate $\bar{h}^{d-1}/(n|\bH|\bh^{2\bnu})$ and consistency of its plug-in estimator. Lemma~\ref{sa-lem: Fuzzy Integral: Bias} and the side condition $n|\bH|\bar{h}^{2p+2}/\bar{h}^{d-1}=o(1)$ make the linearized bias negligible relative to the standard deviation. The final two displayed rate conditions imply the two stochastic pieces of the uniform reduced-form consistency condition, while $\bh^{-\bnu}\bar h^{p+1}=o(1)$ follows from $|\bnu|\le p$ and $\bar h/\underline h\lesssim1$. Moreover, \eqref{sa-eq: second order term -- fuzzy}, integrability of $w$, and the final displayed rate conditions in the theorem imply
\begin{align*}
    \frac{\left|\int_{\B} \mathfrak{R}_{\mathtt{F}}(\bb) w(\bb)\dif\Haus^{d-1}(\bb)\right|}
    {\left(\bar{h}^{d-1}/(n|\bH|\bh^{2\bnu})\right)^{1/2}}
    =o_{\P}(1).
\end{align*}
The Lyapunov condition follows from the same calculation as in the sharp WBATE proof, with the scalar residual replaced by the linearized fuzzy residual $\mathfrak{w}_1(\bb)\varepsilon_{Y,i}+\mathfrak{w}_2(\bb)\varepsilon_{W,i}$. In particular, if $\chi_i$ denotes the normalized integrated linearized fuzzy summand, then the boundedness of $\mathfrak{w}(\bb)$, the boundedness of $\varepsilon_{W,i}$, and Assumption~\ref{sa-assump: Sharp DGP}(v) imply
\begin{align*}
    \sum_{i=1}^n \E[|\chi_i|^3]
    \lesssim
    \left(\frac{\bar{h}^{d-1}}{n|\bH|}\right)^{1/2}
    = o(1).
\end{align*}
Therefore the feasible statistic $\widehat{\Tstat}_{\mathtt{F,WBATE}}^{(\bnu)}$ is asymptotically standard normal.
\qed

\subsection{Proof of Theorem~\ref{sa-thm: Convergence Rate: Fuzzy LBATE}}

Follows by Theorem~\ref{sa-thm: Convergence Rates: Fuzzy BATEC} after noting that
\begin{align*}
    |\widehat{\zeta}_{\mathtt{LBATE}}^{(\bnu)} - \zeta_{\mathtt{LBATE}}^{(\bnu)}|
    \leq \sup_{\bx\in\B} |\widehat{\zeta}^{(\bnu)}(\bx) - \zeta^{(\bnu)}(\bx)|.
\end{align*}
\qed

\subsection{Proof of Theorem~\ref{sa-thm: Confidence Intervals: Fuzzy LBATE}}

Consider the event $E_{\mathtt{F}} = \Big\{\sup_{\bb \in \B} \frac{|\widehat{\zeta}^{(\bnu)}(\bb) - \zeta^{(\bnu)}(\bb)|}{(\widehat{\Omega}_{\mathtt{F},\bb,\bb}^{(\bnu)})^{1/2}} \leq \q_{\alpha}\Big\}$. Theorem~\ref{sa-thm: Confidence Bands: Fuzzy BATEC -- suprema} implies that $\P(E_{\mathtt{F}}) = 1 - \alpha + o(1)$. On the event $E_{\mathtt{F}}$, we have
\begin{align*}
    \widehat{\zeta}^{(\bnu)}(\bb) - \q_{\alpha}(\widehat{\Omega}_{\mathtt{F},\bb,\bb}^{(\bnu)})^{1/2}
    \leq \zeta^{(\bnu)}(\bb)
    \leq \widehat{\zeta}^{(\bnu)}(\bb) + \q_{\alpha}(\widehat{\Omega}_{\mathtt{F},\bb,\bb}^{(\bnu)})^{1/2}, \qquad \forall \; \bb \in \B,
\end{align*}
which implies
\begin{align*}
    \sup_{\bb \in \B}  \widehat{\zeta}^{(\bnu)}(\bb) - \q_{\alpha}(\widehat{\Omega}_{\mathtt{F},\bb,\bb}^{(\bnu)})^{1/2}
    \leq \sup_{\bb \in \B} \zeta^{(\bnu)}(\bb)
    \leq \sup_{\bb \in \B} \widehat{\zeta}^{(\bnu)}(\bb) + \q_{\alpha}(\widehat{\Omega}_{\mathtt{F},\bb,\bb}^{(\bnu)})^{1/2}.
\end{align*}
The stated result then follows.
\qed

\section{Implementation Details}\label{sa-sec: Implementation Details}

This section provides additional details on the practical implementation of the pointwise and uniform inference methods discussed in Section 3.3 of the paper. The main issue for uniform inference is that the target process is indexed by the continuum $\bx\in\B$, whereas computation is necessarily carried out on a finite discretization of the boundary. The following steps describe the default implementation used by the companion software and recommended in applications.

\begin{enumerate}[leftmargin=*]

    \item \textit{Represent the boundary and choose the target set.}
    When $\B$ is piecewise linear, as in the empirical application, it is convenient to parameterize the boundary by arc length. Let $\gamma:[0,\mathfrak{L}(\B)]\to\B$ denote such a parameterization, where $\mathfrak{L}(\B)$ is the total boundary length. If the researcher wants inference only on a subset of the assignment boundary, for example after trimming endpoints, corners, or regions with weak local support, the same construction can be applied to the trimmed target set $\B_0\subseteq\B$. This distinction is important: uniform confidence bands cover the chosen target set, so any trimming changes the estimand from the full boundary to the trimmed boundary.

    \item \textit{Construct a quasi-uniform grid on the boundary.}
    Choose grid points $\bb_1,\ldots,\bb_J\in\B_0$ with approximately equal spacing in arc length. For a single connected component this can be done by setting $\bb_j=\gamma(s_j)$ with $s_j=(j-1)\mathfrak{L}(\B_0)/(J-1)$, $j=1,\ldots,J$. For a piecewise linear boundary with multiple segments, grid points can be allocated across segments in proportion to segment length, while forcing vertices or other substantively important boundary points to be included when desired. In higher-dimensional settings, the analogous requirement is a quasi-uniform mesh on the $(d-1)$-dimensional boundary with respect to Hausdorff measure.

    \item \textit{Choose the mesh size.}
    The grid should be fine enough that the discretized supremum is a good approximation to the continuous supremum. A useful practical rule is to make the maximum spacing between adjacent boundary grid points small relative to the bandwidth scale used for inference. Equivalently, the mesh should be refined until the simulated critical value and the visual shape of the confidence band are stable. In simple two-dimensional applications, values such as $J=40$ or $J=80$ often give similar results, but the appropriate choice depends on the length and geometry of $\B$, the bandwidth, and the degree of treatment effect heterogeneity. We recommend reporting a grid-sensitivity check, for example comparing $J$, $2J$, and, when computationally feasible, $4J$.

    \item \textit{Select bandwidths and polynomial orders.}
    The plug-in selectors described in Section 3.3 of the paper can be implemented pointwise, producing $\widehat{h}_{\mathtt{MSE},\bx}$, or globally, producing $\widehat{h}_{\mathtt{IMSE}}$. For uniform inference and graphical display along the boundary, a global bandwidth is often easier to interpret and tends to produce smoother covariance estimates. Pointwise bandwidths can also be used, provided the covariance matrix of the resulting process is computed using the same bandwidth choices. Robust bias-corrected inference is implemented by using the bandwidth selected for the $p$th-order point estimator and constructing the associated $t$-statistic with a $(p+1)$th-order local polynomial, including the corresponding variance adjustment. When the coordinates of the score have different scales, one may either standardize the score before using a scalar bandwidth or use a diagonal bandwidth matrix $\bH=\diag(h_1,\ldots,h_d)$ as allowed in this supplemental appendix.

    \item \textit{Compute point estimates and covariance estimates on the grid.}
    For each grid point $\bb_j$, compute the local-polynomial estimates, the robust bias-corrected $t$-statistics, and the variance estimates. For uniform inference, also compute the estimated cross-covariances
    \[
        \widehat{\Omega}_{\bb_j,\bb_k}^{(\bnu)},
        \qquad 1\leq j,k\leq J,
    \]
    using the formulas in Sections SA-3 and SA-6. The simulated Gaussian process is then represented by a mean-zero Gaussian vector with correlation matrix
    \[
        \widehat{\bPi}_{j,k}
        =
        \frac{\widehat{\Omega}_{\bb_j,\bb_k}^{(\bnu)}}
        {\sqrt{\widehat{\Omega}_{\bb_j,\bb_j}^{(\bnu)}
               \widehat{\Omega}_{\bb_k,\bb_k}^{(\bnu)}}},
        \qquad 1\leq j,k\leq J.
    \]
    In finite samples, $\widehat{\bPi}$ should be symmetrized before simulation.

    \item \textit{Regularize the covariance matrix if needed.}
    The estimated matrix $\widehat{\bPi}$ may fail to be positive semidefinite because of simulation error, numerical precision, or unstable local covariance estimates. This is a finite-sample numerical issue rather than a change in the target parameter. A simple regularization is to replace $\widehat{\bPi}$ by its nearest positive semidefinite approximation, or equivalently to truncate negative eigenvalues at a small nonnegative tolerance and rescale the diagonal entries to one. Another common approach is to use a small ridge adjustment, $\widehat{\bPi}_{\lambda}=(1-\lambda)\widehat{\bPi}+\lambda \mathbf{I}_J$, with the smallest $\lambda\geq0$ that makes the matrix positive semidefinite. The chosen adjustment should be reported when it is not negligible. See \citet{Cattaneo-Feng-Underwood_2024_JASA} for related discussion and theoretical results on covariance regularization.

    \item \textit{Simulate the critical value.}
    Given the regularized correlation matrix, draw $M$ independent vectors from $N(0,\widehat{\bPi})$ and compute the empirical $(1-\alpha)$ quantile of
    \[
        \max_{1\leq j\leq J}|Z_j|.
    \]
    This quantile is the simulated critical value for the uniform confidence band over the grid. In practice, $M$ should be large enough that simulation error is small relative to the desired reporting precision; increasing $M$ and checking that the critical value is stable is a simple diagnostic.

    \item \textit{Check local support and numerical stability.}
    Uniform inference can be sensitive to boundary regions where one side has little local support or the local design matrix is poorly conditioned. We recommend inspecting, for each grid point, effective local sample sizes or kernel masses on the two sides of the boundary, together with condition numbers of the estimated Gram matrices. Grid points with very small one-sided kernel mass, extremely unbalanced mass across sides, or unstable Gram matrices indicate that the local design is weak at those locations. In such cases, researchers should report the issue, consider sensitivity to bandwidth choices, and, when substantively justified, trim the problematic portion of the boundary and redefine the target set accordingly.

    \item \textit{Report sensitivity analyses.}
    A transparent empirical implementation should report the main estimates together with basic sensitivity checks: alternative grid sizes, alternative bandwidth choices, and whether covariance regularization was needed. For applications with corners or visibly irregular boundaries, it is also useful to report results with and without small neighborhoods of those regions. Researchers should also examine whether the density of the score changes abruptly near the boundary or whether observations appear to bunch on one side of $\B$. In BD designs this type of manipulation or sorting diagnostic is necessarily multivariate and boundary-local: useful checks include scatterplots near $\B$, side-specific kernel masses along the boundary grid, and sensitivity to excluding small neighborhoods where the empirical distribution of $\bX_i$ appears discontinuous or unusually concentrated. These checks help distinguish genuine treatment effect heterogeneity from artifacts of discretization, weak local support, numerical instability, or nonrandom sorting around the assignment boundary.

\end{enumerate}

The companion software package \texttt{rd2d} and article \citep{Cattaneo-Titiunik-Yu_2025_rd2d} discuss additional implementation details.

\section{Empirical Application}

This section reports numerical results for the SPP empirical application that are omitted from the main paper. Replication codes are available at \url{https://rdpackages.github.io/replication/}. The main paper uses the ITT bandwidth selector throughout; this section reports the corresponding full boundary-grid results and then gives the analogous results using the fuzzy bandwidth selector.

\subsection{ITT Bandwidth Selector}

Due to space limitations, the paper reports selected boundary points and aggregate summaries in Table 1. Here we provide the corresponding full boundary-grid results for all $40$ evaluation points.
\begin{itemize}
    \item Table~\ref{tab:sa-empapp_main_itt} reports the reduced-form, or intention-to-treat (ITT), estimates for college enrollment.
    \item Table~\ref{tab:sa-empapp_main_fs} reports the first-stage estimates for treatment take-up.
    \item Table~\ref{tab:sa-empapp_main_fuzzy} reports the fuzzy estimates for college enrollment.
    \item Table~\ref{tab:sa-empapp_covbal_itt} reports the covariate-balance reduced-form estimates using mother's education as the outcome.
\end{itemize}

\begin{figure}
    \centering
    \begin{subfigure}[b]{0.70\textwidth}
        \centering
        \includegraphics[width=\linewidth]{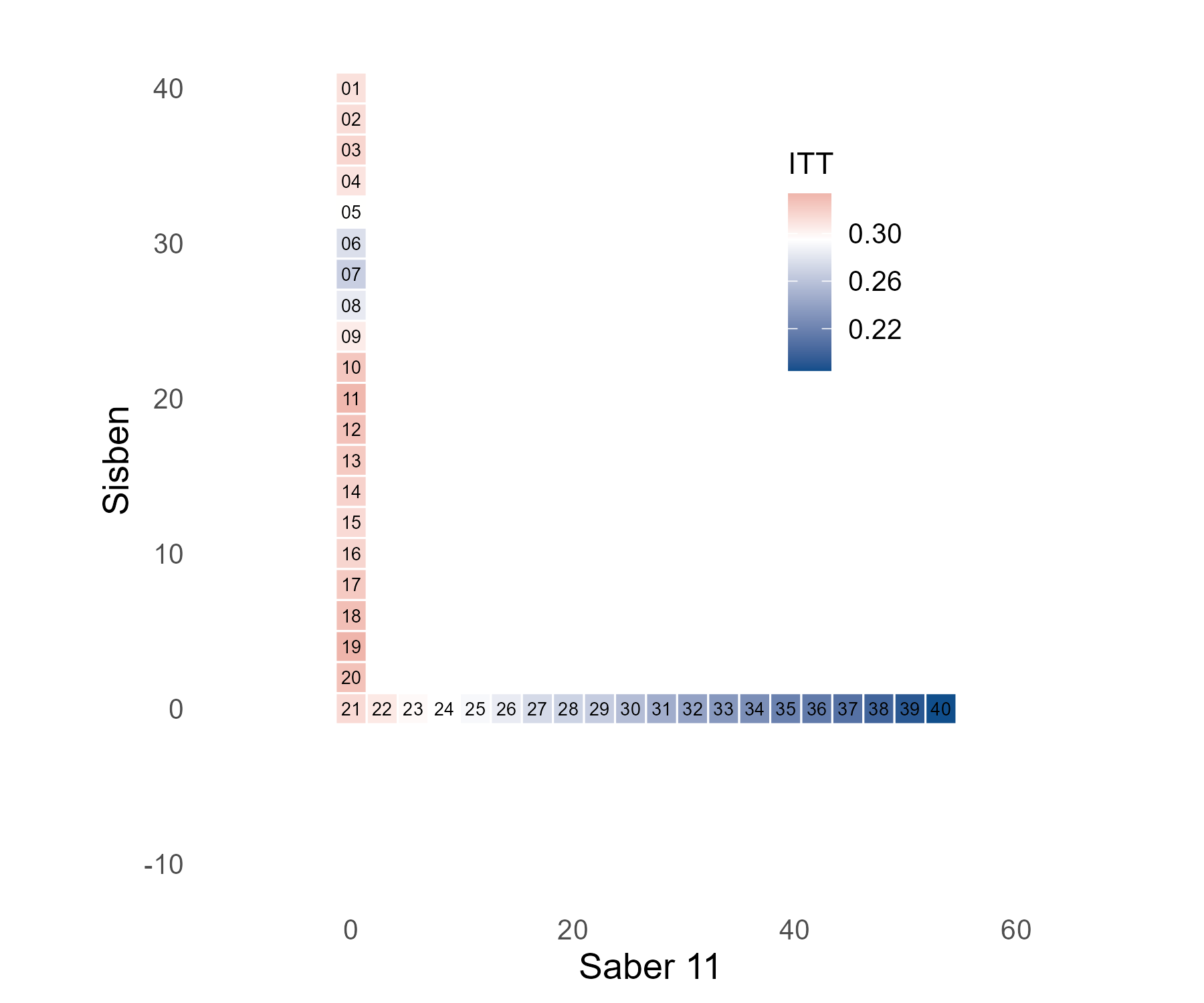}
        \caption{Point estimates.}
    \end{subfigure}

    \begin{subfigure}[b]{0.70\textwidth}
        \centering
        \includegraphics[width=\linewidth]{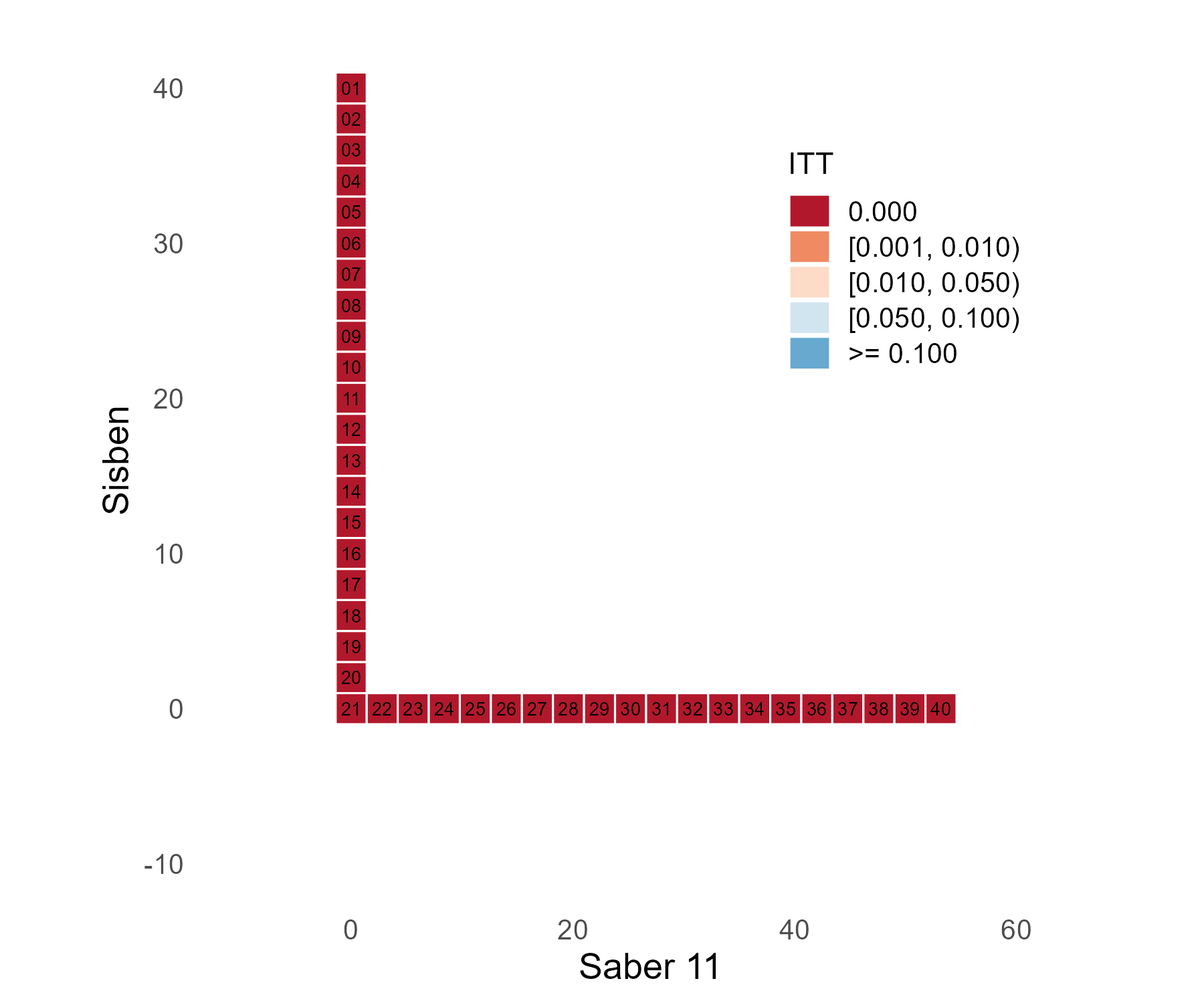}
        \caption{Pointwise $p$-values.}
    \end{subfigure}
    \caption{Intention-to-Treat Effects (SPP Application, ITT Bandwidth Selector).}
    \label{fig:sa-heatmaps-itt-main}
\end{figure}

\begin{figure}
    \centering
    \begin{subfigure}[b]{0.70\textwidth}
        \centering
        \includegraphics[width=\linewidth]{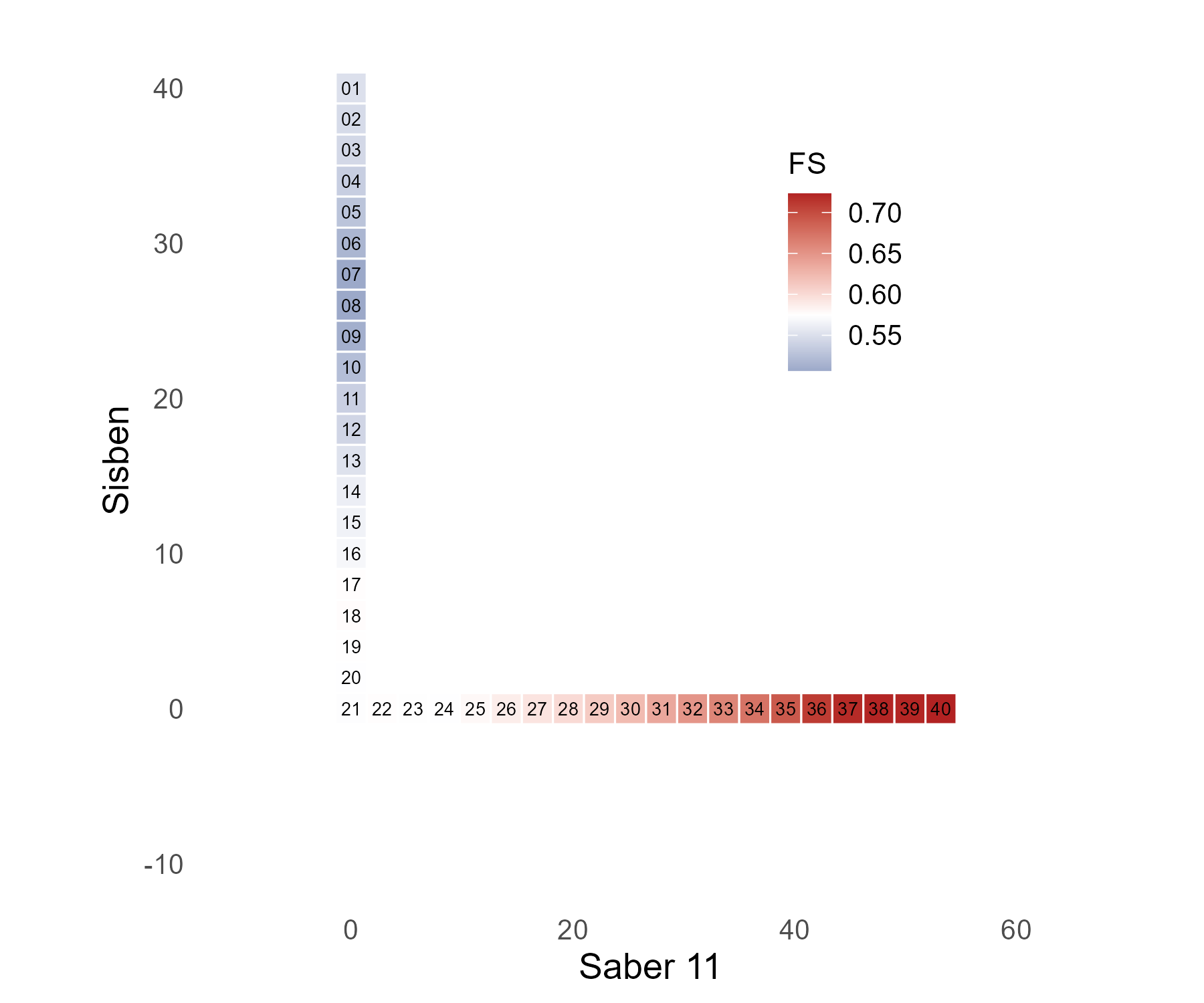}
        \caption{Point estimates.}
    \end{subfigure}

    \begin{subfigure}[b]{0.70\textwidth}
        \centering
        \includegraphics[width=\linewidth]{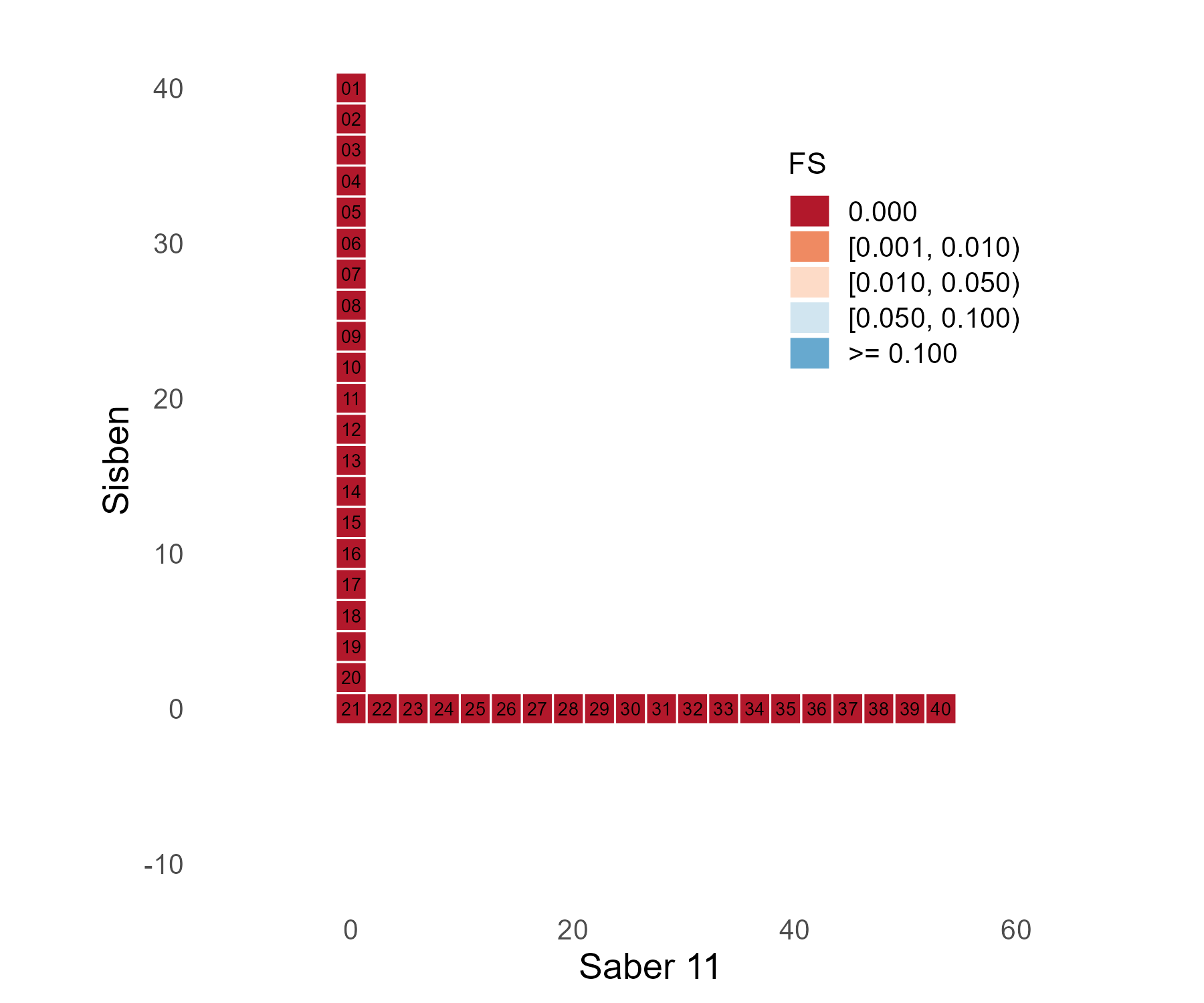}
        \caption{Pointwise $p$-values.}
    \end{subfigure}
    \caption{First-Stage Effects (SPP Application, ITT Bandwidth Selector).}
    \label{fig:sa-heatmaps-fs-main}
\end{figure}

\begin{figure}
    \centering
    \begin{subfigure}[b]{0.70\textwidth}
        \centering
        \includegraphics[width=\linewidth]{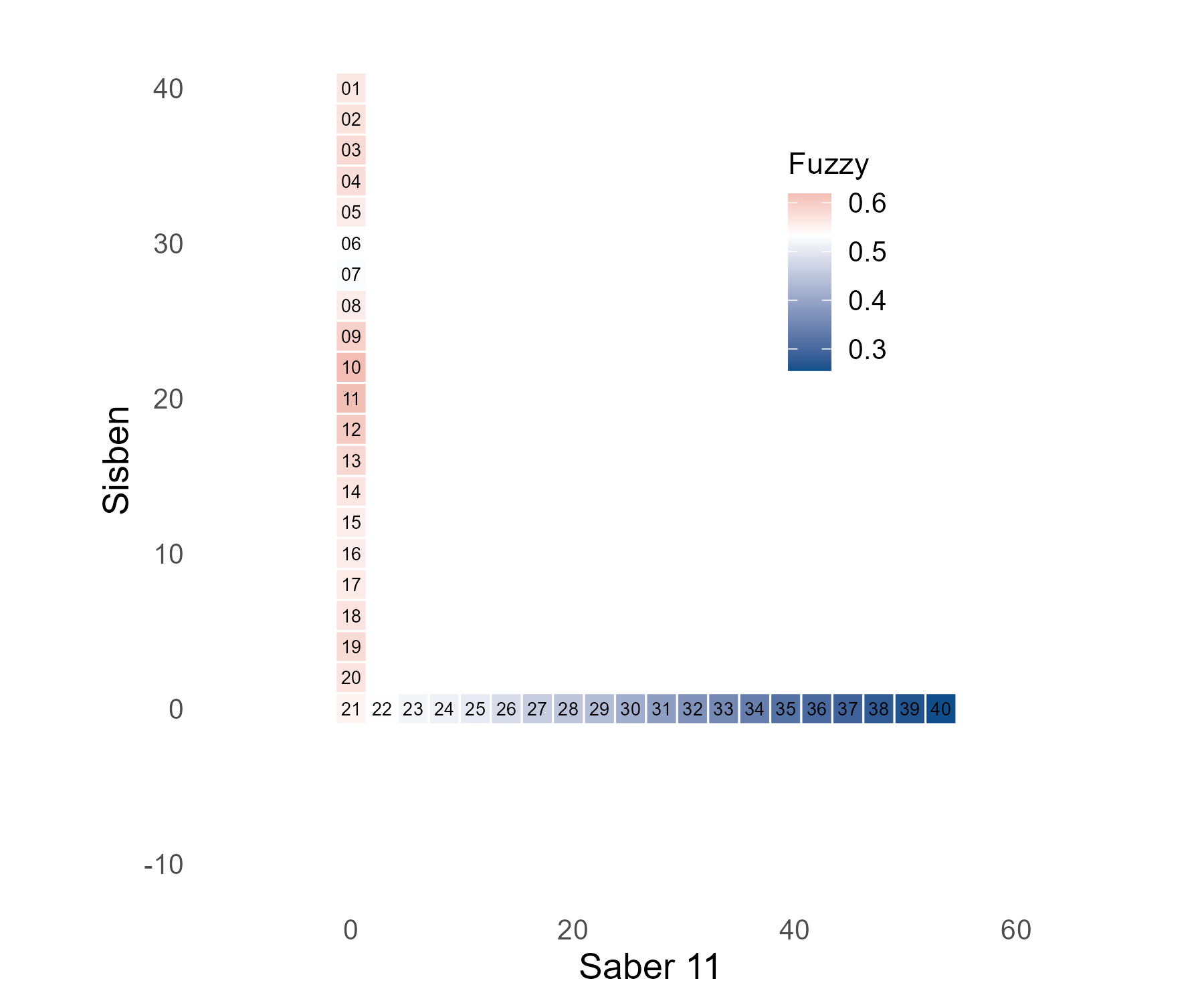}
        \caption{Point estimates.}
    \end{subfigure}

    \begin{subfigure}[b]{0.70\textwidth}
        \centering
        \includegraphics[width=\linewidth]{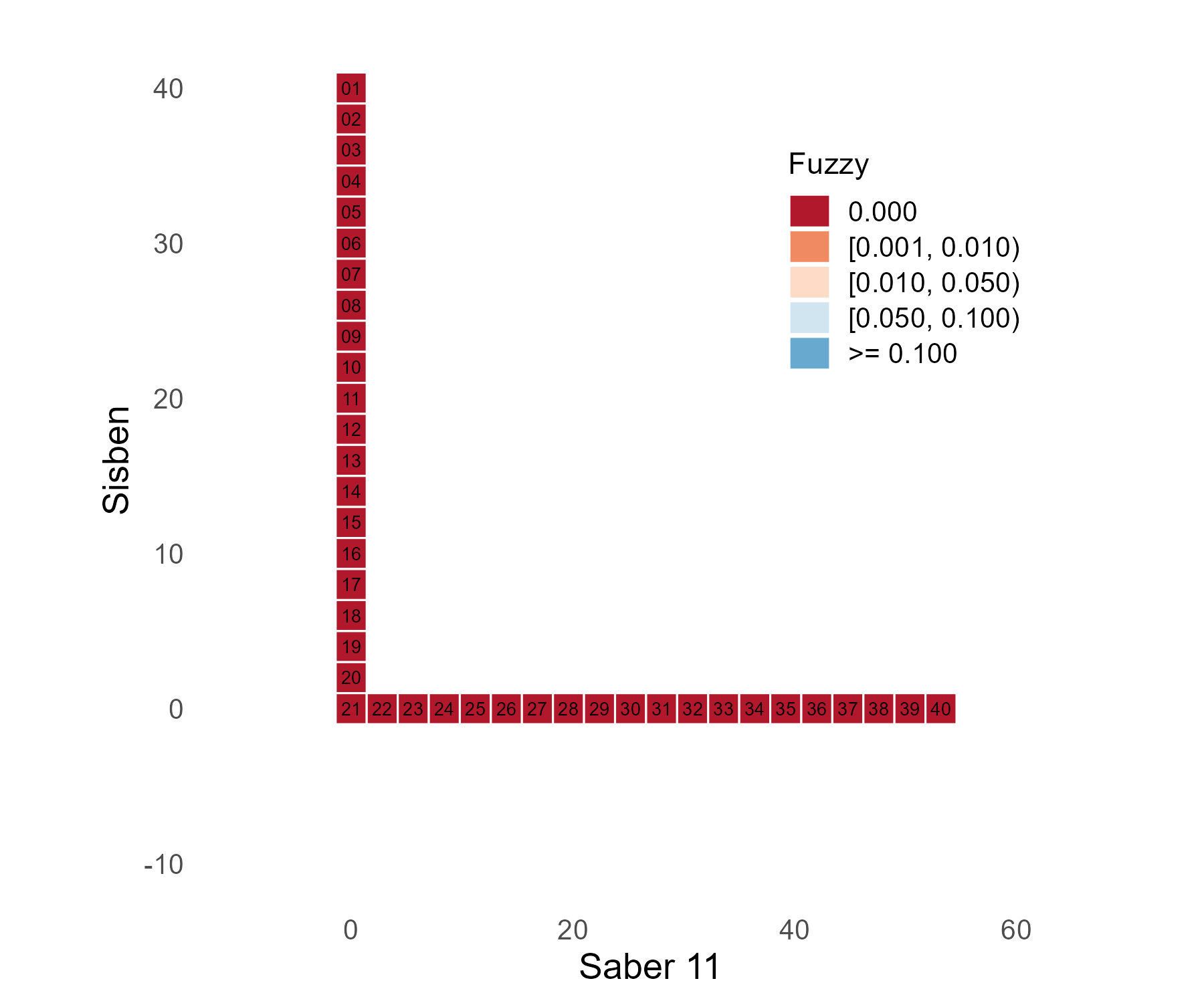}
        \caption{Pointwise $p$-values.}
    \end{subfigure}
    \caption{Fuzzy Effects (SPP Application, ITT Bandwidth Selector).}
    \label{fig:sa-heatmaps-fuzzy-main}
\end{figure}

\begin{figure}
    \centering
    \begin{subfigure}[b]{0.70\textwidth}
        \centering
        \includegraphics[width=\linewidth]{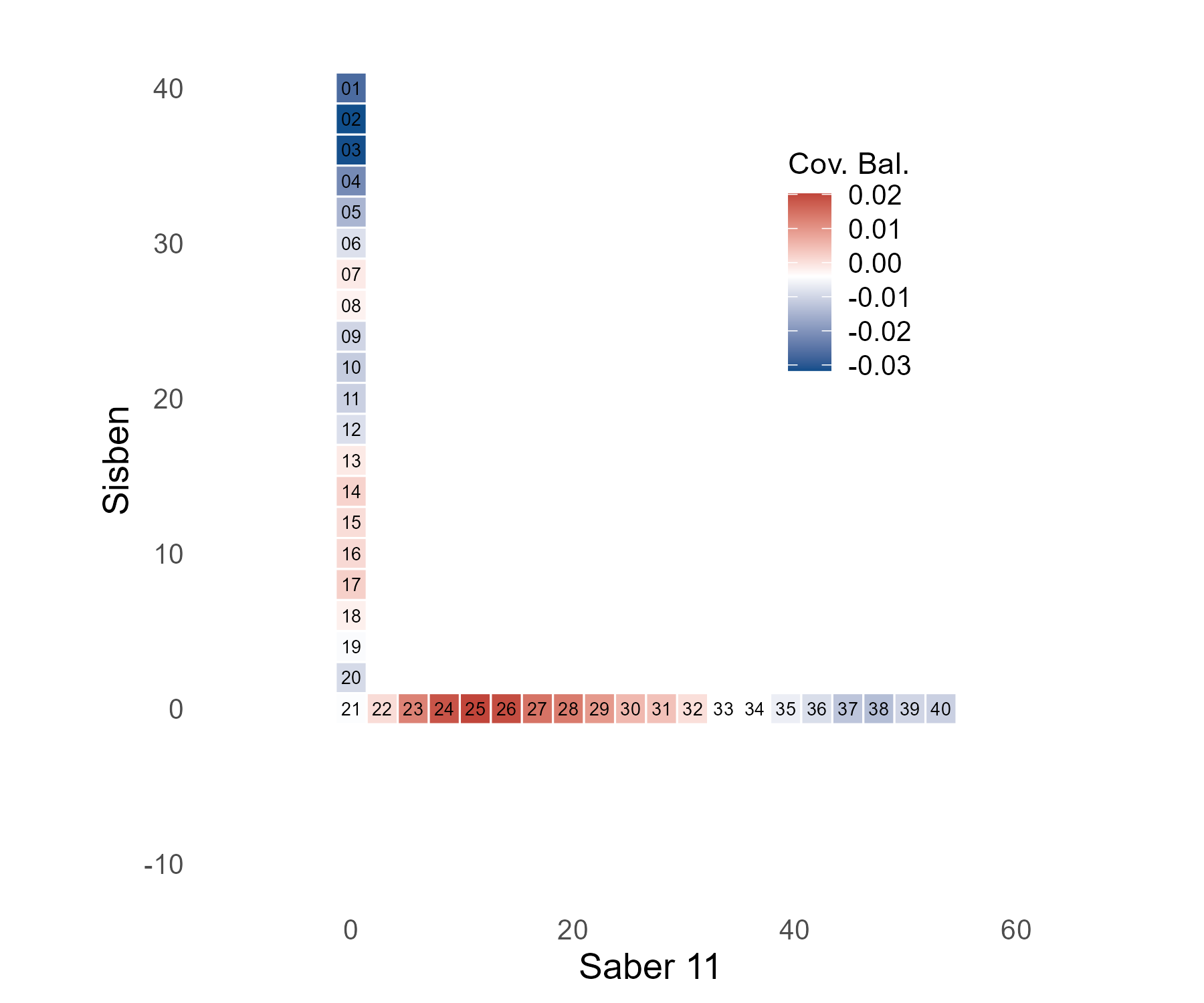}
        \caption{Point estimates.}
    \end{subfigure}

    \begin{subfigure}[b]{0.70\textwidth}
        \centering
        \includegraphics[width=\linewidth]{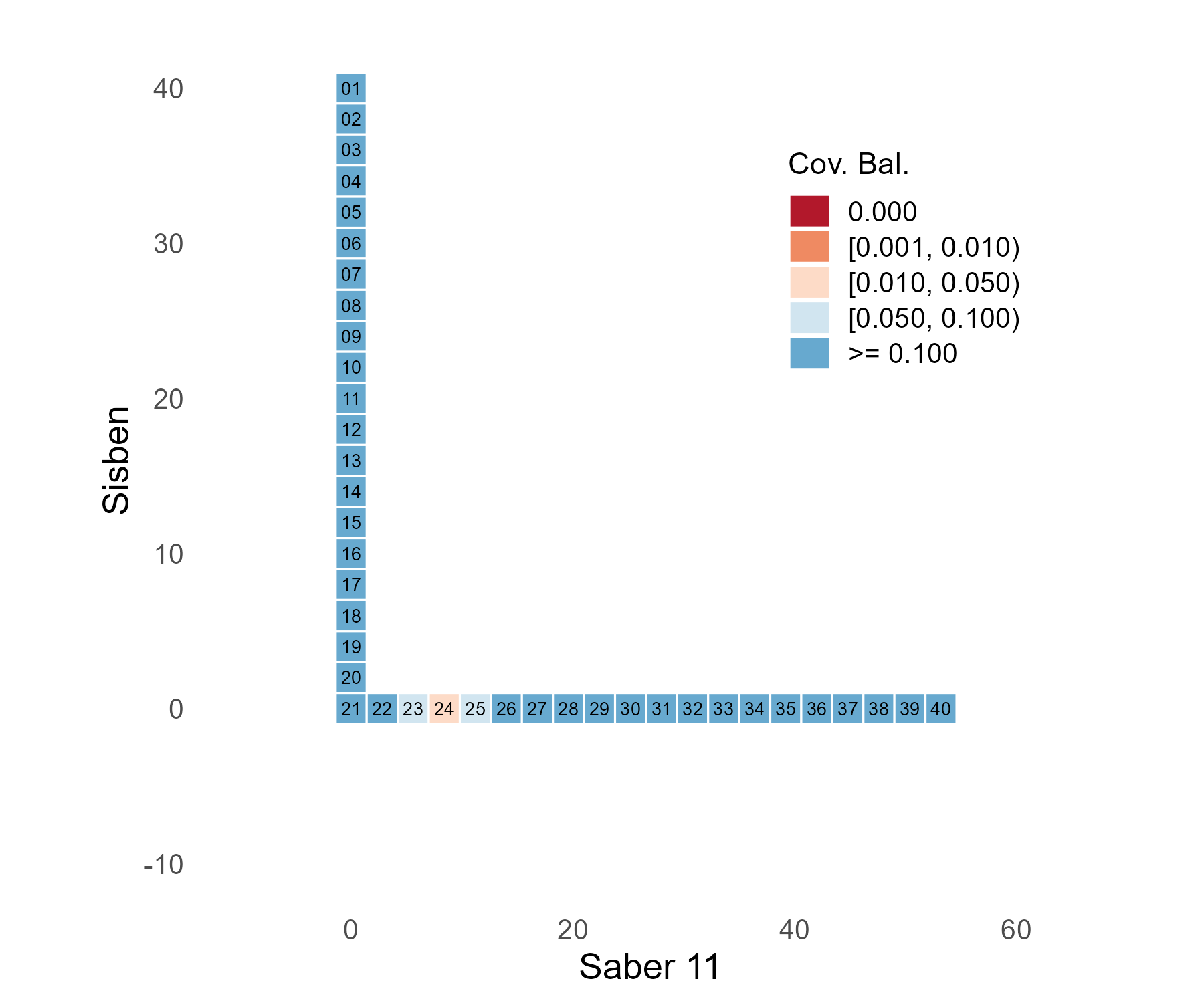}
        \caption{Pointwise $p$-values.}
    \end{subfigure}
    \caption{Covariate Balance for Mother's Education (SPP Application, ITT Bandwidth Selector).}
    \label{fig:sa-heatmaps-covbal}
\end{figure}

The figures in this subsection complement the confidence intervals and bands plots presented in the paper. For each quantity, we report heat maps along the boundary for the point estimates and associated pointwise robust bias-corrected $p$-values. Specifically:
\begin{itemize}
    \item Figure~\ref{fig:sa-heatmaps-itt-main} reports heatmaps for the reduced-form analysis of college enrollment.
    \item Figure~\ref{fig:sa-heatmaps-fs-main} reports heatmaps for the first-stage estimates for treatment take-up.
    \item Figure~\ref{fig:sa-heatmaps-fuzzy-main} reports heatmaps for the fuzzy analysis of college enrollment.
    \item Figure~\ref{fig:sa-heatmaps-covbal} reports heatmaps for the covariate-balance analysis using mother's education as the outcome.
\end{itemize}
These plots make it easier to see where along the boundary the estimated effects are largest and where the evidence against the zero-effect null is strongest.

\begin{figure}
    \centering
    \begin{subfigure}[b]{0.70\textwidth}
        \centering
        \includegraphics[width=\linewidth]{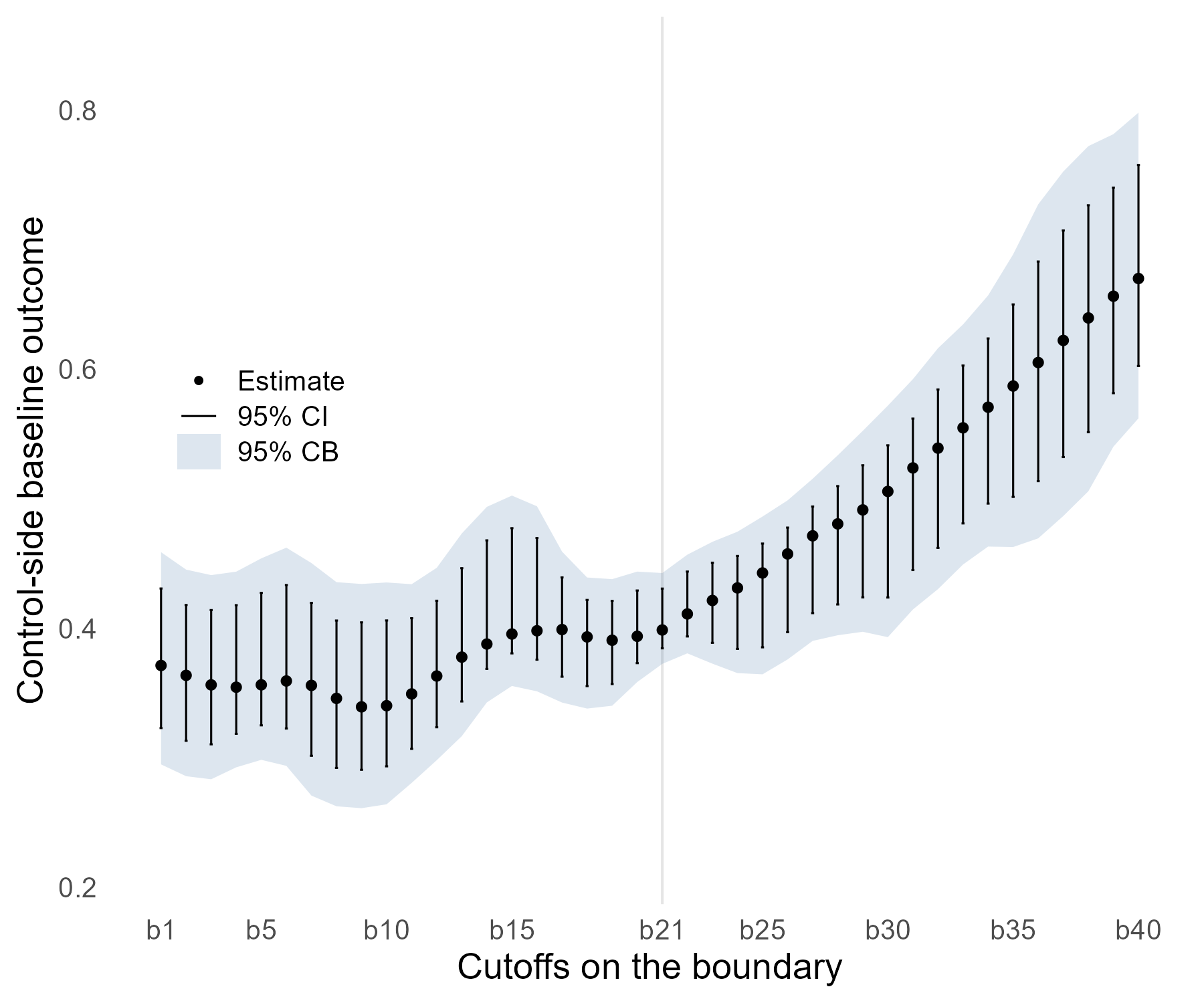}
        \caption{Point Estimates and Uncertainty Quantification.}
    \end{subfigure}

    \begin{subfigure}[b]{0.70\textwidth}
        \centering
        \includegraphics[width=\linewidth]{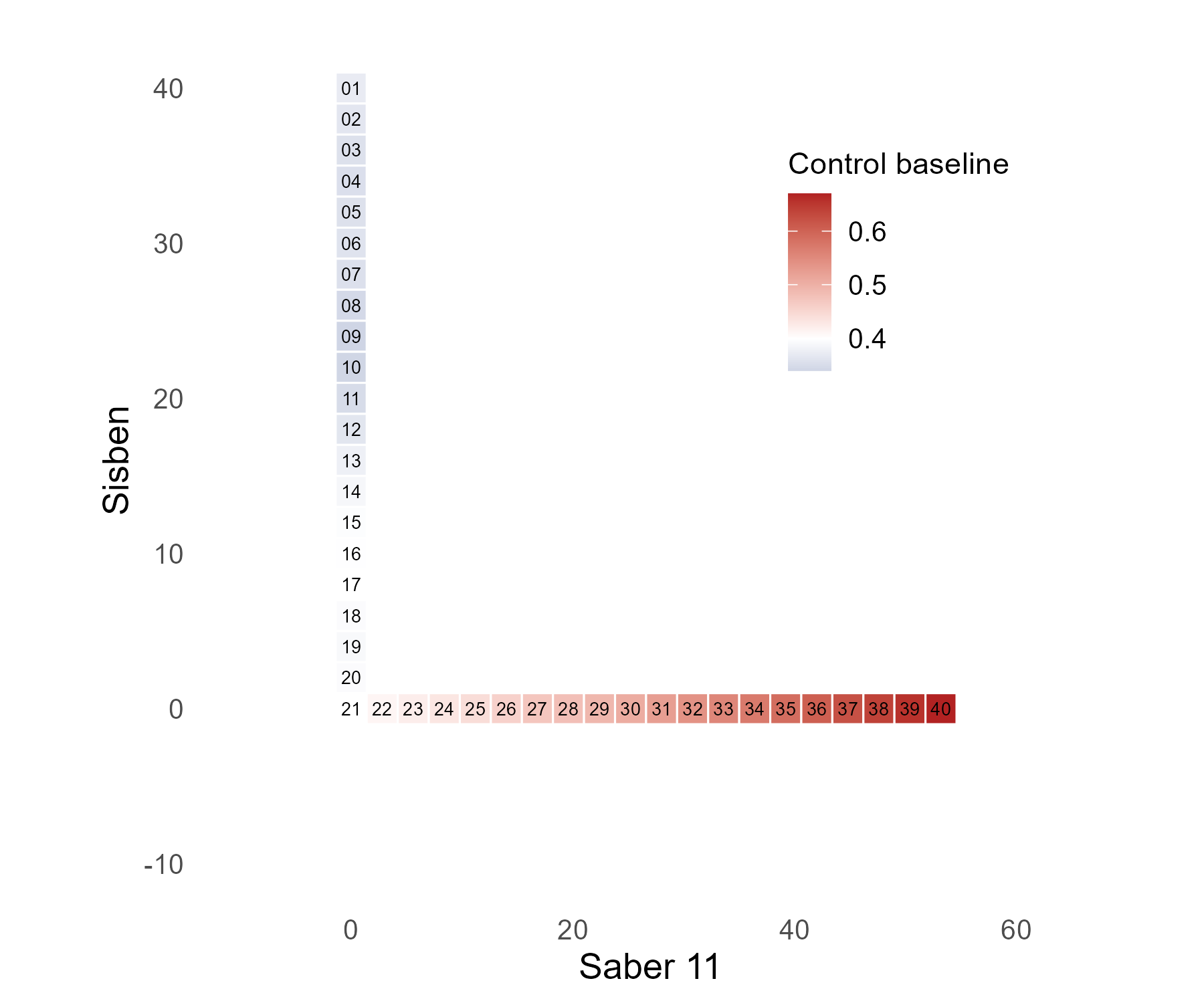}
        \caption{Heatmap of Point Estimates.}
    \end{subfigure}
    \caption{Control-Side Baseline Outcome (SPP Application, ITT Bandwidth Selector).}
    \label{fig:sa-empapp_main_itt0}
\end{figure}

\begin{figure}
    \centering
    \begin{subfigure}[b]{0.70\textwidth}
        \centering
        \includegraphics[width=\linewidth]{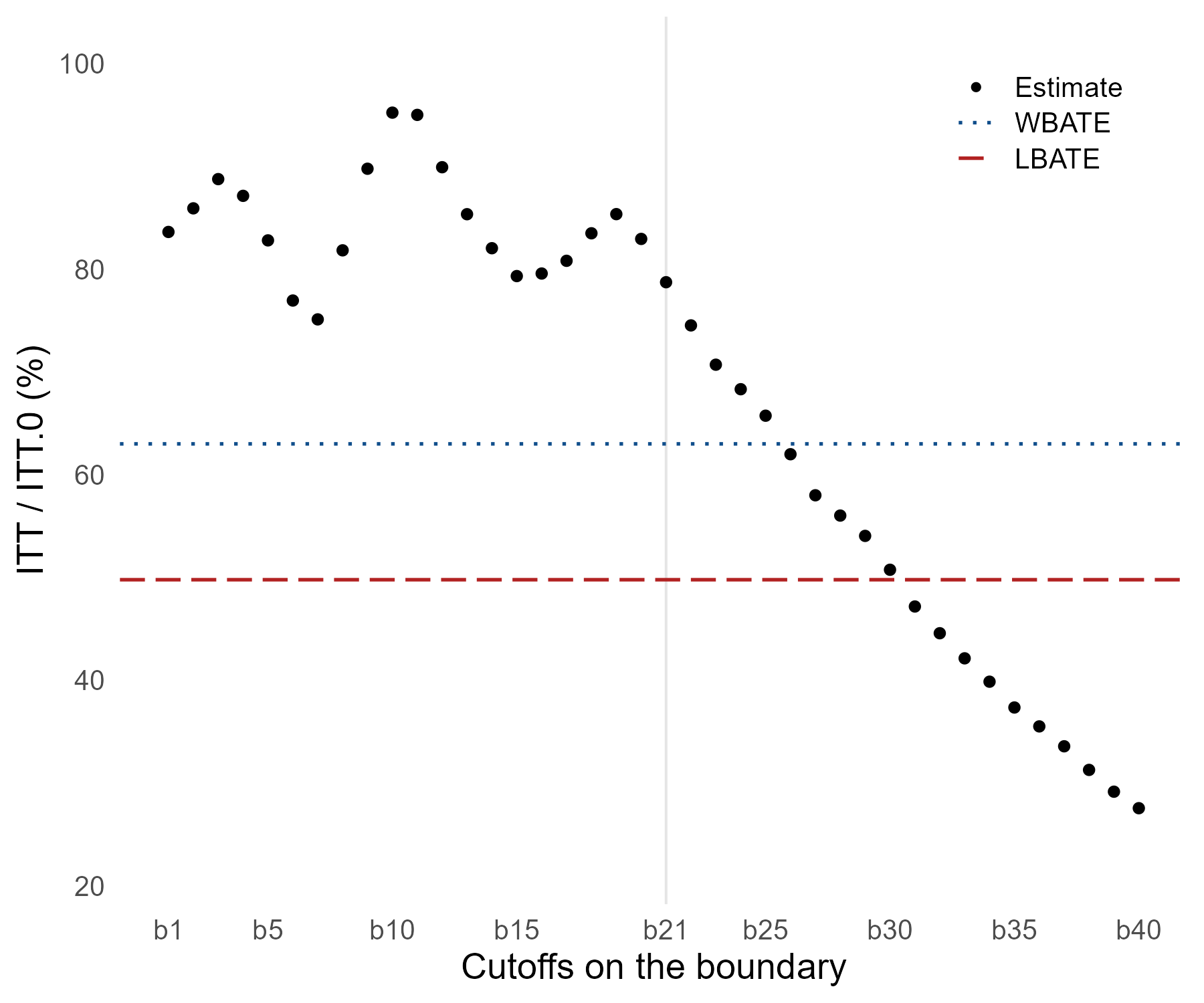}
        \caption{Intention-to-Treat Effect.}
    \end{subfigure}

    \begin{subfigure}[b]{0.70\textwidth}
        \centering
        \includegraphics[width=\linewidth]{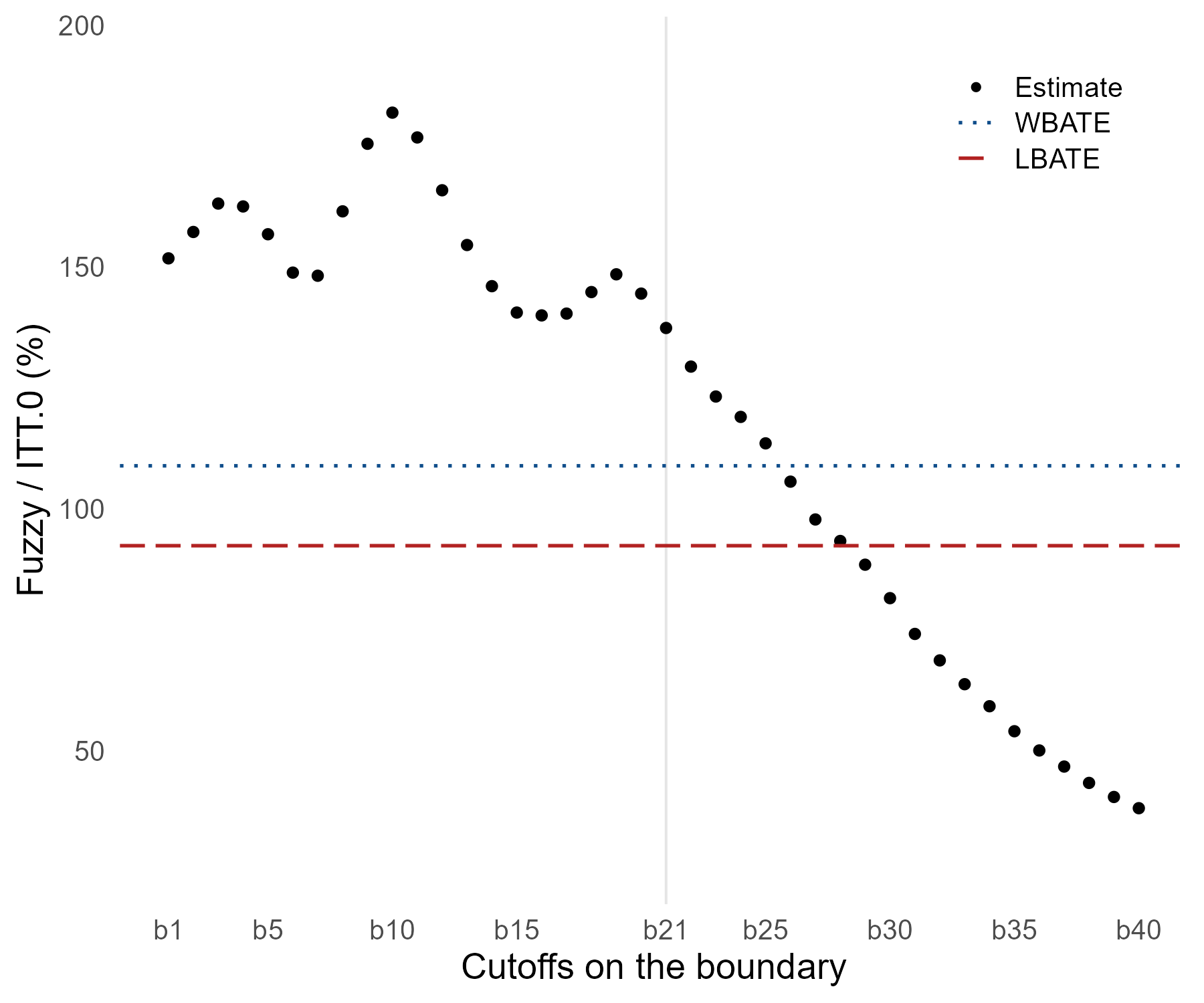}
        \caption{Fuzzy Effect.}
    \end{subfigure}
    \caption{Treatment Effect Relative to Control-Side Baseline (SPP Application, ITT Bandwidth Selector).}
    \label{fig:sa-empapp_main_percent}
\end{figure}

In addition, we report baseline estimates and treatment effects relative to control-side baselines.
\begin{itemize}
    \item Table~\ref{tab:sa-empapp_main_itt0} reports the control-side baseline estimates $\widehat{\mu}_{Y,0}(\bx)$ for college enrollment.
    \item Table~\ref{tab:sa-empapp_main_percent} reports ITT and Fuzzy treatment effects relative to control-side baseline estimates $\widehat{\mu}_{Y,0}(\bx)$ for college enrollment.
    \item Figure~\ref{fig:sa-empapp_main_itt0} reports baseline estimates and uncertainty quantification.
    \item Figure~\ref{fig:sa-empapp_main_percent} reports treatment effects relative to the estimated control-side baseline.
\end{itemize}

The baseline estimates in Table~\ref{tab:sa-empapp_main_itt0} help interpret the magnitude of the treatment effects.

\clearpage

\begin{table}[p]
    \centering
    \resizebox{.8\textwidth}{!}{\begin{tabular}{lccccccc}
  \toprule\toprule
   & $h_1$ & $h_2$ & $N_{\mathrm{Co}}$ & $N_{\mathrm{Tr}}$ & Estimate & $p$-value & 95\% CI \\
  \midrule
   $\bb_{1}$ & 27.6 & 12.2 & 8921 & 3532 & 0.311 & 0.000 & (0.204, 0.423)\\
   $\bb_{2}$ & 26.0 & 11.5 & 8998 & 3677 & 0.313 & 0.000 & (0.217, 0.430)\\
   $\bb_{3}$ & 24.6 & 10.9 & 8510 & 3678 & 0.316 & 0.000 & (0.218, 0.428)\\
   $\bb_{4}$ & 24.5 & 10.9 & 9236 & 3976 & 0.309 & 0.000 & (0.197, 0.399)\\
   $\bb_{5}$ & 23.1 & 10.2 & 9044 & 3997 & 0.295 & 0.000 & (0.164, 0.370)\\
   $\bb_{6}$ & 20.7 & 9.2 & 7176 & 3525 & 0.277 & 0.000 & (0.128, 0.350)\\
   $\bb_{7}$ & 19.0 & 8.4 & 6089 & 3209 & 0.267 & 0.000 & (0.118, 0.356)\\
   $\bb_{8}$ & 19.2 & 8.5 & 6725 & 3422 & 0.283 & 0.000 & (0.137, 0.370)\\
   $\bb_{9}$ & 19.2 & 8.5 & 6803 & 3516 & 0.305 & 0.000 & (0.159, 0.391)\\
   $\bb_{10}$ & 19.3 & 8.6 & 6925 & 3563 & 0.324 & 0.000 & (0.189, 0.417)\\
   $\bb_{11}$ & 21.1 & 9.3 & 8665 & 4206 & 0.332 & 0.000 & (0.219, 0.424)\\
   $\bb_{12}$ & 21.8 & 9.7 & 9017 & 4456 & 0.327 & 0.000 & (0.223, 0.420)\\
   $\bb_{13}$ & 21.1 & 9.3 & 8683 & 4417 & 0.322 & 0.000 & (0.209, 0.412)\\
   $\bb_{14}$ & 21.8 & 9.6 & 9069 & 4646 & 0.318 & 0.000 & (0.194, 0.388)\\
   $\bb_{15}$ & 22.2 & 9.8 & 10063 & 4977 & 0.314 & 0.000 & (0.190, 0.378)\\
   $\bb_{16}$ & 22.5 & 10.0 & 10210 & 5162 & 0.317 & 0.000 & (0.208, 0.390)\\
   $\bb_{17}$ & 21.3 & 9.4 & 9409 & 4444 & 0.323 & 0.000 & (0.251, 0.416)\\
   $\bb_{18}$ & 20.7 & 9.2 & 9182 & 3850 & 0.328 & 0.000 & (0.264, 0.427)\\
   $\bb_{19}$ & 19.3 & 8.5 & 8468 & 3157 & 0.334 & 0.000 & (0.263, 0.439)\\
   $\bb_{20}$ & 19.4 & 8.6 & 8945 & 2680 & 0.327 & 0.000 & (0.231, 0.433)\\
   $\bb_{21}$ & 20.9 & 9.3 & 10677 & 2442 & 0.314 & 0.000 & (0.157, 0.470)\\
   $\bb_{22}$ & 21.5 & 9.5 & 10221 & 2786 & 0.306 & 0.000 & (0.183, 0.418)\\
   $\bb_{23}$ & 20.3 & 9.0 & 7743 & 2679 & 0.298 & 0.000 & (0.186, 0.408)\\
   $\bb_{24}$ & 20.5 & 9.1 & 7097 & 2856 & 0.295 & 0.000 & (0.198, 0.413)\\
   $\bb_{25}$ & 21.0 & 9.3 & 6188 & 3114 & 0.291 & 0.000 & (0.211, 0.423)\\
   $\bb_{26}$ & 23.2 & 10.3 & 6948 & 3635 & 0.284 & 0.000 & (0.211, 0.414)\\
   $\bb_{27}$ & 25.1 & 11.1 & 7232 & 4098 & 0.273 & 0.000 & (0.195, 0.393)\\
   $\bb_{28}$ & 25.2 & 11.2 & 5984 & 4204 & 0.269 & 0.000 & (0.182, 0.391)\\
   $\bb_{29}$ & 25.6 & 11.3 & 5319 & 4343 & 0.266 & 0.000 & (0.171, 0.393)\\
   $\bb_{30}$ & 26.3 & 11.7 & 4698 & 4560 & 0.257 & 0.000 & (0.157, 0.398)\\
   $\bb_{31}$ & 28.5 & 12.6 & 4759 & 4995 & 0.247 & 0.000 & (0.139, 0.377)\\
   $\bb_{32}$ & 28.9 & 12.8 & 4482 & 4655 & 0.240 & 0.000 & (0.123, 0.370)\\
   $\bb_{33}$ & 30.8 & 13.7 & 4559 & 4766 & 0.234 & 0.000 & (0.115, 0.360)\\
   $\bb_{34}$ & 31.2 & 13.8 & 4048 & 4222 & 0.228 & 0.000 & (0.103, 0.357)\\
   $\bb_{35}$ & 28.1 & 12.5 & 2875 & 2950 & 0.219 & 0.000 & (0.077, 0.373)\\
   $\bb_{36}$ & 26.0 & 11.5 & 2214 & 2243 & 0.215 & 0.000 & (0.064, 0.394)\\
   $\bb_{37}$ & 26.8 & 11.9 & 1987 & 2002 & 0.209 & 0.000 & (0.059, 0.391)\\
   $\bb_{38}$ & 28.5 & 12.6 & 2049 & 2068 & 0.200 & 0.000 & (0.052, 0.379)\\
   $\bb_{39}$ & 33.6 & 14.9 & 2758 & 2787 & 0.191 & 0.000 & (0.053, 0.349)\\
   $\bb_{40}$ & 36.3 & 16.1 & 2916 & 2993 & 0.185 & 0.000 & (0.044, 0.335)\\
  \midrule
   $\mathtt{WBATE}$ &  &  &  &  & 0.282 & 0.000 & (0.248, 0.316)\\
  \midrule
   $\mathtt{LBATE}$ &  &  &  &  & 0.334 &  & (0.263, 0.472)\\
  \bottomrule\bottomrule
\end{tabular}
}
    \caption{Intention-to-Treat Effects (SPP Application, ITT Bandwidth Selector).}
    \label{tab:sa-empapp_main_itt}
\end{table}

\begin{table}[p]
    \centering
    \resizebox{.8\textwidth}{!}{\begin{tabular}{lccccccc}
  \toprule\toprule
   & $h_1$ & $h_2$ & $N_{\mathrm{Co}}$ & $N_{\mathrm{Tr}}$ & Estimate & $p$-value & 95\% CI \\
  \midrule
   $\bb_{1}$ & 27.6 & 12.2 & 8921 & 3532 & 0.551 & 0.000 & (0.495, 0.655)\\
   $\bb_{2}$ & 26.0 & 11.5 & 8998 & 3677 & 0.546 & 0.000 & (0.496, 0.651)\\
   $\bb_{3}$ & 24.6 & 10.9 & 8510 & 3678 & 0.544 & 0.000 & (0.494, 0.646)\\
   $\bb_{4}$ & 24.5 & 10.9 & 9236 & 3976 & 0.536 & 0.000 & (0.483, 0.629)\\
   $\bb_{5}$ & 23.1 & 10.2 & 9044 & 3997 & 0.528 & 0.000 & (0.466, 0.612)\\
   $\bb_{6}$ & 20.7 & 9.2 & 7176 & 3525 & 0.517 & 0.000 & (0.441, 0.595)\\
   $\bb_{7}$ & 19.0 & 8.4 & 6089 & 3209 & 0.507 & 0.000 & (0.419, 0.582)\\
   $\bb_{8}$ & 19.2 & 8.5 & 6725 & 3422 & 0.507 & 0.000 & (0.407, 0.570)\\
   $\bb_{9}$ & 19.2 & 8.5 & 6803 & 3516 & 0.512 & 0.000 & (0.406, 0.571)\\
   $\bb_{10}$ & 19.3 & 8.6 & 6925 & 3563 & 0.524 & 0.000 & (0.417, 0.582)\\
   $\bb_{11}$ & 21.1 & 9.3 & 8665 & 4206 & 0.538 & 0.000 & (0.448, 0.599)\\
   $\bb_{12}$ & 21.8 & 9.7 & 9017 & 4456 & 0.542 & 0.000 & (0.463, 0.607)\\
   $\bb_{13}$ & 21.1 & 9.3 & 8683 & 4417 & 0.552 & 0.000 & (0.472, 0.616)\\
   $\bb_{14}$ & 21.8 & 9.6 & 9069 & 4646 & 0.562 & 0.000 & (0.480, 0.617)\\
   $\bb_{15}$ & 22.2 & 9.8 & 10063 & 4977 & 0.564 & 0.000 & (0.481, 0.614)\\
   $\bb_{16}$ & 22.5 & 10.0 & 10210 & 5162 & 0.568 & 0.000 & (0.493, 0.622)\\
   $\bb_{17}$ & 21.3 & 9.4 & 9409 & 4444 & 0.576 & 0.000 & (0.501, 0.636)\\
   $\bb_{18}$ & 20.7 & 9.2 & 9182 & 3850 & 0.577 & 0.000 & (0.498, 0.642)\\
   $\bb_{19}$ & 19.3 & 8.5 & 8468 & 3157 & 0.575 & 0.000 & (0.488, 0.654)\\
   $\bb_{20}$ & 19.4 & 8.6 & 8945 & 2680 & 0.574 & 0.000 & (0.467, 0.674)\\
   $\bb_{21}$ & 20.9 & 9.3 & 10677 & 2442 & 0.573 & 0.000 & (0.424, 0.762)\\
   $\bb_{22}$ & 21.5 & 9.5 & 10221 & 2786 & 0.576 & 0.000 & (0.469, 0.715)\\
   $\bb_{23}$ & 20.3 & 9.0 & 7743 & 2679 & 0.574 & 0.000 & (0.473, 0.696)\\
   $\bb_{24}$ & 20.5 & 9.1 & 7097 & 2856 & 0.574 & 0.000 & (0.472, 0.683)\\
   $\bb_{25}$ & 21.0 & 9.3 & 6188 & 3114 & 0.579 & 0.000 & (0.481, 0.684)\\
   $\bb_{26}$ & 23.2 & 10.3 & 6948 & 3635 & 0.587 & 0.000 & (0.490, 0.679)\\
   $\bb_{27}$ & 25.1 & 11.1 & 7232 & 4098 & 0.593 & 0.000 & (0.494, 0.674)\\
   $\bb_{28}$ & 25.2 & 11.2 & 5984 & 4204 & 0.600 & 0.000 & (0.499, 0.682)\\
   $\bb_{29}$ & 25.6 & 11.3 & 5319 & 4343 & 0.611 & 0.000 & (0.511, 0.697)\\
   $\bb_{30}$ & 26.3 & 11.7 & 4698 & 4560 & 0.623 & 0.000 & (0.522, 0.711)\\
   $\bb_{31}$ & 28.5 & 12.6 & 4759 & 4995 & 0.636 & 0.000 & (0.535, 0.720)\\
   $\bb_{32}$ & 28.9 & 12.8 & 4482 & 4655 & 0.649 & 0.000 & (0.547, 0.738)\\
   $\bb_{33}$ & 30.8 & 13.7 & 4559 & 4766 & 0.661 & 0.000 & (0.564, 0.751)\\
   $\bb_{34}$ & 31.2 & 13.8 & 4048 & 4222 & 0.674 & 0.000 & (0.578, 0.772)\\
   $\bb_{35}$ & 28.1 & 12.5 & 2875 & 2950 & 0.691 & 0.000 & (0.587, 0.811)\\
   $\bb_{36}$ & 26.0 & 11.5 & 2214 & 2243 & 0.710 & 0.000 & (0.601, 0.849)\\
   $\bb_{37}$ & 26.8 & 11.9 & 1987 & 2002 & 0.719 & 0.000 & (0.610, 0.861)\\
   $\bb_{38}$ & 28.5 & 12.6 & 2049 & 2068 & 0.722 & 0.000 & (0.611, 0.863)\\
   $\bb_{39}$ & 33.6 & 14.9 & 2758 & 2787 & 0.721 & 0.000 & (0.618, 0.852)\\
   $\bb_{40}$ & 36.3 & 16.1 & 2916 & 2993 & 0.723 & 0.000 & (0.621, 0.855)\\
  \midrule
   $\mathtt{WBATE}$ &  &  &  &  & 0.592 & 0.000 & (0.567, 0.620)\\
  \midrule
   $\mathtt{LBATE}$ &  &  &  &  & 0.723 &  & (0.624, 0.860)\\
  \bottomrule\bottomrule
\end{tabular}
}
    \caption{First-Stage Effects (SPP Application, ITT Bandwidth Selector).}
    \label{tab:sa-empapp_main_fs}
\end{table}

\begin{table}[p]
    \centering
    \resizebox{.8\textwidth}{!}{\begin{tabular}{lccccccc}
  \toprule\toprule
   & $h_1$ & $h_2$ & $N_{\mathrm{Co}}$ & $N_{\mathrm{Tr}}$ & Estimate & $p$-value & 95\% CI \\
  \midrule
   $\bb_{1}$ & 27.6 & 12.2 & 8921 & 3532 & 0.564 & 0.000 & (0.374, 0.715)\\
   $\bb_{2}$ & 26.0 & 11.5 & 8998 & 3677 & 0.572 & 0.000 & (0.398, 0.730)\\
   $\bb_{3}$ & 24.6 & 10.9 & 8510 & 3678 & 0.581 & 0.000 & (0.403, 0.732)\\
   $\bb_{4}$ & 24.5 & 10.9 & 9236 & 3976 & 0.576 & 0.000 & (0.374, 0.697)\\
   $\bb_{5}$ & 23.1 & 10.2 & 9044 & 3997 & 0.559 & 0.000 & (0.323, 0.666)\\
   $\bb_{6}$ & 20.7 & 9.2 & 7176 & 3525 & 0.535 & 0.000 & (0.268, 0.654)\\
   $\bb_{7}$ & 19.0 & 8.4 & 6089 & 3209 & 0.527 & 0.000 & (0.259, 0.689)\\
   $\bb_{8}$ & 19.2 & 8.5 & 6725 & 3422 & 0.559 & 0.000 & (0.304, 0.734)\\
   $\bb_{9}$ & 19.2 & 8.5 & 6803 & 3516 & 0.595 & 0.000 & (0.350, 0.776)\\
   $\bb_{10}$ & 19.3 & 8.6 & 6925 & 3563 & 0.619 & 0.000 & (0.401, 0.812)\\
   $\bb_{11}$ & 21.1 & 9.3 & 8665 & 4206 & 0.618 & 0.000 & (0.437, 0.791)\\
   $\bb_{12}$ & 21.8 & 9.7 & 9017 & 4456 & 0.602 & 0.000 & (0.433, 0.767)\\
   $\bb_{13}$ & 21.1 & 9.3 & 8683 & 4417 & 0.584 & 0.000 & (0.401, 0.740)\\
   $\bb_{14}$ & 21.8 & 9.6 & 9069 & 4646 & 0.566 & 0.000 & (0.371, 0.692)\\
   $\bb_{15}$ & 22.2 & 9.8 & 10063 & 4977 & 0.556 & 0.000 & (0.362, 0.675)\\
   $\bb_{16}$ & 22.5 & 10.0 & 10210 & 5162 & 0.557 & 0.000 & (0.388, 0.686)\\
   $\bb_{17}$ & 21.3 & 9.4 & 9409 & 4444 & 0.560 & 0.000 & (0.458, 0.715)\\
   $\bb_{18}$ & 20.7 & 9.2 & 9182 & 3850 & 0.570 & 0.000 & (0.484, 0.727)\\
   $\bb_{19}$ & 19.3 & 8.5 & 8468 & 3157 & 0.580 & 0.000 & (0.484, 0.744)\\
   $\bb_{20}$ & 19.4 & 8.6 & 8945 & 2680 & 0.569 & 0.000 & (0.440, 0.723)\\
   $\bb_{21}$ & 20.9 & 9.3 & 10677 & 2442 & 0.548 & 0.000 & (0.338, 0.720)\\
   $\bb_{22}$ & 21.5 & 9.5 & 10221 & 2786 & 0.532 & 0.000 & (0.359, 0.655)\\
   $\bb_{23}$ & 20.3 & 9.0 & 7743 & 2679 & 0.519 & 0.000 & (0.357, 0.659)\\
   $\bb_{24}$ & 20.5 & 9.1 & 7097 & 2856 & 0.513 & 0.000 & (0.373, 0.686)\\
   $\bb_{25}$ & 21.0 & 9.3 & 6188 & 3114 & 0.503 & 0.000 & (0.385, 0.703)\\
   $\bb_{26}$ & 23.2 & 10.3 & 6948 & 3635 & 0.483 & 0.000 & (0.382, 0.689)\\
   $\bb_{27}$ & 25.1 & 11.1 & 7232 & 4098 & 0.461 & 0.000 & (0.352, 0.655)\\
   $\bb_{28}$ & 25.2 & 11.2 & 5984 & 4204 & 0.449 & 0.000 & (0.326, 0.643)\\
   $\bb_{29}$ & 25.6 & 11.3 & 5319 & 4343 & 0.435 & 0.000 & (0.302, 0.632)\\
   $\bb_{30}$ & 26.3 & 11.7 & 4698 & 4560 & 0.412 & 0.000 & (0.272, 0.627)\\
   $\bb_{31}$ & 28.5 & 12.6 & 4759 & 4995 & 0.388 & 0.000 & (0.238, 0.584)\\
   $\bb_{32}$ & 28.9 & 12.8 & 4482 & 4655 & 0.370 & 0.000 & (0.206, 0.561)\\
   $\bb_{33}$ & 30.8 & 13.7 & 4559 & 4766 & 0.354 & 0.000 & (0.188, 0.534)\\
   $\bb_{34}$ & 31.2 & 13.8 & 4048 & 4222 & 0.338 & 0.000 & (0.164, 0.517)\\
   $\bb_{35}$ & 28.1 & 12.5 & 2875 & 2950 & 0.317 & 0.000 & (0.123, 0.521)\\
   $\bb_{36}$ & 26.0 & 11.5 & 2214 & 2243 & 0.303 & 0.000 & (0.100, 0.532)\\
   $\bb_{37}$ & 26.8 & 11.9 & 1987 & 2002 & 0.291 & 0.000 & (0.090, 0.522)\\
   $\bb_{38}$ & 28.5 & 12.6 & 2049 & 2068 & 0.277 & 0.000 & (0.079, 0.506)\\
   $\bb_{39}$ & 33.6 & 14.9 & 2758 & 2787 & 0.265 & 0.000 & (0.079, 0.468)\\
   $\bb_{40}$ & 36.3 & 16.1 & 2916 & 2993 & 0.255 & 0.000 & (0.067, 0.447)\\
  \midrule
   $\mathtt{WBATE}$ &  &  &  &  & 0.487 & 0.000 & (0.435, 0.535)\\
  \midrule
   $\mathtt{LBATE}$ &  &  &  &  & 0.619 &  & (0.487, 0.808)\\
  \bottomrule\bottomrule
\end{tabular}
}
    \caption{Fuzzy Effects (SPP Application, ITT Bandwidth Selector).}
    \label{tab:sa-empapp_main_fuzzy}
\end{table}

\begin{table}[p]
    \centering
    \resizebox{.8\textwidth}{!}{\begin{tabular}{lccccccc}
  \toprule\toprule
   & $h_1$ & $h_2$ & $N_{\mathrm{Co}}$ & $N_{\mathrm{Tr}}$ & Estimate & $p$-value & 95\% CI \\
  \midrule
   $\bb_{1}$ & 27.4 & 12.2 & 8750 & 3469 & -0.027 & 0.170 & (-0.153, 0.058)\\
   $\bb_{2}$ & 25.9 & 11.5 & 8454 & 3560 & -0.032 & 0.115 & (-0.158, 0.050)\\
   $\bb_{3}$ & 24.8 & 11.0 & 8447 & 3647 & -0.031 & 0.115 & (-0.157, 0.050)\\
   $\bb_{4}$ & 24.7 & 11.0 & 9184 & 3947 & -0.022 & 0.213 & (-0.139, 0.058)\\
   $\bb_{5}$ & 24.2 & 10.7 & 9767 & 4156 & -0.015 & 0.398 & (-0.123, 0.069)\\
   $\bb_{6}$ & 22.7 & 10.1 & 8801 & 3951 & -0.008 & 0.739 & (-0.108, 0.086)\\
   $\bb_{7}$ & 21.9 & 9.7 & 8282 & 3901 & -0.001 & 0.984 & (-0.099, 0.098)\\
   $\bb_{8}$ & 21.4 & 9.5 & 8372 & 3993 & -0.002 & 0.678 & (-0.115, 0.087)\\
   $\bb_{9}$ & 21.7 & 9.6 & 8680 & 4130 & -0.010 & 0.367 & (-0.130, 0.071)\\
   $\bb_{10}$ & 21.0 & 9.3 & 8085 & 3981 & -0.011 & 0.272 & (-0.139, 0.065)\\
   $\bb_{11}$ & 20.6 & 9.1 & 7894 & 3964 & -0.011 & 0.444 & (-0.129, 0.077)\\
   $\bb_{12}$ & 21.6 & 9.6 & 8832 & 4358 & -0.009 & 0.931 & (-0.100, 0.095)\\
   $\bb_{13}$ & 21.0 & 9.3 & 8098 & 4270 & -0.001 & 0.851 & (-0.094, 0.106)\\
   $\bb_{14}$ & 20.9 & 9.3 & 8189 & 4348 & 0.002 & 0.849 & (-0.091, 0.104)\\
   $\bb_{15}$ & 20.3 & 9.0 & 8051 & 4329 & 0.000 & 0.909 & (-0.093, 0.100)\\
   $\bb_{16}$ & 21.8 & 9.7 & 9040 & 4809 & 0.001 & 0.560 & (-0.071, 0.105)\\
   $\bb_{17}$ & 22.7 & 10.0 & 10699 & 4678 & 0.002 & 0.516 & (-0.057, 0.088)\\
   $\bb_{18}$ & 20.1 & 8.9 & 8788 & 3746 & -0.002 & 0.472 & (-0.058, 0.095)\\
   $\bb_{19}$ & 18.7 & 8.3 & 7730 & 3014 & -0.005 & 0.815 & (-0.085, 0.073)\\
   $\bb_{20}$ & 20.2 & 8.9 & 9828 & 2825 & -0.009 & 0.613 & (-0.103, 0.073)\\
   $\bb_{21}$ & 19.8 & 8.8 & 9415 & 2203 & -0.004 & 0.503 & (-0.118, 0.185)\\
   $\bb_{22}$ & 21.0 & 9.3 & 9846 & 2645 & 0.001 & 0.350 & (-0.076, 0.144)\\
   $\bb_{23}$ & 17.9 & 7.9 & 5971 & 2207 & 0.012 & 0.065 & (-0.043, 0.176)\\
   $\bb_{24}$ & 17.9 & 7.9 & 4992 & 2339 & 0.018 & 0.037 & (-0.034, 0.180)\\
   $\bb_{25}$ & 18.7 & 8.3 & 4697 & 2598 & 0.020 & 0.058 & (-0.039, 0.168)\\
   $\bb_{26}$ & 20.6 & 9.1 & 4929 & 3087 & 0.019 & 0.112 & (-0.046, 0.148)\\
   $\bb_{27}$ & 22.3 & 9.9 & 5182 & 3532 & 0.015 & 0.233 & (-0.057, 0.130)\\
   $\bb_{28}$ & 22.5 & 10.0 & 4222 & 3677 & 0.013 & 0.312 & (-0.065, 0.130)\\
   $\bb_{29}$ & 23.0 & 10.2 & 3755 & 3831 & 0.009 & 0.412 & (-0.077, 0.134)\\
   $\bb_{30}$ & 24.5 & 10.9 & 3842 & 3945 & 0.005 & 0.556 & (-0.085, 0.126)\\
   $\bb_{31}$ & 29.4 & 13.0 & 5150 & 5113 & 0.004 & 0.678 & (-0.079, 0.105)\\
   $\bb_{32}$ & 30.0 & 13.3 & 4746 & 4961 & 0.000 & 0.961 & (-0.094, 0.097)\\
   $\bb_{33}$ & 27.7 & 12.3 & 3589 & 3723 & -0.004 & 0.798 & (-0.117, 0.099)\\
   $\bb_{34}$ & 26.1 & 11.6 & 2715 & 2794 & -0.004 & 0.665 & (-0.135, 0.101)\\
   $\bb_{35}$ & 23.1 & 10.3 & 1825 & 1842 & -0.006 & 0.639 & (-0.156, 0.114)\\
   $\bb_{36}$ & 21.2 & 9.4 & 1396 & 1393 & -0.009 & 0.528 & (-0.172, 0.113)\\
   $\bb_{37}$ & 21.7 & 9.6 & 1252 & 1223 & -0.012 & 0.423 & (-0.177, 0.103)\\
   $\bb_{38}$ & 22.7 & 10.1 & 1304 & 1274 & -0.014 & 0.372 & (-0.173, 0.094)\\
   $\bb_{39}$ & 29.3 & 13.0 & 1944 & 1969 & -0.010 & 0.434 & (-0.134, 0.079)\\
   $\bb_{40}$ & 27.8 & 12.3 & 1560 & 1511 & -0.011 & 0.516 & (-0.134, 0.087)\\
  \midrule
   $\mathtt{WBATE}$ &  &  &  &  & -0.004 & 0.938 & (-0.030, 0.028)\\
  \midrule
   $\mathtt{LBATE}$ &  &  &  &  & 0.020 &  & (-0.031, 0.182)\\
  \bottomrule\bottomrule
\end{tabular}
}
    \caption{Covariate Balance for Mother's Education (SPP Application, ITT Bandwidth Selector).}
    \label{tab:sa-empapp_covbal_itt}
\end{table}

\clearpage

\clearpage

\begin{table}[p]
    \centering
    \resizebox{.8\textwidth}{!}{\begin{tabular}{lccccccc}
  \toprule\toprule
   & $h_1$ & $h_2$ & $N_{\mathrm{Co}}$ & $N_{\mathrm{Tr}}$ & Estimate & $p$-value & 95\% CI \\
  \midrule
   $\bb_{1}$ & 27.6 & 12.2 & 8921 &  & 0.371 & 0.000 & (0.295, 0.459)\\
   $\bb_{2}$ & 26.0 & 11.5 & 8998 &  & 0.364 & 0.000 & (0.286, 0.445)\\
   $\bb_{3}$ & 24.6 & 10.9 & 8510 &  & 0.356 & 0.000 & (0.283, 0.441)\\
   $\bb_{4}$ & 24.5 & 10.9 & 9236 &  & 0.355 & 0.000 & (0.293, 0.444)\\
   $\bb_{5}$ & 23.1 & 10.2 & 9044 &  & 0.356 & 0.000 & (0.298, 0.454)\\
   $\bb_{6}$ & 20.7 & 9.2 & 7176 &  & 0.359 & 0.000 & (0.294, 0.462)\\
   $\bb_{7}$ & 19.0 & 8.4 & 6089 &  & 0.356 & 0.000 & (0.271, 0.450)\\
   $\bb_{8}$ & 19.2 & 8.5 & 6725 &  & 0.346 & 0.000 & (0.263, 0.436)\\
   $\bb_{9}$ & 19.2 & 8.5 & 6803 &  & 0.339 & 0.000 & (0.261, 0.434)\\
   $\bb_{10}$ & 19.3 & 8.6 & 6925 &  & 0.340 & 0.000 & (0.264, 0.435)\\
   $\bb_{11}$ & 21.1 & 9.3 & 8665 &  & 0.349 & 0.000 & (0.281, 0.434)\\
   $\bb_{12}$ & 21.8 & 9.7 & 9017 &  & 0.363 & 0.000 & (0.298, 0.447)\\
   $\bb_{13}$ & 21.1 & 9.3 & 8683 &  & 0.378 & 0.000 & (0.317, 0.473)\\
   $\bb_{14}$ & 21.8 & 9.6 & 9069 &  & 0.388 & 0.000 & (0.343, 0.494)\\
   $\bb_{15}$ & 22.2 & 9.8 & 10063 &  & 0.396 & 0.000 & (0.355, 0.502)\\
   $\bb_{16}$ & 22.5 & 10.0 & 10210 &  & 0.398 & 0.000 & (0.351, 0.494)\\
   $\bb_{17}$ & 21.3 & 9.4 & 9409 &  & 0.399 & 0.000 & (0.343, 0.459)\\
   $\bb_{18}$ & 20.7 & 9.2 & 9182 &  & 0.393 & 0.000 & (0.338, 0.439)\\
   $\bb_{19}$ & 19.3 & 8.5 & 8468 &  & 0.391 & 0.000 & (0.340, 0.438)\\
   $\bb_{20}$ & 19.4 & 8.6 & 8945 &  & 0.394 & 0.000 & (0.359, 0.444)\\
   $\bb_{21}$ & 20.9 & 9.3 & 10677 &  & 0.399 & 0.000 & (0.373, 0.443)\\
   $\bb_{22}$ & 21.5 & 9.5 & 10221 &  & 0.411 & 0.000 & (0.381, 0.457)\\
   $\bb_{23}$ & 20.3 & 9.0 & 7743 &  & 0.421 & 0.000 & (0.373, 0.467)\\
   $\bb_{24}$ & 20.5 & 9.1 & 7097 &  & 0.431 & 0.000 & (0.365, 0.475)\\
   $\bb_{25}$ & 21.0 & 9.3 & 6188 &  & 0.443 & 0.000 & (0.364, 0.486)\\
   $\bb_{26}$ & 23.2 & 10.3 & 6948 &  & 0.457 & 0.000 & (0.376, 0.499)\\
   $\bb_{27}$ & 25.1 & 11.1 & 7232 &  & 0.471 & 0.000 & (0.390, 0.515)\\
   $\bb_{28}$ & 25.2 & 11.2 & 5984 &  & 0.481 & 0.000 & (0.395, 0.534)\\
   $\bb_{29}$ & 25.6 & 11.3 & 5319 &  & 0.491 & 0.000 & (0.397, 0.552)\\
   $\bb_{30}$ & 26.3 & 11.7 & 4698 &  & 0.506 & 0.000 & (0.393, 0.572)\\
   $\bb_{31}$ & 28.5 & 12.6 & 4759 &  & 0.524 & 0.000 & (0.415, 0.592)\\
   $\bb_{32}$ & 28.9 & 12.8 & 4482 &  & 0.539 & 0.000 & (0.430, 0.616)\\
   $\bb_{33}$ & 30.8 & 13.7 & 4559 &  & 0.555 & 0.000 & (0.449, 0.635)\\
   $\bb_{34}$ & 31.2 & 13.8 & 4048 &  & 0.571 & 0.000 & (0.463, 0.657)\\
   $\bb_{35}$ & 28.1 & 12.5 & 2875 &  & 0.587 & 0.000 & (0.463, 0.689)\\
   $\bb_{36}$ & 26.0 & 11.5 & 2214 &  & 0.605 & 0.000 & (0.469, 0.727)\\
   $\bb_{37}$ & 26.8 & 11.9 & 1987 &  & 0.622 & 0.000 & (0.487, 0.753)\\
   $\bb_{38}$ & 28.5 & 12.6 & 2049 &  & 0.640 & 0.000 & (0.506, 0.772)\\
   $\bb_{39}$ & 33.6 & 14.9 & 2758 &  & 0.656 & 0.000 & (0.540, 0.782)\\
   $\bb_{40}$ & 36.3 & 16.1 & 2916 &  & 0.670 & 0.000 & (0.562, 0.798)\\
  \midrule
   $\mathtt{WBATE}$ &  &  &  &  & 0.447 & 0.000 & (0.423, 0.473)\\
  \midrule
   $\mathtt{LBATE}$ &  &  &  &  & 0.670 &  & (0.561, 0.800)\\
  \bottomrule\bottomrule
\end{tabular}
}
    \caption{Control-Side Baseline Outcome (SPP Application, ITT Bandwidth Selector).}
    \label{tab:sa-empapp_main_itt0}
\end{table}

\begin{table}[p]
    \centering
    \resizebox{.8\textwidth}{!}{\begin{tabular}{lcccccc}
  \toprule\toprule
   & $h_1$ & $h_2$ & $N_{\mathrm{Co}}$ & $N_{\mathrm{Tr}}$ & ITT / ITT.0 (\%) & Fuzzy / ITT.0 (\%) \\
  \midrule
   $\bb_{1}$ & 27.6 & 12.2 & 8921 & 3532 & 83.631 & 151.785\\
   $\bb_{2}$ & 26.0 & 11.5 & 8998 & 3677 & 85.929 & 157.241\\
   $\bb_{3}$ & 24.6 & 10.9 & 8510 & 3678 & 88.773 & 163.114\\
   $\bb_{4}$ & 24.5 & 10.9 & 9236 & 3976 & 87.145 & 162.517\\
   $\bb_{5}$ & 23.1 & 10.2 & 9044 & 3997 & 82.811 & 156.772\\
   $\bb_{6}$ & 20.7 & 9.2 & 7176 & 3525 & 76.957 & 148.838\\
   $\bb_{7}$ & 19.0 & 8.4 & 6089 & 3209 & 75.122 & 148.193\\
   $\bb_{8}$ & 19.2 & 8.5 & 6725 & 3422 & 81.842 & 161.505\\
   $\bb_{9}$ & 19.2 & 8.5 & 6803 & 3516 & 89.783 & 175.477\\
   $\bb_{10}$ & 19.3 & 8.6 & 6925 & 3563 & 95.242 & 181.913\\
   $\bb_{11}$ & 21.1 & 9.3 & 8665 & 4206 & 95.023 & 176.783\\
   $\bb_{12}$ & 21.8 & 9.7 & 9017 & 4456 & 89.931 & 165.864\\
   $\bb_{13}$ & 21.1 & 9.3 & 8683 & 4417 & 85.359 & 154.546\\
   $\bb_{14}$ & 21.8 & 9.6 & 9069 & 4646 & 82.049 & 146.036\\
   $\bb_{15}$ & 22.2 & 9.8 & 10063 & 4977 & 79.337 & 140.579\\
   $\bb_{16}$ & 22.5 & 10.0 & 10210 & 5162 & 79.586 & 139.992\\
   $\bb_{17}$ & 21.3 & 9.4 & 9409 & 4444 & 80.823 & 140.364\\
   $\bb_{18}$ & 20.7 & 9.2 & 9182 & 3850 & 83.501 & 144.816\\
   $\bb_{19}$ & 19.3 & 8.5 & 8468 & 3157 & 85.366 & 148.486\\
   $\bb_{20}$ & 19.4 & 8.6 & 8945 & 2680 & 82.952 & 144.496\\
   $\bb_{21}$ & 20.9 & 9.3 & 10677 & 2442 & 78.739 & 137.393\\
   $\bb_{22}$ & 21.5 & 9.5 & 10221 & 2786 & 74.537 & 129.406\\
   $\bb_{23}$ & 20.3 & 9.0 & 7743 & 2679 & 70.713 & 123.220\\
   $\bb_{24}$ & 20.5 & 9.1 & 7097 & 2856 & 68.327 & 119.007\\
   $\bb_{25}$ & 21.0 & 9.3 & 6188 & 3114 & 65.748 & 113.536\\
   $\bb_{26}$ & 23.2 & 10.3 & 6948 & 3635 & 62.000 & 105.621\\
   $\bb_{27}$ & 25.1 & 11.1 & 7232 & 4098 & 58.005 & 97.800\\
   $\bb_{28}$ & 25.2 & 11.2 & 5984 & 4204 & 56.030 & 93.353\\
   $\bb_{29}$ & 25.6 & 11.3 & 5319 & 4343 & 54.053 & 88.449\\
   $\bb_{30}$ & 26.3 & 11.7 & 4698 & 4560 & 50.756 & 81.532\\
   $\bb_{31}$ & 28.5 & 12.6 & 4759 & 4995 & 47.181 & 74.136\\
   $\bb_{32}$ & 28.9 & 12.8 & 4482 & 4655 & 44.580 & 68.662\\
   $\bb_{33}$ & 30.8 & 13.7 & 4559 & 4766 & 42.138 & 63.743\\
   $\bb_{34}$ & 31.2 & 13.8 & 4048 & 4222 & 39.864 & 59.187\\
   $\bb_{35}$ & 28.1 & 12.5 & 2875 & 2950 & 37.353 & 54.018\\
   $\bb_{36}$ & 26.0 & 11.5 & 2214 & 2243 & 35.513 & 50.042\\
   $\bb_{37}$ & 26.8 & 11.9 & 1987 & 2002 & 33.577 & 46.713\\
   $\bb_{38}$ & 28.5 & 12.6 & 2049 & 2068 & 31.279 & 43.332\\
   $\bb_{39}$ & 33.6 & 14.9 & 2758 & 2787 & 29.159 & 40.427\\
   $\bb_{40}$ & 36.3 & 16.1 & 2916 & 2993 & 27.554 & 38.108\\
  \midrule
   $\mathtt{WBATE}$ &  &  &  &  & 63.008 & 108.897\\
  \midrule
   $\mathtt{LBATE}$ &  &  &  &  & 49.787 & 92.377\\
  \bottomrule\bottomrule
\end{tabular}
}
    \caption{Treatment Effect Relative to Control-Side Baseline (SPP Application, ITT Bandwidth Selector).}
    \label{tab:sa-empapp_main_percent}
\end{table}

\clearpage

\subsection{Fuzzy Bandwidth Selector}

This subsection reports the same empirical quantities using the bandwidth selected for the fuzzy Wald estimand. These results provide a bandwidth-sensitivity check for the main ITT-bandwidth empirical analysis. We provide the corresponding full boundary-grid results for all $40$ evaluation points.
\begin{itemize}
    \item Table~\ref{tab:sa-empapp_main_itt_bwmain} reports the reduced-form, or intention-to-treat (ITT), estimates for college enrollment.
    \item Table~\ref{tab:sa-empapp_main_fs_bwmain} reports the first-stage estimates for treatment take-up.
    \item Table~\ref{tab:sa-empapp_main_fuzzy_bwmain} reports the fuzzy estimates for college enrollment.
    \item Table~\ref{tab:sa-empapp_covbal_itt_bwmain} reports the covariate-balance reduced-form estimates using mother's education as the outcome.
\end{itemize}

\begin{figure}
    \centering
    \begin{subfigure}[b]{0.70\textwidth}
        \centering
        \includegraphics[width=\linewidth]{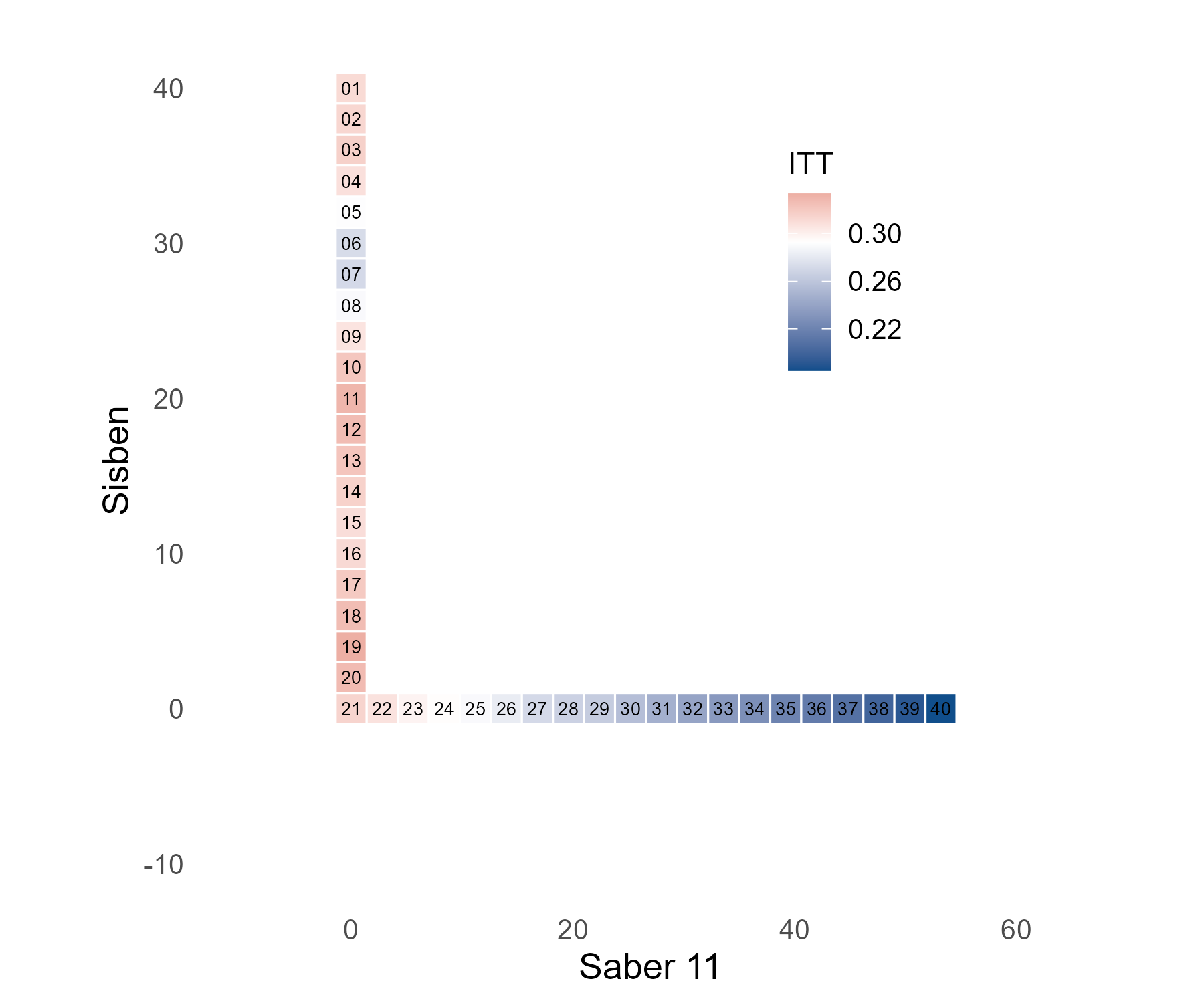}
        \caption{Point estimates.}
    \end{subfigure}

    \begin{subfigure}[b]{0.70\textwidth}
        \centering
        \includegraphics[width=\linewidth]{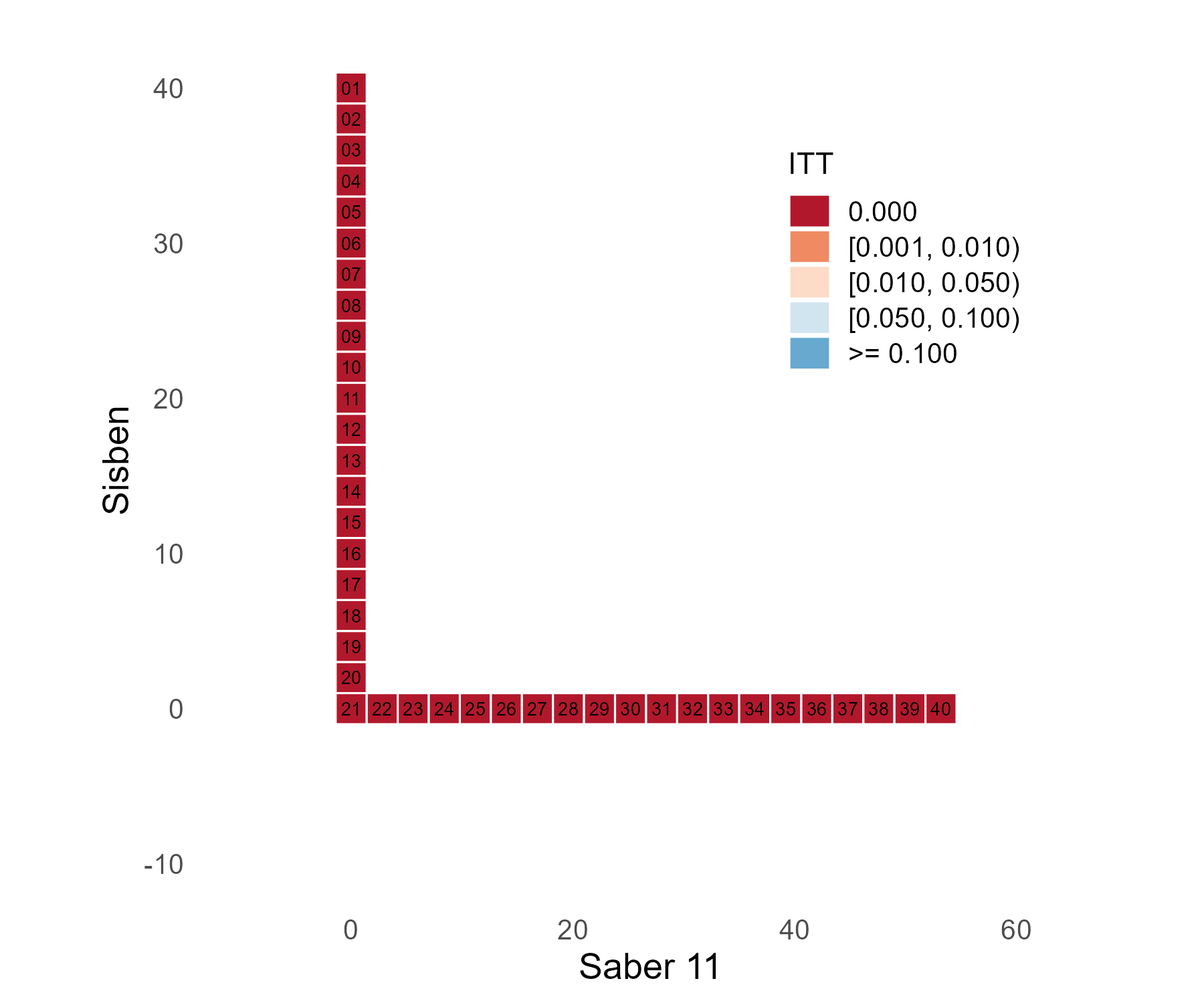}
        \caption{Pointwise $p$-values.}
    \end{subfigure}
    \caption{Intention-to-Treat Effects (SPP Application, Fuzzy Bandwidth Selector).}
    \label{fig:sa-heatmaps-itt-main-bwmain}
\end{figure}

\begin{figure}
    \centering
    \begin{subfigure}[b]{0.70\textwidth}
        \centering
        \includegraphics[width=\linewidth]{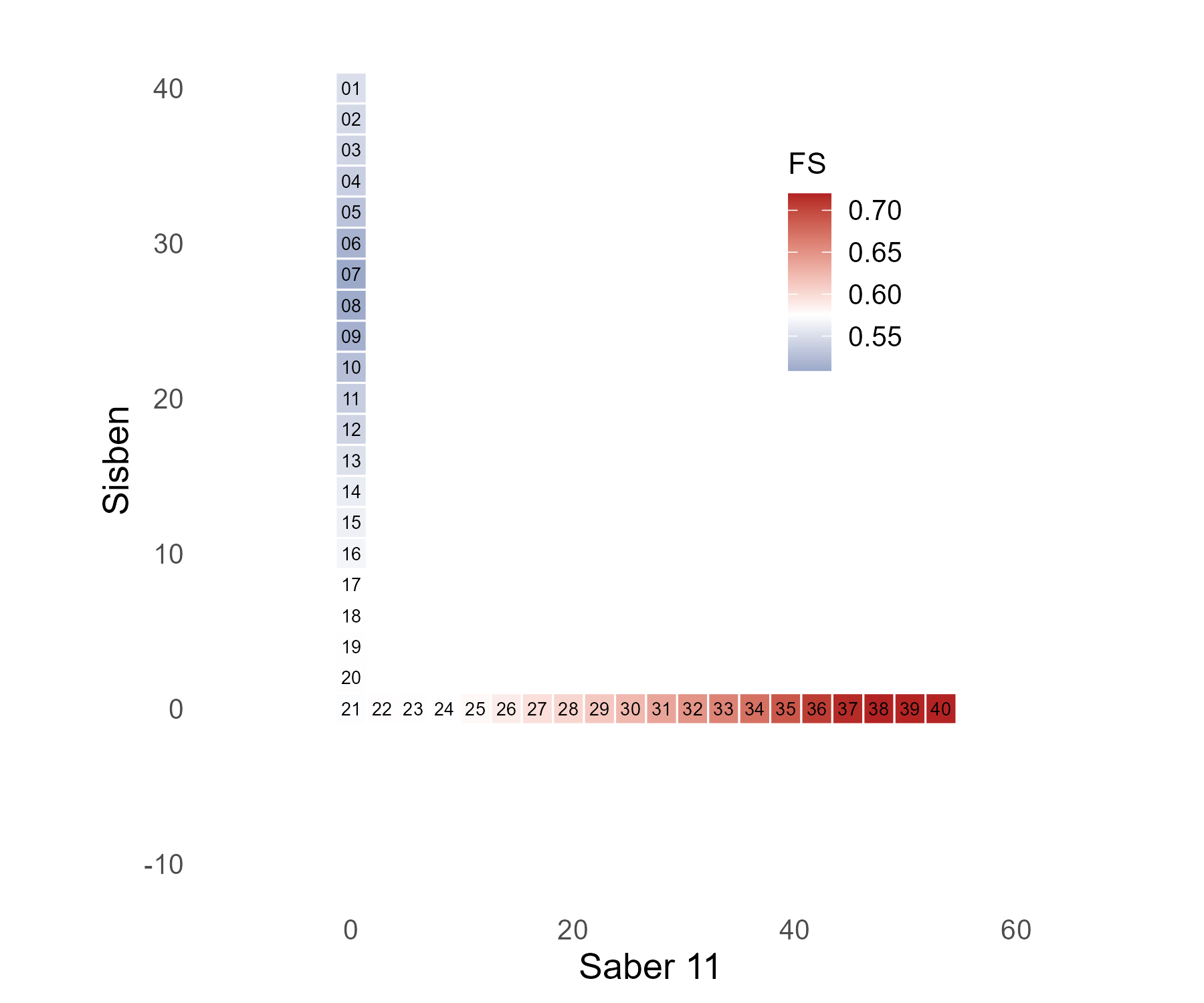}
        \caption{Point estimates.}
    \end{subfigure}

    \begin{subfigure}[b]{0.70\textwidth}
        \centering
        \includegraphics[width=\linewidth]{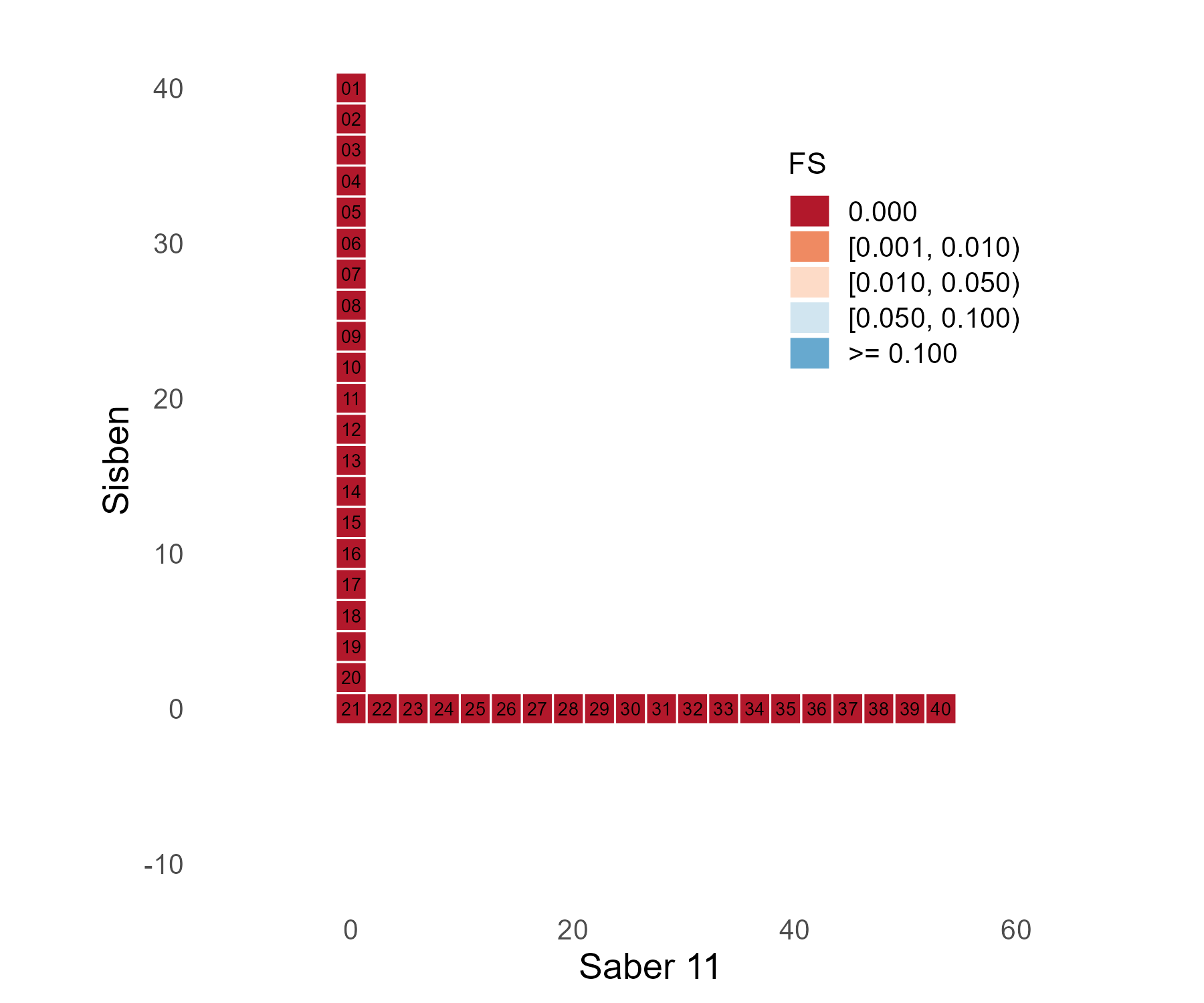}
        \caption{Pointwise $p$-values.}
    \end{subfigure}
    \caption{First-Stage Effects (SPP Application, Fuzzy Bandwidth Selector).}
    \label{fig:sa-heatmaps-fs-main-bwmain}
\end{figure}

\begin{figure}
    \centering
    \begin{subfigure}[b]{0.70\textwidth}
        \centering
        \includegraphics[width=\linewidth]{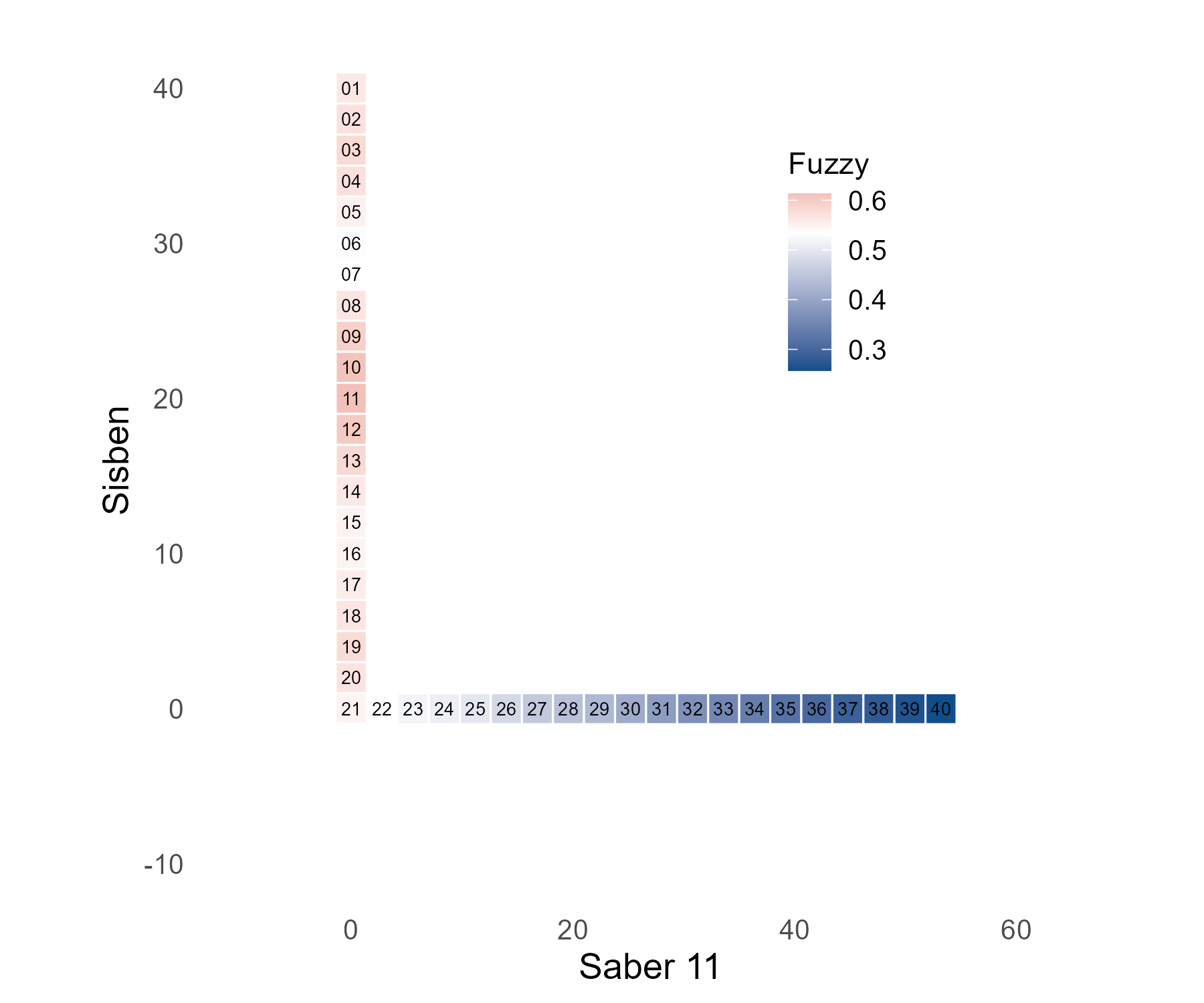}
        \caption{Point estimates.}
    \end{subfigure}

    \begin{subfigure}[b]{0.70\textwidth}
        \centering
        \includegraphics[width=\linewidth]{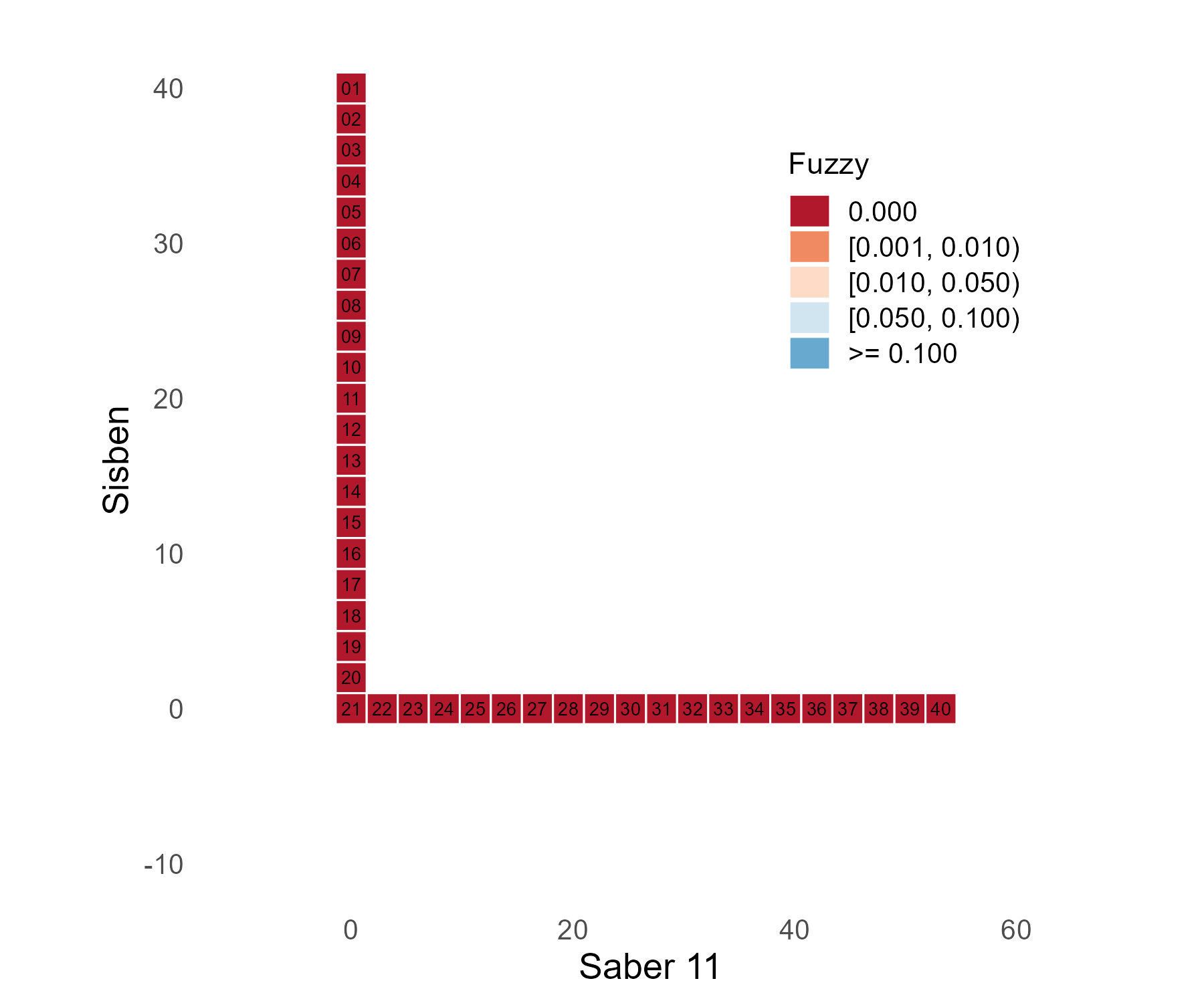}
        \caption{Pointwise $p$-values.}
    \end{subfigure}
    \caption{Fuzzy Effects (SPP Application, Fuzzy Bandwidth Selector).}
    \label{fig:sa-heatmaps-fuzzy-main-bwmain}
\end{figure}

\begin{figure}
    \centering
    \begin{subfigure}[b]{0.70\textwidth}
        \centering
        \includegraphics[width=\linewidth]{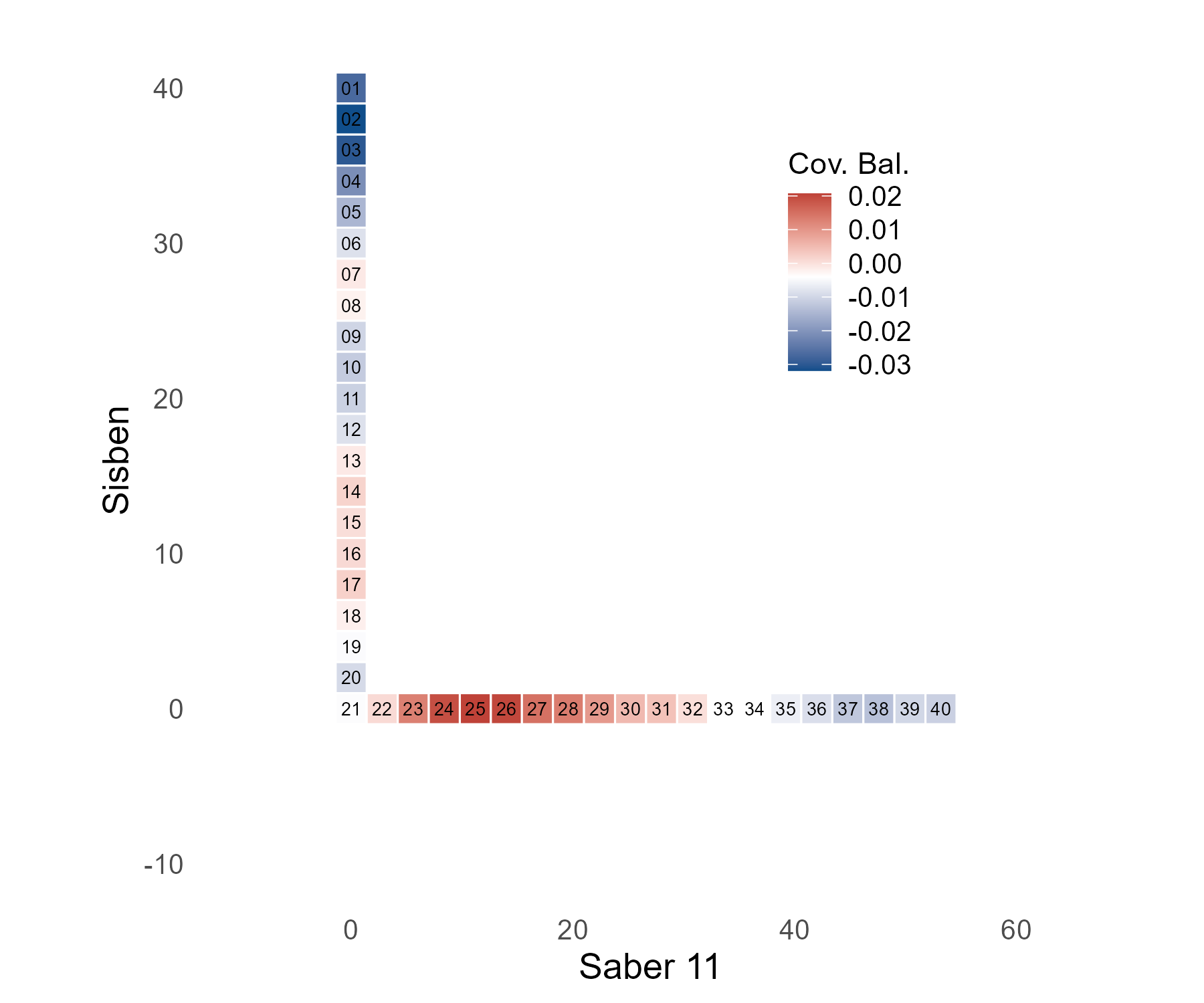}
        \caption{Point estimates.}
    \end{subfigure}

    \begin{subfigure}[b]{0.70\textwidth}
        \centering
        \includegraphics[width=\linewidth]{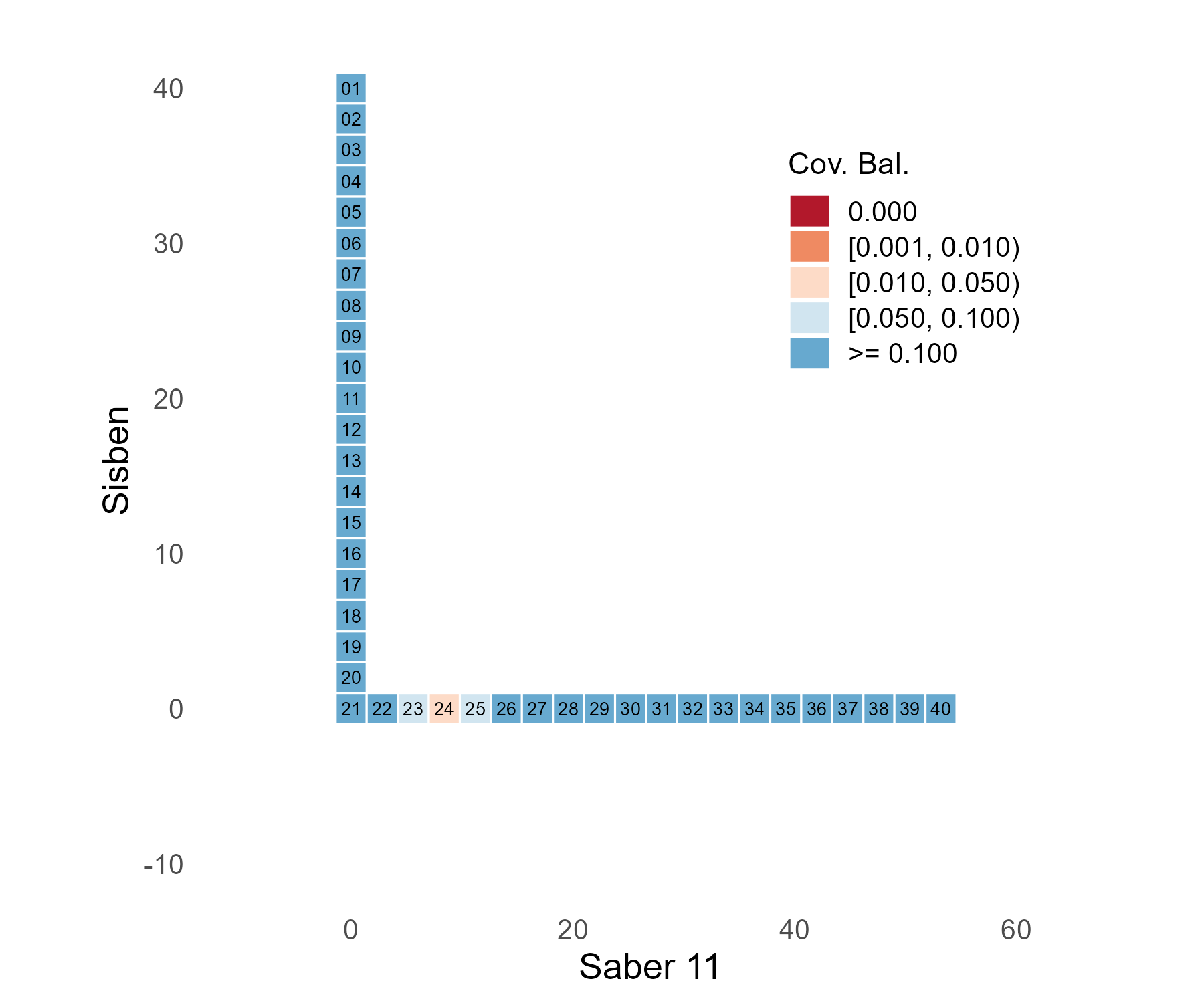}
        \caption{Pointwise $p$-values.}
    \end{subfigure}
    \caption{Covariate Balance for Mother's Education (SPP Application, Fuzzy Bandwidth Selector).}
    \label{fig:sa-heatmaps-covbal-bwmain}
\end{figure}

The figures in this subsection complement the corresponding confidence intervals and bands plots for the fuzzy bandwidth selector. For each quantity, we report heat maps along the boundary for the point estimates and associated pointwise robust bias-corrected $p$-values. Specifically:
\begin{itemize}
    \item Figure~\ref{fig:sa-heatmaps-itt-main-bwmain} reports the reduced-form, or ITT, estimates for college enrollment.
    \item Figure~\ref{fig:sa-heatmaps-fs-main-bwmain} reports the first-stage estimates for treatment take-up.
    \item Figure~\ref{fig:sa-heatmaps-fuzzy-main-bwmain} reports the fuzzy estimates for college enrollment.
    \item Figure~\ref{fig:sa-heatmaps-covbal-bwmain} reports the covariate-balance reduced-form estimates using mother's education as the outcome.
\end{itemize}
These plots make it easier to inspect how the estimated effects and pointwise significance patterns vary along the boundary.

\begin{figure}
    \centering
    \begin{subfigure}[b]{0.70\textwidth}
        \centering
        \includegraphics[width=\linewidth]{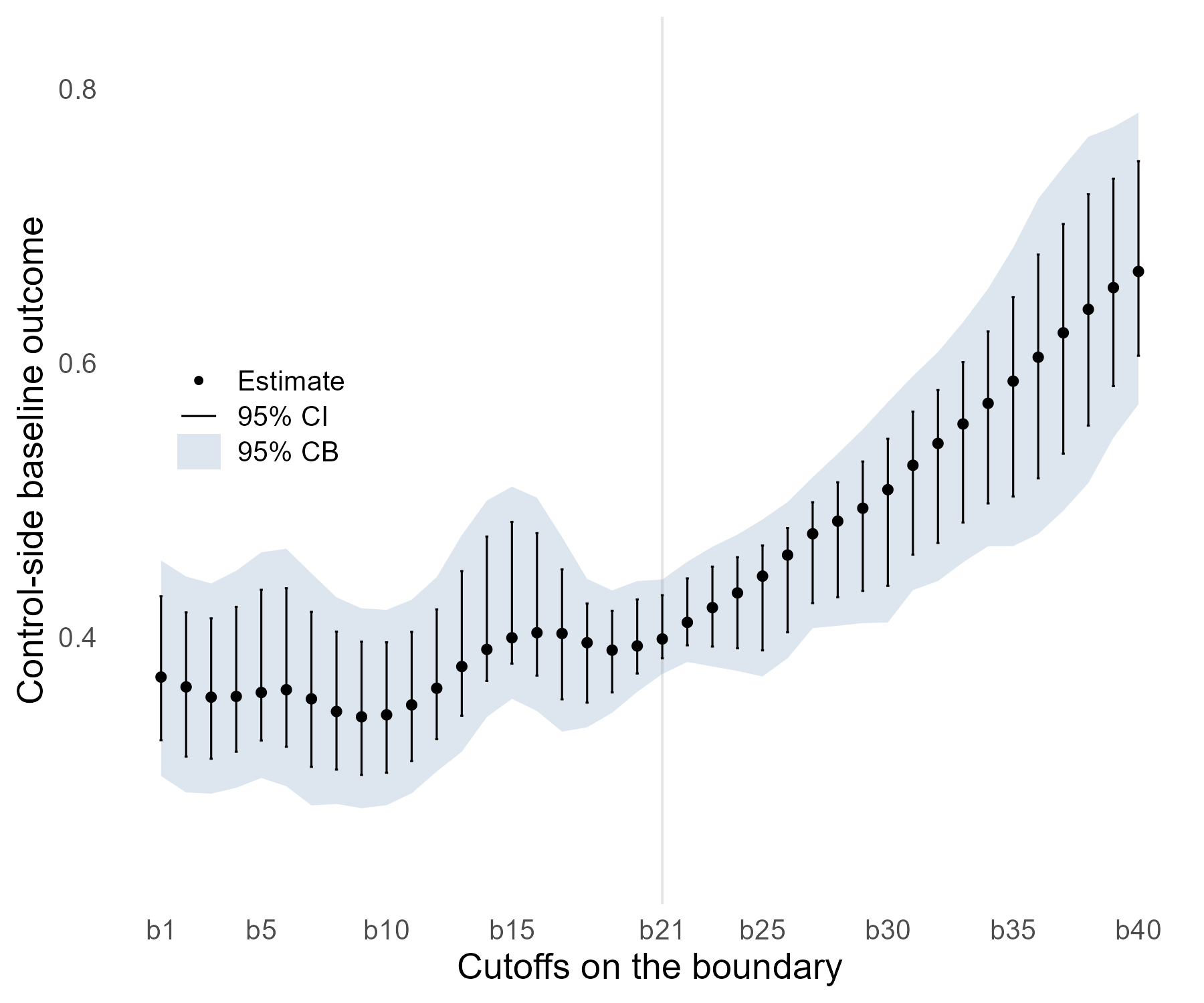}
        \caption{Point Estimates and Uncertainty Quantification.}
    \end{subfigure}

    \begin{subfigure}[b]{0.70\textwidth}
        \centering
        \includegraphics[width=\linewidth]{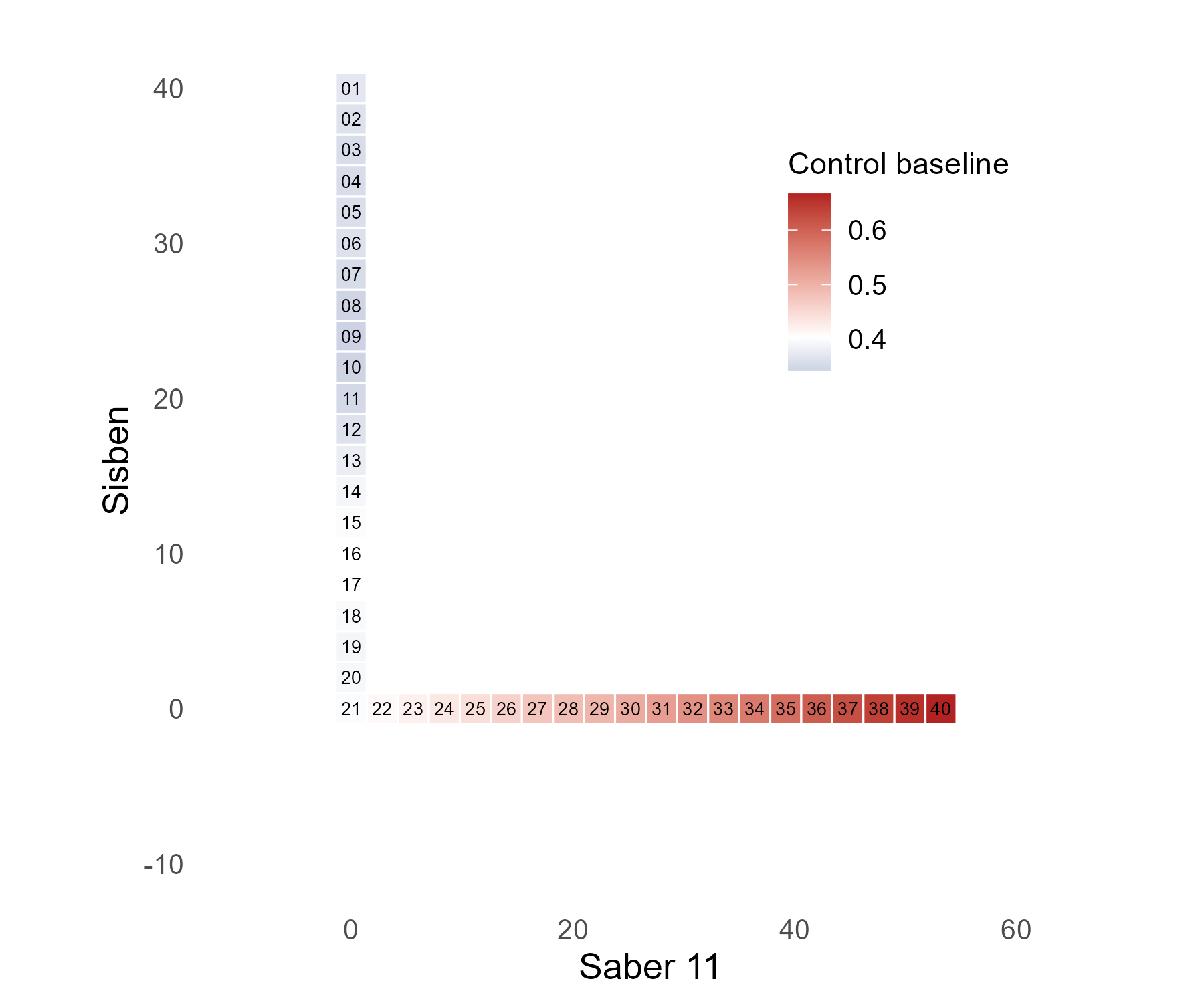}
        \caption{Heatmap of Point Estimates.}
    \end{subfigure}
    \caption{Control-Side Baseline Outcome (SPP Application, Fuzzy Bandwidth Selector).}
    \label{fig:sa-empapp_main_itt0_bwmain}
\end{figure}

\begin{figure}
    \centering
    \begin{subfigure}[b]{0.70\textwidth}
        \centering
        \includegraphics[width=\linewidth]{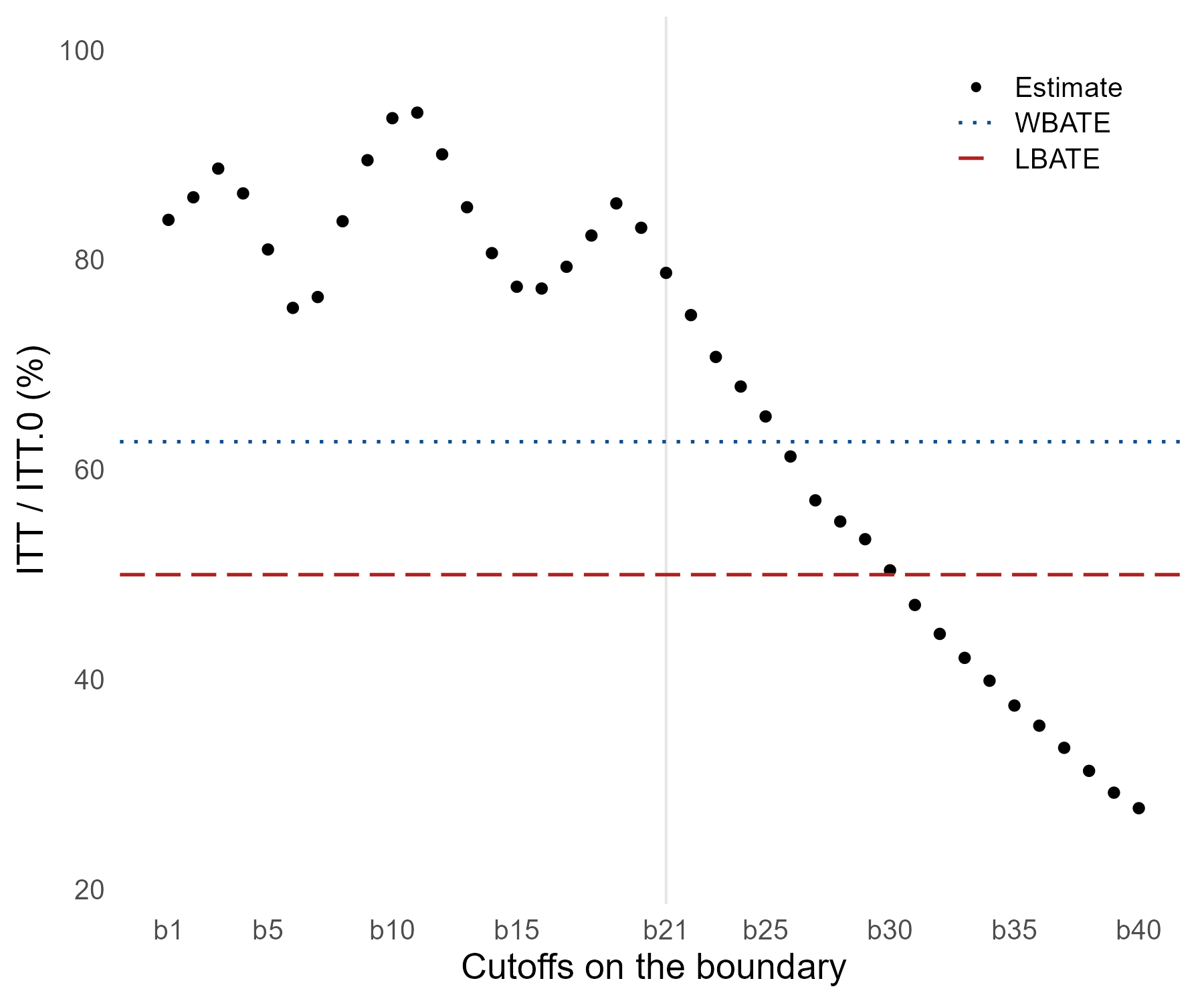}
        \caption{Intention-to-Treat Effect.}
    \end{subfigure}

    \begin{subfigure}[b]{0.70\textwidth}
        \centering
        \includegraphics[width=\linewidth]{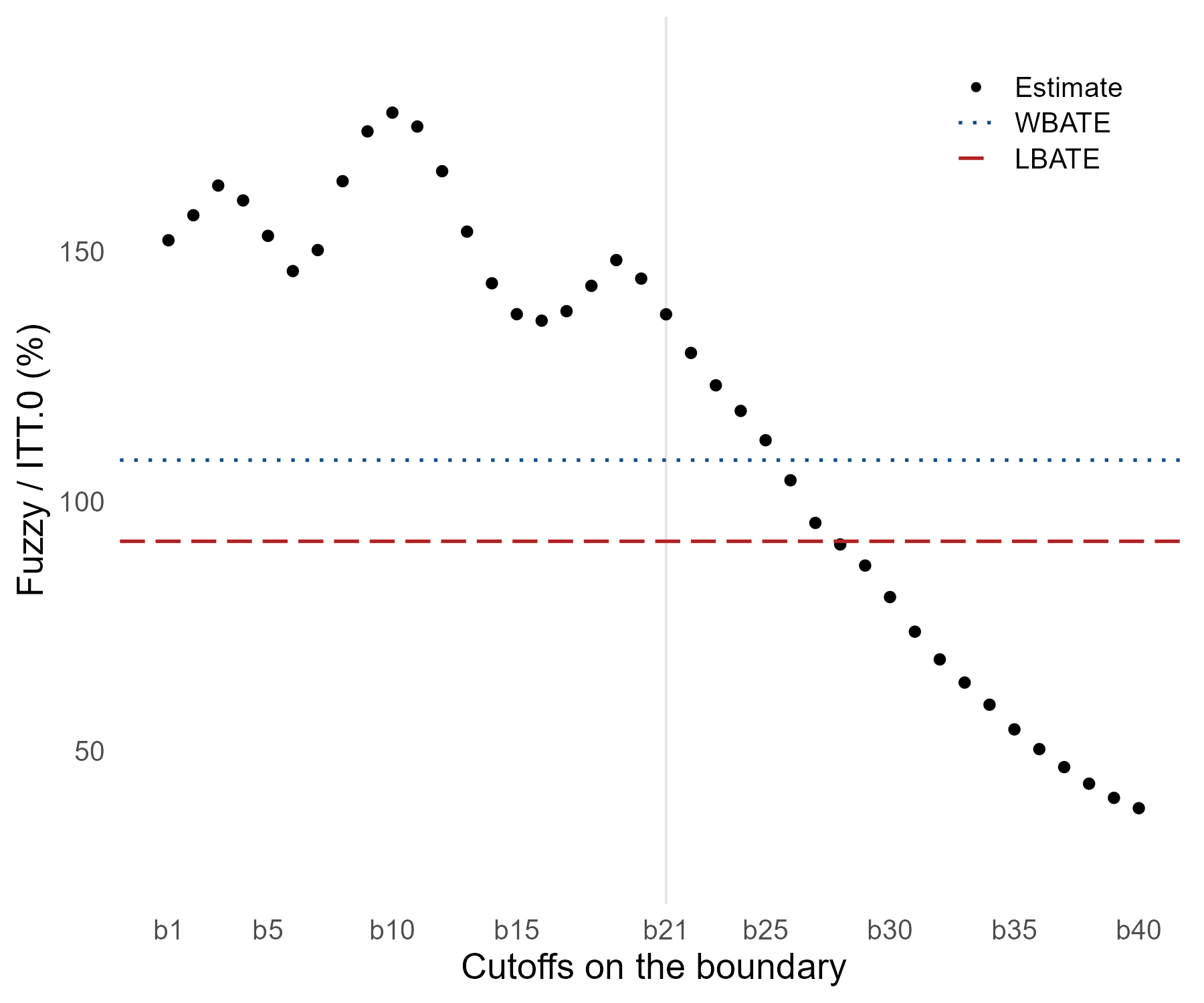}
        \caption{Fuzzy Effect.}
    \end{subfigure}
    \caption{Treatment Effect Relative to Control-Side Baseline (SPP Application, Fuzzy Bandwidth Selector).}
    \label{fig:sa-empapp_main_percent_bwmain}
\end{figure}

In addition, we report baseline estimates and treatment effects relative to control-side baselines.
\begin{itemize}
    \item Table~\ref{tab:sa-empapp_main_itt0_bwmain} reports the control-side baseline estimates $\widehat{\mu}_{Y,0}(\bx)$ for college enrollment.
    \item Table~\ref{tab:sa-empapp_main_percent_bwmain} reports ITT and Fuzzy treatment effects relative to control-side baseline estimates $\widehat{\mu}_{Y,0}(\bx)$ for college enrollment.
    \item Figure~\ref{fig:sa-empapp_main_itt0_bwmain} reports baseline estimates and uncertainty quantification.
    \item Figure~\ref{fig:sa-empapp_main_percent_bwmain} reports treatment effects relative to the estimated control-side baseline.
\end{itemize}
The baseline estimates in Table~\ref{tab:sa-empapp_main_itt0_bwmain} help interpret the economic magnitude of the effects, while Table~\ref{tab:sa-empapp_main_percent_bwmain} reports the same treatment effects in percentage terms relative to the local control-side baseline.

\clearpage

\begin{table}[p]
    \centering
    \resizebox{.8\textwidth}{!}{\begin{tabular}{lccccccc}
  \toprule\toprule
   & $h_1$ & $h_2$ & $N_{\mathrm{Co}}$ & $N_{\mathrm{Tr}}$ & Estimate & $p$-value & 95\% CI \\
  \midrule
   $\bb_{1}$ & 28.2 & 12.5 & 9704 & 3687 & 0.311 & 0.000 & (0.205, 0.418)\\
   $\bb_{2}$ & 26.0 & 11.5 & 8572 & 3612 & 0.313 & 0.000 & (0.218, 0.429)\\
   $\bb_{3}$ & 24.9 & 11.0 & 8590 & 3720 & 0.316 & 0.000 & (0.220, 0.425)\\
   $\bb_{4}$ & 23.2 & 10.3 & 8397 & 3724 & 0.308 & 0.000 & (0.190, 0.401)\\
   $\bb_{5}$ & 21.8 & 9.6 & 7403 & 3568 & 0.291 & 0.000 & (0.154, 0.371)\\
   $\bb_{6}$ & 20.0 & 8.9 & 6977 & 3427 & 0.273 & 0.000 & (0.122, 0.351)\\
   $\bb_{7}$ & 19.7 & 8.7 & 6633 & 3404 & 0.271 & 0.000 & (0.126, 0.352)\\
   $\bb_{8}$ & 21.3 & 9.4 & 8437 & 4032 & 0.289 & 0.000 & (0.156, 0.362)\\
   $\bb_{9}$ & 21.9 & 9.7 & 8859 & 4218 & 0.306 & 0.000 & (0.178, 0.378)\\
   $\bb_{10}$ & 22.2 & 9.8 & 9828 & 4469 & 0.321 & 0.000 & (0.204, 0.399)\\
   $\bb_{11}$ & 22.3 & 9.9 & 9906 & 4584 & 0.330 & 0.000 & (0.223, 0.415)\\
   $\bb_{12}$ & 22.3 & 9.9 & 9949 & 4710 & 0.327 & 0.000 & (0.224, 0.414)\\
   $\bb_{13}$ & 20.6 & 9.1 & 7995 & 4229 & 0.322 & 0.000 & (0.208, 0.412)\\
   $\bb_{14}$ & 20.7 & 9.2 & 8168 & 4350 & 0.315 & 0.000 & (0.188, 0.390)\\
   $\bb_{15}$ & 21.0 & 9.3 & 8370 & 4493 & 0.309 & 0.000 & (0.182, 0.380)\\
   $\bb_{16}$ & 20.8 & 9.2 & 8278 & 4560 & 0.312 & 0.000 & (0.201, 0.398)\\
   $\bb_{17}$ & 19.7 & 8.7 & 7544 & 4057 & 0.319 & 0.000 & (0.241, 0.428)\\
   $\bb_{18}$ & 19.6 & 8.7 & 7997 & 3623 & 0.326 & 0.000 & (0.262, 0.431)\\
   $\bb_{19}$ & 20.4 & 9.0 & 9592 & 3367 & 0.333 & 0.000 & (0.266, 0.432)\\
   $\bb_{20}$ & 20.0 & 8.9 & 9225 & 2755 & 0.327 & 0.000 & (0.234, 0.430)\\
   $\bb_{21}$ & 20.9 & 9.3 & 10677 & 2442 & 0.314 & 0.000 & (0.159, 0.469)\\
   $\bb_{22}$ & 22.0 & 9.8 & 10830 & 2843 & 0.307 & 0.000 & (0.186, 0.414)\\
   $\bb_{23}$ & 21.3 & 9.4 & 8679 & 2871 & 0.298 & 0.000 & (0.190, 0.401)\\
   $\bb_{24}$ & 21.7 & 9.6 & 8114 & 3140 & 0.294 & 0.000 & (0.201, 0.403)\\
   $\bb_{25}$ & 21.8 & 9.6 & 6858 & 3223 & 0.289 & 0.000 & (0.212, 0.416)\\
   $\bb_{26}$ & 24.3 & 10.8 & 7784 & 3867 & 0.282 & 0.000 & (0.209, 0.401)\\
   $\bb_{27}$ & 27.4 & 12.1 & 8856 & 4564 & 0.271 & 0.000 & (0.190, 0.370)\\
   $\bb_{28}$ & 26.9 & 11.9 & 7294 & 4540 & 0.267 & 0.000 & (0.179, 0.373)\\
   $\bb_{29}$ & 26.7 & 11.8 & 5916 & 4593 & 0.264 & 0.000 & (0.170, 0.377)\\
   $\bb_{30}$ & 27.3 & 12.1 & 5250 & 4759 & 0.256 & 0.000 & (0.155, 0.378)\\
   $\bb_{31}$ & 30.1 & 13.3 & 5758 & 5285 & 0.247 & 0.000 & (0.141, 0.356)\\
   $\bb_{32}$ & 31.8 & 14.1 & 5266 & 5607 & 0.240 & 0.000 & (0.136, 0.359)\\
   $\bb_{33}$ & 32.1 & 14.2 & 4925 & 5218 & 0.234 & 0.000 & (0.122, 0.355)\\
   $\bb_{34}$ & 31.7 & 14.0 & 4323 & 4517 & 0.228 & 0.000 & (0.107, 0.355)\\
   $\bb_{35}$ & 28.8 & 12.7 & 3022 & 3135 & 0.220 & 0.000 & (0.083, 0.368)\\
   $\bb_{36}$ & 27.0 & 12.0 & 2452 & 2467 & 0.215 & 0.000 & (0.072, 0.386)\\
   $\bb_{37}$ & 28.1 & 12.4 & 2318 & 2326 & 0.208 & 0.000 & (0.069, 0.383)\\
   $\bb_{38}$ & 29.7 & 13.1 & 2330 & 2353 & 0.200 & 0.000 & (0.058, 0.370)\\
   $\bb_{39}$ & 35.2 & 15.6 & 2954 & 3036 & 0.191 & 0.000 & (0.061, 0.341)\\
   $\bb_{40}$ & 39.6 & 17.5 & 3538 & 3725 & 0.185 & 0.000 & (0.057, 0.322)\\
  \midrule
   $\mathtt{WBATE}$ &  &  &  &  & 0.281 & 0.000 & (0.246, 0.313)\\
  \midrule
   $\mathtt{LBATE}$ &  &  &  &  & 0.333 &  & (0.264, 0.473)\\
  \bottomrule\bottomrule
\end{tabular}
}
    \caption{Intention-to-Treat Effects (SPP Application, Fuzzy Bandwidth Selector).}
    \label{tab:sa-empapp_main_itt_bwmain}
\end{table}

\begin{table}[p]
    \centering
    \resizebox{.8\textwidth}{!}{\begin{tabular}{lccccccc}
  \toprule\toprule
   & $h_1$ & $h_2$ & $N_{\mathrm{Co}}$ & $N_{\mathrm{Tr}}$ & Estimate & $p$-value & 95\% CI \\
  \midrule
   $\bb_{1}$ & 28.2 & 12.5 & 9704 & 3687 & 0.550 & 0.000 & (0.496, 0.650)\\
   $\bb_{2}$ & 26.0 & 11.5 & 8572 & 3612 & 0.547 & 0.000 & (0.498, 0.650)\\
   $\bb_{3}$ & 24.9 & 11.0 & 8590 & 3720 & 0.543 & 0.000 & (0.495, 0.644)\\
   $\bb_{4}$ & 23.2 & 10.3 & 8397 & 3724 & 0.539 & 0.000 & (0.481, 0.630)\\
   $\bb_{5}$ & 21.8 & 9.6 & 7403 & 3568 & 0.529 & 0.000 & (0.463, 0.613)\\
   $\bb_{6}$ & 20.0 & 8.9 & 6977 & 3427 & 0.516 & 0.000 & (0.440, 0.594)\\
   $\bb_{7}$ & 19.7 & 8.7 & 6633 & 3404 & 0.509 & 0.000 & (0.424, 0.579)\\
   $\bb_{8}$ & 21.3 & 9.4 & 8437 & 4032 & 0.510 & 0.000 & (0.422, 0.569)\\
   $\bb_{9}$ & 21.9 & 9.7 & 8859 & 4218 & 0.514 & 0.000 & (0.419, 0.564)\\
   $\bb_{10}$ & 22.2 & 9.8 & 9828 & 4469 & 0.526 & 0.000 & (0.431, 0.575)\\
   $\bb_{11}$ & 22.3 & 9.9 & 9906 & 4584 & 0.537 & 0.000 & (0.452, 0.594)\\
   $\bb_{12}$ & 22.3 & 9.9 & 9949 & 4710 & 0.542 & 0.000 & (0.465, 0.604)\\
   $\bb_{13}$ & 20.6 & 9.1 & 7995 & 4229 & 0.552 & 0.000 & (0.471, 0.615)\\
   $\bb_{14}$ & 20.7 & 9.2 & 8168 & 4350 & 0.561 & 0.000 & (0.477, 0.616)\\
   $\bb_{15}$ & 21.0 & 9.3 & 8370 & 4493 & 0.563 & 0.000 & (0.478, 0.614)\\
   $\bb_{16}$ & 20.8 & 9.2 & 8278 & 4560 & 0.567 & 0.000 & (0.489, 0.624)\\
   $\bb_{17}$ & 19.7 & 8.7 & 7544 & 4057 & 0.575 & 0.000 & (0.499, 0.638)\\
   $\bb_{18}$ & 19.6 & 8.7 & 7997 & 3623 & 0.575 & 0.000 & (0.497, 0.643)\\
   $\bb_{19}$ & 20.4 & 9.0 & 9592 & 3367 & 0.576 & 0.000 & (0.493, 0.651)\\
   $\bb_{20}$ & 20.0 & 8.9 & 9225 & 2755 & 0.574 & 0.000 & (0.471, 0.670)\\
   $\bb_{21}$ & 20.9 & 9.3 & 10677 & 2442 & 0.573 & 0.000 & (0.428, 0.759)\\
   $\bb_{22}$ & 22.0 & 9.8 & 10830 & 2843 & 0.576 & 0.000 & (0.470, 0.709)\\
   $\bb_{23}$ & 21.3 & 9.4 & 8679 & 2871 & 0.574 & 0.000 & (0.477, 0.689)\\
   $\bb_{24}$ & 21.7 & 9.6 & 8114 & 3140 & 0.575 & 0.000 & (0.478, 0.676)\\
   $\bb_{25}$ & 21.8 & 9.6 & 6858 & 3223 & 0.580 & 0.000 & (0.486, 0.679)\\
   $\bb_{26}$ & 24.3 & 10.8 & 7784 & 3867 & 0.588 & 0.000 & (0.494, 0.672)\\
   $\bb_{27}$ & 27.4 & 12.1 & 8856 & 4564 & 0.597 & 0.000 & (0.498, 0.663)\\
   $\bb_{28}$ & 26.9 & 11.9 & 7294 & 4540 & 0.603 & 0.000 & (0.504, 0.675)\\
   $\bb_{29}$ & 26.7 & 11.8 & 5916 & 4593 & 0.613 & 0.000 & (0.515, 0.691)\\
   $\bb_{30}$ & 27.3 & 12.1 & 5250 & 4759 & 0.624 & 0.000 & (0.526, 0.705)\\
   $\bb_{31}$ & 30.1 & 13.3 & 5758 & 5285 & 0.637 & 0.000 & (0.541, 0.712)\\
   $\bb_{32}$ & 31.8 & 14.1 & 5266 & 5607 & 0.649 & 0.000 & (0.557, 0.726)\\
   $\bb_{33}$ & 32.1 & 14.2 & 4925 & 5218 & 0.660 & 0.000 & (0.568, 0.744)\\
   $\bb_{34}$ & 31.7 & 14.0 & 4323 & 4517 & 0.673 & 0.000 & (0.581, 0.768)\\
   $\bb_{35}$ & 28.8 & 12.7 & 3022 & 3135 & 0.691 & 0.000 & (0.591, 0.805)\\
   $\bb_{36}$ & 27.0 & 12.0 & 2452 & 2467 & 0.707 & 0.000 & (0.606, 0.840)\\
   $\bb_{37}$ & 28.1 & 12.4 & 2318 & 2326 & 0.716 & 0.000 & (0.617, 0.853)\\
   $\bb_{38}$ & 29.7 & 13.1 & 2330 & 2353 & 0.720 & 0.000 & (0.618, 0.856)\\
   $\bb_{39}$ & 35.2 & 15.6 & 2954 & 3036 & 0.719 & 0.000 & (0.626, 0.845)\\
   $\bb_{40}$ & 39.6 & 17.5 & 3538 & 3725 & 0.719 & 0.000 & (0.635, 0.845)\\
  \midrule
   $\mathtt{WBATE}$ &  &  &  &  & 0.593 & 0.000 & (0.567, 0.619)\\
  \midrule
   $\mathtt{LBATE}$ &  &  &  &  & 0.720 &  & (0.636, 0.855)\\
  \bottomrule\bottomrule
\end{tabular}
}
    \caption{First-Stage Effects (SPP Application, Fuzzy Bandwidth Selector).}
    \label{tab:sa-empapp_main_fs_bwmain}
\end{table}

\begin{table}[p]
    \centering
    \resizebox{.8\textwidth}{!}{\begin{tabular}{lccccccc}
  \toprule\toprule
   & $h_1$ & $h_2$ & $N_{\mathrm{Co}}$ & $N_{\mathrm{Tr}}$ & Estimate & $p$-value & 95\% CI \\
  \midrule
   $\bb_{1}$ & 28.2 & 12.5 & 9704 & 3687 & 0.565 & 0.000 & (0.381, 0.708)\\
   $\bb_{2}$ & 26.0 & 11.5 & 8572 & 3612 & 0.572 & 0.000 & (0.401, 0.727)\\
   $\bb_{3}$ & 24.9 & 11.0 & 8590 & 3720 & 0.581 & 0.000 & (0.407, 0.726)\\
   $\bb_{4}$ & 23.2 & 10.3 & 8397 & 3724 & 0.572 & 0.000 & (0.365, 0.700)\\
   $\bb_{5}$ & 21.8 & 9.6 & 7403 & 3568 & 0.551 & 0.000 & (0.308, 0.667)\\
   $\bb_{6}$ & 20.0 & 8.9 & 6977 & 3427 & 0.528 & 0.000 & (0.261, 0.655)\\
   $\bb_{7}$ & 19.7 & 8.7 & 6633 & 3404 & 0.533 & 0.000 & (0.275, 0.679)\\
   $\bb_{8}$ & 21.3 & 9.4 & 8437 & 4032 & 0.567 & 0.000 & (0.338, 0.708)\\
   $\bb_{9}$ & 21.9 & 9.7 & 8859 & 4218 & 0.595 & 0.000 & (0.387, 0.746)\\
   $\bb_{10}$ & 22.2 & 9.8 & 9828 & 4469 & 0.610 & 0.000 & (0.428, 0.771)\\
   $\bb_{11}$ & 22.3 & 9.9 & 9906 & 4584 & 0.613 & 0.000 & (0.446, 0.772)\\
   $\bb_{12}$ & 22.3 & 9.9 & 9949 & 4710 & 0.602 & 0.000 & (0.438, 0.756)\\
   $\bb_{13}$ & 20.6 & 9.1 & 7995 & 4229 & 0.583 & 0.000 & (0.401, 0.741)\\
   $\bb_{14}$ & 20.7 & 9.2 & 8168 & 4350 & 0.562 & 0.000 & (0.363, 0.695)\\
   $\bb_{15}$ & 21.0 & 9.3 & 8370 & 4493 & 0.549 & 0.000 & (0.351, 0.677)\\
   $\bb_{16}$ & 20.8 & 9.2 & 8278 & 4560 & 0.549 & 0.000 & (0.378, 0.698)\\
   $\bb_{17}$ & 19.7 & 8.7 & 7544 & 4057 & 0.556 & 0.000 & (0.441, 0.736)\\
   $\bb_{18}$ & 19.6 & 8.7 & 7997 & 3623 & 0.567 & 0.000 & (0.482, 0.734)\\
   $\bb_{19}$ & 20.4 & 9.0 & 9592 & 3367 & 0.579 & 0.000 & (0.490, 0.731)\\
   $\bb_{20}$ & 20.0 & 8.9 & 9225 & 2755 & 0.569 & 0.000 & (0.446, 0.717)\\
   $\bb_{21}$ & 20.9 & 9.3 & 10677 & 2442 & 0.548 & 0.000 & (0.343, 0.715)\\
   $\bb_{22}$ & 22.0 & 9.8 & 10830 & 2843 & 0.533 & 0.000 & (0.365, 0.652)\\
   $\bb_{23}$ & 21.3 & 9.4 & 8679 & 2871 & 0.519 & 0.000 & (0.364, 0.649)\\
   $\bb_{24}$ & 21.7 & 9.6 & 8114 & 3140 & 0.510 & 0.000 & (0.378, 0.667)\\
   $\bb_{25}$ & 21.8 & 9.6 & 6858 & 3223 & 0.499 & 0.000 & (0.389, 0.689)\\
   $\bb_{26}$ & 24.3 & 10.8 & 7784 & 3867 & 0.479 & 0.000 & (0.380, 0.666)\\
   $\bb_{27}$ & 27.4 & 12.1 & 8856 & 4564 & 0.455 & 0.000 & (0.347, 0.618)\\
   $\bb_{28}$ & 26.9 & 11.9 & 7294 & 4540 & 0.443 & 0.000 & (0.324, 0.613)\\
   $\bb_{29}$ & 26.7 & 11.8 & 5916 & 4593 & 0.430 & 0.000 & (0.301, 0.606)\\
   $\bb_{30}$ & 27.3 & 12.1 & 5250 & 4759 & 0.410 & 0.000 & (0.271, 0.596)\\
   $\bb_{31}$ & 30.1 & 13.3 & 5758 & 5285 & 0.388 & 0.000 & (0.242, 0.551)\\
   $\bb_{32}$ & 31.8 & 14.1 & 5266 & 5607 & 0.370 & 0.000 & (0.228, 0.544)\\
   $\bb_{33}$ & 32.1 & 14.2 & 4925 & 5218 & 0.354 & 0.000 & (0.201, 0.526)\\
   $\bb_{34}$ & 31.7 & 14.0 & 4323 & 4517 & 0.338 & 0.000 & (0.172, 0.512)\\
   $\bb_{35}$ & 28.8 & 12.7 & 3022 & 3135 & 0.319 & 0.000 & (0.133, 0.513)\\
   $\bb_{36}$ & 27.0 & 12.0 & 2452 & 2467 & 0.304 & 0.000 & (0.113, 0.520)\\
   $\bb_{37}$ & 28.1 & 12.4 & 2318 & 2326 & 0.291 & 0.000 & (0.105, 0.510)\\
   $\bb_{38}$ & 29.7 & 13.1 & 2330 & 2353 & 0.278 & 0.000 & (0.089, 0.492)\\
   $\bb_{39}$ & 35.2 & 15.6 & 2954 & 3036 & 0.266 & 0.000 & (0.091, 0.455)\\
   $\bb_{40}$ & 39.6 & 17.5 & 3538 & 3725 & 0.257 & 0.000 & (0.086, 0.426)\\
  \midrule
   $\mathtt{WBATE}$ &  &  &  &  & 0.485 & 0.000 & (0.433, 0.531)\\
  \midrule
   $\mathtt{LBATE}$ &  &  &  &  & 0.613 &  & (0.484, 0.781)\\
  \bottomrule\bottomrule
\end{tabular}
}
    \caption{Fuzzy Effects (SPP Application, Fuzzy Bandwidth Selector).}
    \label{tab:sa-empapp_main_fuzzy_bwmain}
\end{table}

\begin{table}[p]
    \centering
    \resizebox{.8\textwidth}{!}{\begin{tabular}{lccccccc}
  \toprule\toprule
   & $h_1$ & $h_2$ & $N_{\mathrm{Co}}$ & $N_{\mathrm{Tr}}$ & Estimate & $p$-value & 95\% CI \\
  \midrule
   $\bb_{1}$ & 27.2 & 12.1 & 8688 & 3446 & -0.027 & 0.172 & (-0.153, 0.058)\\
   $\bb_{2}$ & 25.8 & 11.5 & 8428 & 3552 & -0.032 & 0.116 & (-0.157, 0.049)\\
   $\bb_{3}$ & 25.4 & 11.3 & 9246 & 3841 & -0.030 & 0.108 & (-0.153, 0.046)\\
   $\bb_{4}$ & 25.4 & 11.3 & 10066 & 4154 & -0.021 & 0.215 & (-0.134, 0.056)\\
   $\bb_{5}$ & 24.4 & 10.8 & 9822 & 4176 & -0.015 & 0.402 & (-0.121, 0.068)\\
   $\bb_{6}$ & 22.7 & 10.1 & 8801 & 3951 & -0.008 & 0.739 & (-0.107, 0.086)\\
   $\bb_{7}$ & 21.9 & 9.7 & 8262 & 3891 & -0.001 & 0.985 & (-0.098, 0.097)\\
   $\bb_{8}$ & 21.4 & 9.5 & 8372 & 3993 & -0.002 & 0.678 & (-0.114, 0.086)\\
   $\bb_{9}$ & 21.6 & 9.6 & 8604 & 4098 & -0.010 & 0.356 & (-0.131, 0.070)\\
   $\bb_{10}$ & 20.9 & 9.3 & 8049 & 3966 & -0.012 & 0.262 & (-0.140, 0.064)\\
   $\bb_{11}$ & 20.7 & 9.2 & 7952 & 3986 & -0.011 & 0.456 & (-0.127, 0.076)\\
   $\bb_{12}$ & 21.5 & 9.5 & 8766 & 4328 & -0.008 & 0.935 & (-0.100, 0.095)\\
   $\bb_{13}$ & 21.0 & 9.3 & 8090 & 4265 & -0.001 & 0.850 & (-0.093, 0.106)\\
   $\bb_{14}$ & 21.0 & 9.3 & 8207 & 4358 & 0.002 & 0.848 & (-0.090, 0.103)\\
   $\bb_{15}$ & 20.4 & 9.0 & 8059 & 4332 & 0.000 & 0.908 & (-0.092, 0.099)\\
   $\bb_{16}$ & 21.6 & 9.6 & 8936 & 4750 & 0.001 & 0.550 & (-0.071, 0.106)\\
   $\bb_{17}$ & 22.5 & 10.0 & 10627 & 4667 & 0.002 & 0.510 & (-0.057, 0.088)\\
   $\bb_{18}$ & 20.1 & 8.9 & 8788 & 3746 & -0.002 & 0.472 & (-0.058, 0.094)\\
   $\bb_{19}$ & 18.7 & 8.3 & 7730 & 3014 & -0.005 & 0.816 & (-0.084, 0.072)\\
   $\bb_{20}$ & 20.1 & 8.9 & 9828 & 2825 & -0.009 & 0.613 & (-0.102, 0.073)\\
   $\bb_{21}$ & 19.9 & 8.8 & 9436 & 2206 & -0.004 & 0.505 & (-0.117, 0.184)\\
   $\bb_{22}$ & 20.8 & 9.2 & 9760 & 2625 & 0.001 & 0.337 & (-0.075, 0.145)\\
   $\bb_{23}$ & 17.8 & 7.9 & 5890 & 2194 & 0.013 & 0.061 & (-0.041, 0.177)\\
   $\bb_{24}$ & 17.7 & 7.9 & 4911 & 2316 & 0.019 & 0.035 & (-0.032, 0.182)\\
   $\bb_{25}$ & 18.5 & 8.2 & 4641 & 2566 & 0.021 & 0.056 & (-0.038, 0.169)\\
   $\bb_{26}$ & 20.2 & 9.0 & 4845 & 3018 & 0.020 & 0.114 & (-0.046, 0.149)\\
   $\bb_{27}$ & 22.0 & 9.8 & 5110 & 3458 & 0.015 & 0.238 & (-0.057, 0.130)\\
   $\bb_{28}$ & 22.5 & 10.0 & 4215 & 3666 & 0.013 & 0.312 & (-0.065, 0.130)\\
   $\bb_{29}$ & 23.0 & 10.2 & 3754 & 3828 & 0.009 & 0.413 & (-0.076, 0.133)\\
   $\bb_{30}$ & 24.4 & 10.8 & 3836 & 3936 & 0.005 & 0.556 & (-0.084, 0.125)\\
   $\bb_{31}$ & 29.3 & 13.0 & 5139 & 5095 & 0.004 & 0.683 & (-0.080, 0.104)\\
   $\bb_{32}$ & 30.0 & 13.3 & 4744 & 4958 & 0.000 & 0.963 & (-0.093, 0.096)\\
   $\bb_{33}$ & 27.7 & 12.3 & 3587 & 3720 & -0.004 & 0.799 & (-0.116, 0.098)\\
   $\bb_{34}$ & 26.1 & 11.6 & 2714 & 2786 & -0.004 & 0.667 & (-0.134, 0.101)\\
   $\bb_{35}$ & 23.1 & 10.3 & 1825 & 1842 & -0.006 & 0.639 & (-0.155, 0.113)\\
   $\bb_{36}$ & 21.2 & 9.4 & 1396 & 1393 & -0.009 & 0.528 & (-0.171, 0.112)\\
   $\bb_{37}$ & 21.9 & 9.7 & 1350 & 1316 & -0.012 & 0.429 & (-0.173, 0.101)\\
   $\bb_{38}$ & 23.1 & 10.2 & 1313 & 1295 & -0.013 & 0.382 & (-0.168, 0.093)\\
   $\bb_{39}$ & 29.5 & 13.1 & 2021 & 2033 & -0.010 & 0.423 & (-0.133, 0.077)\\
   $\bb_{40}$ & 27.8 & 12.3 & 1556 & 1506 & -0.011 & 0.516 & (-0.133, 0.086)\\
  \midrule
   $\mathtt{WBATE}$ &  &  &  &  & -0.004 & 0.951 & (-0.030, 0.028)\\
  \midrule
   $\mathtt{LBATE}$ &  &  &  &  & 0.021 &  & (-0.033, 0.185)\\
  \bottomrule\bottomrule
\end{tabular}
}
    \caption{Covariate Balance for Mother's Education (SPP Application, Fuzzy Bandwidth Selector).}
    \label{tab:sa-empapp_covbal_itt_bwmain}
\end{table}

\clearpage

\clearpage

\begin{table}[p]
    \centering
    \resizebox{.8\textwidth}{!}{\begin{tabular}{lccccccc}
  \toprule\toprule
   & $h_1$ & $h_2$ & $N_{\mathrm{Co}}$ & $N_{\mathrm{Tr}}$ & Estimate & $p$-value & 95\% CI \\
  \midrule
   $\bb_{1}$ & 28.2 & 12.5 & 9704 &  & 0.371 & 0.000 & (0.299, 0.456)\\
   $\bb_{2}$ & 26.0 & 11.5 & 8572 &  & 0.364 & 0.000 & (0.287, 0.444)\\
   $\bb_{3}$ & 24.9 & 11.0 & 8590 &  & 0.356 & 0.000 & (0.286, 0.439)\\
   $\bb_{4}$ & 23.2 & 10.3 & 8397 &  & 0.357 & 0.000 & (0.290, 0.448)\\
   $\bb_{5}$ & 21.8 & 9.6 & 7403 &  & 0.360 & 0.000 & (0.297, 0.462)\\
   $\bb_{6}$ & 20.0 & 8.9 & 6977 &  & 0.362 & 0.000 & (0.291, 0.465)\\
   $\bb_{7}$ & 19.7 & 8.7 & 6633 &  & 0.355 & 0.000 & (0.277, 0.447)\\
   $\bb_{8}$ & 21.3 & 9.4 & 8437 &  & 0.346 & 0.000 & (0.278, 0.429)\\
   $\bb_{9}$ & 21.9 & 9.7 & 8859 &  & 0.342 & 0.000 & (0.275, 0.421)\\
   $\bb_{10}$ & 22.2 & 9.8 & 9828 &  & 0.343 & 0.000 & (0.277, 0.420)\\
   $\bb_{11}$ & 22.3 & 9.9 & 9906 &  & 0.351 & 0.000 & (0.286, 0.427)\\
   $\bb_{12}$ & 22.3 & 9.9 & 9949 &  & 0.363 & 0.000 & (0.302, 0.444)\\
   $\bb_{13}$ & 20.6 & 9.1 & 7995 &  & 0.379 & 0.000 & (0.316, 0.474)\\
   $\bb_{14}$ & 20.7 & 9.2 & 8168 &  & 0.391 & 0.000 & (0.342, 0.500)\\
   $\bb_{15}$ & 21.0 & 9.3 & 8370 &  & 0.400 & 0.000 & (0.355, 0.510)\\
   $\bb_{16}$ & 20.8 & 9.2 & 8278 &  & 0.403 & 0.000 & (0.346, 0.502)\\
   $\bb_{17}$ & 19.7 & 8.7 & 7544 &  & 0.403 & 0.000 & (0.331, 0.473)\\
   $\bb_{18}$ & 19.6 & 8.7 & 7997 &  & 0.396 & 0.000 & (0.334, 0.443)\\
   $\bb_{19}$ & 20.4 & 9.0 & 9592 &  & 0.391 & 0.000 & (0.345, 0.434)\\
   $\bb_{20}$ & 20.0 & 8.9 & 9225 &  & 0.394 & 0.000 & (0.360, 0.441)\\
   $\bb_{21}$ & 20.9 & 9.3 & 10677 &  & 0.399 & 0.000 & (0.373, 0.442)\\
   $\bb_{22}$ & 22.0 & 9.8 & 10830 &  & 0.411 & 0.000 & (0.382, 0.455)\\
   $\bb_{23}$ & 21.3 & 9.4 & 8679 &  & 0.422 & 0.000 & (0.379, 0.466)\\
   $\bb_{24}$ & 21.7 & 9.6 & 8114 &  & 0.432 & 0.000 & (0.375, 0.475)\\
   $\bb_{25}$ & 21.8 & 9.6 & 6858 &  & 0.445 & 0.000 & (0.371, 0.486)\\
   $\bb_{26}$ & 24.3 & 10.8 & 7784 &  & 0.460 & 0.000 & (0.385, 0.499)\\
   $\bb_{27}$ & 27.4 & 12.1 & 8856 &  & 0.475 & 0.000 & (0.407, 0.517)\\
   $\bb_{28}$ & 26.9 & 11.9 & 7294 &  & 0.485 & 0.000 & (0.408, 0.534)\\
   $\bb_{29}$ & 26.7 & 11.8 & 5916 &  & 0.494 & 0.000 & (0.410, 0.552)\\
   $\bb_{30}$ & 27.3 & 12.1 & 5250 &  & 0.508 & 0.000 & (0.411, 0.572)\\
   $\bb_{31}$ & 30.1 & 13.3 & 5758 &  & 0.525 & 0.000 & (0.434, 0.591)\\
   $\bb_{32}$ & 31.8 & 14.1 & 5266 &  & 0.541 & 0.000 & (0.441, 0.608)\\
   $\bb_{33}$ & 32.1 & 14.2 & 4925 &  & 0.556 & 0.000 & (0.455, 0.630)\\
   $\bb_{34}$ & 31.7 & 14.0 & 4323 &  & 0.571 & 0.000 & (0.466, 0.654)\\
   $\bb_{35}$ & 28.8 & 12.7 & 3022 &  & 0.587 & 0.000 & (0.466, 0.684)\\
   $\bb_{36}$ & 27.0 & 12.0 & 2452 &  & 0.604 & 0.000 & (0.475, 0.720)\\
   $\bb_{37}$ & 28.1 & 12.4 & 2318 &  & 0.622 & 0.000 & (0.492, 0.743)\\
   $\bb_{38}$ & 29.7 & 13.1 & 2330 &  & 0.639 & 0.000 & (0.512, 0.765)\\
   $\bb_{39}$ & 35.2 & 15.6 & 2954 &  & 0.655 & 0.000 & (0.546, 0.772)\\
   $\bb_{40}$ & 39.6 & 17.5 & 3538 &  & 0.667 & 0.000 & (0.570, 0.783)\\
  \midrule
   $\mathtt{WBATE}$ &  &  &  &  & 0.448 & 0.000 & (0.425, 0.474)\\
  \midrule
   $\mathtt{LBATE}$ &  &  &  &  & 0.667 &  & (0.568, 0.784)\\
  \bottomrule\bottomrule
\end{tabular}
}
    \caption{Control-Side Baseline Outcome (SPP Application, Fuzzy Bandwidth Selector).}
    \label{tab:sa-empapp_main_itt0_bwmain}
\end{table}

\begin{table}[p]
    \centering
    \resizebox{.8\textwidth}{!}{\begin{tabular}{lcccccc}
  \toprule\toprule
   & $h_1$ & $h_2$ & $N_{\mathrm{Co}}$ & $N_{\mathrm{Tr}}$ & ITT / ITT.0 (\%) & Fuzzy / ITT.0 (\%) \\
  \midrule
   $\bb_{1}$ & 28.2 & 12.5 & 9704 & 3687 & 83.801 & 152.231\\
   $\bb_{2}$ & 26.0 & 11.5 & 8572 & 3612 & 85.950 & 157.238\\
   $\bb_{3}$ & 24.9 & 11.0 & 8590 & 3720 & 88.686 & 163.182\\
   $\bb_{4}$ & 23.2 & 10.3 & 8397 & 3724 & 86.320 & 160.201\\
   $\bb_{5}$ & 21.8 & 9.6 & 7403 & 3568 & 80.977 & 153.111\\
   $\bb_{6}$ & 20.0 & 8.9 & 6977 & 3427 & 75.410 & 146.062\\
   $\bb_{7}$ & 19.7 & 8.7 & 6633 & 3404 & 76.445 & 150.264\\
   $\bb_{8}$ & 21.3 & 9.4 & 8437 & 4032 & 83.668 & 164.054\\
   $\bb_{9}$ & 21.9 & 9.7 & 8859 & 4218 & 89.489 & 174.020\\
   $\bb_{10}$ & 22.2 & 9.8 & 9828 & 4469 & 93.492 & 177.775\\
   $\bb_{11}$ & 22.3 & 9.9 & 9906 & 4584 & 94.019 & 174.994\\
   $\bb_{12}$ & 22.3 & 9.9 & 9949 & 4710 & 90.043 & 166.056\\
   $\bb_{13}$ & 20.6 & 9.1 & 7995 & 4229 & 85.004 & 153.958\\
   $\bb_{14}$ & 20.7 & 9.2 & 8168 & 4350 & 80.630 & 143.619\\
   $\bb_{15}$ & 21.0 & 9.3 & 8370 & 4493 & 77.426 & 137.434\\
   $\bb_{16}$ & 20.8 & 9.2 & 8278 & 4560 & 77.256 & 136.138\\
   $\bb_{17}$ & 19.7 & 8.7 & 7544 & 4057 & 79.328 & 138.034\\
   $\bb_{18}$ & 19.6 & 8.7 & 7997 & 3623 & 82.308 & 143.110\\
   $\bb_{19}$ & 20.4 & 9.0 & 9592 & 3367 & 85.368 & 148.263\\
   $\bb_{20}$ & 20.0 & 8.9 & 9225 & 2755 & 83.046 & 144.554\\
   $\bb_{21}$ & 20.9 & 9.3 & 10677 & 2442 & 78.747 & 137.406\\
   $\bb_{22}$ & 22.0 & 9.8 & 10830 & 2843 & 74.724 & 129.677\\
   $\bb_{23}$ & 21.3 & 9.4 & 8679 & 2871 & 70.735 & 123.194\\
   $\bb_{24}$ & 21.7 & 9.6 & 8114 & 3140 & 67.914 & 118.071\\
   $\bb_{25}$ & 21.8 & 9.6 & 6858 & 3223 & 65.063 & 112.204\\
   $\bb_{26}$ & 24.3 & 10.8 & 7784 & 3867 & 61.248 & 104.183\\
   $\bb_{27}$ & 27.4 & 12.1 & 8856 & 4564 & 57.067 & 95.654\\
   $\bb_{28}$ & 26.9 & 11.9 & 7294 & 4540 & 55.052 & 91.338\\
   $\bb_{29}$ & 26.7 & 11.8 & 5916 & 4593 & 53.365 & 87.115\\
   $\bb_{30}$ & 27.3 & 12.1 & 5250 & 4759 & 50.399 & 80.804\\
   $\bb_{31}$ & 30.1 & 13.3 & 5758 & 5285 & 47.095 & 73.880\\
   $\bb_{32}$ & 31.8 & 14.1 & 5266 & 5607 & 44.339 & 68.325\\
   $\bb_{33}$ & 32.1 & 14.2 & 4925 & 5218 & 42.053 & 63.694\\
   $\bb_{34}$ & 31.7 & 14.0 & 4323 & 4517 & 39.874 & 59.262\\
   $\bb_{35}$ & 28.8 & 12.7 & 3022 & 3135 & 37.509 & 54.321\\
   $\bb_{36}$ & 27.0 & 12.0 & 2452 & 2467 & 35.590 & 50.375\\
   $\bb_{37}$ & 28.1 & 12.4 & 2318 & 2326 & 33.480 & 46.767\\
   $\bb_{38}$ & 29.7 & 13.1 & 2330 & 2353 & 31.284 & 43.456\\
   $\bb_{39}$ & 35.2 & 15.6 & 2954 & 3036 & 29.205 & 40.612\\
   $\bb_{40}$ & 39.6 & 17.5 & 3538 & 3725 & 27.726 & 38.547\\
  \midrule
   $\mathtt{WBATE}$ &  &  &  &  & 62.655 & 108.222\\
  \midrule
   $\mathtt{LBATE}$ &  &  &  &  & 49.984 & 91.972\\
  \bottomrule\bottomrule
\end{tabular}
}
    \caption{Treatment Effect Relative to Control-Side Baseline (SPP Application, Fuzzy Bandwidth Selector).}
    \label{tab:sa-empapp_main_percent_bwmain}
\end{table}

\clearpage

\section{Simulation Study}

The score $\bX_i = (X_{1i}, X_{2i})^\top$ is generated from the distribution $(100\cdot\mathsf{Beta}(3,4)-25,\;100\cdot\mathsf{Beta}(3,4)-25)^\top$ with independent components, which approximately reproduces salient features of the SPP dataset. We employ the same L-shaped treatment assignment boundary as in the empirical application, featuring a kink at $\bx=(0,0)^\top$. To reduce computation burden we consider $21$ evaluation points with the kink corresponding to the evaluation grid point $\bb_{11}$.

The simulation design sets $n=10,000$ and uses $5,000$ Monte Carlo replications. We evaluate performance in terms of point estimation and uncertainty quantification, both pointwise and uniformly over the boundary $\B$. The Monte Carlo design jointly generates the reduced-form outcome and the treatment status, so that the sharp intention-to-treat (ITT), first-stage (FS), and fuzzy analyses can all be performed on the same simulated samples.

\subsection{Data-Generating Process}

For each Monte Carlo replication, draw $\bX_i=(X_{1i},X_{2i})^\top$ independently from the distribution described above and set $T_i=\Indicator(\bX_i\in\A_1)$. We consider DGP polynomial specifications of order $s\in\{1,2,3\}$, using the monomial vector $\br_s(\cdot)$, evaluated at the unscaled score. Thus, for $d=2$, $\br_1(\bx)=(1,x_1,x_2)^\top$ and $\br_2(\bx)=(1,x_1,x_2,x_1^2,x_1x_2,x_2^2)^\top$, with $\br_3(\bx)$ defined analogously by adding all cubic monomials. For each $t\in\{0,1\}$ and $i=1,\ldots,n$, define
\begin{align*}
    \mu_{Y,t}(\bX_i)
    &= \br_s(\bX_i)^\top\bbeta_{Y,t},\\
    \mu_{W,t}(\bX_i)
    &= \Lambda(\eta_{W,t}(\bX_i)),
    \qquad
    \eta_{W,t}(\bX_i) = \br_s(\bX_i)^\top\bbeta_{W,t},
\end{align*}
where $\Lambda(a)=1/(1+\exp(-a))$. Given these conditional mean functions, draw $U_{W,t,i}\thicksim\mathsf{Uniform}(0,1)$ and $\varepsilon_{Y,t,i}\thicksim\mathsf{Normal}(0,\sigma_{Y,t}^2)$ independently across $i$ and $t$ and independently of $\bX_i$. The potential treatment statuses and treatment-indexed potential outcomes are
\begin{align*}
    W_i(t)
    &= \Indicator\{U_{W,t,i}\leq \mu_{W,t}(\bX_i)\},\\
    Y_i(t,r)
    &= \mu_{Y,t}(\bX_i) + \lambda_t\{r-\mu_{W,t}(\bX_i)\}
    + \varepsilon_{Y,t,i},
    \qquad r\in\{0,1\}.
\end{align*}
The reduced-form potential outcome induced by assignment $t$ is $Y_i(t,W_i(t))$. Equivalently,
\begin{align*}
    Y_i(t,W_i(t))
    &=
    \mu_{Y,t}(\bX_i)
    +\lambda_t\{W_i(t)-\mu_{W,t}(\bX_i)\}
    +\varepsilon_{Y,t,i}.
\end{align*}
The observed outcomes and treatment status are
\begin{align*}
    Y_i &= Y_i(0,W_i(0)) (1-T_i) + Y_i(1,W_i(1)) T_i,\\
    W_i &= W_i(0) (1-T_i) + W_i(1) T_i, \qquad  T_i = \Indicator(\bX_i \in \A_1),
\end{align*}
for $i=1,2,\ldots,n$.

In the reduced-form notation used for sharp estimation and inference, $Y_i(t)$ is shorthand for $Y_i(t,W_i(t))$. This centered specification preserves the reduced-form outcome surface because $\E[Y_i(t,W_i(t))|\bX_i=\bx]=\mu_{Y,t}(\bx)$, while allowing the potential outcome and potential treatment status equations to be correlated:
\begin{align*}
    \E[W_i(t)|\bX_i=\bx] = \mu_{W,t}(\bx)
    \qquad\text{and}\qquad
    \Cov(Y_i(t,W_i(t)),W_i(t)|\bX_i=\bx) = \lambda_t\mu_{W,t}(\bx)\{1-\mu_{W,t}(\bx)\}.
\end{align*}
Consequently, for each boundary point $\bb\in\B$,
\begin{align*}
    \tau_W(\bb)
    &=
    \E[W_i(1)-W_i(0)|\bX_i=\bb]
    =
    \mu_{W,1}(\bb)-\mu_{W,0}(\bb),\\
    \tau_Y(\bb)
    &=
    \E[Y_i(1,W_i(1))-Y_i(0,W_i(0))|\bX_i=\bb]
    =
    \mu_{Y,1}(\bb)-\mu_{Y,0}(\bb),\\
    \zeta(\bb)
    &=
    \frac{\tau_Y(\bb)}{\tau_W(\bb)}.
\end{align*}
In each Monte Carlo replication, the sharp analysis applied to $Y_i$ targets the reduced-form or ITT curve $\tau_Y(\bb)$, the same analysis applied to $W_i$ targets the first-stage curve $\tau_W(\bb)$, and the fuzzy analysis targets the Wald curve $\zeta(\bb)$. Thus, the ITT and fuzzy simulation results are different summaries of the same simulated samples rather than results from separate sharp and fuzzy DGPs.

\subsection{Calibration from the SPP data}

The DGP parameters are calibrated once from the SPP data and then held fixed across Monte Carlo replications. We use the following steps separately for $s\in\{1,2,3\}$.

\begin{enumerate}[label=\normalfont Step \arabic*.,leftmargin=*]
    \item Fix the DGP polynomial order $s\in\{1,2,3\}$ and define the side-specific calibration samples $\mathcal I_t=\{i:T_i=t\}$ and $N_t=|\mathcal I_t|$, for $t\in\{0,1\}$.

    \item For each side $t$, estimate the outcome regression coefficients by side-specific least squares:
    \[
        \widehat{\bbeta}_{Y,t}
        =
        \operatorname*{arg\,min}_{\bbeta\in\mathbb R^{\mathfrak p_s}}
        \sum_{i\in\mathcal I_t}\{Y_i-\br_s(\bX_i)^\top\bbeta\}^2.
    \]
    This gives the fitted reduced-form outcome surface $\widehat{\mu}_{Y,t}(\bx)=\br_s(\bx)^\top\widehat{\bbeta}_{Y,t}$.

    \item For each side $t$, estimate the treatment-status regression coefficients by side-specific logit:
    \[
        \widehat{\bbeta}_{W,t}
        =
        \operatorname*{arg\,max}_{\bbeta\in\mathbb R^{\mathfrak p_s}}
        \sum_{i\in\mathcal I_t}
        \left[
            W_i \br_s(\bX_i)^\top\bbeta
            -\log\{1+\exp(\br_s(\bX_i)^\top\bbeta)\}
        \right].
    \]
    This gives the fitted first-stage surface $\widehat{\mu}_{W,t}(\bx)=\Lambda(\br_s(\bx)^\top\widehat{\bbeta}_{W,t})$.

    \item Construct side-specific residuals in the calibration sample:
    \[
        \widehat{\varepsilon}_{Y,i}=Y_i-\widehat{\mu}_{Y,T_i}(\bX_i),
        \qquad
        \widehat{\varepsilon}_{W,i}=W_i-\widehat{\mu}_{W,T_i}(\bX_i).
    \]

    \item Estimate the residual dependence parameter and outcome-shock variance on each side:
\begin{align*}
    \widehat{\lambda}_t
    &=
    \frac{\sum_{i\in\mathcal I_t}\widehat{\varepsilon}_{Y,i}\widehat{\varepsilon}_{W,i}}
    {\sum_{i\in\mathcal I_t}\widehat{\varepsilon}_{W,i}^2},
    \qquad
    \widehat{\sigma}_{Y,t}^2
    =
    \frac{1}{N_t}\sum_{i\in\mathcal I_t}
    \{\widehat{\varepsilon}_{Y,i}-\widehat{\lambda}_t\widehat{\varepsilon}_{W,i}\}^2,
\end{align*}

    \item Set the Monte Carlo DGP parameters equal to the calibration estimates: $\bbeta_{Y,t}=\widehat{\bbeta}_{Y,t}$, $\bbeta_{W,t}=\widehat{\bbeta}_{W,t}$, $\lambda_t=\widehat{\lambda}_t$, and $\sigma_{Y,t}^2=\widehat{\sigma}_{Y,t}^2$ for $t\in\{0,1\}$.

    \item Before running the Monte Carlo replications, compute the true target functions analytically from the calibrated conditional mean surfaces. For each evaluation point $\bb\in\B$,
    \[
        \tau_Y(\bb)
        =
        \br_s(\bb)^\top(\bbeta_{Y,1}-\bbeta_{Y,0}),
        \qquad
        \tau_W(\bb)
        =
        \Lambda(\br_s(\bb)^\top\bbeta_{W,1})
        -
        \Lambda(\br_s(\bb)^\top\bbeta_{W,0}),
        \qquad
        \zeta(\bb)=\frac{\tau_Y(\bb)}{\tau_W(\bb)}.
    \]
\end{enumerate}

This calibration reproduces the empirical residual covariance between the outcome and take-up equations on each side of the boundary, while keeping the marginal reduced-form and first-stage surfaces fixed by the fitted SPP models.

\subsection{Monte Carlo Implementation}

The following steps describe how to simulate from the calibrated model. They are repeated separately for each DGP polynomial order $s\in\{1,2,3\}$.

\begin{enumerate}[label=\normalfont Step \arabic*.,leftmargin=*]
    \item Fix the Monte Carlo inputs: the sample size $n$, the number of replications $M$, the evaluation grid $\B=\{\bb_1,\ldots,\bb_{21}\}$ with $\bb_{11}$ at the kink, the assignment regions $\A_0$ and $\A_1$, and the calibrated parameters
    \[
        \{\bbeta_{Y,t},\bbeta_{W,t},\lambda_t,\sigma_{Y,t}^2:t\in\{0,1\}\}.
    \]

    \item Before entering the Monte Carlo loop, evaluate the true target functions on the grid:
    \[
        \tau_Y(\bb_j),\qquad
        \tau_W(\bb_j),\qquad
        \zeta(\bb_j),
        \qquad j=1,\ldots,21,
    \]
    using the analytic formulas in the calibration step above. These target values are fixed across Monte Carlo replications.

    \item For replication $m=1,\ldots,M$, draw independent score vectors
    \[
        \bX_i=(X_{1i},X_{2i})^\top,
        \qquad
        X_{\ell i}=100B_{\ell i}-25,\quad
        B_{\ell i}\thicksim\mathsf{Beta}(3,4),
        \quad \ell\in\{1,2\},
    \]
    with $B_{1i}$ and $B_{2i}$ independent. Set $T_i=\Indicator(\bX_i\in\A_1)$.

    \item Form the design vectors $\br_s(\bX_i)$ and compute the side-specific conditional means
    \[
        \mu_{Y,t}(\bX_i)=\br_s(\bX_i)^\top\bbeta_{Y,t},
        \qquad
        \mu_{W,t}(\bX_i)=\Lambda(\br_s(\bX_i)^\top\bbeta_{W,t}),
        \qquad t\in\{0,1\}.
    \]

    \item Draw the treatment-status shocks and outcome shocks independently:
    \[
        U_{W,t,i}\thicksim\mathsf{Uniform}(0,1),
        \qquad
        \varepsilon_{Y,t,i}\thicksim\mathsf{Normal}(0,\sigma_{Y,t}^2),
        \qquad t\in\{0,1\}.
    \]
    Then construct the potential treatment statuses and treatment-indexed potential outcomes:
    \begin{align*}
        W_i(t)
        &=
        \Indicator\{U_{W,t,i}\leq \mu_{W,t}(\bX_i)\},\\
        Y_i(t,r)
        &=
        \mu_{Y,t}(\bX_i)
        +\lambda_t\{r-\mu_{W,t}(\bX_i)\}
        +\varepsilon_{Y,t,i},
        \qquad r\in\{0,1\}.
    \end{align*}
    Evaluate the reduced-form potential outcome at the simulated treatment status:
    \begin{align*}
        Y_i(t,W_i(t))
        &=
        \mu_{Y,t}(\bX_i)
        +\lambda_t\{W_i(t)-\mu_{W,t}(\bX_i)\}
        +\varepsilon_{Y,t,i}.
    \end{align*}

    \item Select the observed variables using the assignment indicator:
    \begin{align*}
        W_i
        &=
        W_i(0)(1-T_i)+W_i(1)T_i,\\
        Y_i
        &=
        Y_i(0,W_i(0))(1-T_i)+Y_i(1,W_i(1))T_i.
    \end{align*}
    The simulated dataset for replication $m$ is therefore $\{Y_i,W_i,T_i,\bX_i:i=1,\ldots,n\}$.

    \item \textbf{ITT Analysis}. Apply the sharp estimator to $Y_i$ using the simulated assignment rule and score vectors for the target $\tau_Y(\bb_j)$.

    \item \textbf{FS Analysis}. Apply the sharp estimator to $W_i$ using the simulated assignment rule and score vectors for the target $\tau_W(\bb_j)$.

    \item \textbf{Fuzzy Analysis}. Apply the fuzzy estimator using $Y_i$ as the outcome, $W_i$ as the treatment status, and using the simulated assignment rule and score vectors for the Wald target $\zeta(\bb_j)$.

    \item After all $M$ replications, compute Monte Carlo summaries by comparing the stored estimates and confidence sets with the fixed analytic targets. For each target and each grid point, report bias, standard deviation, root mean squared error, pointwise coverage, and average interval length. For uniform inference, report simultaneous coverage and average band length over the grid $\B$.
    Finally, also recover WBATE and LBATE and report their associated performance.
\end{enumerate}

We use the fuzzy simulations to study the finite-sample behavior of the reduced-form/ITT estimators, the first-stage estimators, and the fuzzy Wald estimator. A complier-average causal interpretation of $\zeta(\bb)$ would require additional assumptions that are not imposed in the simulations. In particular, one would need a local exclusion restriction ruling out a direct effect of assignment on the outcome except through treatment take-up, a monotonicity condition ruling out defiers near the boundary, and a nonzero first-stage effect. Under those conditions, the Wald curve $\zeta(\bb)$ can be interpreted as a local average treatment effect for compliers at the boundary point $\bb$.

\subsection{Results}

The following tables report Monte Carlo summaries separately for the reduced-form/ITT, first-stage, and fuzzy estimators, and separately by DGP polynomial order $s\in\{1,2,3\}$. We report two sets of results: one using the ITT bandwidth selector, matching the main empirical specification, and one using the fuzzy bandwidth selector.

The first set of simulation results uses the ITT bandwidth selector. Specifically:
\begin{itemize}
    \item Table~\ref{tab:sa-simuls-itt-s1} reports the reduced-form (ITT) simulation results for $s=1$.
    \item Table~\ref{tab:sa-simuls-itt-s2} reports the reduced-form (ITT) simulation results for $s=2$.
    \item Table~\ref{tab:sa-simuls-itt-s3} reports the reduced-form (ITT) simulation results for $s=3$.
    \item Table~\ref{tab:sa-simuls-fs-s1} reports the take-up (FS) simulation results for $s=1$.
    \item Table~\ref{tab:sa-simuls-fs-s2} reports the take-up (FS) simulation results for $s=2$.
    \item Table~\ref{tab:sa-simuls-fs-s3} reports the take-up (FS) simulation results for $s=3$.
    \item Table~\ref{tab:sa-simuls-fuzzy-s1} reports the fuzzy simulation results for $s=1$.
    \item Table~\ref{tab:sa-simuls-fuzzy-s2} reports the fuzzy simulation results for $s=2$.
    \item Table~\ref{tab:sa-simuls-fuzzy-s3} reports the fuzzy simulation results for $s=3$.
\end{itemize}

The second set of simulation results uses the fuzzy bandwidth selector. Specifically:
\begin{itemize}
    \item Table~\ref{tab:sa-simuls-itt-s1-bwmain} reports the reduced-form (ITT) simulation results for $s=1$.
    \item Table~\ref{tab:sa-simuls-itt-s2-bwmain} reports the reduced-form (ITT) simulation results for $s=2$.
    \item Table~\ref{tab:sa-simuls-itt-s3-bwmain} reports the reduced-form (ITT) simulation results for $s=3$.
    \item Table~\ref{tab:sa-simuls-fs-s1-bwmain} reports the take-up (FS) simulation results for $s=1$.
    \item Table~\ref{tab:sa-simuls-fs-s2-bwmain} reports the take-up (FS) simulation results for $s=2$.
    \item Table~\ref{tab:sa-simuls-fs-s3-bwmain} reports the take-up (FS) simulation results for $s=3$.
    \item Table~\ref{tab:sa-simuls-fuzzy-s1-bwmain} reports the fuzzy simulation results for $s=1$.
    \item Table~\ref{tab:sa-simuls-fuzzy-s2-bwmain} reports the fuzzy simulation results for $s=2$.
    \item Table~\ref{tab:sa-simuls-fuzzy-s3-bwmain} reports the fuzzy simulation results for $s=3$.
\end{itemize}
The columns report the average bandwidths, average bias, standard deviation, root mean squared error, empirical coverage, and average interval or band length. For the Uniform row, empirical coverage and average length refer to uniform coverage and average uniform band length.

\clearpage

\begin{table}[p]
    \centering
    \resizebox{.8\textwidth}{!}{\begin{tabular}{lccccrrrrr}
\toprule\toprule
 & $h_1$ & $h_2$ & $N_{\mathrm{Co}}$ & $N_{\mathrm{Tr}}$ & Bias & SD & RMSE & EC & IL \\
\midrule
$\bb_{1}$ & 13.220 & 13.220 & 732 & 1267 & 0.001 & 0.045 & 0.045 & 0.950 & 0.263 \\
$\bb_{2}$ & 13.525 & 13.525 & 765 & 1343 & 0.001 & 0.044 & 0.044 & 0.949 & 0.255 \\
$\bb_{3}$ & 13.740 & 13.740 & 787 & 1393 & 0.001 & 0.043 & 0.043 & 0.950 & 0.251 \\
$\bb_{4}$ & 13.857 & 13.857 & 808 & 1400 & 0.001 & 0.042 & 0.042 & 0.949 & 0.246 \\
$\bb_{5}$ & 13.904 & 13.904 & 869 & 1329 & 0.001 & 0.041 & 0.041 & 0.953 & 0.234 \\
$\bb_{6}$ & 13.890 & 13.890 & 924 & 1229 & 0.001 & 0.040 & 0.040 & 0.955 & 0.232 \\
$\bb_{7}$ & 13.811 & 13.811 & 959 & 1116 & 0.000 & 0.040 & 0.040 & 0.950 & 0.239 \\
$\bb_{8}$ & 13.639 & 13.639 & 968 & 989 & 0.001 & 0.042 & 0.042 & 0.949 & 0.253 \\
$\bb_{9}$ & 13.993 & 13.993 & 1024 & 928 & 0.000 & 0.046 & 0.046 & 0.948 & 0.271 \\
$\bb_{10}$ & 15.131 & 15.131 & 1147 & 961 & 0.000 & 0.051 & 0.051 & 0.948 & 0.306 \\
$\bb_{11}$ & 14.782 & 14.782 & 1091 & 803 & 0.000 & 0.066 & 0.066 & 0.936 & 0.425 \\
$\bb_{12}$ & 14.679 & 14.679 & 1101 & 949 & 0.000 & 0.048 & 0.048 & 0.941 & 0.288 \\
$\bb_{13}$ & 13.614 & 13.614 & 969 & 965 & 0.000 & 0.043 & 0.043 & 0.949 & 0.257 \\
$\bb_{14}$ & 13.811 & 13.811 & 951 & 1137 & 0.000 & 0.040 & 0.040 & 0.953 & 0.237 \\
$\bb_{15}$ & 13.870 & 13.870 & 889 & 1286 & 0.001 & 0.041 & 0.041 & 0.954 & 0.232 \\
$\bb_{16}$ & 13.837 & 13.837 & 805 & 1397 & 0.001 & 0.043 & 0.043 & 0.949 & 0.246 \\
$\bb_{17}$ & 13.678 & 13.678 & 781 & 1379 & 0.001 & 0.044 & 0.044 & 0.945 & 0.251 \\
$\bb_{18}$ & 13.289 & 13.289 & 739 & 1284 & 0.001 & 0.045 & 0.045 & 0.945 & 0.260 \\
$\bb_{19}$ & 13.011 & 13.011 & 697 & 1195 & 0.001 & 0.047 & 0.047 & 0.947 & 0.269 \\
$\bb_{20}$ & 13.077 & 13.077 & 671 & 1154 & 0.001 & 0.048 & 0.048 & 0.949 & 0.274 \\
$\bb_{21}$ & 13.330 & 13.330 & 651 & 1130 & 0.000 & 0.049 & 0.049 & 0.944 & 0.278 \\
\midrule
$\mathtt{Uniform}$ &  &  &  &  &  &  &  & 0.930 & 0.381 \\
\midrule
$\mathtt{WBATE}$ &  &  &  &  & 0.001 & 0.026 & 0.026 & 0.943 & 0.138 \\
\midrule
$\mathtt{LBATE}$ &  &  &  &  & 0.053 & 0.038 & 0.065 & 0.971 & 0.433 \\
\bottomrule\bottomrule
\end{tabular}
}
    \caption{Intention-to-Treat Effects (Simulations, $s=1$, ITT Bandwidth Selector).}
    \label{tab:sa-simuls-itt-s1}
\end{table}

\begin{table}[p]
    \centering
    \resizebox{.8\textwidth}{!}{\begin{tabular}{lccccrrrrr}
\toprule\toprule
 & $h_1$ & $h_2$ & $N_{\mathrm{Co}}$ & $N_{\mathrm{Tr}}$ & Bias & SD & RMSE & EC & IL \\
\midrule
$\bb_{1}$ & 13.191 & 13.191 & 728 & 1262 & 0.001 & 0.045 & 0.045 & 0.949 & 0.263 \\
$\bb_{2}$ & 13.494 & 13.494 & 762 & 1338 & 0.001 & 0.044 & 0.044 & 0.951 & 0.256 \\
$\bb_{3}$ & 13.717 & 13.717 & 784 & 1390 & 0.002 & 0.043 & 0.043 & 0.951 & 0.251 \\
$\bb_{4}$ & 13.830 & 13.830 & 804 & 1397 & 0.002 & 0.042 & 0.042 & 0.954 & 0.247 \\
$\bb_{5}$ & 13.870 & 13.870 & 864 & 1325 & 0.001 & 0.041 & 0.041 & 0.955 & 0.235 \\
$\bb_{6}$ & 13.846 & 13.846 & 919 & 1223 & 0.000 & 0.040 & 0.040 & 0.950 & 0.233 \\
$\bb_{7}$ & 13.777 & 13.777 & 954 & 1112 & -0.001 & 0.040 & 0.040 & 0.947 & 0.239 \\
$\bb_{8}$ & 13.630 & 13.630 & 966 & 988 & -0.001 & 0.043 & 0.043 & 0.949 & 0.253 \\
$\bb_{9}$ & 14.017 & 14.017 & 1026 & 932 & -0.002 & 0.046 & 0.046 & 0.944 & 0.271 \\
$\bb_{10}$ & 15.156 & 15.156 & 1148 & 965 & -0.003 & 0.051 & 0.051 & 0.942 & 0.305 \\
$\bb_{11}$ & 14.778 & 14.778 & 1090 & 803 & -0.003 & 0.067 & 0.067 & 0.939 & 0.423 \\
$\bb_{12}$ & 14.646 & 14.646 & 1097 & 945 & -0.003 & 0.049 & 0.049 & 0.939 & 0.287 \\
$\bb_{13}$ & 13.581 & 13.581 & 965 & 961 & -0.002 & 0.044 & 0.044 & 0.946 & 0.256 \\
$\bb_{14}$ & 13.768 & 13.768 & 945 & 1131 & -0.001 & 0.041 & 0.041 & 0.947 & 0.236 \\
$\bb_{15}$ & 13.834 & 13.834 & 883 & 1281 & -0.001 & 0.042 & 0.042 & 0.949 & 0.231 \\
$\bb_{16}$ & 13.800 & 13.800 & 801 & 1391 & 0.000 & 0.043 & 0.043 & 0.945 & 0.246 \\
$\bb_{17}$ & 13.629 & 13.629 & 776 & 1370 & 0.000 & 0.044 & 0.044 & 0.945 & 0.251 \\
$\bb_{18}$ & 13.237 & 13.237 & 734 & 1275 & 0.000 & 0.045 & 0.045 & 0.951 & 0.260 \\
$\bb_{19}$ & 12.978 & 12.978 & 693 & 1189 & -0.000 & 0.047 & 0.047 & 0.952 & 0.269 \\
$\bb_{20}$ & 13.065 & 13.065 & 669 & 1153 & -0.001 & 0.047 & 0.047 & 0.947 & 0.273 \\
$\bb_{21}$ & 13.327 & 13.327 & 649 & 1131 & -0.001 & 0.048 & 0.048 & 0.947 & 0.277 \\
\midrule
$\mathtt{Uniform}$ &  &  &  &  &  &  &  & 0.932 & 0.381 \\
\midrule
$\mathtt{WBATE}$ &  &  &  &  & -0.001 & 0.027 & 0.027 & 0.944 & 0.138 \\
\midrule
$\mathtt{LBATE}$ &  &  &  &  & 0.034 & 0.046 & 0.057 & 0.982 & 0.448 \\
\bottomrule\bottomrule
\end{tabular}
}
    \caption{Intention-to-Treat Effects (Simulations, $s=2$, ITT Bandwidth Selector).}
    \label{tab:sa-simuls-itt-s2}
\end{table}

\begin{table}[p]
    \centering
    \resizebox{.8\textwidth}{!}{\begin{tabular}{lccccrrrrr}
\toprule\toprule
 & $h_1$ & $h_2$ & $N_{\mathrm{Co}}$ & $N_{\mathrm{Tr}}$ & Bias & SD & RMSE & EC & IL \\
\midrule
$\bb_{1}$ & 13.216 & 13.216 & 731 & 1266 & 0.002 & 0.045 & 0.045 & 0.948 & 0.262 \\
$\bb_{2}$ & 13.505 & 13.505 & 763 & 1339 & 0.001 & 0.044 & 0.044 & 0.950 & 0.255 \\
$\bb_{3}$ & 13.716 & 13.716 & 784 & 1389 & 0.001 & 0.043 & 0.043 & 0.948 & 0.251 \\
$\bb_{4}$ & 13.816 & 13.816 & 803 & 1393 & 0.000 & 0.043 & 0.043 & 0.946 & 0.247 \\
$\bb_{5}$ & 13.843 & 13.843 & 861 & 1320 & -0.001 & 0.042 & 0.042 & 0.951 & 0.235 \\
$\bb_{6}$ & 13.821 & 13.821 & 915 & 1219 & -0.001 & 0.041 & 0.041 & 0.951 & 0.233 \\
$\bb_{7}$ & 13.740 & 13.740 & 950 & 1107 & -0.002 & 0.041 & 0.041 & 0.951 & 0.239 \\
$\bb_{8}$ & 13.566 & 13.566 & 959 & 980 & -0.002 & 0.043 & 0.043 & 0.950 & 0.253 \\
$\bb_{9}$ & 13.909 & 13.909 & 1014 & 919 & -0.002 & 0.046 & 0.046 & 0.952 & 0.272 \\
$\bb_{10}$ & 15.049 & 15.049 & 1138 & 951 & -0.002 & 0.051 & 0.051 & 0.949 & 0.306 \\
$\bb_{11}$ & 14.775 & 14.775 & 1090 & 803 & -0.001 & 0.064 & 0.064 & 0.944 & 0.423 \\
$\bb_{12}$ & 14.642 & 14.642 & 1097 & 945 & -0.000 & 0.048 & 0.048 & 0.949 & 0.288 \\
$\bb_{13}$ & 13.564 & 13.564 & 963 & 959 & 0.000 & 0.044 & 0.044 & 0.946 & 0.257 \\
$\bb_{14}$ & 13.758 & 13.758 & 944 & 1130 & 0.001 & 0.041 & 0.041 & 0.940 & 0.237 \\
$\bb_{15}$ & 13.839 & 13.839 & 885 & 1281 & 0.003 & 0.042 & 0.042 & 0.949 & 0.232 \\
$\bb_{16}$ & 13.799 & 13.799 & 802 & 1390 & 0.004 & 0.044 & 0.044 & 0.948 & 0.246 \\
$\bb_{17}$ & 13.634 & 13.634 & 777 & 1371 & 0.004 & 0.044 & 0.044 & 0.948 & 0.251 \\
$\bb_{18}$ & 13.260 & 13.260 & 737 & 1279 & 0.004 & 0.045 & 0.045 & 0.949 & 0.260 \\
$\bb_{19}$ & 13.014 & 13.014 & 697 & 1195 & 0.003 & 0.046 & 0.047 & 0.951 & 0.269 \\
$\bb_{20}$ & 13.095 & 13.095 & 672 & 1158 & 0.003 & 0.047 & 0.047 & 0.948 & 0.274 \\
$\bb_{21}$ & 13.341 & 13.341 & 651 & 1133 & 0.003 & 0.048 & 0.048 & 0.948 & 0.277 \\
\midrule
$\mathtt{Uniform}$ &  &  &  &  &  &  &  & 0.932 & 0.381 \\
\midrule
$\mathtt{WBATE}$ &  &  &  &  & 0.001 & 0.026 & 0.026 & 0.941 & 0.138 \\
\midrule
$\mathtt{LBATE}$ &  &  &  &  & 0.035 & 0.044 & 0.056 & 0.980 & 0.444 \\
\bottomrule\bottomrule
\end{tabular}
}
    \caption{Intention-to-Treat Effects (Simulations, $s=3$, ITT Bandwidth Selector).}
    \label{tab:sa-simuls-itt-s3}
\end{table}

\begin{table}[p]
    \centering
    \resizebox{.8\textwidth}{!}{\begin{tabular}{lccccrrrrr}
\toprule\toprule
 & $h_1$ & $h_2$ & $N_{\mathrm{Co}}$ & $N_{\mathrm{Tr}}$ & Bias & SD & RMSE & EC & IL \\
\midrule
$\bb_{1}$ & 13.220 & 13.220 & 732 & 1267 & 0.000 & 0.038 & 0.038 & 0.950 & 0.230 \\
$\bb_{2}$ & 13.525 & 13.525 & 765 & 1343 & 0.000 & 0.037 & 0.037 & 0.947 & 0.223 \\
$\bb_{3}$ & 13.740 & 13.740 & 787 & 1393 & 0.000 & 0.036 & 0.036 & 0.945 & 0.219 \\
$\bb_{4}$ & 13.857 & 13.857 & 808 & 1400 & 0.000 & 0.036 & 0.036 & 0.946 & 0.217 \\
$\bb_{5}$ & 13.904 & 13.904 & 869 & 1329 & 0.000 & 0.036 & 0.036 & 0.944 & 0.219 \\
$\bb_{6}$ & 13.890 & 13.890 & 924 & 1229 & 0.000 & 0.037 & 0.037 & 0.946 & 0.226 \\
$\bb_{7}$ & 13.811 & 13.811 & 959 & 1116 & 0.001 & 0.040 & 0.040 & 0.944 & 0.239 \\
$\bb_{8}$ & 13.639 & 13.639 & 968 & 989 & 0.001 & 0.044 & 0.044 & 0.945 & 0.261 \\
$\bb_{9}$ & 13.993 & 13.993 & 1024 & 928 & 0.001 & 0.048 & 0.048 & 0.944 & 0.289 \\
$\bb_{10}$ & 15.131 & 15.131 & 1147 & 961 & 0.001 & 0.054 & 0.054 & 0.939 & 0.338 \\
$\bb_{11}$ & 14.782 & 14.782 & 1091 & 803 & 0.001 & 0.070 & 0.070 & 0.931 & 0.483 \\
$\bb_{12}$ & 14.679 & 14.679 & 1101 & 949 & 0.000 & 0.051 & 0.051 & 0.942 & 0.313 \\
$\bb_{13}$ & 13.614 & 13.614 & 969 & 965 & 0.000 & 0.044 & 0.044 & 0.947 & 0.266 \\
$\bb_{14}$ & 13.811 & 13.811 & 951 & 1137 & 0.000 & 0.039 & 0.039 & 0.950 & 0.235 \\
$\bb_{15}$ & 13.870 & 13.870 & 889 & 1286 & 0.000 & 0.036 & 0.036 & 0.948 & 0.220 \\
$\bb_{16}$ & 13.837 & 13.837 & 805 & 1397 & 0.000 & 0.036 & 0.036 & 0.951 & 0.215 \\
$\bb_{17}$ & 13.678 & 13.678 & 781 & 1379 & 0.000 & 0.036 & 0.036 & 0.949 & 0.217 \\
$\bb_{18}$ & 13.289 & 13.289 & 739 & 1284 & 0.001 & 0.037 & 0.037 & 0.947 & 0.224 \\
$\bb_{19}$ & 13.011 & 13.011 & 697 & 1195 & 0.001 & 0.038 & 0.038 & 0.947 & 0.232 \\
$\bb_{20}$ & 13.077 & 13.077 & 671 & 1154 & 0.001 & 0.039 & 0.039 & 0.947 & 0.236 \\
$\bb_{21}$ & 13.330 & 13.330 & 651 & 1130 & 0.001 & 0.039 & 0.039 & 0.945 & 0.238 \\
\midrule
$\mathtt{Uniform}$ &  &  &  &  &  &  &  & 0.924 & 0.364 \\
\midrule
$\mathtt{WBATE}$ &  &  &  &  & 0.001 & 0.025 & 0.025 & 0.944 & 0.134 \\
\midrule
$\mathtt{LBATE}$ &  &  &  &  & 0.017 & 0.035 & 0.039 & 0.984 & 0.421 \\
\bottomrule\bottomrule
\end{tabular}
}
    \caption{First-Stage Effects (Simulations, $s=1$, ITT Bandwidth Selector).}
    \label{tab:sa-simuls-fs-s1}
\end{table}

\begin{table}[p]
    \centering
    \resizebox{.8\textwidth}{!}{\begin{tabular}{lccccrrrrr}
\toprule\toprule
 & $h_1$ & $h_2$ & $N_{\mathrm{Co}}$ & $N_{\mathrm{Tr}}$ & Bias & SD & RMSE & EC & IL \\
\midrule
$\bb_{1}$ & 13.191 & 13.191 & 728 & 1262 & 0.002 & 0.039 & 0.039 & 0.948 & 0.231 \\
$\bb_{2}$ & 13.494 & 13.494 & 762 & 1338 & 0.002 & 0.038 & 0.038 & 0.943 & 0.224 \\
$\bb_{3}$ & 13.717 & 13.717 & 784 & 1390 & 0.002 & 0.037 & 0.037 & 0.944 & 0.219 \\
$\bb_{4}$ & 13.830 & 13.830 & 804 & 1397 & 0.002 & 0.036 & 0.037 & 0.948 & 0.218 \\
$\bb_{5}$ & 13.870 & 13.870 & 864 & 1325 & 0.002 & 0.037 & 0.037 & 0.944 & 0.220 \\
$\bb_{6}$ & 13.846 & 13.846 & 919 & 1223 & 0.001 & 0.038 & 0.038 & 0.941 & 0.227 \\
$\bb_{7}$ & 13.777 & 13.777 & 954 & 1112 & 0.001 & 0.040 & 0.040 & 0.942 & 0.239 \\
$\bb_{8}$ & 13.630 & 13.630 & 966 & 988 & 0.000 & 0.043 & 0.043 & 0.936 & 0.259 \\
$\bb_{9}$ & 14.017 & 14.017 & 1026 & 932 & -0.000 & 0.048 & 0.048 & 0.937 & 0.286 \\
$\bb_{10}$ & 15.156 & 15.156 & 1148 & 965 & -0.002 & 0.054 & 0.054 & 0.936 & 0.333 \\
$\bb_{11}$ & 14.778 & 14.778 & 1090 & 803 & -0.002 & 0.069 & 0.069 & 0.932 & 0.476 \\
$\bb_{12}$ & 14.646 & 14.646 & 1097 & 945 & -0.002 & 0.051 & 0.051 & 0.942 & 0.309 \\
$\bb_{13}$ & 13.581 & 13.581 & 965 & 961 & -0.001 & 0.044 & 0.044 & 0.944 & 0.262 \\
$\bb_{14}$ & 13.768 & 13.768 & 945 & 1131 & -0.001 & 0.039 & 0.039 & 0.949 & 0.232 \\
$\bb_{15}$ & 13.834 & 13.834 & 883 & 1281 & -0.001 & 0.036 & 0.036 & 0.945 & 0.218 \\
$\bb_{16}$ & 13.800 & 13.800 & 801 & 1391 & -0.001 & 0.035 & 0.035 & 0.940 & 0.213 \\
$\bb_{17}$ & 13.629 & 13.629 & 776 & 1370 & -0.001 & 0.036 & 0.036 & 0.943 & 0.215 \\
$\bb_{18}$ & 13.237 & 13.237 & 734 & 1275 & -0.001 & 0.037 & 0.037 & 0.941 & 0.222 \\
$\bb_{19}$ & 12.978 & 12.978 & 693 & 1189 & -0.002 & 0.038 & 0.038 & 0.940 & 0.229 \\
$\bb_{20}$ & 13.065 & 13.065 & 669 & 1153 & -0.002 & 0.039 & 0.039 & 0.941 & 0.233 \\
$\bb_{21}$ & 13.327 & 13.327 & 649 & 1131 & -0.002 & 0.039 & 0.039 & 0.939 & 0.235 \\
\midrule
$\mathtt{Uniform}$ &  &  &  &  &  &  &  & 0.917 & 0.361 \\
\midrule
$\mathtt{WBATE}$ &  &  &  &  & -0.000 & 0.024 & 0.024 & 0.938 & 0.133 \\
\midrule
$\mathtt{LBATE}$ &  &  &  &  & 0.021 & 0.038 & 0.043 & 0.978 & 0.427 \\
\bottomrule\bottomrule
\end{tabular}
}
    \caption{First-Stage Effects (Simulations, $s=2$, ITT Bandwidth Selector).}
    \label{tab:sa-simuls-fs-s2}
\end{table}

\begin{table}[p]
    \centering
    \resizebox{.8\textwidth}{!}{\begin{tabular}{lccccrrrrr}
\toprule\toprule
 & $h_1$ & $h_2$ & $N_{\mathrm{Co}}$ & $N_{\mathrm{Tr}}$ & Bias & SD & RMSE & EC & IL \\
\midrule
$\bb_{1}$ & 13.216 & 13.216 & 731 & 1266 & 0.001 & 0.038 & 0.038 & 0.951 & 0.230 \\
$\bb_{2}$ & 13.505 & 13.505 & 763 & 1339 & 0.001 & 0.037 & 0.037 & 0.949 & 0.223 \\
$\bb_{3}$ & 13.716 & 13.716 & 784 & 1389 & 0.001 & 0.036 & 0.036 & 0.950 & 0.219 \\
$\bb_{4}$ & 13.816 & 13.816 & 803 & 1393 & 0.000 & 0.036 & 0.036 & 0.950 & 0.217 \\
$\bb_{5}$ & 13.843 & 13.843 & 861 & 1320 & -0.000 & 0.036 & 0.036 & 0.951 & 0.220 \\
$\bb_{6}$ & 13.821 & 13.821 & 915 & 1219 & -0.000 & 0.037 & 0.037 & 0.949 & 0.226 \\
$\bb_{7}$ & 13.740 & 13.740 & 950 & 1107 & -0.000 & 0.039 & 0.039 & 0.947 & 0.239 \\
$\bb_{8}$ & 13.566 & 13.566 & 959 & 980 & 0.000 & 0.043 & 0.043 & 0.942 & 0.260 \\
$\bb_{9}$ & 13.909 & 13.909 & 1014 & 919 & 0.001 & 0.048 & 0.048 & 0.941 & 0.288 \\
$\bb_{10}$ & 15.049 & 15.049 & 1138 & 951 & 0.002 & 0.053 & 0.053 & 0.940 & 0.336 \\
$\bb_{11}$ & 14.775 & 14.775 & 1090 & 803 & 0.003 & 0.068 & 0.068 & 0.930 & 0.478 \\
$\bb_{12}$ & 14.642 & 14.642 & 1097 & 945 & 0.002 & 0.051 & 0.051 & 0.939 & 0.311 \\
$\bb_{13}$ & 13.564 & 13.564 & 963 & 959 & 0.002 & 0.044 & 0.044 & 0.938 & 0.264 \\
$\bb_{14}$ & 13.758 & 13.758 & 944 & 1130 & 0.002 & 0.039 & 0.039 & 0.943 & 0.235 \\
$\bb_{15}$ & 13.839 & 13.839 & 885 & 1281 & 0.002 & 0.037 & 0.037 & 0.943 & 0.220 \\
$\bb_{16}$ & 13.799 & 13.799 & 802 & 1390 & 0.003 & 0.036 & 0.036 & 0.943 & 0.215 \\
$\bb_{17}$ & 13.634 & 13.634 & 777 & 1371 & 0.002 & 0.036 & 0.036 & 0.945 & 0.217 \\
$\bb_{18}$ & 13.260 & 13.260 & 737 & 1279 & 0.002 & 0.037 & 0.037 & 0.947 & 0.224 \\
$\bb_{19}$ & 13.014 & 13.014 & 697 & 1195 & 0.002 & 0.038 & 0.038 & 0.948 & 0.231 \\
$\bb_{20}$ & 13.095 & 13.095 & 672 & 1158 & 0.002 & 0.038 & 0.038 & 0.947 & 0.235 \\
$\bb_{21}$ & 13.341 & 13.341 & 651 & 1133 & 0.002 & 0.039 & 0.039 & 0.943 & 0.237 \\
\midrule
$\mathtt{Uniform}$ &  &  &  &  &  &  &  & 0.920 & 0.363 \\
\midrule
$\mathtt{WBATE}$ &  &  &  &  & 0.001 & 0.024 & 0.024 & 0.944 & 0.133 \\
\midrule
$\mathtt{LBATE}$ &  &  &  &  & 0.024 & 0.036 & 0.043 & 0.980 & 0.431 \\
\bottomrule\bottomrule
\end{tabular}
}
    \caption{First-Stage Effects (Simulations, $s=3$, ITT Bandwidth Selector).}
    \label{tab:sa-simuls-fs-s3}
\end{table}

\begin{table}[p]
    \centering
    \resizebox{.8\textwidth}{!}{\begin{tabular}{lccccrrrrr}
\toprule\toprule
 & $h_1$ & $h_2$ & $N_{\mathrm{Co}}$ & $N_{\mathrm{Tr}}$ & Bias & SD & RMSE & EC & IL \\
\midrule
$\bb_{1}$ & 13.220 & 13.220 & 732 & 1267 & 0.001 & 0.072 & 0.072 & 0.954 & 0.433 \\
$\bb_{2}$ & 13.525 & 13.525 & 765 & 1343 & 0.001 & 0.069 & 0.069 & 0.958 & 0.419 \\
$\bb_{3}$ & 13.740 & 13.740 & 787 & 1393 & 0.001 & 0.067 & 0.067 & 0.958 & 0.411 \\
$\bb_{4}$ & 13.857 & 13.857 & 808 & 1400 & 0.001 & 0.066 & 0.066 & 0.958 & 0.402 \\
$\bb_{5}$ & 13.904 & 13.904 & 869 & 1329 & 0.001 & 0.064 & 0.064 & 0.957 & 0.373 \\
$\bb_{6}$ & 13.890 & 13.890 & 924 & 1229 & 0.001 & 0.061 & 0.061 & 0.959 & 0.364 \\
$\bb_{7}$ & 13.811 & 13.811 & 959 & 1116 & 0.001 & 0.060 & 0.060 & 0.955 & 0.369 \\
$\bb_{8}$ & 13.639 & 13.639 & 968 & 989 & 0.001 & 0.063 & 0.063 & 0.955 & 0.387 \\
$\bb_{9}$ & 13.993 & 13.993 & 1024 & 928 & 0.001 & 0.067 & 0.067 & 0.958 & 0.411 \\
$\bb_{10}$ & 15.131 & 15.131 & 1147 & 961 & 0.000 & 0.072 & 0.072 & 0.959 & 0.463 \\
$\bb_{11}$ & 14.782 & 14.782 & 1091 & 803 & -0.000 & 0.092 & 0.092 & 0.964 & 0.719 \\
$\bb_{12}$ & 14.679 & 14.679 & 1101 & 949 & 0.000 & 0.068 & 0.068 & 0.951 & 0.425 \\
$\bb_{13}$ & 13.614 & 13.614 & 969 & 965 & 0.000 & 0.061 & 0.061 & 0.954 & 0.378 \\
$\bb_{14}$ & 13.811 & 13.811 & 951 & 1137 & 0.000 & 0.057 & 0.057 & 0.959 & 0.348 \\
$\bb_{15}$ & 13.870 & 13.870 & 889 & 1286 & 0.001 & 0.059 & 0.059 & 0.955 & 0.342 \\
$\bb_{16}$ & 13.837 & 13.837 & 805 & 1397 & 0.001 & 0.062 & 0.062 & 0.954 & 0.369 \\
$\bb_{17}$ & 13.678 & 13.678 & 781 & 1379 & 0.001 & 0.063 & 0.063 & 0.951 & 0.374 \\
$\bb_{18}$ & 13.289 & 13.289 & 739 & 1284 & 0.001 & 0.065 & 0.065 & 0.951 & 0.382 \\
$\bb_{19}$ & 13.011 & 13.011 & 697 & 1195 & 0.001 & 0.066 & 0.066 & 0.952 & 0.390 \\
$\bb_{20}$ & 13.077 & 13.077 & 671 & 1154 & 0.000 & 0.066 & 0.066 & 0.954 & 0.393 \\
$\bb_{21}$ & 13.330 & 13.330 & 651 & 1130 & -0.000 & 0.067 & 0.067 & 0.952 & 0.394 \\
\midrule
$\mathtt{Uniform}$ &  &  &  &  &  &  &  & 0.962 & 0.586 \\
\midrule
$\mathtt{WBATE}$ &  &  &  &  & 0.001 & 0.039 & 0.039 & 0.949 & 0.214 \\
\midrule
$\mathtt{LBATE}$ &  &  &  &  & 0.053 & 0.058 & 0.078 & 0.997 & 0.753 \\
\bottomrule\bottomrule
\end{tabular}
}
    \caption{Fuzzy Effects (Simulations, $s=1$, ITT Bandwidth Selector).}
    \label{tab:sa-simuls-fuzzy-s1}
\end{table}

\begin{table}[p]
    \centering
    \resizebox{.8\textwidth}{!}{\begin{tabular}{lccccrrrrr}
\toprule\toprule
 & $h_1$ & $h_2$ & $N_{\mathrm{Co}}$ & $N_{\mathrm{Tr}}$ & Bias & SD & RMSE & EC & IL \\
\midrule
$\bb_{1}$ & 13.191 & 13.191 & 728 & 1262 & 0.000 & 0.073 & 0.073 & 0.958 & 0.445 \\
$\bb_{2}$ & 13.494 & 13.494 & 762 & 1338 & 0.001 & 0.070 & 0.070 & 0.957 & 0.429 \\
$\bb_{3}$ & 13.717 & 13.717 & 784 & 1390 & 0.001 & 0.068 & 0.068 & 0.959 & 0.417 \\
$\bb_{4}$ & 13.830 & 13.830 & 804 & 1397 & 0.001 & 0.067 & 0.067 & 0.959 & 0.405 \\
$\bb_{5}$ & 13.870 & 13.870 & 864 & 1325 & 0.000 & 0.064 & 0.064 & 0.959 & 0.372 \\
$\bb_{6}$ & 13.846 & 13.846 & 919 & 1223 & -0.001 & 0.061 & 0.061 & 0.954 & 0.359 \\
$\bb_{7}$ & 13.777 & 13.777 & 954 & 1112 & -0.002 & 0.059 & 0.059 & 0.951 & 0.359 \\
$\bb_{8}$ & 13.630 & 13.630 & 966 & 988 & -0.003 & 0.061 & 0.061 & 0.953 & 0.369 \\
$\bb_{9}$ & 14.017 & 14.017 & 1026 & 932 & -0.003 & 0.064 & 0.064 & 0.952 & 0.383 \\
$\bb_{10}$ & 15.156 & 15.156 & 1148 & 965 & -0.003 & 0.068 & 0.068 & 0.949 & 0.421 \\
$\bb_{11}$ & 14.778 & 14.778 & 1090 & 803 & -0.004 & 0.085 & 0.085 & 0.961 & 1.058 \\
$\bb_{12}$ & 14.646 & 14.646 & 1097 & 945 & -0.003 & 0.063 & 0.064 & 0.952 & 0.384 \\
$\bb_{13}$ & 13.581 & 13.581 & 965 & 961 & -0.002 & 0.058 & 0.058 & 0.951 & 0.345 \\
$\bb_{14}$ & 13.768 & 13.768 & 945 & 1131 & -0.001 & 0.054 & 0.054 & 0.953 & 0.321 \\
$\bb_{15}$ & 13.834 & 13.834 & 883 & 1281 & -0.000 & 0.056 & 0.056 & 0.950 & 0.318 \\
$\bb_{16}$ & 13.800 & 13.800 & 801 & 1391 & 0.001 & 0.059 & 0.059 & 0.948 & 0.345 \\
$\bb_{17}$ & 13.629 & 13.629 & 776 & 1370 & 0.001 & 0.060 & 0.060 & 0.951 & 0.351 \\
$\bb_{18}$ & 13.237 & 13.237 & 734 & 1275 & 0.001 & 0.061 & 0.061 & 0.952 & 0.361 \\
$\bb_{19}$ & 12.978 & 12.978 & 693 & 1189 & 0.000 & 0.063 & 0.063 & 0.951 & 0.372 \\
$\bb_{20}$ & 13.065 & 13.065 & 669 & 1153 & -0.001 & 0.064 & 0.064 & 0.951 & 0.376 \\
$\bb_{21}$ & 13.327 & 13.327 & 649 & 1131 & -0.001 & 0.065 & 0.065 & 0.949 & 0.379 \\
\midrule
$\mathtt{Uniform}$ &  &  &  &  &  &  &  & 0.961 & 0.588 \\
\midrule
$\mathtt{WBATE}$ &  &  &  &  & -0.001 & 0.038 & 0.038 & 0.947 & 0.224 \\
\midrule
$\mathtt{LBATE}$ &  &  &  &  & 0.052 & 0.060 & 0.079 & 0.994 & 0.993 \\
\bottomrule\bottomrule
\end{tabular}
}
    \caption{Fuzzy Effects (Simulations, $s=2$, ITT Bandwidth Selector).}
    \label{tab:sa-simuls-fuzzy-s2}
\end{table}

\begin{table}[p]
    \centering
    \resizebox{.8\textwidth}{!}{\begin{tabular}{lccccrrrrr}
\toprule\toprule
 & $h_1$ & $h_2$ & $N_{\mathrm{Co}}$ & $N_{\mathrm{Tr}}$ & Bias & SD & RMSE & EC & IL \\
\midrule
$\bb_{1}$ & 13.216 & 13.216 & 731 & 1266 & 0.001 & 0.071 & 0.071 & 0.958 & 0.428 \\
$\bb_{2}$ & 13.505 & 13.505 & 763 & 1339 & 0.001 & 0.069 & 0.069 & 0.954 & 0.412 \\
$\bb_{3}$ & 13.716 & 13.716 & 784 & 1389 & 0.001 & 0.067 & 0.067 & 0.955 & 0.400 \\
$\bb_{4}$ & 13.816 & 13.816 & 803 & 1393 & 0.000 & 0.066 & 0.066 & 0.952 & 0.388 \\
$\bb_{5}$ & 13.843 & 13.843 & 861 & 1320 & -0.001 & 0.063 & 0.063 & 0.954 & 0.358 \\
$\bb_{6}$ & 13.821 & 13.821 & 915 & 1219 & -0.002 & 0.059 & 0.059 & 0.953 & 0.346 \\
$\bb_{7}$ & 13.740 & 13.740 & 950 & 1107 & -0.003 & 0.058 & 0.058 & 0.954 & 0.349 \\
$\bb_{8}$ & 13.566 & 13.566 & 959 & 980 & -0.003 & 0.060 & 0.060 & 0.960 & 0.363 \\
$\bb_{9}$ & 13.909 & 13.909 & 1014 & 919 & -0.004 & 0.063 & 0.063 & 0.961 & 0.383 \\
$\bb_{10}$ & 15.049 & 15.049 & 1138 & 951 & -0.004 & 0.067 & 0.067 & 0.958 & 0.427 \\
$\bb_{11}$ & 14.775 & 14.775 & 1090 & 803 & -0.005 & 0.084 & 0.084 & 0.964 & 0.636 \\
$\bb_{12}$ & 14.642 & 14.642 & 1097 & 945 & -0.003 & 0.064 & 0.064 & 0.952 & 0.399 \\
$\bb_{13}$ & 13.564 & 13.564 & 963 & 959 & -0.002 & 0.059 & 0.059 & 0.954 & 0.361 \\
$\bb_{14}$ & 13.758 & 13.758 & 944 & 1130 & -0.000 & 0.056 & 0.056 & 0.950 & 0.336 \\
$\bb_{15}$ & 13.839 & 13.839 & 885 & 1281 & 0.002 & 0.059 & 0.059 & 0.951 & 0.332 \\
$\bb_{16}$ & 13.799 & 13.799 & 802 & 1390 & 0.003 & 0.062 & 0.062 & 0.949 & 0.360 \\
$\bb_{17}$ & 13.634 & 13.634 & 777 & 1371 & 0.003 & 0.062 & 0.062 & 0.949 & 0.366 \\
$\bb_{18}$ & 13.260 & 13.260 & 737 & 1279 & 0.004 & 0.064 & 0.064 & 0.956 & 0.376 \\
$\bb_{19}$ & 13.014 & 13.014 & 697 & 1195 & 0.003 & 0.065 & 0.065 & 0.954 & 0.386 \\
$\bb_{20}$ & 13.095 & 13.095 & 672 & 1158 & 0.003 & 0.066 & 0.066 & 0.951 & 0.390 \\
$\bb_{21}$ & 13.341 & 13.341 & 651 & 1133 & 0.002 & 0.066 & 0.066 & 0.954 & 0.392 \\
\midrule
$\mathtt{Uniform}$ &  &  &  &  &  &  &  & 0.961 & 0.561 \\
\midrule
$\mathtt{WBATE}$ &  &  &  &  & -0.000 & 0.037 & 0.037 & 0.950 & 0.203 \\
\midrule
$\mathtt{LBATE}$ &  &  &  &  & 0.050 & 0.059 & 0.078 & 0.995 & 0.687 \\
\bottomrule\bottomrule
\end{tabular}
}
    \caption{Fuzzy Effects (Simulations, $s=3$, ITT Bandwidth Selector).}
    \label{tab:sa-simuls-fuzzy-s3}
\end{table}

\clearpage

\begin{table}[p]
    \centering
    \resizebox{.8\textwidth}{!}{\begin{tabular}{lccccrrrrr}
\toprule\toprule
 & $h_1$ & $h_2$ & $N_{\mathrm{Co}}$ & $N_{\mathrm{Tr}}$ & Bias & SD & RMSE & EC & IL \\
\midrule
$\bb_{1}$ & 13.404 & 13.404 & 746 & 1303 & 0.000 & 0.044 & 0.044 & 0.951 & 0.260 \\
$\bb_{2}$ & 13.823 & 13.823 & 788 & 1402 & 0.000 & 0.042 & 0.042 & 0.951 & 0.251 \\
$\bb_{3}$ & 14.104 & 14.104 & 815 & 1466 & 0.000 & 0.041 & 0.041 & 0.949 & 0.245 \\
$\bb_{4}$ & 14.257 & 14.257 & 849 & 1470 & 0.001 & 0.040 & 0.040 & 0.951 & 0.238 \\
$\bb_{5}$ & 14.318 & 14.318 & 919 & 1394 & 0.001 & 0.039 & 0.039 & 0.951 & 0.227 \\
$\bb_{6}$ & 14.306 & 14.306 & 975 & 1291 & 0.001 & 0.038 & 0.038 & 0.951 & 0.226 \\
$\bb_{7}$ & 14.235 & 14.235 & 1009 & 1176 & 0.000 & 0.038 & 0.038 & 0.949 & 0.233 \\
$\bb_{8}$ & 13.991 & 13.991 & 1009 & 1036 & 0.000 & 0.041 & 0.041 & 0.948 & 0.248 \\
$\bb_{9}$ & 14.133 & 14.133 & 1040 & 946 & 0.000 & 0.045 & 0.045 & 0.948 & 0.270 \\
$\bb_{10}$ & 15.100 & 15.100 & 1144 & 957 & -0.000 & 0.050 & 0.050 & 0.947 & 0.308 \\
$\bb_{11}$ & 14.722 & 14.722 & 1085 & 796 & -0.001 & 0.063 & 0.063 & 0.935 & 0.431 \\
$\bb_{12}$ & 14.698 & 14.698 & 1103 & 951 & -0.000 & 0.047 & 0.047 & 0.941 & 0.288 \\
$\bb_{13}$ & 13.915 & 13.915 & 1004 & 1004 & -0.000 & 0.042 & 0.042 & 0.949 & 0.252 \\
$\bb_{14}$ & 14.252 & 14.252 & 1004 & 1200 & 0.000 & 0.038 & 0.038 & 0.952 & 0.231 \\
$\bb_{15}$ & 14.307 & 14.307 & 941 & 1352 & 0.000 & 0.039 & 0.039 & 0.954 & 0.225 \\
$\bb_{16}$ & 14.254 & 14.254 & 849 & 1469 & 0.001 & 0.041 & 0.041 & 0.948 & 0.237 \\
$\bb_{17}$ & 14.044 & 14.044 & 809 & 1453 & 0.001 & 0.042 & 0.042 & 0.945 & 0.245 \\
$\bb_{18}$ & 13.534 & 13.534 & 758 & 1332 & 0.001 & 0.044 & 0.044 & 0.943 & 0.256 \\
$\bb_{19}$ & 13.137 & 13.137 & 707 & 1218 & 0.001 & 0.046 & 0.046 & 0.947 & 0.267 \\
$\bb_{20}$ & 13.188 & 13.188 & 680 & 1174 & 0.000 & 0.047 & 0.047 & 0.947 & 0.272 \\
$\bb_{21}$ & 13.422 & 13.422 & 657 & 1146 & -0.000 & 0.047 & 0.047 & 0.946 & 0.276 \\
\midrule
$\mathtt{Uniform}$ &  &  &  &  &  &  &  & 0.934 & 0.375 \\
\midrule
$\mathtt{WBATE}$ &  &  &  &  & 0.000 & 0.026 & 0.026 & 0.945 & 0.137 \\
\midrule
$\mathtt{LBATE}$ &  &  &  &  & 0.049 & 0.034 & 0.060 & 0.976 & 0.430 \\
\bottomrule\bottomrule
\end{tabular}
}
    \caption{Intention-to-Treat Effects (Simulations, $s=1$, Fuzzy Bandwidth Selector).}
    \label{tab:sa-simuls-itt-s1-bwmain}
\end{table}

\begin{table}[p]
    \centering
    \resizebox{.8\textwidth}{!}{\begin{tabular}{lccccrrrrr}
\toprule\toprule
 & $h_1$ & $h_2$ & $N_{\mathrm{Co}}$ & $N_{\mathrm{Tr}}$ & Bias & SD & RMSE & EC & IL \\
\midrule
$\bb_{1}$ & 13.386 & 13.386 & 744 & 1300 & 0.001 & 0.044 & 0.044 & 0.949 & 0.260 \\
$\bb_{2}$ & 13.791 & 13.791 & 785 & 1397 & 0.002 & 0.042 & 0.042 & 0.951 & 0.251 \\
$\bb_{3}$ & 14.069 & 14.069 & 811 & 1460 & 0.002 & 0.041 & 0.041 & 0.949 & 0.245 \\
$\bb_{4}$ & 14.218 & 14.218 & 844 & 1464 & 0.002 & 0.040 & 0.040 & 0.953 & 0.239 \\
$\bb_{5}$ & 14.287 & 14.287 & 914 & 1389 & 0.001 & 0.039 & 0.039 & 0.953 & 0.227 \\
$\bb_{6}$ & 14.284 & 14.284 & 972 & 1288 & 0.000 & 0.038 & 0.038 & 0.949 & 0.226 \\
$\bb_{7}$ & 14.225 & 14.225 & 1008 & 1176 & -0.000 & 0.039 & 0.039 & 0.947 & 0.233 \\
$\bb_{8}$ & 13.995 & 13.995 & 1009 & 1037 & -0.001 & 0.041 & 0.041 & 0.951 & 0.248 \\
$\bb_{9}$ & 14.126 & 14.126 & 1038 & 946 & -0.002 & 0.045 & 0.045 & 0.945 & 0.270 \\
$\bb_{10}$ & 15.108 & 15.108 & 1144 & 958 & -0.003 & 0.050 & 0.050 & 0.942 & 0.307 \\
$\bb_{11}$ & 14.748 & 14.748 & 1087 & 799 & -0.005 & 0.063 & 0.064 & 0.936 & 0.429 \\
$\bb_{12}$ & 14.726 & 14.726 & 1106 & 955 & -0.004 & 0.048 & 0.048 & 0.940 & 0.287 \\
$\bb_{13}$ & 13.933 & 13.933 & 1006 & 1007 & -0.003 & 0.043 & 0.043 & 0.946 & 0.251 \\
$\bb_{14}$ & 14.255 & 14.255 & 1003 & 1201 & -0.002 & 0.039 & 0.039 & 0.946 & 0.229 \\
$\bb_{15}$ & 14.312 & 14.312 & 941 & 1353 & -0.001 & 0.039 & 0.039 & 0.947 & 0.223 \\
$\bb_{16}$ & 14.250 & 14.250 & 847 & 1469 & -0.001 & 0.041 & 0.041 & 0.944 & 0.236 \\
$\bb_{17}$ & 14.019 & 14.019 & 806 & 1449 & -0.000 & 0.042 & 0.042 & 0.944 & 0.244 \\
$\bb_{18}$ & 13.498 & 13.498 & 754 & 1326 & -0.000 & 0.044 & 0.044 & 0.949 & 0.255 \\
$\bb_{19}$ & 13.115 & 13.115 & 703 & 1214 & -0.001 & 0.046 & 0.046 & 0.950 & 0.266 \\
$\bb_{20}$ & 13.180 & 13.180 & 678 & 1173 & -0.001 & 0.046 & 0.047 & 0.946 & 0.271 \\
$\bb_{21}$ & 13.415 & 13.415 & 655 & 1146 & -0.002 & 0.047 & 0.047 & 0.946 & 0.275 \\
\midrule
$\mathtt{Uniform}$ &  &  &  &  &  &  &  & 0.934 & 0.374 \\
\midrule
$\mathtt{WBATE}$ &  &  &  &  & -0.001 & 0.026 & 0.026 & 0.947 & 0.136 \\
\midrule
$\mathtt{LBATE}$ &  &  &  &  & 0.030 & 0.041 & 0.051 & 0.984 & 0.445 \\
\bottomrule\bottomrule
\end{tabular}
}
    \caption{Intention-to-Treat Effects (Simulations, $s=2$, Fuzzy Bandwidth Selector).}
    \label{tab:sa-simuls-itt-s2-bwmain}
\end{table}

\begin{table}[p]
    \centering
    \resizebox{.8\textwidth}{!}{\begin{tabular}{lccccrrrrr}
\toprule\toprule
 & $h_1$ & $h_2$ & $N_{\mathrm{Co}}$ & $N_{\mathrm{Tr}}$ & Bias & SD & RMSE & EC & IL \\
\midrule
$\bb_{1}$ & 13.416 & 13.416 & 746 & 1305 & 0.001 & 0.044 & 0.044 & 0.949 & 0.259 \\
$\bb_{2}$ & 13.823 & 13.823 & 788 & 1403 & 0.001 & 0.042 & 0.042 & 0.950 & 0.250 \\
$\bb_{3}$ & 14.093 & 14.093 & 813 & 1465 & 0.000 & 0.041 & 0.041 & 0.947 & 0.245 \\
$\bb_{4}$ & 14.229 & 14.229 & 845 & 1465 & -0.000 & 0.041 & 0.041 & 0.945 & 0.238 \\
$\bb_{5}$ & 14.282 & 14.282 & 913 & 1388 & -0.001 & 0.040 & 0.040 & 0.948 & 0.227 \\
$\bb_{6}$ & 14.281 & 14.281 & 971 & 1288 & -0.002 & 0.038 & 0.039 & 0.950 & 0.226 \\
$\bb_{7}$ & 14.214 & 14.214 & 1006 & 1174 & -0.003 & 0.039 & 0.039 & 0.952 & 0.232 \\
$\bb_{8}$ & 13.955 & 13.955 & 1005 & 1031 & -0.003 & 0.041 & 0.041 & 0.948 & 0.248 \\
$\bb_{9}$ & 14.080 & 14.080 & 1034 & 940 & -0.003 & 0.045 & 0.045 & 0.952 & 0.270 \\
$\bb_{10}$ & 15.038 & 15.038 & 1137 & 949 & -0.002 & 0.049 & 0.049 & 0.949 & 0.308 \\
$\bb_{11}$ & 14.667 & 14.667 & 1079 & 790 & -0.002 & 0.062 & 0.062 & 0.941 & 0.430 \\
$\bb_{12}$ & 14.687 & 14.687 & 1102 & 951 & -0.001 & 0.047 & 0.047 & 0.949 & 0.288 \\
$\bb_{13}$ & 13.902 & 13.902 & 1003 & 1003 & 0.000 & 0.042 & 0.042 & 0.946 & 0.252 \\
$\bb_{14}$ & 14.233 & 14.233 & 1001 & 1197 & 0.001 & 0.039 & 0.039 & 0.941 & 0.230 \\
$\bb_{15}$ & 14.292 & 14.292 & 940 & 1350 & 0.003 & 0.040 & 0.040 & 0.948 & 0.225 \\
$\bb_{16}$ & 14.217 & 14.217 & 845 & 1463 & 0.004 & 0.042 & 0.042 & 0.946 & 0.237 \\
$\bb_{17}$ & 13.995 & 13.995 & 805 & 1443 & 0.004 & 0.042 & 0.043 & 0.945 & 0.245 \\
$\bb_{18}$ & 13.492 & 13.492 & 755 & 1324 & 0.004 & 0.044 & 0.044 & 0.947 & 0.256 \\
$\bb_{19}$ & 13.126 & 13.126 & 705 & 1216 & 0.003 & 0.046 & 0.046 & 0.952 & 0.267 \\
$\bb_{20}$ & 13.183 & 13.183 & 679 & 1173 & 0.003 & 0.046 & 0.046 & 0.947 & 0.272 \\
$\bb_{21}$ & 13.408 & 13.408 & 655 & 1145 & 0.003 & 0.047 & 0.047 & 0.951 & 0.276 \\
\midrule
$\mathtt{Uniform}$ &  &  &  &  &  &  &  & 0.934 & 0.375 \\
\midrule
$\mathtt{WBATE}$ &  &  &  &  & 0.000 & 0.026 & 0.026 & 0.943 & 0.136 \\
\midrule
$\mathtt{LBATE}$ &  &  &  &  & 0.032 & 0.041 & 0.052 & 0.984 & 0.443 \\
\bottomrule\bottomrule
\end{tabular}
}
    \caption{Intention-to-Treat Effects (Simulations, $s=3$, Fuzzy Bandwidth Selector).}
    \label{tab:sa-simuls-itt-s3-bwmain}
\end{table}

\begin{table}[p]
    \centering
    \resizebox{.8\textwidth}{!}{\begin{tabular}{lccccrrrrr}
\toprule\toprule
 & $h_1$ & $h_2$ & $N_{\mathrm{Co}}$ & $N_{\mathrm{Tr}}$ & Bias & SD & RMSE & EC & IL \\
\midrule
$\bb_{1}$ & 13.404 & 13.404 & 746 & 1303 & -0.000 & 0.037 & 0.037 & 0.949 & 0.227 \\
$\bb_{2}$ & 13.823 & 13.823 & 788 & 1402 & -0.000 & 0.035 & 0.035 & 0.948 & 0.218 \\
$\bb_{3}$ & 14.104 & 14.104 & 815 & 1466 & -0.000 & 0.035 & 0.035 & 0.949 & 0.213 \\
$\bb_{4}$ & 14.257 & 14.257 & 849 & 1470 & -0.000 & 0.034 & 0.034 & 0.946 & 0.211 \\
$\bb_{5}$ & 14.318 & 14.318 & 919 & 1394 & -0.000 & 0.035 & 0.035 & 0.945 & 0.214 \\
$\bb_{6}$ & 14.306 & 14.306 & 975 & 1291 & -0.000 & 0.036 & 0.036 & 0.946 & 0.221 \\
$\bb_{7}$ & 14.235 & 14.235 & 1009 & 1176 & 0.000 & 0.038 & 0.038 & 0.945 & 0.234 \\
$\bb_{8}$ & 13.991 & 13.991 & 1009 & 1036 & 0.000 & 0.042 & 0.042 & 0.946 & 0.256 \\
$\bb_{9}$ & 14.133 & 14.133 & 1040 & 946 & 0.000 & 0.047 & 0.047 & 0.941 & 0.287 \\
$\bb_{10}$ & 15.100 & 15.100 & 1144 & 957 & 0.000 & 0.053 & 0.053 & 0.939 & 0.338 \\
$\bb_{11}$ & 14.722 & 14.722 & 1085 & 796 & -0.001 & 0.069 & 0.069 & 0.930 & 0.486 \\
$\bb_{12}$ & 14.698 & 14.698 & 1103 & 951 & 0.000 & 0.050 & 0.050 & 0.944 & 0.313 \\
$\bb_{13}$ & 13.915 & 13.915 & 1004 & 1004 & -0.000 & 0.043 & 0.043 & 0.948 & 0.261 \\
$\bb_{14}$ & 14.252 & 14.252 & 1004 & 1200 & -0.000 & 0.037 & 0.037 & 0.949 & 0.229 \\
$\bb_{15}$ & 14.307 & 14.307 & 941 & 1352 & -0.000 & 0.034 & 0.034 & 0.948 & 0.215 \\
$\bb_{16}$ & 14.254 & 14.254 & 849 & 1469 & -0.000 & 0.034 & 0.034 & 0.948 & 0.209 \\
$\bb_{17}$ & 14.044 & 14.044 & 809 & 1453 & -0.000 & 0.034 & 0.034 & 0.948 & 0.212 \\
$\bb_{18}$ & 13.534 & 13.534 & 758 & 1332 & 0.000 & 0.036 & 0.036 & 0.947 & 0.221 \\
$\bb_{19}$ & 13.137 & 13.137 & 707 & 1218 & 0.000 & 0.037 & 0.037 & 0.947 & 0.230 \\
$\bb_{20}$ & 13.188 & 13.188 & 680 & 1174 & 0.000 & 0.038 & 0.038 & 0.943 & 0.234 \\
$\bb_{21}$ & 13.422 & 13.422 & 657 & 1146 & 0.000 & 0.038 & 0.038 & 0.944 & 0.237 \\
\midrule
$\mathtt{Uniform}$ &  &  &  &  &  &  &  & 0.929 & 0.359 \\
\midrule
$\mathtt{WBATE}$ &  &  &  &  & -0.000 & 0.024 & 0.024 & 0.946 & 0.133 \\
\midrule
$\mathtt{LBATE}$ &  &  &  &  & 0.014 & 0.033 & 0.036 & 0.988 & 0.416 \\
\bottomrule\bottomrule
\end{tabular}
}
    \caption{First-Stage Effects (Simulations, $s=1$, Fuzzy Bandwidth Selector).}
    \label{tab:sa-simuls-fs-s1-bwmain}
\end{table}

\begin{table}[p]
    \centering
    \resizebox{.8\textwidth}{!}{\begin{tabular}{lccccrrrrr}
\toprule\toprule
 & $h_1$ & $h_2$ & $N_{\mathrm{Co}}$ & $N_{\mathrm{Tr}}$ & Bias & SD & RMSE & EC & IL \\
\midrule
$\bb_{1}$ & 13.386 & 13.386 & 744 & 1300 & 0.002 & 0.037 & 0.037 & 0.945 & 0.228 \\
$\bb_{2}$ & 13.791 & 13.791 & 785 & 1397 & 0.002 & 0.036 & 0.036 & 0.942 & 0.219 \\
$\bb_{3}$ & 14.069 & 14.069 & 811 & 1460 & 0.002 & 0.035 & 0.035 & 0.943 & 0.214 \\
$\bb_{4}$ & 14.218 & 14.218 & 844 & 1464 & 0.002 & 0.035 & 0.035 & 0.947 & 0.212 \\
$\bb_{5}$ & 14.287 & 14.287 & 914 & 1389 & 0.002 & 0.035 & 0.035 & 0.945 & 0.214 \\
$\bb_{6}$ & 14.284 & 14.284 & 972 & 1288 & 0.001 & 0.036 & 0.036 & 0.940 & 0.221 \\
$\bb_{7}$ & 14.225 & 14.225 & 1008 & 1176 & 0.001 & 0.038 & 0.038 & 0.941 & 0.233 \\
$\bb_{8}$ & 13.995 & 13.995 & 1009 & 1037 & 0.000 & 0.042 & 0.042 & 0.936 & 0.254 \\
$\bb_{9}$ & 14.126 & 14.126 & 1038 & 946 & -0.001 & 0.047 & 0.047 & 0.938 & 0.285 \\
$\bb_{10}$ & 15.108 & 15.108 & 1144 & 958 & -0.003 & 0.052 & 0.052 & 0.937 & 0.334 \\
$\bb_{11}$ & 14.748 & 14.748 & 1087 & 799 & -0.005 & 0.067 & 0.067 & 0.932 & 0.479 \\
$\bb_{12}$ & 14.726 & 14.726 & 1106 & 955 & -0.003 & 0.049 & 0.049 & 0.942 & 0.308 \\
$\bb_{13}$ & 13.933 & 13.933 & 1006 & 1007 & -0.002 & 0.042 & 0.042 & 0.945 & 0.257 \\
$\bb_{14}$ & 14.255 & 14.255 & 1003 & 1201 & -0.002 & 0.037 & 0.037 & 0.947 & 0.226 \\
$\bb_{15}$ & 14.312 & 14.312 & 941 & 1353 & -0.002 & 0.034 & 0.034 & 0.943 & 0.212 \\
$\bb_{16}$ & 14.250 & 14.250 & 847 & 1469 & -0.002 & 0.033 & 0.033 & 0.943 & 0.206 \\
$\bb_{17}$ & 14.019 & 14.019 & 806 & 1449 & -0.002 & 0.034 & 0.034 & 0.944 & 0.209 \\
$\bb_{18}$ & 13.498 & 13.498 & 754 & 1326 & -0.002 & 0.036 & 0.036 & 0.939 & 0.218 \\
$\bb_{19}$ & 13.115 & 13.115 & 703 & 1214 & -0.002 & 0.037 & 0.037 & 0.939 & 0.227 \\
$\bb_{20}$ & 13.180 & 13.180 & 678 & 1173 & -0.003 & 0.038 & 0.038 & 0.940 & 0.231 \\
$\bb_{21}$ & 13.415 & 13.415 & 655 & 1146 & -0.003 & 0.038 & 0.038 & 0.938 & 0.234 \\
\midrule
$\mathtt{Uniform}$ &  &  &  &  &  &  &  & 0.921 & 0.355 \\
\midrule
$\mathtt{WBATE}$ &  &  &  &  & -0.001 & 0.024 & 0.024 & 0.941 & 0.132 \\
\midrule
$\mathtt{LBATE}$ &  &  &  &  & 0.017 & 0.034 & 0.038 & 0.982 & 0.422 \\
\bottomrule\bottomrule
\end{tabular}
}
    \caption{First-Stage Effects (Simulations, $s=2$, Fuzzy Bandwidth Selector).}
    \label{tab:sa-simuls-fs-s2-bwmain}
\end{table}

\begin{table}[p]
    \centering
    \resizebox{.8\textwidth}{!}{\begin{tabular}{lccccrrrrr}
\toprule\toprule
 & $h_1$ & $h_2$ & $N_{\mathrm{Co}}$ & $N_{\mathrm{Tr}}$ & Bias & SD & RMSE & EC & IL \\
\midrule
$\bb_{1}$ & 13.416 & 13.416 & 746 & 1305 & 0.001 & 0.037 & 0.037 & 0.949 & 0.227 \\
$\bb_{2}$ & 13.823 & 13.823 & 788 & 1403 & 0.000 & 0.035 & 0.035 & 0.947 & 0.219 \\
$\bb_{3}$ & 14.093 & 14.093 & 813 & 1465 & 0.000 & 0.034 & 0.034 & 0.949 & 0.213 \\
$\bb_{4}$ & 14.229 & 14.229 & 845 & 1465 & -0.001 & 0.034 & 0.034 & 0.949 & 0.211 \\
$\bb_{5}$ & 14.282 & 14.282 & 913 & 1388 & -0.001 & 0.034 & 0.034 & 0.951 & 0.214 \\
$\bb_{6}$ & 14.281 & 14.281 & 971 & 1288 & -0.001 & 0.035 & 0.035 & 0.947 & 0.220 \\
$\bb_{7}$ & 14.214 & 14.214 & 1006 & 1174 & -0.001 & 0.037 & 0.037 & 0.945 & 0.233 \\
$\bb_{8}$ & 13.955 & 13.955 & 1005 & 1031 & -0.001 & 0.041 & 0.041 & 0.945 & 0.254 \\
$\bb_{9}$ & 14.080 & 14.080 & 1034 & 940 & -0.000 & 0.047 & 0.047 & 0.940 & 0.286 \\
$\bb_{10}$ & 15.038 & 15.038 & 1137 & 949 & -0.000 & 0.052 & 0.052 & 0.939 & 0.336 \\
$\bb_{11}$ & 14.667 & 14.667 & 1079 & 790 & 0.001 & 0.067 & 0.067 & 0.929 & 0.483 \\
$\bb_{12}$ & 14.687 & 14.687 & 1102 & 951 & 0.002 & 0.050 & 0.050 & 0.939 & 0.310 \\
$\bb_{13}$ & 13.902 & 13.902 & 1003 & 1003 & 0.002 & 0.043 & 0.043 & 0.938 & 0.260 \\
$\bb_{14}$ & 14.233 & 14.233 & 1001 & 1197 & 0.002 & 0.037 & 0.037 & 0.943 & 0.228 \\
$\bb_{15}$ & 14.292 & 14.292 & 940 & 1350 & 0.003 & 0.035 & 0.035 & 0.944 & 0.214 \\
$\bb_{16}$ & 14.217 & 14.217 & 845 & 1463 & 0.003 & 0.034 & 0.034 & 0.943 & 0.209 \\
$\bb_{17}$ & 13.995 & 13.995 & 805 & 1443 & 0.002 & 0.034 & 0.035 & 0.944 & 0.211 \\
$\bb_{18}$ & 13.492 & 13.492 & 755 & 1324 & 0.002 & 0.036 & 0.036 & 0.946 & 0.221 \\
$\bb_{19}$ & 13.126 & 13.126 & 705 & 1216 & 0.002 & 0.037 & 0.037 & 0.946 & 0.229 \\
$\bb_{20}$ & 13.183 & 13.183 & 679 & 1173 & 0.002 & 0.037 & 0.037 & 0.948 & 0.233 \\
$\bb_{21}$ & 13.408 & 13.408 & 655 & 1145 & 0.002 & 0.038 & 0.038 & 0.944 & 0.236 \\
\midrule
$\mathtt{Uniform}$ &  &  &  &  &  &  &  & 0.925 & 0.357 \\
\midrule
$\mathtt{WBATE}$ &  &  &  &  & 0.001 & 0.024 & 0.024 & 0.943 & 0.132 \\
\midrule
$\mathtt{LBATE}$ &  &  &  &  & 0.021 & 0.033 & 0.040 & 0.984 & 0.429 \\
\bottomrule\bottomrule
\end{tabular}
}
    \caption{First-Stage Effects (Simulations, $s=3$, Fuzzy Bandwidth Selector).}
    \label{tab:sa-simuls-fs-s3-bwmain}
\end{table}

\begin{table}[p]
    \centering
    \resizebox{.8\textwidth}{!}{\begin{tabular}{lccccrrrrr}
\toprule\toprule
 & $h_1$ & $h_2$ & $N_{\mathrm{Co}}$ & $N_{\mathrm{Tr}}$ & Bias & SD & RMSE & EC & IL \\
\midrule
$\bb_{1}$ & 13.404 & 13.404 & 746 & 1303 & 0.001 & 0.073 & 0.073 & 0.955 & 0.427 \\
$\bb_{2}$ & 13.823 & 13.823 & 788 & 1402 & 0.001 & 0.069 & 0.069 & 0.958 & 0.411 \\
$\bb_{3}$ & 14.104 & 14.104 & 815 & 1466 & 0.002 & 0.067 & 0.067 & 0.958 & 0.401 \\
$\bb_{4}$ & 14.257 & 14.257 & 849 & 1470 & 0.002 & 0.066 & 0.066 & 0.958 & 0.387 \\
$\bb_{5}$ & 14.318 & 14.318 & 919 & 1394 & 0.002 & 0.064 & 0.064 & 0.959 & 0.360 \\
$\bb_{6}$ & 14.306 & 14.306 & 975 & 1291 & 0.002 & 0.061 & 0.061 & 0.959 & 0.353 \\
$\bb_{7}$ & 14.235 & 14.235 & 1009 & 1176 & 0.001 & 0.060 & 0.060 & 0.955 & 0.359 \\
$\bb_{8}$ & 13.991 & 13.991 & 1009 & 1036 & 0.001 & 0.062 & 0.062 & 0.955 & 0.378 \\
$\bb_{9}$ & 14.133 & 14.133 & 1040 & 946 & 0.001 & 0.067 & 0.067 & 0.958 & 0.407 \\
$\bb_{10}$ & 15.100 & 15.100 & 1144 & 957 & 0.001 & 0.074 & 0.074 & 0.959 & 0.462 \\
$\bb_{11}$ & 14.722 & 14.722 & 1085 & 796 & 0.001 & 0.098 & 0.098 & 0.965 & 0.964 \\
$\bb_{12}$ & 14.698 & 14.698 & 1103 & 951 & 0.000 & 0.070 & 0.070 & 0.953 & 0.424 \\
$\bb_{13}$ & 13.915 & 13.915 & 1004 & 1004 & 0.001 & 0.061 & 0.061 & 0.956 & 0.370 \\
$\bb_{14}$ & 14.252 & 14.252 & 1004 & 1200 & 0.001 & 0.057 & 0.057 & 0.959 & 0.338 \\
$\bb_{15}$ & 14.307 & 14.307 & 941 & 1352 & 0.001 & 0.058 & 0.058 & 0.953 & 0.331 \\
$\bb_{16}$ & 14.254 & 14.254 & 849 & 1469 & 0.001 & 0.061 & 0.061 & 0.956 & 0.355 \\
$\bb_{17}$ & 14.044 & 14.044 & 809 & 1453 & 0.001 & 0.062 & 0.062 & 0.950 & 0.365 \\
$\bb_{18}$ & 13.534 & 13.534 & 758 & 1332 & 0.001 & 0.064 & 0.064 & 0.950 & 0.375 \\
$\bb_{19}$ & 13.137 & 13.137 & 707 & 1218 & 0.001 & 0.066 & 0.066 & 0.952 & 0.387 \\
$\bb_{20}$ & 13.188 & 13.188 & 680 & 1174 & 0.000 & 0.067 & 0.067 & 0.953 & 0.389 \\
$\bb_{21}$ & 13.422 & 13.422 & 657 & 1146 & -0.000 & 0.067 & 0.067 & 0.952 & 0.391 \\
\midrule
$\mathtt{Uniform}$ &  &  &  &  &  &  &  & 0.964 & 0.591 \\
\midrule
$\mathtt{WBATE}$ &  &  &  &  & 0.001 & 0.039 & 0.039 & 0.950 & 0.222 \\
\midrule
$\mathtt{LBATE}$ &  &  &  &  & 0.055 & 0.065 & 0.085 & 0.998 & 0.931 \\
\bottomrule\bottomrule
\end{tabular}
}
    \caption{Fuzzy Effects (Simulations, $s=1$, Fuzzy Bandwidth Selector).}
    \label{tab:sa-simuls-fuzzy-s1-bwmain}
\end{table}

\begin{table}[p]
    \centering
    \resizebox{.8\textwidth}{!}{\begin{tabular}{lccccrrrrr}
\toprule\toprule
 & $h_1$ & $h_2$ & $N_{\mathrm{Co}}$ & $N_{\mathrm{Tr}}$ & Bias & SD & RMSE & EC & IL \\
\midrule
$\bb_{1}$ & 13.386 & 13.386 & 744 & 1300 & 0.001 & 0.074 & 0.074 & 0.959 & 0.439 \\
$\bb_{2}$ & 13.791 & 13.791 & 785 & 1397 & 0.001 & 0.071 & 0.071 & 0.957 & 0.420 \\
$\bb_{3}$ & 14.069 & 14.069 & 811 & 1460 & 0.002 & 0.068 & 0.068 & 0.959 & 0.407 \\
$\bb_{4}$ & 14.218 & 14.218 & 844 & 1464 & 0.002 & 0.067 & 0.067 & 0.960 & 0.390 \\
$\bb_{5}$ & 14.287 & 14.287 & 914 & 1389 & 0.001 & 0.063 & 0.063 & 0.962 & 0.358 \\
$\bb_{6}$ & 14.284 & 14.284 & 972 & 1288 & -0.001 & 0.060 & 0.060 & 0.956 & 0.347 \\
$\bb_{7}$ & 14.225 & 14.225 & 1008 & 1176 & -0.001 & 0.059 & 0.059 & 0.953 & 0.348 \\
$\bb_{8}$ & 13.995 & 13.995 & 1009 & 1037 & -0.002 & 0.061 & 0.061 & 0.952 & 0.360 \\
$\bb_{9}$ & 14.126 & 14.126 & 1038 & 946 & -0.002 & 0.065 & 0.065 & 0.954 & 0.380 \\
$\bb_{10}$ & 15.108 & 15.108 & 1144 & 958 & -0.003 & 0.070 & 0.070 & 0.952 & 0.422 \\
$\bb_{11}$ & 14.748 & 14.748 & 1087 & 799 & -0.003 & 0.089 & 0.089 & 0.963 & 1.656 \\
$\bb_{12}$ & 14.726 & 14.726 & 1106 & 955 & -0.003 & 0.064 & 0.064 & 0.951 & 0.382 \\
$\bb_{13}$ & 13.933 & 13.933 & 1006 & 1007 & -0.002 & 0.057 & 0.057 & 0.950 & 0.337 \\
$\bb_{14}$ & 14.255 & 14.255 & 1003 & 1201 & -0.002 & 0.053 & 0.053 & 0.954 & 0.311 \\
$\bb_{15}$ & 14.312 & 14.312 & 941 & 1353 & -0.001 & 0.055 & 0.055 & 0.952 & 0.306 \\
$\bb_{16}$ & 14.250 & 14.250 & 847 & 1469 & 0.000 & 0.058 & 0.058 & 0.951 & 0.330 \\
$\bb_{17}$ & 14.019 & 14.019 & 806 & 1449 & 0.001 & 0.059 & 0.059 & 0.950 & 0.342 \\
$\bb_{18}$ & 13.498 & 13.498 & 754 & 1326 & 0.001 & 0.061 & 0.061 & 0.950 & 0.355 \\
$\bb_{19}$ & 13.115 & 13.115 & 703 & 1214 & 0.000 & 0.063 & 0.063 & 0.953 & 0.368 \\
$\bb_{20}$ & 13.180 & 13.180 & 678 & 1173 & -0.001 & 0.064 & 0.064 & 0.953 & 0.373 \\
$\bb_{21}$ & 13.415 & 13.415 & 655 & 1146 & -0.001 & 0.065 & 0.065 & 0.950 & 0.377 \\
\midrule
$\mathtt{Uniform}$ &  &  &  &  &  &  &  & 0.961 & 0.619 \\
\midrule
$\mathtt{WBATE}$ &  &  &  &  & -0.001 & 0.038 & 0.038 & 0.943 & 0.249 \\
\midrule
$\mathtt{LBATE}$ &  &  &  &  & 0.054 & 0.064 & 0.084 & 0.995 & 1.461 \\
\bottomrule\bottomrule
\end{tabular}
}
    \caption{Fuzzy Effects (Simulations, $s=2$, Fuzzy Bandwidth Selector).}
    \label{tab:sa-simuls-fuzzy-s2-bwmain}
\end{table}

\begin{table}[p]
    \centering
    \resizebox{.8\textwidth}{!}{\begin{tabular}{lccccrrrrr}
\toprule\toprule
 & $h_1$ & $h_2$ & $N_{\mathrm{Co}}$ & $N_{\mathrm{Tr}}$ & Bias & SD & RMSE & EC & IL \\
\midrule
$\bb_{1}$ & 13.416 & 13.416 & 746 & 1305 & 0.002 & 0.072 & 0.072 & 0.959 & 0.422 \\
$\bb_{2}$ & 13.823 & 13.823 & 788 & 1403 & 0.001 & 0.069 & 0.069 & 0.955 & 0.403 \\
$\bb_{3}$ & 14.093 & 14.093 & 813 & 1465 & 0.001 & 0.067 & 0.067 & 0.956 & 0.390 \\
$\bb_{4}$ & 14.229 & 14.229 & 845 & 1465 & 0.000 & 0.065 & 0.065 & 0.952 & 0.373 \\
$\bb_{5}$ & 14.282 & 14.282 & 913 & 1388 & -0.001 & 0.062 & 0.062 & 0.954 & 0.344 \\
$\bb_{6}$ & 14.281 & 14.281 & 971 & 1288 & -0.003 & 0.059 & 0.059 & 0.954 & 0.335 \\
$\bb_{7}$ & 14.214 & 14.214 & 1006 & 1174 & -0.003 & 0.057 & 0.057 & 0.954 & 0.338 \\
$\bb_{8}$ & 13.955 & 13.955 & 1005 & 1031 & -0.004 & 0.059 & 0.059 & 0.960 & 0.354 \\
$\bb_{9}$ & 14.080 & 14.080 & 1034 & 940 & -0.004 & 0.063 & 0.063 & 0.962 & 0.379 \\
$\bb_{10}$ & 15.038 & 15.038 & 1137 & 949 & -0.004 & 0.069 & 0.069 & 0.960 & 0.428 \\
$\bb_{11}$ & 14.667 & 14.667 & 1079 & 790 & -0.003 & 0.090 & 0.090 & 0.966 & 0.675 \\
$\bb_{12}$ & 14.687 & 14.687 & 1102 & 951 & -0.003 & 0.065 & 0.065 & 0.952 & 0.398 \\
$\bb_{13}$ & 13.902 & 13.902 & 1003 & 1003 & -0.001 & 0.059 & 0.059 & 0.957 & 0.352 \\
$\bb_{14}$ & 14.233 & 14.233 & 1001 & 1197 & -0.000 & 0.055 & 0.055 & 0.952 & 0.325 \\
$\bb_{15}$ & 14.292 & 14.292 & 940 & 1350 & 0.002 & 0.058 & 0.058 & 0.952 & 0.320 \\
$\bb_{16}$ & 14.217 & 14.217 & 845 & 1463 & 0.003 & 0.061 & 0.061 & 0.950 & 0.345 \\
$\bb_{17}$ & 13.995 & 13.995 & 805 & 1443 & 0.004 & 0.062 & 0.062 & 0.948 & 0.357 \\
$\bb_{18}$ & 13.492 & 13.492 & 755 & 1324 & 0.004 & 0.064 & 0.064 & 0.954 & 0.370 \\
$\bb_{19}$ & 13.126 & 13.126 & 705 & 1216 & 0.003 & 0.065 & 0.065 & 0.954 & 0.383 \\
$\bb_{20}$ & 13.183 & 13.183 & 679 & 1173 & 0.003 & 0.066 & 0.066 & 0.951 & 0.387 \\
$\bb_{21}$ & 13.408 & 13.408 & 655 & 1145 & 0.002 & 0.066 & 0.066 & 0.954 & 0.390 \\
\midrule
$\mathtt{Uniform}$ &  &  &  &  &  &  &  & 0.962 & 0.552 \\
\midrule
$\mathtt{WBATE}$ &  &  &  &  & 0.000 & 0.038 & 0.038 & 0.946 & 0.202 \\
\midrule
$\mathtt{LBATE}$ &  &  &  &  & 0.053 & 0.065 & 0.083 & 0.994 & 0.717 \\
\bottomrule\bottomrule
\end{tabular}
}
    \caption{Fuzzy Effects (Simulations, $s=3$, Fuzzy Bandwidth Selector).}
    \label{tab:sa-simuls-fuzzy-s3-bwmain}
\end{table}

\clearpage

\clearpage
\bibliography{CTY_2026_JASA--bib}
\bibliographystyle{plainnat}